\documentclass[structabstract]{aa}  
\usepackage{subfig}
\usepackage{xspace}
\usepackage{amsmath}
\usepackage{graphicx}
\usepackage{natbib}
\usepackage{lscape}
\usepackage{rotating} 
\usepackage{txfonts}
\begin{document}
\newcommand{\um}{\ensuremath{\mu\mathrm{m}}\xspace}
\newcommand{\nsources}{23~}
   \title{A diversity of dusty AGN tori\thanks{Based on observations collected at the European Organisation for Astronomical Research in the Southern Hemisphere, Chile, program numbers 184.B-0832, 087.B-0266 (PI: K. Meisenheimer), 086.B-0919 (PI: K. Tristram). Based on data obtained from the ESO Science Archive Facility.}}

   \subtitle{Data release for the VLTI/MIDI AGN Large Program and first results for \nsources galaxies}

   \author{L. Burtscher \inst{1,2}
          \and
          K. Meisenheimer \inst{1}
          \and
          K. R. W. Tristram \inst{3}
          \and
          W. Jaffe \inst{4}
          \and
          S. F. H\"onig \inst{5}
          \and
          R. I. Davies \inst{2}
          \and
          M. Kishimoto \inst{3}
          \and
          J.-U. Pott \inst{1}
          \and
          H. R\"ottgering \inst{4}
          \and
          M. Schartmann \inst{2,6}
          \and
          G. Weigelt \inst{3}
          \and
          S. Wolf \inst{7}
}

\institute{Max-Planck-Institut f\"ur Astronomie,
		K\"onigstuhl 17, 69117 Heidelberg, Germany
	\and
		Max-Planck-Institut f\"ur extraterrestrische Physik,
		Postfach 1312, Gie\ss enbachstr., 85741 Garching, Germany\\
		\email{burtscher@mpe.mpg.de}
	\and
		Max-Planck-Institut f\"ur Radioastronomie,
		Auf dem H\"ugel 69, 53121 Bonn, Germany
	\and
		Sterrewacht Leiden, Universiteit Leiden,
		Niels-Bohr-Weg 2, 2300 CA Leiden, The Netherlands
	\and
		University of California Santa Barbara, Department of Physics, Broida Hall, Santa Barbara, CA 93106-9530, USA
	\and
		Universit\"ats-Sternwarte M\"unchen, Scheinerstra{\ss}e 1, D-81679 M\"unchen, Germany
	\and
		Institut f\"ur Theoretische Physik und Astrophysik, Universit\"at zu Kiel, Leibnizstr. 15, 24118 Kiel,
Germany
             }

   \date{Received ; accepted }

 
  \abstract
   {The AGN-heated dust distribution (the ``torus") is increasingly recognized not only as the absorber required in unifying models, but as a tracer for the reservoir that feeds the nuclear Super-Massive Black Hole. Yet, even its most basic structural properties (such as its extent, geometry and elongation) are unknown for all but a few archetypal objects.}
   {In order to understand how the properties of AGN tori are related to feeding and obscuration, we need to resolve the matter distribution on parsec scales.}
   {Since most AGNs are unresolved in the mid-infrared, even with the largest telescopes, we utilize the MID-infrared interferometric Instrument (MIDI) at the Very Large Telescope Interferometer (VLTI) that is sensitive to structures as small as a few milli-arcseconds (mas). We present here an extensive amount of new interferometric observations from the MIDI AGN Large Program (2009 -- 2011) and add data from the archive to give a complete view of the existing MIDI observations of AGNs. Additionally, we have obtained high-quality mid-infrared spectra from VLT/VISIR to provide a precise total flux reference for the interferometric data.}
  {We present correlated and total fluxes for \nsources AGNs (16 of which with new data) and derive flux and size estimates at 12 \um using simple axisymmetric geometrical models. Perhaps the most surprising result is the relatively high level of unresolved flux and its large scatter: The median ``point source fraction'' is 70\% for type 1 and 47 \% for type 2 AGNs meaning that a large part of the flux is concentrated on scales $<$~5~mas (0.1 -- 10 pc). Among sources observed with similar spatial resolution, it varies from 20 \% -- 100 \%.  
  For 18 of the sources, two nuclear components can be distinguished in the radial fits. While these models provide good fits to all but the brightest sources, significant elongations are detected in eight sources.}
   {The half-light radii of the fainter sources are smaller than expected from the size $\propto L^{0.5}$ scaling of the bright sources and show a large scatter, especially when compared to the relatively tight size--luminosity relation in the near-infrared. It is likely that a common ``size-luminosity'' relation does not exist for AGN tori, but that they are dominated by intrinsic differences in their dust structures. Variations in the relative contribution of extended dust in the narrow line region or heated by star formation vs. compact AGN-heated dust and non-thermal emission also have to be taken into account.}


   \keywords{galaxies: active -- galaxies: nuclei -- galaxies: Seyfert -- techniques: interferometric}

   \maketitle
%
%
%
\section{Introduction}

The parsec-scale dust distribution in Active Galactic Nuclei (AGNs), that is often referred to as the ``torus'', is increasingly recognized not only as the absorber required for unification \citep{antonucci1993}, but as a tracer for the reservoir that feeds the nuclear Super-Massive Black Hole (SMBH). While the gas transport on scales of hundreds of parsecs is now well studied \citep[e.g.][]{dumas2007} and theoretical models of nuclear accretion disks have been well developed \citep[e.g.][]{frank2002}, the link between these scales is not clear at all. Current dynamical models of the torus see it either made up from outflowing \citep[e.g.][]{elitzur2006b,czerny2011,gallagher2013} material that is part of a wind blown off from the accretion disk or as an inflowing structure whose scale-height may be provided by exploding stars from nuclear star clusters on tens of parsec scale that would also be able to provide the dust required for the obscuration \citep[e.g.][]{schartmann2009}. It is not clear however whether nuclear star clusters would rather hinder or help accretion or if there is a time-delay between the onset of nuclear star-formation and powerful accretion onto the AGN \citep[e.g.][]{davies2007, vollmer2008, schartmann2009, wada2009, davies2012}. 

Direct observations of the torus are therefore required to reveal its size and its role in the feeding and in the nuclear stellar evolution of AGNs. They can most favorably be done in the infrared since the spectral energy distribution (SED, in $\nu F_{\nu}$) peaks at $\sim 12 \um$ for a typical torus temperature of 400 K. But even the largest optical-infrared telescopes are not able to resolve the torus in the nearest galaxies and we need to employ interferometry to achieve the resolution required to resolve it.

The first near- and mid-infrared (mid-IR) observations of AGNs using long-baseline ``stellar'' interferometry have revealed that nuclear dust structures indeed exist and that they can be observed with the extremely high angular resolution of a few milli-arcseconds (mas) provided by optical/infrared interferometry \citep{swain2003, wittkowski2004, jaffe2004}. In the near-infrared, where AGNs are only marginally resolved by optical interferometers, the radius of the hot dust $r_K$ has now been measured in nine sources and a tight correlation with AGN luminosity $L_{\rm UV}$ has been found of the form $r_K \propto \sqrt{L_{\rm UV}}$ \citep{kishimoto2011,weigelt2012}.

In the mid-IR, the first detailed studies focussed on the brightest AGNs (Circinus galaxy, \citet{tristram2007b,tristram2012} and NGC 1068, \citet{raban2009}) and have resolved the ``tori'' into a large, almost round, structure that contains most of the nuclear mid-IR flux, and a compact disk component with an axis ratio $\approx$ 4. Both of these torus components have to be clumpy to explain the low surface brightnesses. A study of the brightest type 1 Seyfert galaxy, NGC~4151, found its torus to be of similar size, temperature and clumpiness as the bright type 2 objects \citep{burtscher2009}. In the nearest radio galaxy, Centaurus~A, the nuclear mid-infrared emission was found to be dominated by non-thermal radiation \citep{meisenheimer2007}. The nuclear disk that was seen in the first data of this source, could not be clearly identified in a more extended search \citep{burtscher2010}. Recently, very well sampled $(u,v)$ coverages have been achieved for two fainter sources, the Seyfert~2 galaxy NGC~424 and the Seyfert~1.5 galaxy NGC~3783, and their extended components have been found to be slightly elongated with axis ratios of 1.3 in NGC~424 \citep{hoenig2012} and 1.5 in NGC~3783 \citep{hoenig2013}. Interestingly, the elongation is along the polar direction in both cases, placing the emission in the narrow-line region, since neither smooth nor clumpy torus models cannot explain elongations along the polar axis \citep[e.g.][, Figs. 18,19]{schartmann2008}.

In addition to studies of individual objects, \citet{kishimoto2009} analyzed the published and unpublished data of four AGNs and claimed evidence for a ``common radial structure'' in the sense that the radial structure of tori is self-similar if normalized to the inner torus radius. About a dozen, weaker, galaxies had been observed using MIDI GTO time in a ``snapshot'' survey \citep{tristram2009} that served to prove the observability of weaker targets. The size--luminosity relation of AGN tori was examined and it was found that the mid-IR size of tori $r$ scales with mid-IR luminosity $\nu L_{\nu}$ as $r \propto \sqrt{\nu L_{\nu}}$. However, only one or two $(u,v)$-points were observed per source and the data quality was very low. Apart from that, the number of successfully observed type 1 sources, which are rare in the local universe, was very low compared to type 2 sources. \citet{kishimoto2011b} independently built a mid-infrared size--luminosity relation from six type 1 objects and found a weaker dependence on (optical) luminosity at 8.5 \um and a nearly constant size (independent of the luminosity) at 13 \um.

It was clear that a larger and more systematic observational campaign was needed to collect the basic observational information necessary to understand dusty tori on a statistical basis. This is the aim of the MIDI AGN Large Program (LP).

The questions to be addressed in the LP are:

\begin{enumerate}
	\item How does the measured mid-IR morphology of AGN tori depend on wavelength, nuclear orientation and luminosity?  This is the key to understanding the radiation transfer effects in the dust structure.  Is there a common torus size -- AGN luminosity relation for all types of AGNs as suggested by the GTO study \citep{tristram2009}? Understanding and properly calibrating this relation, if it exists, with local AGNs is the only way to safely apply the relation to the wealth of distant spatially unresolved AGNs.
	\item Are other ``parameters" important for the morphology, such as mass feeding rate from circum-stellar star clusters, or dust chemistry?
	\item Is the Seyfert 1 and Seyfert 2 dichotomy only an orientation effect or a simplified view of a wide variety of different intrinsic morphologies?  Are 10 \um silicate features always seen in absorption in Seyfert 2s and in emission in Seyfert 1s?
	\item Is the apparent ``two component" structure of a compact dense disk and an extended almost round (spherical?) distribution a general property? Is the dust distribution patchy / clumpy (as found in Circinus)? In which AGNs does maser emission coincide with the dust emission?
\end{enumerate}

Some of these questions (e.g. on the size--luminosity relation) can be answered directly from the interferometric data, others, such as questions on the structural properties of the dust (e.g. clump sizes and distributions), require a more detailed analysis through radiative transfer models. The most far-reaching astrophysical questions, involving accretion mechanisms from the tens of parsec scale of the nuclear star cluster to the parsec-scale dust and further in, require comparisons with hydrodynamical models \citep[e.g.][]{schartmann2010}, or at least a study of the physical mechanisms responsible on these scales.

In this paper, we present the observations and the first full data analysis of the MIDI AGN Large Program observations. We include in this analysis also all published data from the ESO archive in an attempt to give a complete overview of mid-infrared interferometric AGN observations to date. We analyze all data with the same data reduction method, apply uniform selection criteria for the reduced data and derive model-independent size estimates for each source using the same prescription for all sources to allow for an easy comparison between the targets.

The paper is structured as follows: In Section 2 we describe the sample selection, give an overview about the observations and data reduction method. In Section 3 we present our results: $(u,v)$ coverage plots, correlated and total flux spectra as well as additional, new high-resolution mid-infrared spectrophotometry from our supplementary VISIR observations. We describe the interferometric data with simple one-dimensional models (Section 4) and derive a model-independent size estimate (or a limit to it) for all sources. In Section 5 we compare the expected scaling relations for AGN tori with the observed ones. Our conclusions and summary are given in Section 6.

A number of additional material can be found in the appendix: 
\begin{itemize}
	\item Tests and quality checks from our data analysis
	\item The total and correlated flux spectra as well as photometries for all targets
	\item Plots of the differential phases for all targets
	\item Observing logs and data selection for all targets
\end{itemize}

%
%
%
%
%
%
%
%
%

\section{Sample selection, observations and Data Reduction}

%
%

\subsection{Sample selection}
\label{sec:obs:sample}

\begin{figure}
	{\includegraphics[trim=3cm 3cm 3cm 3cm, width=0.95\hsize]{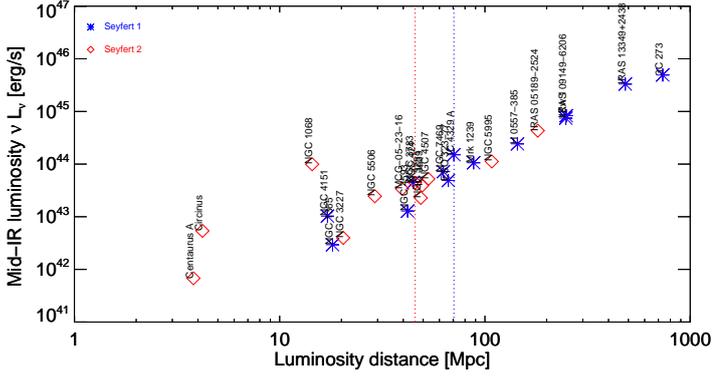}}
	\caption{\label{fig:obs:lum_dist}Luminosity as a function of luminosity distance: Most of the sources are in a very narrow range of observed flux. Note, however, that both at $\sim$ 4 and at $\sim$ 15 Mpc there are two (four) sources that differ significantly in luminosity. The blue (red) vertical dotted lines indicate the median distance of the type 1 (type 2) sub-samples.}
\end{figure}

Mid-infrared interferometric observations typically exceed the resolution achievable with single telescopes by at least a factor of 20. On the other hand, interferometers are typically not very sensitive and interferometric observations are time-consuming. For these reasons, our sample was built up from objects about which we had good prior information, either from our own pre-observations or from the literature. Our selection criteria were that the source is

\begin{itemize}
	\item reliably classified as an AGN,
	\item a point source in high-resolution single-dish observations with a flux $\gtrsim$ 300 mJy at $\sim 12 \um$ and
	\item observable in the southern hemisphere, $\delta \lesssim 40^{\circ}$.
\end{itemize}

A large part of the target list was drawn from the source selection for the study that preceded this project \citep[][in the following GTO survey]{tristram2009}. This was based on near-diffraction limited observations using TIMMI2, the mid-infrared instrument mounted on ESO's 3.6m telescope on Cerro La Silla, Chile. All sources with $\delta < 22$ and $F_{\nu, {\rm N}} >$ 400 mJy had been selected from catalogues of Seyfert galaxies plus a few well-known hand-picked candidates, see \citet{raban2008}. For the MIDI AGN Large Program we selected from this sample all sources except the brightest ones (NGC~1068, the Circinus~galaxy, Centaurus~A, NGC~4151, NGC~3783) on which dedicated projects had been started. We also omitted two sources that are not actually AGNs (NGC~253, NGC~3256) and NGC~7582 on which fringes had been searched in vain before.

Additionally we collected sources that met our criteria from the literature available at the time of proposal writing \citep{krabbe2001, siebenmorgen2004, gorjian2004, galliano2005, haas2007, raban2008, horst2009, gandhi2009, prieto2010, reunanen2010}. Some of these sources have been observed from subsets of this team; we have included all published archive data on MIDI AGNs in this publication to give a complete picture of the status quo.\footnote{Apart from the observations and sources reported here, further MIDI observations exist on the following extragalactic targets: the Circinus~galaxy, NGC~1068, Centaurus~A and the faint -- but observable -- sources 3C~120, Mrk~1095, MCG+06-30-015, 3C~273, ESO383-35, ESO 138-1, Fairall 49 and Fairall 51. Publications on these sources are in preparation by various members of this group.} Most of our sources are detected in the BAT 70 month hard X-ray survey, proving that they contain a strong AGN. The four BAT non-detections (I Zw 1, Mrk 1239, IC 3639, IRAS 13349+2438) show other clear signs for AGN activity.

The median luminosity distance of our sources is 53 Mpc for the complete sample and 70.6 Mpc (45.7 Mpc) for the type 1 (type 2) sub-samples. As can be seen from Fig.~\ref{fig:obs:lum_dist}, most of the sources, except NGC~1068 and the Circinus galaxy, show a very tight correlation between luminosity and distance, i.e. are in a very narrow range of observed flux.

The properties of the individual sources are collected in Table~\ref{tab:sources}. Apart from basic information (co-ordinates, source type, distance, redshift and scale), we give information about the expected AO performance on each source (a combination of distance and magnitude of the Coud\'e guide star) and the number of successful observations at (stacked) $(u,v)$ positions $n_{(u,v)}$ (see below for our definition of stacking).

\begin{table*}
\caption{\label{tab:sources}The \nsources Large Program and archive targets included in this paper. In addition the two sources for which only upper limits have been derived are included in this table. Explanations of the columns are given below the table.}
\centering
\begin{tabular}{r l r r l l l l l l l l l}
\hline
 & Name & RA & DEC & type & type & $D$ & $z$ & scale & \multicolumn{2}{c}{Coud\'e star} & $n_{(u,v)}$ & Ref.\\
     & [h m s] & [$^{\circ}$ ' ''] & (1) & (2) & [Mpc] & & [mas/pc] & $\Psi$[''] & V [mag] & & \\
\hline\hline
 1 & I Zw 1 & 00 53 35.1 & +12 41 34 & NL Sy 1 & Sy 1 & 222 & 0.0589 &  0.9 & 0 & 14.0 & 10 &         12\\
 2 & NGC 424 & 01 11 27.6 & -38 05 00 & Sy 1h & Sy 2 & 44.7 & 0.0110 &  4.6 & 0 & 14.1 & 38 &         11\\
 3 & NGC 1068 & 02 42 40.7 & -00 00 48 & Sy 1h & Sy 2 & \emph{14.4} & 0.0038 & 14.3 & 0 & 10.8 & 32 &     0,5,12\\
 4 & NGC 1365 & 03 33 36.4 & -36 08 25 & Sy 1.8 & Sy 1 & \emph{18.1} & 0.0055 & 11.4 & 0 & 12.9 & 19 &       6,12\\
 5 & IRAS 05189-2524 & 05 21 01.7 & -25 21 45 & Sy 1h & Sy 2 & 167 & 0.0426 &  1.2 & 0 & 14.8 &  7 &         12\\
 6 & H 0557-385 & 05 58 02.1 & -38 20 04 & Sy 1.2 & Sy 1 & 135 & 0.0339 &  1.5 & 0 & 15.0 &  4 &         10\\
 7 & IRAS 09149-6206 & 09 16 09.4 & -62 19 30 & Sy 1 & Sy 1 & 222 & 0.0579 &  0.9 & 0 & 13.6 &  5 &         10\\
 8 & MCG-05-23-16 & 09 47 40.2 & -30 56 54 & Sy 1i & Sy 2 & 38.8 & 0.0085 &  5.3 & 11 & 13.5 & 18 &     6,9,12\\
 9 & Mrk 1239 & 09 52 19.1 & -01 36 43 & NL Sy 1 & Sy 1 & 84.5 & 0.0200 &  2.4 & 0 & 14.4 &  8 &     6,9,12\\
10 & NGC 3281 & 10 31 52.1 & -34 51 13 & Sy 2 & Sy 2 & 47.6 & 0.0107 &  4.3 & 15 & 15.7 &  7 &         12\\
11 & NGC 3783 & 11 39 01.8 & -37 44 19 & Sy 1.5 & Sy 1 & 43.8 & 0.0097 &  4.7 & 0 & 13.4 & 12 &     3,4,10\\
12 & NGC 4151 & 12 10 32.6 & +39 24 21 & Sy 1.5 & Sy 1 & 16.9 & 0.0033 & 12.2 & 0 & 11.8 & 14 &     7,9,10\\
13 & 3C 273 & 12 29 06.7 & +02 03 08 & Sy 1.0 & Quasar & 546 & 0.1580 &  0.4 & 0 & 12.8 &  3 &       6,12\\
14 & NGC 4507 & 12 35 36.6 & -39 54 33 & Sy 1h & Sy 2 & 51.7 & 0.0118 &  4.0 & 28 & 16.0 & 10 &       6,12\\
15 & NGC 4593 & 12 39 39.4 & -05 20 39 & Sy 1.0 & Sy 1 & 41.2 & 0.0090 &  5.0 & 0 & 13.2 &  9 &         12\\
16 & ESO 323-77 & 13 06 26.1 & -40 24 52 & Sy 1.2 & Sy 1 & 64.2 & 0.0150 &  3.2 & 0 & 13.4 &  9 &         10\\
17 & Centaurus A & 13 25 26.6 & -43 01 09 & ? & Sy 2 & \emph{3.8} & 0.0018 & 54.3 & 45 & 12.5 & 12 &        1,8\\
18 & IRAS 13349+2438 & 13 37 18.7 & +24 23 03 & NL Sy 1 & Sy 1 & 393 & 0.1076 &  0.5 & 0 & 15.0 &  6 &         10\\
19 & IC 4329 A & 13 49 19.3 & -30 18 34 & Sy 1.2 & Sy 1 & 68.3 & 0.0161 &  3.0 & 0 & 13.7 & 11 &     6,9,12\\
20 & Circinus & 14 13 09.9 & -65 20 21 & Sy 1h & Sy 2 & \emph{4.2} & 0.0014 & 49.1 & 50 & 12.5 & 28 &          2\\
21 & NGC 5506 & 14 13 15.0 & -03 12 27 & Sy 1i & Sy 2 & 28.7 & 0.0062 &  7.2 & 0 & 14.4 & 12 &       6,12\\
22 & NGC 5995 & 15 48 25.0 & -13 45 28 & Sy 1.9 & Sy 2 & 102 & 0.0252 &  2.0 & 0 & 13.7 &  4 &         12\\
23 & NGC 7469 & 23 03 15.6 & +08 52 26 & Sy 1.5 & Sy 1 & 60.9 & 0.0163 &  3.4 & 0 & 13.0 & 13 &       6,12\\
\hline
& NGC 3227 & 10 23 30.6 & +19 51 54 & Sy 2 & Sy 2 & 20.2 & 0.0039 & 10.2 & 0 & 11.8 &  0 &         12\\
& IC 3639 & 12 40 52.8 & -36 45 21 & Sy 1h & Sy 2 & 48.3 & 0.0109 &  4.3 & 37 & 14.0 &  0 &         12\\
\end{tabular}
\vspace{0.2cm}
\raggedright\\
{\scriptsize \begin{tabular}{p{0.07\hsize}p{0.9\hsize}}
RA, DEC & J2000 coordinates\\
type & (1) AGN type as classified by \citet{veroncetty2010}: The classification of intermediate Seyfert types (1 $<$ type $<$ 2) is defined by the ratio of H$\beta$ to [OIII]$\lambda$5007 fluxes. Sy 1h are 'hidden' Seyfert 1 galaxies (i.e. where broad lines have been detected in polarized light), In the Sy 1i type, broad Pa $\beta$ lines are detected ``indicating the presence of a highly reddened broad line region''; (2) AGN type from SIMBAD\\
$D$ & angular-size distance derived from redshift using the CMB reference frame and a concordance $\Lambda$CDM cosmology (plain typeface); redshift-independent distance measurement / average over several observations, see NED (italic typeface).\\
$z$ & redshift (from NED)\\
$\Psi$ & Coud\'e (MACAO) guide star separation from science target (0: guiding on nucleus)\\
V[mag] & V magnitude of guide star; from \citet{veroncetty2010} if guide star is AGN\\
$n_{(u,v)}$ & number of good fringe tracks at unique $(u,v)$ positions (see text for explanation)\\
References & 0: \citet{jaffe2004}; 1: \citet{meisenheimer2007}; 2: \citet{tristram2007b}; 3: \citet{beckert2008}; 4: \citet{kishimoto2009}; 5:\citet{raban2009}; 6: \citet{tristram2009}; 7: \citet{burtscher2009}; 8: \citet{burtscher2010}; 9: \citet{tristram2011}; 10: \citet{kishimoto2011b}; 11: \citet{hoenig2012}; 12: this work
\end{tabular}}
\end{table*}

%
%

\subsection{MIDI observations}
\label{sec:obs:MIDI}

All interferometric observations were carried out with MIDI, the MID-infrared interferometric Instrument \citep{leinert2003b} at the European Southern Observatory's (ESO's) Very Large Telescope Interferometer (VLTI) \citep[e.g.][]{haguenauer2010} on Cerro Paranal, Chile. MIDI is a two-telescope Michelson-type beam combiner that operates in the atmospherical $N$ band ($\lambda \sim$ 8 -- 13 \um). Its principal observable is the ``correlated flux'', i.e. the coherent flux of the target with a given projected baseline and position angle \citep[see e.g.][for an introduction to interferometry]{haniff2007}. To maximize sensitivity we used MIDI in its {\tt HIGH\_SENS} mode, where all the light is used to produce fringes and calibration follows later, and chose the lower of the two dispersing elements, i.e. the prism with spectral resolution $R \equiv \lambda/\Delta \lambda \approx 30$. The minimum projected baseline lengths that can be achieved with the large (diameter $D=8.2$ m) Unit Telescopes (UTs) is about 30m. Shorter baselines are possible when observing with the smaller Auxiliary Telescopes (ATs, $D=1.8$m). However, due to the smaller light collecting area, currently only the two mid-IR brightest AGNs, NGC~1068 and the Circinus~galaxy, have been observed successfully with the ATs. The longest baseline used for this work was UT1--UT4 with a baseline length of 130 m.

A typical MIDI observing sequence consists of target acquisition, fringe search, fringe track and ``photometry'' observation on both the science target and a near-by standard star and takes about 50 minutes of time. In the course of the Large Program, we adopted a different observing strategy \citep{burtscher2012b} that allowed us to obtain calibrated $(u,v)$ points up to three times faster than this, both due to improvements at the VLTI and because of a change in observing paradigm: We multiply the correlated flux of the target -- the principal outcome of any MIDI ``fringe track'' observation -- directly with a conversion factor, derived from the correlated flux of the calibrator. This allows us to omit the time-consuming ``photometry'' observations. Thanks to the VLTI lab guiding facility IRIS, MIDI acquisition frames were no longer needed and the VLTI baseline models were sufficiently accurate to predict the position of Zero Optical Path Difference (ZOPD) to within the coherence length of MIDI so that the fringe search step could also be omitted in many cases. In effect, we actually obtained much denser sampled $(u,v)$ coverages for most sources, although not for all sources an optimal configuration as described above was achievable. The obtainable $(u,v)$ coverage depends on the source's declination and is limited by the fixed positions of the VLT Unit Telescopes (UTs) as well as timing constraints, weather loss etc.

The MIDI AGN Large Program (ESO program number 184.B-0832) consisted of 13.1 nights of Visitor Mode observations. Between December 2009 and August 2011, in total 228 science fringe track observations of 15 AGNs have been observed in this program.\footnote{A significant fraction of the originally allocated time was lost due to bad weather and the program was completed under program number 087.B-0266 (2.5 nights), also in Visitor Mode.} For this paper, we also include from the archive 159 previously observed tracks for these sources, 156 fringe tracks of other weak AGNs and 132 tracks for the two mid-IR brightest AGNs (NGC~1068 and the Circinus~galaxy).\footnote{For NGC~1068 and the Circinus~galaxy we show some unpublished data collected on AT~baselines (N. Lopez-Gonzaga, K.R.W. Tristram in prep., respectively). For NGC~5128 (Centaurus~A) we restrict ourselves to the data from the 2008~epoch since the source is variable. For NGC~3783 only data published before 2013 is included, i.e. not the most recent data published in \citet{hoenig2013}.} Each track typically consists of 8000 frames and, including the dominating overheads, takes about eight minutes of observing time. In total, about 7.1 million science frames have been analyzed for this publication. Observing logs for each source with a detailed list of all observations, observing conditions and our most relevant quality check criteria, can be found in the appendix in Tab.~\ref{tab:obs:log:IZw1} and the following.

Our goal was to cover the $(u,v)$ plane for each source at least as good as to probe two perpendicular directions with two different baseline lengths each, i.e. we would consider a minimum good $(u,v)$ coverage to consist of four observations, e.g. at projected baseline lengths of 40m and 120m at position angles (PAs) 30$^{\circ}$ and 120$^{\circ}$ each. Our conclusion from the GTO study and other early studies was that with such a $(u,v)$ coverage we could determine a characteristic size of the emitter in two directions, thus resolving a potential (expected) elongation of the emitter. Additional to this ``extended snapshot survey'' we wanted to observe ``detailed maps'' on three sources which had been successfully observed before.

\subsection{Total fluxes: MIDI photometries; VISIR observations and data reduction}
\label{sec:obs:phot}
The MIDI photometry reduction has been described in detail in \citet{tristram2007}. We have followed this standard procedure except for the masks which have been fitted on the calibrator correlated flux image. We produce a mask for each calibrator by fitting the trace and width ($\times 1.2$) of the calibrator fringe image and use this mask for both interferometric and non-interferometric observations of the calibrator and all science targets associated with it. The width of the mask is slightly enlarged with respect to the calibrator PSF to account for the lower AO performance on the weak science targets. Unfortunately, even with optimal flux extraction, the MIDI photometries are very noisy for weak sources since the background cannot be removed accurately through chopping\footnote{The ``photometry'' errors can be described as a superposition of photon noise and background flux variations \citep{burtscher2011}.} . The reason for the high background in MIDI single-dish observations is unclear but likely related to the complex beam relay through the VLTI. Even by averaging many of single-dish observations together and after binning in wavelength, the errors were on the order of 30 -- 40 \% for the weakest targets and severely limited the interpretation of our correlated flux data. We therefore left out the photometry step in later MIDI observations and, instead, used high spatial resolution spectra or photometry obtained with the VLT spectrometer and imager for the mid--infrared (VISIR) on UT3.

For some of the weak targets such spectra or photometry had already been observed \citep{hoenig2010,kishimoto2011b}, for the remaining six we acquired the data under program ID 086.B-0919 in Service Mode between October 2010 and March 2011. The observing logs for these observations can be found in Tab.~\ref{tab:obs:visir} in the appendix. To cover the full $N$-band from $8-13\,\um$, four different wavelength settings centered at 8.5, 9.8, 11.4, and 12.4\,$\um$ had to be observed. Each setting was extracted using the standard \textit{esorex} pipeline, and further reduction and 2D flux calibration were performed with a dedicated VISIR software presented in \citet{hoenig2010} to minimize the sky background.

Our goal is to obtain a total flux reference that closely matches the setup of MIDI. The MIDI flux extraction described below uses a window of size $0\farcs5 \times 0\farcs52$. All VISIR data has been taken with a $0\farcs75$ slit and we used a $0\farcs53$ window in spatial direction to extract the VISIR fluxes.

%
%

\subsection{Correlated fluxes: MIDI data reduction}
\label{sec:obs:dr}

MIDI data reduction was performed with the Expert WorkStation {\tt EWS} 2.0 \footnote{{\tt EWS} is available for download from {\tt http://home.strw.leidenuniv.nl/$\sim$jaffe/ews/index.html}.} and basically consists of two steps: First the group delay of each (packet of) frame(s) is determined from a Fourier transform of the raw data after extraction under a mask and noise reduction. This group delay is subsequently removed from the fringe data to coherently estimate the correlated flux, i.e. co-phase all frames and average them in the complex plane and compute the modulus. Its principle has been described in detail in \citep{jaffe2004b}.

For the data analysis of weak sources (``weak'' meaning: band-averaged correlated flux $F_{\rm corr,N} < 1$ Jy), a number of biases need to be controlled and the calibration method needs to be changed from visibility calibration to correlated flux calibration involving a more elaborate error analysis. We have published our entire data reduction procedure for weak sources including calibration and quality control in \citet{burtscher2012b}. We repeat the key steps and provide updated plots describing the data analysis in the appendix.

In summary one can conclude that MIDI correlated fluxes can be absolutely calibrated to about 5 -- 15 \% accuracy (at 12 \um, down to $\approx$ 150 mJy), depending on observing conditions.

Given the large number of fringe tracks (675 in total) to be analyzed for this project, we also developed a set of high-level {\tt miditools}\footnote{{\tt http://code.google.com/p/miditools/}} to efficiently reduce and calibrate the data and to check their quality (see example plots in the appendix). The basic idea is to store all relevant information about each observation in a database allowing to link observations together (to find the nearest calibrator) and to quickly reduce a night of data or all data of a specific target etc. The latest release of these routines is described at its download location.

%
%
%
%
%
%
%
%
%

\section{Results}
For each of the \nsources sources we recorded both total flux and correlated flux spectra with MIDI. For most sources, especially the very faint ones for which the total flux MIDI spectrum is very uncertain, we also obtained VISIR spectra and / or photometries. The main observable, however, is the correlated flux at different baselines and position angles.

\subsection{$(u,v)$ coverages}

The $(u,v)$ coverages that we obtained for all Large~Program and archive sources are displayed in Fig.~\ref{fig:rad:IZwicky1} -- \ref{fig:rad:NGC7469} in the top left panel. For most sources, they show all observations that have been obtained with MIDI with the exceptions stated in Section~\ref{sec:obs:sample}. For all sources the requirements for the ``extended snapshot survey'' (see Section~\ref{sec:obs:MIDI}) have been met; the ``detailed maps'' observing strategy on NGC~1365, MCG-05-23-16 and IC~4329~A has not been carried out fully, however, since first results had indicated that a more detailed mapping of these sources would not be very useful. We therefore spent this time on a better sampling of the other sources.

\subsection{Correlated flux spectra}
\label{sec:results:corrflux}

Plots of all photometries and spectra -- correlated flux as well as total flux -- are given in the appendix, see Fig.~\ref{fig:spectra} and following. A pronounced silicate absorption feature can be seen in many type 2 AGNs (e.g. MCG-5-23-16, NGC 3281, NGC 5506) and some type 1 sources show silicates in emission (e.g. I Zw 1 and Mrk 1239). The silicate feature is broad (several microns) and thus generally adds a broad dip or broad peak to the spectrum. In the narrow atmospheric $N$ band the feature is sometimes hard to discern; for a clearer example of the silicate feature in absorption and emission, see the space-based spectra from Spitzer IRS \citep[e.g.][]{weedman2005}.

Other than the silicate feature, there are no spectral features in the correlated flux spectra. We hence focus here on the dependence of correlated flux on $(u,v)$ position which encodes the structure of the object.

In principle, three cases can be distinguished:

\begin{itemize}
	\item sources where the correlated fluxes and the total flux scatter around the same values;
	\item sources where the correlated fluxes scatter around a value different from the total flux;
	\item sources where there is both a difference between the total flux and the correlated fluxes and a dependence of correlated flux on baseline.
\end{itemize}

The level of resolved emission can be appreciated by looking at the difference between the total flux and the correlated fluxes. For a resolved source, the correlated flux is expected to decrease with baseline length. This can be seen in some sources (e.g. MCG-5-23-16) when the spectra align in a rainbow-like manner. If no order is detectable in the colors, the correlated fluxes indicate an unresolved component. If they are further identical to the total flux, the source consists only of this unresolved component. Otherwise the source consists of both an unresolved and an ``over-resolved'' component.

\subsubsection{Sources without fringe tracks}
\label{sec:notrack_limits}

On several sources fringes were searched but either not found or not tracked, see Table~\ref{tab:results:notrack}.

\begin{table*}
\caption{\label{tab:results:notrack}List of all unsuccessful attempts to track an extragalactic target with MIDI; Total flux references: 1: \citet{raban2008}, 2: \citet{ramosalmeida2009}, 3: \citet{gandhi2009}, 4: \citet{haas2007}, 5: \citet{mason2012b}}
\centering
\begin{tabular}{l | l | l | l | l | l}
\hline\hline\
source name & $F_{\rm tot}$ at 12 \um [Jy] & ref. for $F_{\rm tot}$& $F_p$ (12 \um) [Jy] & prog. ID & notes \\
NGC 253     & 2.04                         & 1                          & --                           & 079.B-0398             & not an AGN\\
NGC 3227    & 0.401 (at 11.29 \um) & 2                          & $<0.15$                       & 084.B-0366             & Sy 1.5 \\
GQ Com      &  --                          & --                         & --                           & 083.B-0452             & Sy 1.2\\
IC 3639     & 0.542 (at 13.04 \um) & 3                          & $<0.15$                       & 184.B-0832             & Sy 1h\\
MCG-03-34-064 & 0.674 (at 11.25 \um)  &  4                      & --                           & 075.B-0697, 184.B-0832 & Sy 1h\\
Mrk 463E    & 0.338                        & 1                          & --                           & 184.B-0832             & Sy 1h; tried both E and W nucleus \\
IC 5063     & 0.752 (at 11.25 \um) &  4                        & --                           & 079.B-0398, 080.B-0332 & Sy 1h\\
NGC 7479    & 0.16                          & 5                         & --                           & 079.B-0398, 080.B-0332 & Sy 1.9? AGN classification uncertain\\
NGC 7582    & 0.67                         & 1                          & --                           & 079.B-0398             & Sy 1i\\
\hline
\end{tabular}
\end{table*}

Only two sources that have no fringe detections can be considered firm non-detections and only for them we derive upper limits on the correlated flux. These two sources were observed in very stable nights where the zero OPD position did not change significantly between the fringe tracks as judged from calibrator observations shortly before and after the failed science fringe track. Since no fringe was detected, we assign to these sources an upper limit of the minimum detected flux on this baseline in this night. The two sources are

\begin{itemize}
	\item {\bf NGC 3227} \quad attempted on 2010-02-28 on the UT1-UT2 baseline (projected baseline length 34 m), $F_{\nu} < 150 mJy$
	\item {\bf IC 3639} \quad attempted on 2010-05-27 on the UT1-UT2 baseline (projected baseline length 56 m), $F_{\nu} < 150 mJy$
\end{itemize}

More recent MIDI observations of IC 3639 using the experimental MIDI ``no-track'' mode suggest a correlated flux of $70 \pm 20$ mJy (Fern\'andez-Ontiveros et al. in prep.).

For the other non-detections we can not derive robust limits for the correlated fluxes as it is unclear where the position of zero OPD was during the observation, i.e. the source might have been bright enough for a detection but the fringe might have been searched at the wrong OPD position.

\subsubsection{A note about differential phases}

In principle a MIDI observation reveals not only the correlated flux of the target at a given baseline and position angle, but also its {\em differential phase}, i.e. the phase with 0$^{\rm th}$ and $1^{\rm st}$ order polynomials removed. More direct estimates of the phase are not obtainable in a two-telescope interferometer without phase referencing since the atmospheric turbulence destroys this information.
In all of the Large~Program and other weak targets the differential phase (sometimes referred to as ``chromatic phase'') is consistent with 0. Only the two mid-IR brightest sources NGC~1068 and the Circinus~galaxy show a significant signal in differential phase which is being analyzed by sub-groups of this team. We show the differential phases for all targets in the appendix, Fig.~\ref{fig:dphases}.

\subsubsection{Notes on individual sources}

We include important notes on individual sources as captions to their spectra and model fits, see Fig. \ref{fig:rad:IZwicky1} and following and Fig. \ref{fig:spectra} and following. For a brief introduction to the individual Large~Programme sources, please see Chapter 5 of \citet{burtscher2011}.

\subsection{Total flux photometry and spectra\label{sec:spectra}}

Both the MIDI and VISIR single dish spectra and photometries are displayed in the appendix (Figs.~\ref{fig:spectra} and the following). We have indicated the wavelengths of bright mid-IR lines for reference. The bright [NeII] line is easily detected in a number of sources both in MIDI and VISIR spectra, while the weaker [ArIII] and [SIV] lines are only clearly seen in sources where we have VISIR spectra.

The [ArIII] 8.99 \um line is detected in MCG-05-23-016 and in IC~4329~A; the [SIV] 10.5 \um line is clearly seen in I~Zw~1, NGC~3281, NGC~4507, IC~4329~A and NGC~7469 and we have tentative detections of this line in NGC~424 and NGC~5506 (the latter can be confirmed by comparison with the Gemini spectrum of \citet{roche2007}). The [NeII] 12.81 \um line is detected in all sources with [ArIII] or [SIV] detections and additionally in NGC~3783, NGC~4151, Centaurus~A, NGC~5995. It is not seen in I~Zw~1 since the location of the line is shifted outside the atmospheric $N$ band for this source.

Otherwise the mid-IR total flux spectra of these sources show a featureless continuum with the $\approx$ 9.7 \um silicate emission/absorption feature imprinted on it. In many cases the silicate feature is hard to discern due to the limited bandpass of the atmospheric $N$ band. None of the lines is seen in the correlated flux spectra, meaning that they originate from larger-scale regions.

Our best estimates for the total flux of all the targets at $\sim 12 \um$ are collected in Tab.~\ref{tab:results:phot}. In general, the VISIR fluxes, both from photometry and from spectroscopy data, are very similar to the MIDI single-dish fluxes, where we can get a good estimate from the latter. For comparison we also reduced available mid-IR images taken close in time to the spectra and compare the nuclear photometry to the spectra. Both photometry and spectra are generally in good agreement, which illustrates well the reliability of the VISIR spectra as total flux references. None of the AGN shows significant extension beyond a point source, although NGC~7469 and NGC~5995 appear $\ga$20\% larger than the calibrator in some filters \citep[see also][]{hoenig2010}. Only four of the sources in this paper are partly resolved with 8m telescopes; for eleven sources upper limits on the source size have been published. They are collected in Table~\ref{tab:results:single_dish_limits}.

\begin{table*}
\caption{\label{tab:results:phot}Total flux references from MIDI and VISIR mid-IR spectroscopy and photometry for all targets. The table lists only those VISIR observations that are close in time to the MIDI observations. References: 1: \citet{tristram2009}; 2: \citet{gandhi2009}, 3: \citet{hoenig2010}, 4: \citet{kishimoto2011b}, 5: Sebastian F. H\"onig (priv. comm.), 6: this work}
\centering
\begin{tabular}{l | l l | l l l | l l l l}
\hline\hline\
 & \multicolumn{2}{c}{MIDI spectroscopy} & \multicolumn{3}{c}{VISIR spectroscopy} & \multicolumn{4}{c}{VISIR photometry}\\
Object & $\lambda [\um]$ & $F_{\rm tot}^{\rm MIDI}$ [Jy] & $F_{\rm spec}^{\rm VISIR}$ [Jy] & obsdate & ref & $\lambda [\um]$ & $F_{\rm phot}^{\rm VISIR}$ [Jy] & obsdate & ref\\
\hline
              I Zw 1&12.71$\pm$ 0.20& 0.44$\pm$ 0.04& 0.43$\pm$ 0.02&2010-10-16&      6&
11.88$\pm$ 0.37& 0.44$\pm$ 0.01&2010-10-15&     5\\
             NGC 424&12.13$\pm$ 0.20& 0.72$\pm$ 0.03&---&---&---&---&---&---&---\\
            NGC 1068&12.05$\pm$ 0.20&16.08$\pm$ 0.16&---&---&---&---&---&---&---\\
            NGC 1365&12.07$\pm$ 0.20& 0.38$\pm$ 0.06& 0.30$\pm$ 0.01&2010-12-25&      6&---&---&---&---\\
     IRAS 05189-2524&12.51$\pm$ 0.20& 0.37$\pm$ 0.07& 0.46$\pm$ 0.03&2007-01-30&      6&---&---&---&---\\
          H 0557-385&12.41$\pm$ 0.20& 0.35$\pm$ 0.05&---&---&---&11.88$\pm$ 0.37& 0.40$\pm$ 0.03&2009-09-06& 4\\
     IRAS 09149-6206&12.70$\pm$ 0.20& 0.40$\pm$ 0.05&---&---&---&11.88$\pm$ 0.37& 0.47$\pm$ 0.03&2009-06-07& 4\\
        MCG-05-23-16&12.10$\pm$ 0.20& 0.60$\pm$ 0.04& 0.73$\pm$ 0.01&2007-01-30&     3&---&---&---&---\\
            Mrk 1239&12.24$\pm$ 0.20& 0.47$\pm$ 0.05& 0.36$\pm$ 0.01&2011-03-12&      6&---&---&---&---\\
            NGC 3281&12.13$\pm$ 0.20& --& 0.32$\pm$ 0.01&2011-02-18&      6&---&---&---&---\\
            NGC 3783&12.12$\pm$ 0.20& 0.68$\pm$ 0.03&---&---&---&11.88$\pm$ 0.37& 0.76$\pm$ 0.02&2009-06-07& 4\\
            NGC 4151&12.04$\pm$ 0.20& 1.16$\pm$ 0.03&---&---&---&11.88$\pm$ 0.37& 1.36$\pm$ 0.08&2009-06-07&4,5\\
              3C 273&12.75$\pm$ 0.20& 0.35$\pm$ 0.14&---&---&---&11.85$\pm$ 2.34& 0.32$\pm$ 0.04&2007-03-01&   1\\
            NGC 4507&12.14$\pm$ 0.20& 0.68$\pm$ 0.09& 0.63$\pm$ 0.02&2008-04-11&     3&---&---&---&---\\
            NGC 4593&12.11$\pm$ 0.20& 0.22$\pm$ 0.08& 0.25$\pm$ 0.01&2011-02-18&      6&11.88$\pm$ 0.37& 0.25$\pm$ 0.01&2010-07-12&     5\\
          ESO 323-77&12.18$\pm$ 0.20& 0.44$\pm$ 0.07&---&---&---&11.88$\pm$ 0.37& 0.38$\pm$ 0.01&2009-05-09&     3\\
         Centaurus A&12.02$\pm$ 0.20& 1.57$\pm$ 0.03&---&---&---&---&---&---&---\\
     IRAS 13349+2438&12.75$\pm$ 0.20& 0.49$\pm$ 0.06&---&---&---&11.88$\pm$ 0.37& 0.53$\pm$ 0.03&2009-06-07& 4\\
           IC 4329 A&12.19$\pm$ 0.20& 0.96$\pm$ 0.03& 1.03$\pm$ 0.02&2009-05-01&     3&11.88$\pm$ 0.37& 1.01$\pm$ 0.02&2009-05-09&     3\\
            Circinus&12.02$\pm$ 0.20&10.23$\pm$ 0.05&---&---&---&---&---&---&---\\
            NGC 5506&12.07$\pm$ 0.20& 0.97$\pm$ 0.08&---&---&---&---&---&---&---\\
            NGC 5995&12.30$\pm$ 0.20& 0.40$\pm$ 0.14& 0.33$\pm$ 0.01&2009-04-28&     3&---&---&---&---\\
            NGC 7469&12.20$\pm$ 0.20& 0.47$\pm$ 0.06& 0.62$\pm$ 0.02&2008-09-18&     3&---&---&---&---\\
             NGC3227& ---           & ---           & ---           & ---      & ---  &11.88 $\pm$ 0.37 & 0.3201 $\pm$ 0.0219 & 2008-03-21 & 3\\
             IC3639 & ---           & ---           & ---           & ---      & ---  &13.04 $\pm$ 0.22 & 0.542 $\pm$ 0.029   & 2008-03-21 & 2\\

\end{tabular}
\end{table*}

\begin{table*}
\caption{\label{tab:results:single_dish_limits}Limits to the size of the sample targets from high-resolution observations with 8m class telescopes in the mid-infrared (at $\approx$ 12 \um). References: 1: \citet{radomski2003}, 2: \citet{bock2000}, 3: \citet{radomski2008}, 4: \citet{hoenig2010}, 5: \citet{reunanen2010}}
\centering
\begin{tabular}{l | l | l}
\hline\hline\
source name & Limit [mas] & Reference\\
NGC~1068    & compact core + extended    & 2 \\
MCG-5-23-16 & 0\farcs33       & 4 \\
Mrk~1239    & 0\farcs32       & 5 \\
NGC~3227    & 0\farcs37       & 4 \\
NGC~3783    & 0\farcs35       & 4 \\
NGC~4151    & 0\farcs27       & 1 \\
NGC~4507    & 0\farcs34       & 4 \\
NGC~4593    & 0\farcs36       & 4 \\
NGC~5128    & 0\farcs20        & 3 \\
IC~4329A    & 0\farcs36       & 4 \\
Circinus    & compact core + extended    & 5 \\
NGC~5506    & 0\farcs33       & 5 \\
NGC~5995    & 0\farcs40       & 4 \\
NGC~7469    & 0\farcs33       & 5 \\
\hline
\end{tabular}
\end{table*}

\subsection{Data in electronic format}
All data are also available in electronic form from CDS.

%
%
%
%
%
%
%
%
%

\section{Radial models}
\subsection{Considerations for model fitting}
Due to the limited coverage of $(u,v)$ space and the lack of phase information, we cannot directly reconstruct images from our interferometric data. We therefore describe the source brightness distribution (the ``image'') by a point source and a Gaussian because these are building blocks of parametric imaging in interferometry that have simple analytical forms both in real and in Fourier space: The Fourier transforms of a Gaussian and a point source are simply a Gaussian and a constant. There is some discussion in the interferometry literature as to which is the best parametric description of the mid-IR radial profile of an AGN. The currently employed models are (in order of increasing complexity):

\begin{itemize}
	\item The {\em single Gaussian} has the advantage of being the simple most description of the interferometric data with just one parameter -- full-width at half maximum -- and allows for a rough determination of the size of the emitter \citep{tristram2009,tristram2011}. However, it fails to provide proper fits to most sources with good $(u,v)$ coverage. Most sources in this sample are clearly not well described by a single Gaussian.
	\item The {\em point source + single Gaussian} description has one extra parameter (two extra parameters if the total flux is fitted as well) and can be considered as the minimum extension to the first description. Like the single Gaussian, its Fourier transform is a simple analytic function. It seems to provide good fits (both ``visually'' and in terms of $\chi^2_{\rm red}$) to the mid-IR data of all but the two brightest AGNs and can be easily extended to incorporate a physical model for the underlying temperature distribution, as in \citet{burtscher2009}.
	\item {\em Power-law parameterizations} of the intensity \citep[e.g.][]{kishimoto2009} provide the maximum freedom to fit the radial profiles but do no longer have analytic Fourier transforms. They assume an innermost radius for the otherwise diverging power-law and often imply that the nuclear dust can be described with a single radial profile, from the hottest dust at the innermost radius (as probed by K band observations) to the warm dust on parsec scale (as probed by MIDI). However, it has been shown for NGC~4151 that the mid-IR and near-IR structures likely do {\em not} belong to the same structure \citep{burtscher2009,kishimoto2011}. Apart from that, some of the sources show a clear ``point source'' in the radial profiles (e.g. NGC~4151, NGC~4507, NGC~4593, NGC~5506). It is unclear whether a single power-law would still result in a good fit for these sources.
\end{itemize}

We would like to stress here that our approach of fitting a point source and a Gaussian is intended to describe the brightness distribution as simply as possible without assuming any physical connection (power-law or other) to unobserved scales. For most sources in this sample, the sublimation radius of dust is about ten times smaller than our resolution limit.

In this section we present model fits for all \nsources sources with the purpose of determining their geometry in the context of such an elementary model and give sizes and fluxes of its components. We assume axisymmetric source-brightness distributions so that the modeled intensity distributions only have a radial co-ordinate; the position angles are ignored. For the purpose of this paper -- to present a first analysis of all sources and derive reliable half-light radii -- this approach is justified as we discuss in Section~\ref{sec:radial:pa}.

\subsection{Resolution limits of an interferometer}
An interferometer has resolution limits towards both small and large angles that depend on the spatial frequencies $BL_{\lambda} = BL/\lambda$ that the interferometer samples ($BL$ is the projected baseline length). The smallest structures that can be resolved with an interferometer can be estimated by applying the Rayleigh criterion to the interferometer. It describes the resolution at which the correlated flux of a binary star has dropped to zero, which happens at an angle of $1/(2 BL_{\lambda})$. In reality, the resolution depends on the signal to noise (S/N) of the observations, however. If the S/N is high, a slight decrease of correlated flux can already be detected and angular sizes much smaller than $1/(2 BL_{\lambda})$ can be determined although the source is only {\em marginally} resolved, i.e. the flux has not vanished. We estimate our resolution to be $\sim 1/(4 BL_{\lambda})$ or about 5 mas at $\lambda = 12 \um$. A source (component) that is smaller than this is {\em unresolved} and we give an upper limit to its size.

On the other hand, the minimum baseline of the interferometer and the S/N of the data determine the largest angles that can be resolved. In our case, structures of about 50 mas (100 mas for the two sources observed with the ATs) and larger are {\em over-resolved} at the shortest baselines, i.e. the correlated flux of this component has dropped below the measurement noise already at the shortest projected baselines. We can then only give a lower limit to the size of that component. In those cases, the source is only observable because the unresolved flux from a smaller component is larger than our sensitivity limit. Since most of our sources are essentially unresolved with 8m-class telescopes, we can thus give both an upper and a lower limit in those cases. For a compilation of upper limits from single-dish observations see Tab.~\ref{tab:results:single_dish_limits}.

\subsection{Model parameters}
\label{sec:radial:model}
In principle we could therefore give the flux densities (in the following `fluxes') of an unresolved, a resolved and an over-resolved component. However, in most sources we cannot constrain the properties of both the over-resolved and the resolved emitter, due to the low S/N of our data and the ``gap'' in spatial frequencies between the single-dish resolution (8 m) and the shortest projected VLTI/UT baseline of $\approx$ 25 m. We therefore fit a two-component model with an unresolved ``point source'' (with flux $F_p$) and a ``Gaussian'' component with flux $F_g$ and full width at half maximum (FWHM) $\Theta_g$ and decide, based on the reduced $\chi^2$ ($\chi_{\rm red}^2$) contours, whether or not we can constrain the size of the Gaussian. If it is constrained, the Gaussian is resolved; if we can only give a lower limit to the Gaussian, it is over-resolved.

In real space, the modeled intensity distribution is:
\begin{align}
	\label{eq:imagespace}
	I_{\nu}(r) &= F_p \cdot \delta(r) + F_g \cdot \frac{4 \ln(2)}{\pi \Theta_g^2} \cdot \exp\left(-\frac{4 \ln(2) r^2}{\Theta_g^2}\right)
\end{align}

where $r$ is a radial angular co-ordinate on sky.

In Fourier space, the correlated flux at spatial frequency $BL_{\lambda}$, $F_{\nu}(BL_{\lambda})$, depends on the parameters of the two components as follows:

\begin{equation}
	\label{eq:radial}
	F_{\nu}(BL_{\lambda}) = F_p + F_g \cdot \exp\left(- \frac{(\pi \Theta_g BL_{\lambda})^2}{4 \ln(2)}\right)
\end{equation}

Since we fit only two components, we treat the total flux like a correlated flux measurement at baseline 0, i.e.

\begin{equation}
	F_{\rm \nu, tot} = F_{\nu}(0) = F_p + F_g = \int_0^{\infty} I_{\nu}(r) {\rm d}\Omega
\end{equation}

The total flux is very well constrained for most sources and we therefore approximate its error with zero in order to reduce the number of model parameters to two: $\Theta_g$ and point source flux fraction

\begin{equation}
	\label{eq:fp}
	f_p \equiv F_p / F_{\rm tot}.
\end{equation}

The flux of the resolved (or over-resolved) component is then $F_g = (1-f_p) \cdot F_{\rm tot}$. This two-parameter fit has the advantage that the goodness of fit can be displayed in a single $\chi^2$ plane (see next subsection).

We perform the fits at (12.0 $\pm$ 0.2) \um restframe since the sources are best resolved and brightest at this wavelength range and the atmospheric transmission is good there. Also, there are no bright mid-IR features in this range, allowing for an accurate determination of the continuum flux. For the two most distant sources 3C 273 and IRAS 13349+2438 this restframe wavelength is shifted beyond the atmospheric $N$ band. For these two sources the fit is performed at 12.5 $\um$ {\em observed} frame, corresponding to 10.8 and 11.3 $\um$ restframe respectively.

A full analysis of the spectral information will appear in a subsequent publication.

We show the radial models and the $\chi^2$ plots of the two-component model for all sources in Figs.~\ref{fig:rad:IZwicky1} -- \ref{fig:rad:NGC7469} and the best fitting parameters are summarized in Tab. \ref{tab:radialfit} on page \pageref{tab:radialfit}. Since we approximate the total flux errors with 0 for the radial fitting, the correlated flux can be understood in the same way as the visibility with the only difference that our normalization is the total flux of the target instead of 1. The point source fraction $f_p$ is then essentially identical to the minimum visibility.

For a graphical comparison of the fit results, we show model images with constant pixel scale for all sources in Fig.~\ref{fig:modelimages}, sorted by increasing resolution. These images are simply super-positions of two Gaussians of FWHM = 5 mas (the ``point source'') and FWHM = $\Theta_g$; elongations (see below, Section~\ref{sec:radial:pa}) are not shown. Their relative intensity is given by the point source fraction $f_p$. The point source is encircled with blue dots and in the cases where we can only derive a limit to the resolved emitter, this limit is indicated by a red circle.

\subsection{$\chi^2$ plots and component identification}

\begin{figure}
	{\includegraphics[trim=0cm 13cm 4cm 0cm, width=\hsize]{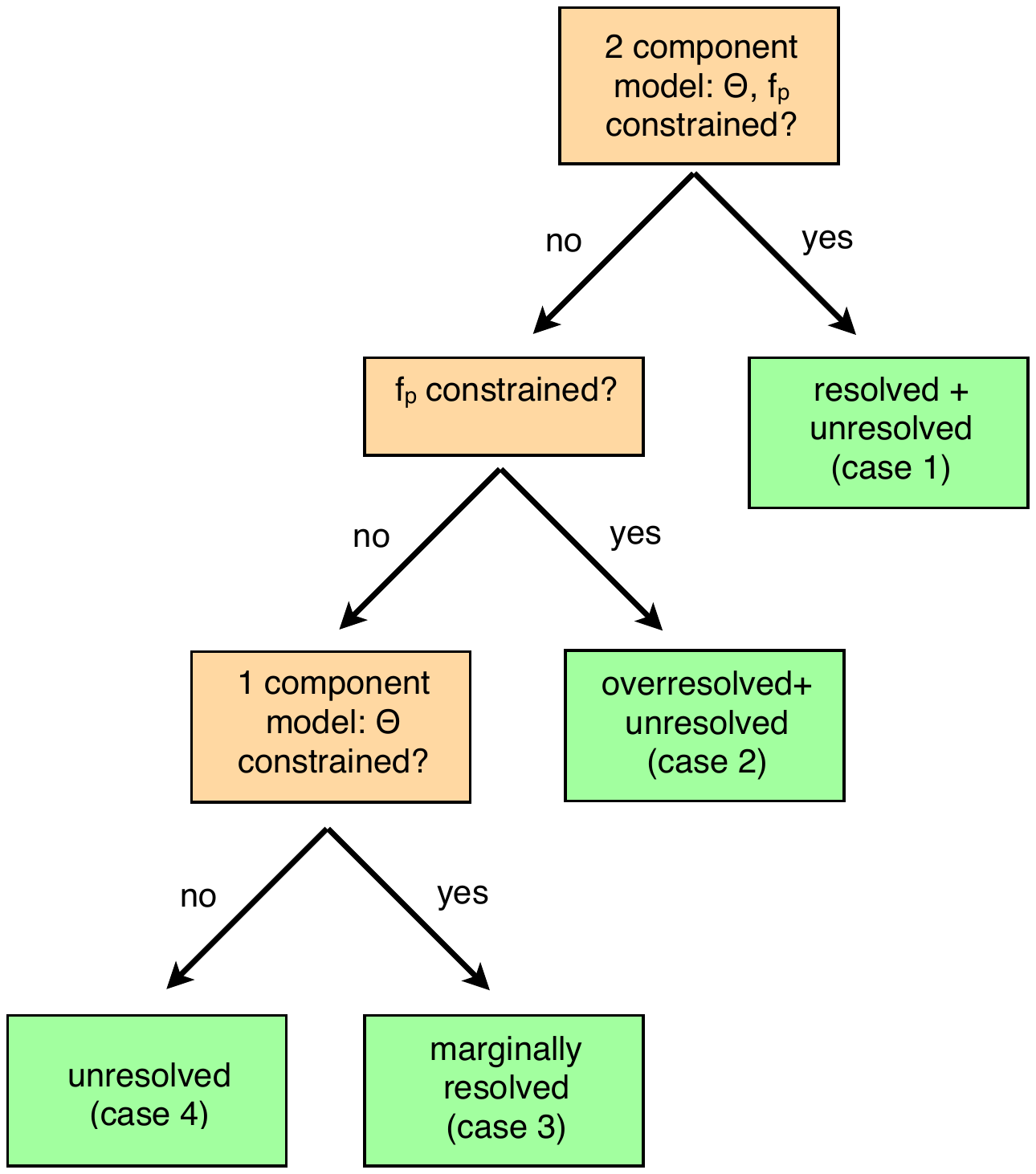}}
	\caption{\label{fig:rad:models} Model selection strategy: If both parameters ($\Theta_g, f_p$) are constrained on the $\chi^2_{\rm red}$ plane, we see a resolved and an unresolved emitter (case 1). If only $f_p$ is constrained, there are two emitters but we can give only limits to their sizes (case 2). If neither $\Theta_g$ nor $f_p$ are constrained, there is only one component which is either marginally resolved (case 3) or entirely unresolved (case 4).}
\end{figure}

\begin{figure}
	{\includegraphics[trim=0cm 10cm 0cm 0cm, width=\hsize]{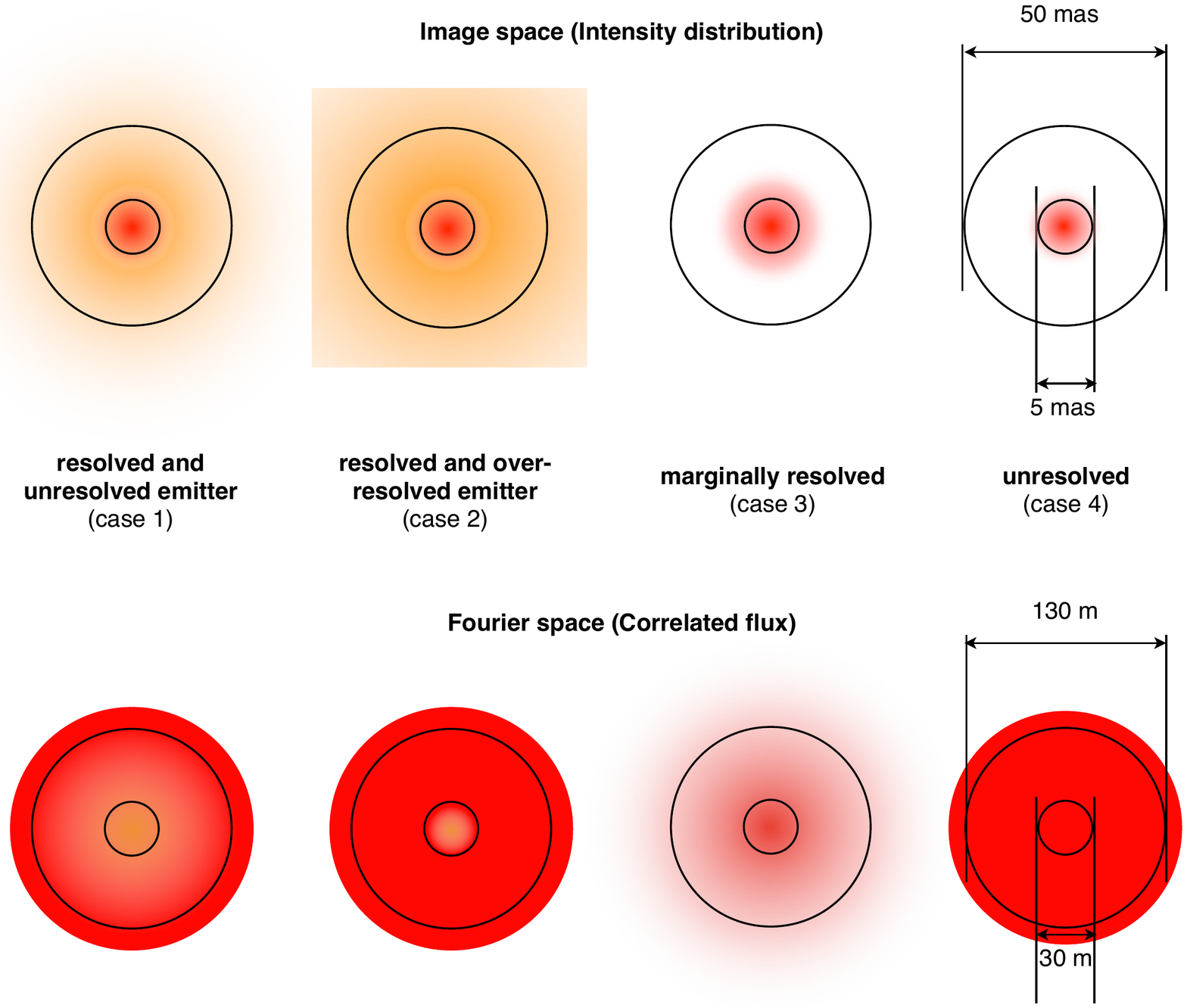}}
	\caption{\label{fig:rad:geo} Model source geometries and their Fourier transforms. In real space (top row), the two black rings stand for the minimum and maximum angular resolution of the interferometer (5 and 50 mas, respectively), the orange and red Gaussians stand for the (over-)resolved and unresolved mid-IR emitters, respectively. In case 1, both the compact unresolved as well as the larger, interferometrically resolved component are constrained by our model fit. Model 2 consists of an essentially unresolved source (red) with over-resolved flux (orange). The size of the over-resolved emitter cannot be determined and could be as large as the single-dish PSF of about 300 mas. Models 3 and 4 contain only one, essentially unresolved, component which in model 3 is marginally resolved and in model 4 entirely unresolved.\\
	In the bottom row, the Fourier transforms of these models are shown. Here the two black rings denote the minimum and maximum projected baseline using the 8m Unit Telescopes of the VLTI. Note how the compact source (red) contributes to the correlated flux at all baselines while the extended source (orange) is only observable at short baselines.}
\end{figure}

In the $\chi^2$ plots, three cases can be distinguished depending on the appearance of the $\min(\chi^2_{\rm red}) + 1$ contour which delimits the 1 $\sigma$ confidence interval. It is either \ldots

\begin{itemize}
	\item \ldots entirely on the plot, or
	\item open to large values of $\Theta_g$ but constrained in $f_p$, or
	\item degenerate with both $\Theta_g$ and $f_p$.
\end{itemize}

In the first two cases, both the point source and the resolved or overresolved emitter have non-zero flux. In the third case, where the two components are degenerate, we test whether the source is marginally resolved or entirely unresolved by fitting a single Gaussian component to the fluxes:

\begin{equation}
	\label{eq:radial1}
	\tilde{F}_{\nu}(BL_{\lambda}) = F_{\rm tot} \cdot \exp\left(- \frac{(\pi \Theta_g BL_{\lambda})^2}{4 \ln(2)}\right)
\end{equation}

In this case, we additionally plot the $\chi^2$ curve of the one-component model as an inset to the $\chi^2$ plots.

In summary, we classify each source into one of the following four cases following straight-forward, hierarchical steps (see Fig.~\ref{fig:rad:models}):

\begin{enumerate}
	\item Both the size of the 'Gaussian', $\Theta_g$, and the fluxes of the two components, $F_g, F_p$, can be derived (12 sources, e.g. I Zw 1).
	\item The fluxes of the over-resolved and unresolved components, $F_g, F_p$ are constrained, but only a lower limit for $\Theta_g$ can be given (6 sources, e.g. NGC~3281).
	\item $f_p$ is degenerate, but $\Theta_g$ is constrained in a one-component fit (4 sources, e.g. 3C 273).
	\item $f_p$ is degenerate and, in a one-component fit, only an upper limit for $\Theta_g$ can be given (1 source: IRAS~13349+2438).
\end{enumerate}

\subsection{Half-light radii}

In order to derive a single size estimate that characterizes the nuclear light distribution, we determine the half-light radius for each source. It is defined by

\begin{equation}
	\label{eq:r12}
	\frac{1}{2} = \frac{\int_0^{r_{1/2}} I_{\nu} 2 \pi r {\rm d}r}{F_{\nu, {\rm tot}}}
\end{equation}

For sources of case '1', Eqs.~\ref{eq:imagespace} and \ref{eq:fp} substituted into Eq.~\ref{eq:r12} give
\begin{equation}
	\label{eq:r12_1}
	\frac{1}{2} = f_p + (1-f_p) \cdot \left(1-\exp\left(-\frac{4 \ln(2) r_{1/2}^2}{\Theta_g^2}\right)\right)
\end{equation}

which leads to the following relation between half light radius and Gaussian FWHM $\Theta_g$:
\begin{equation}
	\label{eq:r12theta}
	\frac{r_{1/2}}{\Theta_g} = \frac{1}{2} \sqrt{\frac{\ln(2-2f_p)}{\ln2}}.
\end{equation}

\begin{figure}
	{\includegraphics[trim=3cm 3cm 3cm 3cm, width=0.9\hsize]{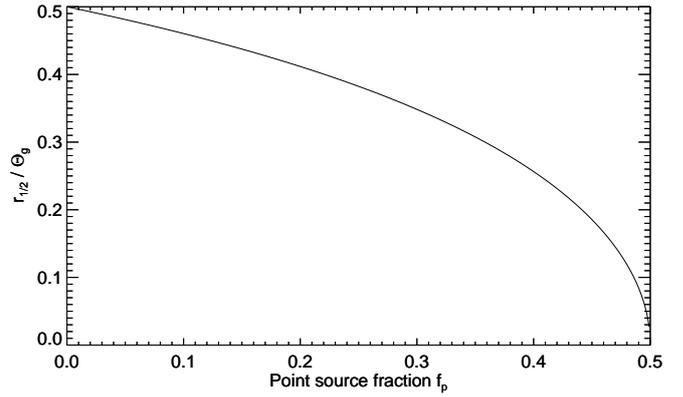}}
	\caption{\label{fig:r12fwhm} Ratio of half-light radius $r_{1/2}$ to Gaussian full-width at half maximum $\Theta_g$ as a function of point source fraction $f_p$.}
\end{figure}

Its dependency on $f_p$ is plotted in Fig.~\ref{fig:r12fwhm}. For $f_p = 0$, it is 1/2 and it vanishes for $f_p = 0.5$. For $f_p > 0.5$ it is not defined and we assign such sources, where more than half of the emission is unresolved, an upper limit to the half-light radius that we set to our approximate resolution limit of $\Theta_{g, {\rm min}}/2 \approx$ 2.5 mas.

For over-resolved sources (case '2'), we can only derive limits to the half-light radius. We give an upper limit, $r_{1/2} < \Theta_{g {\rm min}}/2$ if $f_p > 0.5$ and a lower limit, given by Eq.~\ref{eq:r12theta}, otherwise. For sources where we only model one component, the half-light radius is equal to $\Theta_g/2$ if the source is marginally resolved (case `3'); if it is unresolved (case `4'), the upper limit to $r_{1/2}$ is again given by our resolution limit.

At $f_p = 0.5$ the half-light radius has a discontinuity for over-resolved sources (case `2') since it is unknown whether most emission is un- or over-resolved ($r_{1/2}$ jumps from an upper to a lower limit). We therefore only derive a half-light radius for sources which are clearly off this border, i.e. where either $f_p < 0.5 - \sigma(f_p)$ or $f_p > 0.5 + \sigma(f_p)$. The half-light radii in both angular and linear units are given in Tab.~\ref{tab:radialfit}.

Applying these criteria, we arrive at a total of eleven measured half-light radii and estimate very compact upper limits to $r_{1/2}$ for ten additional sources and one lower limit. For only one source (NGC~4593) we do not derive a half-light radius since it is over-resolved and its point source fraction is $\approx$ 50 \%.

%
%

\subsection{Axial symmetry and elongations}
\label{sec:radial:pa}

\begin{figure*}
	\centering
	\subfloat{\includegraphics[trim=7cm 4cm 7cm 4cm, width=0.5\hsize]{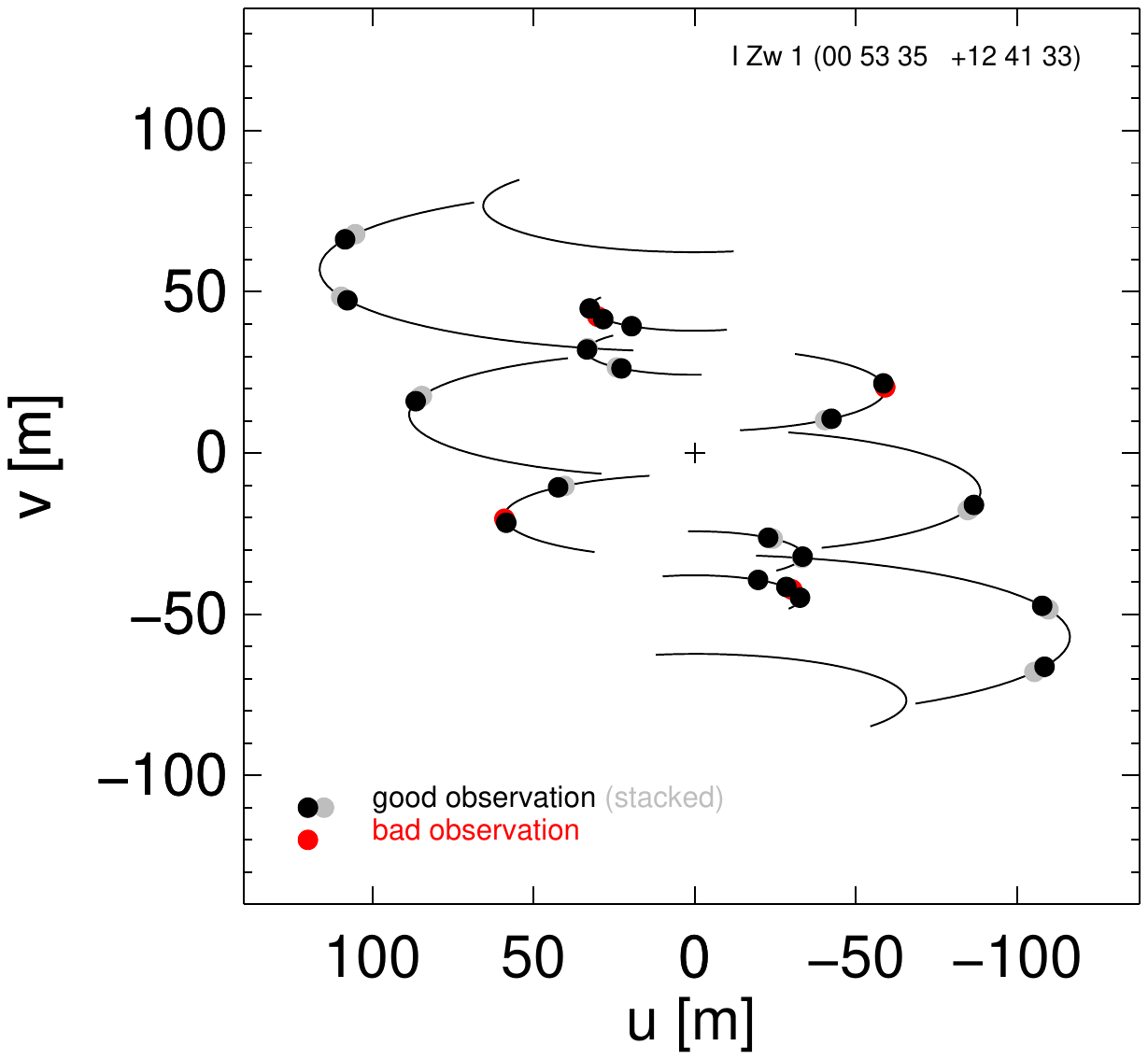}}
	~~~~~~~~~
	\subfloat{\includegraphics[trim=7cm 4cm 7cm 4cm, width=0.5\hsize]{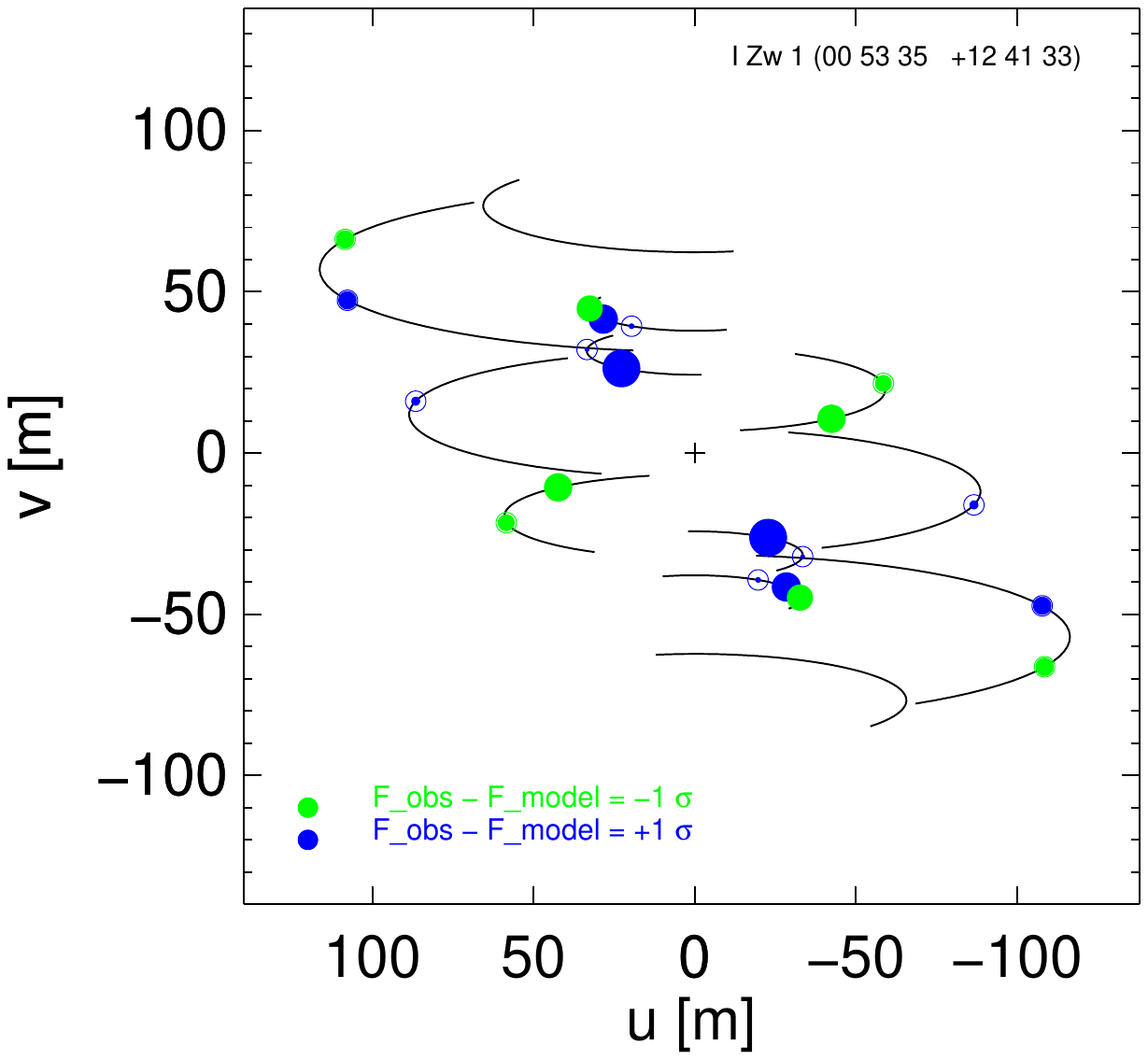}}\\
	\subfloat{\includegraphics[trim=3cm 0cm 3cm 0cm, width=0.5\hsize]{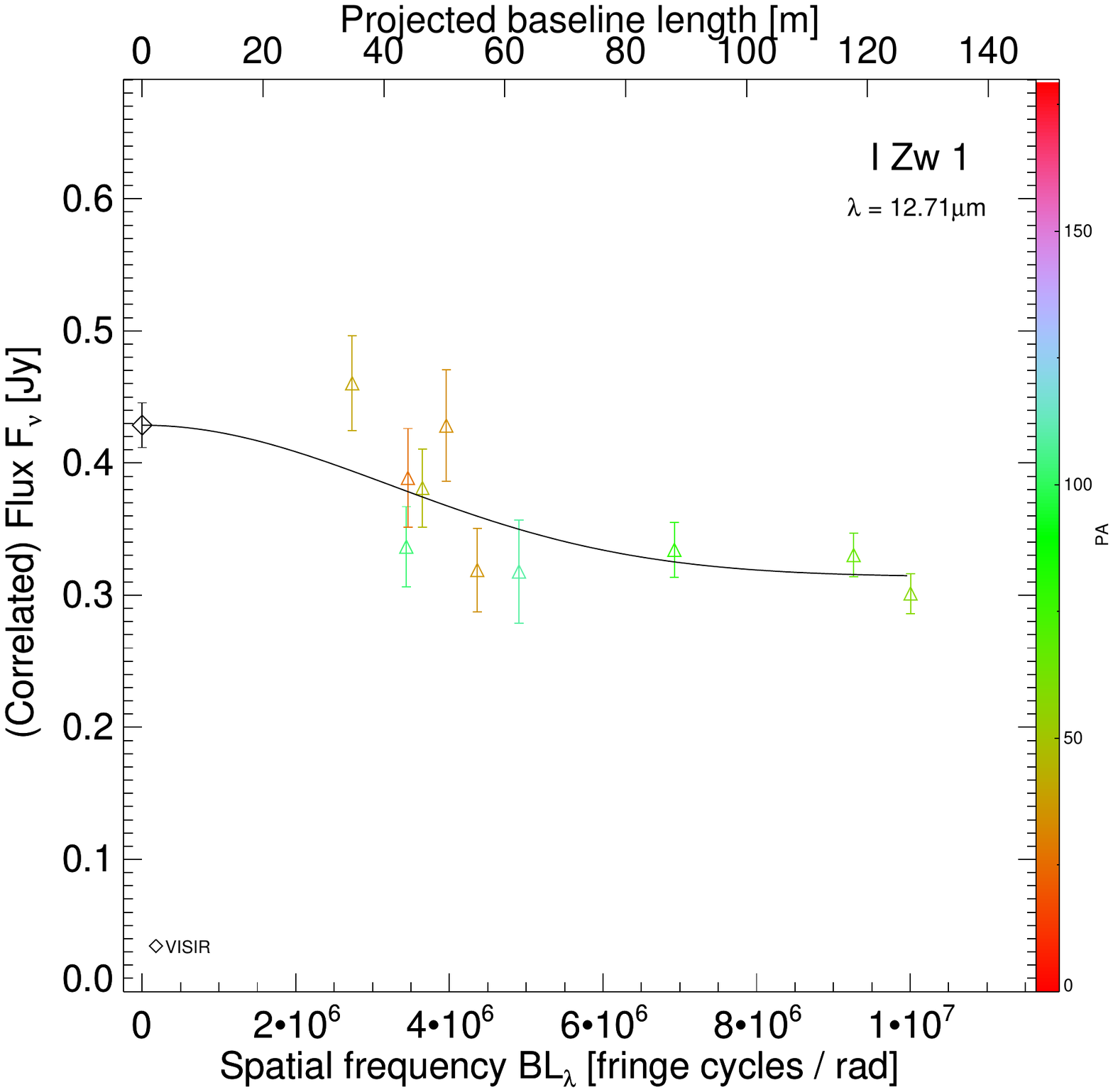}}
	~~~~~~~~~
	\subfloat{\includegraphics[trim=3cm 0cm 3cm 0cm, width=0.5\hsize]{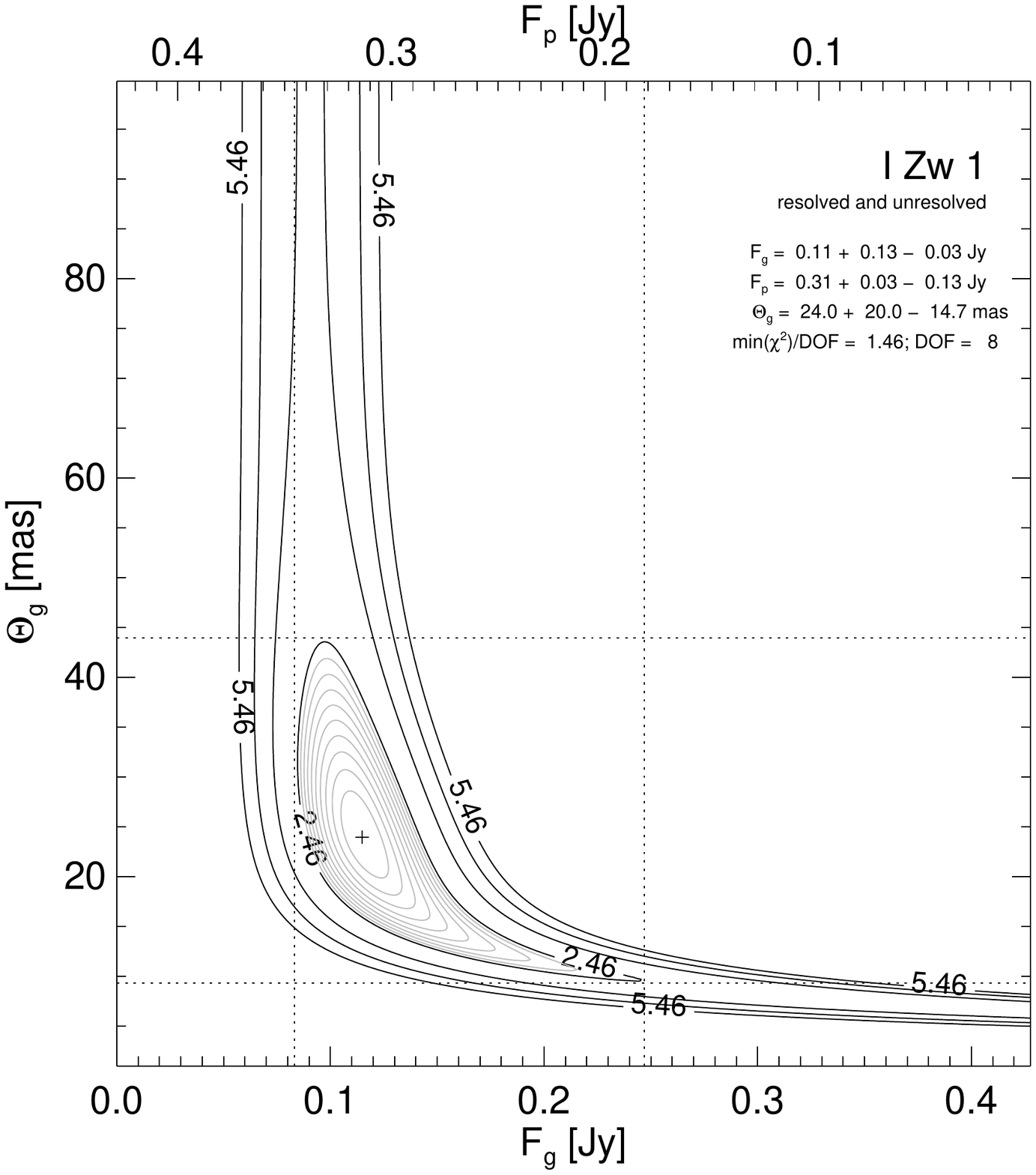}}
	\caption{\label{fig:rad:IZwicky1}I Zw 1 \quad {\em Top left panel:} $(u,v)$ coverage. The first observation of a stack is shown in black, other good observations (that are stacked to the preceding ``black'' one) in gray, bad observations in red. In the upper right hand the co-ordinates of the target are given for reference. {\em Top right:} model residuals on the $(u,v)$ plane. The diameter of the filled circle indicates, in units of the measurement error, the difference between the observed correlated flux at that $(u,v)$ position and the model predicted flux (see bottom left panel). The difference is displayed in green if the observed flux is lower than the model predicted one and in blue otherwise.	 Measurements that lie very close to the model predictions have very small filled circles and are therefore additionally encircled. {\em Bottom left:} Radial plot with position angle (PA) color-coded. Note that PA is only unique modulo 180 degrees. {\em Bottom right:} $\chi^2$ plane. In it, the best fit is marked with a black cross, the $1 \sigma$ limits to the best fit are marked with dotted lines. Since the total flux, $F_{\nu, \rm tot}$, is well constrained, we approximate its error with zero in order to reduce the model to two parameters: Gaussian FWHM of the resolved source, $\Theta_g$, and point source fraction $f_p$. In the $\chi^2$ plot, the abscissae are the fluxes of the two modelled components, $F_g \equiv (1-f_p) \cdot F_{\nu, \rm tot}$ and $F_p \equiv f_p \cdot F_{\nu, \rm tot}$.}
\end{figure*}

\begin{figure*}
	\centering
	\subfloat{\includegraphics[trim=7cm 4cm 7cm 4cm, width=0.5\hsize]{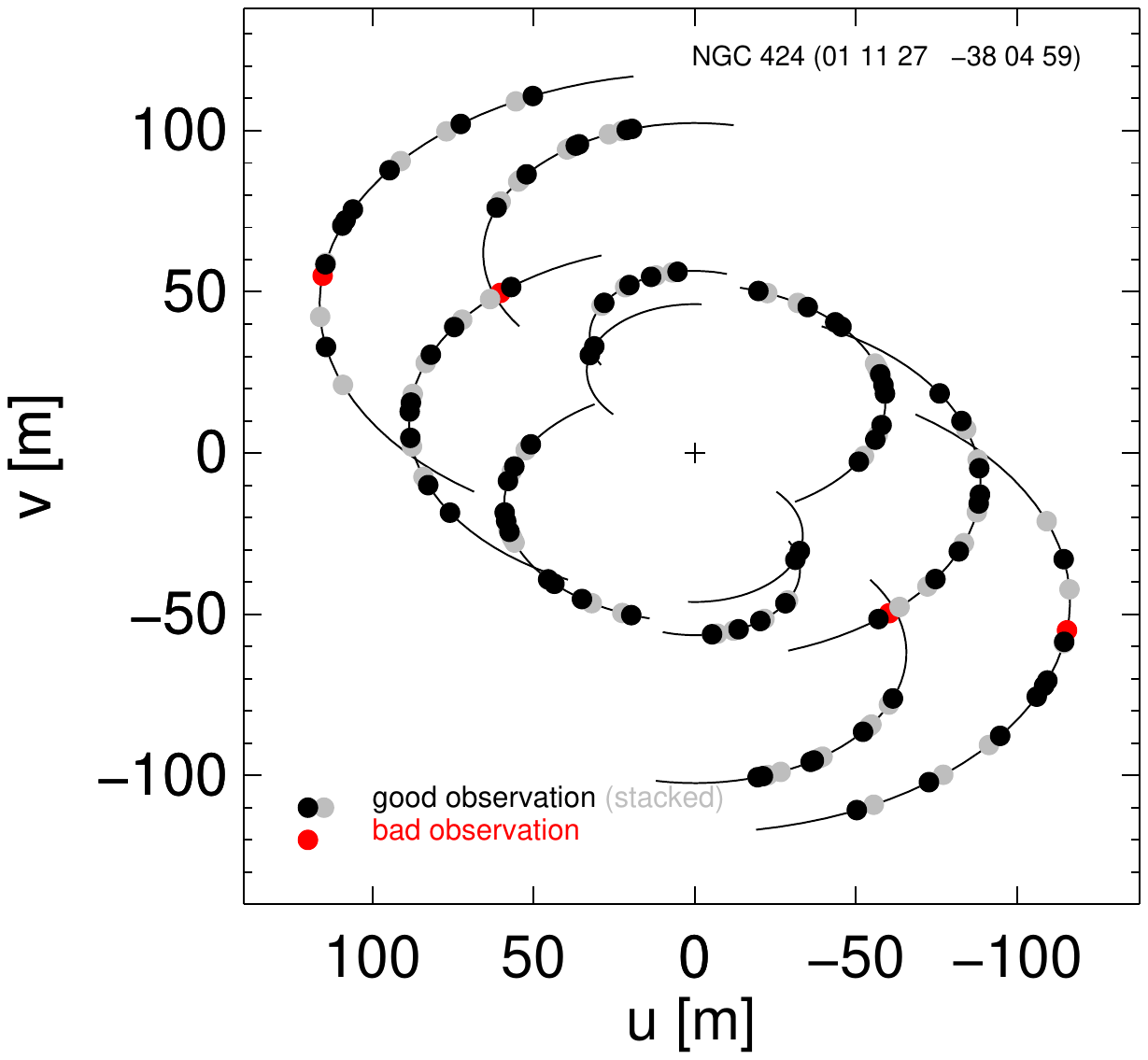}}
	~~~~~~~~~
	\subfloat{\includegraphics[trim=7cm 4cm 7cm 4cm, width=0.5\hsize]{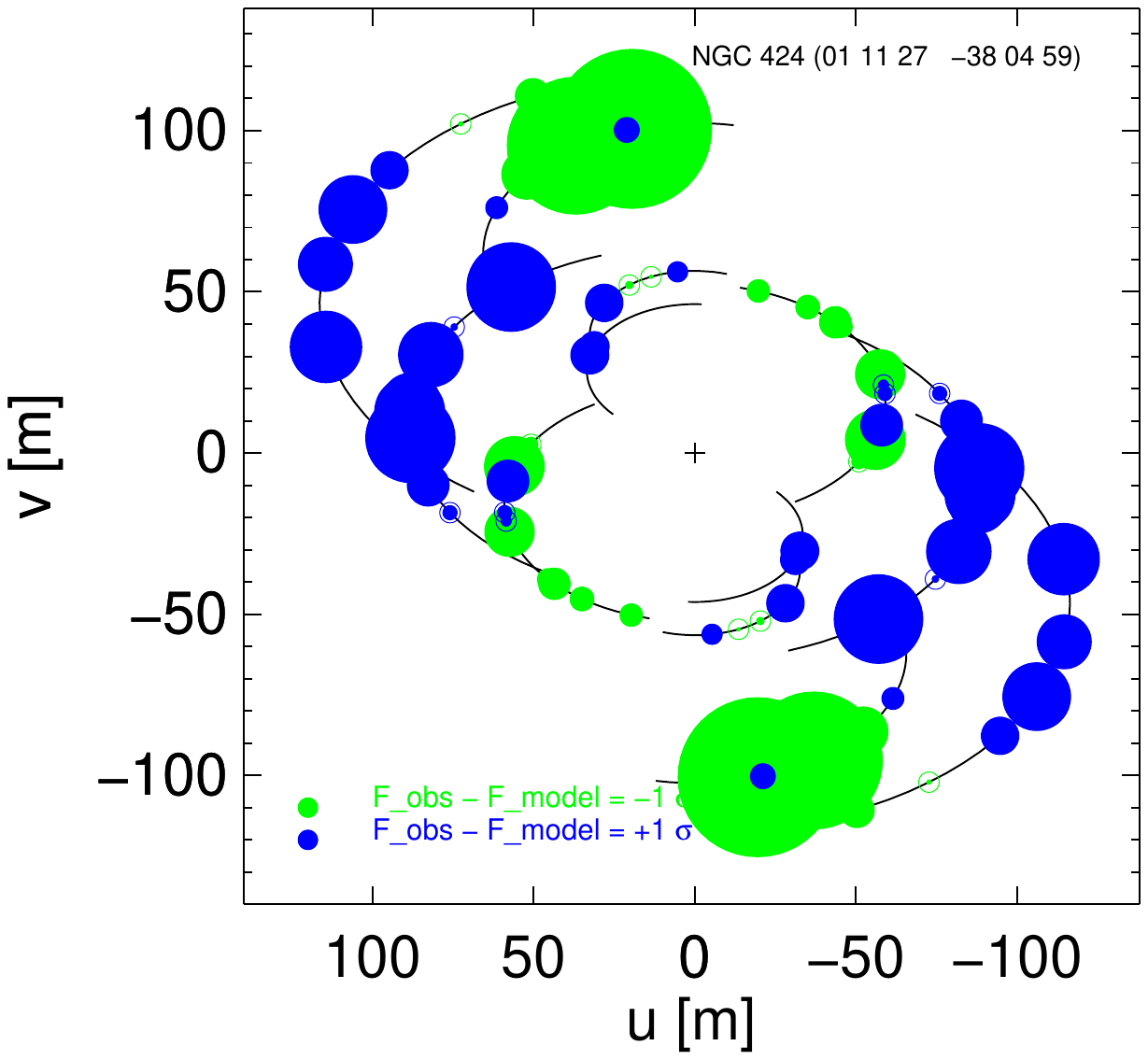}}\\
	\subfloat{\includegraphics[trim=3cm 0cm 3cm 0cm, width=0.5\hsize]{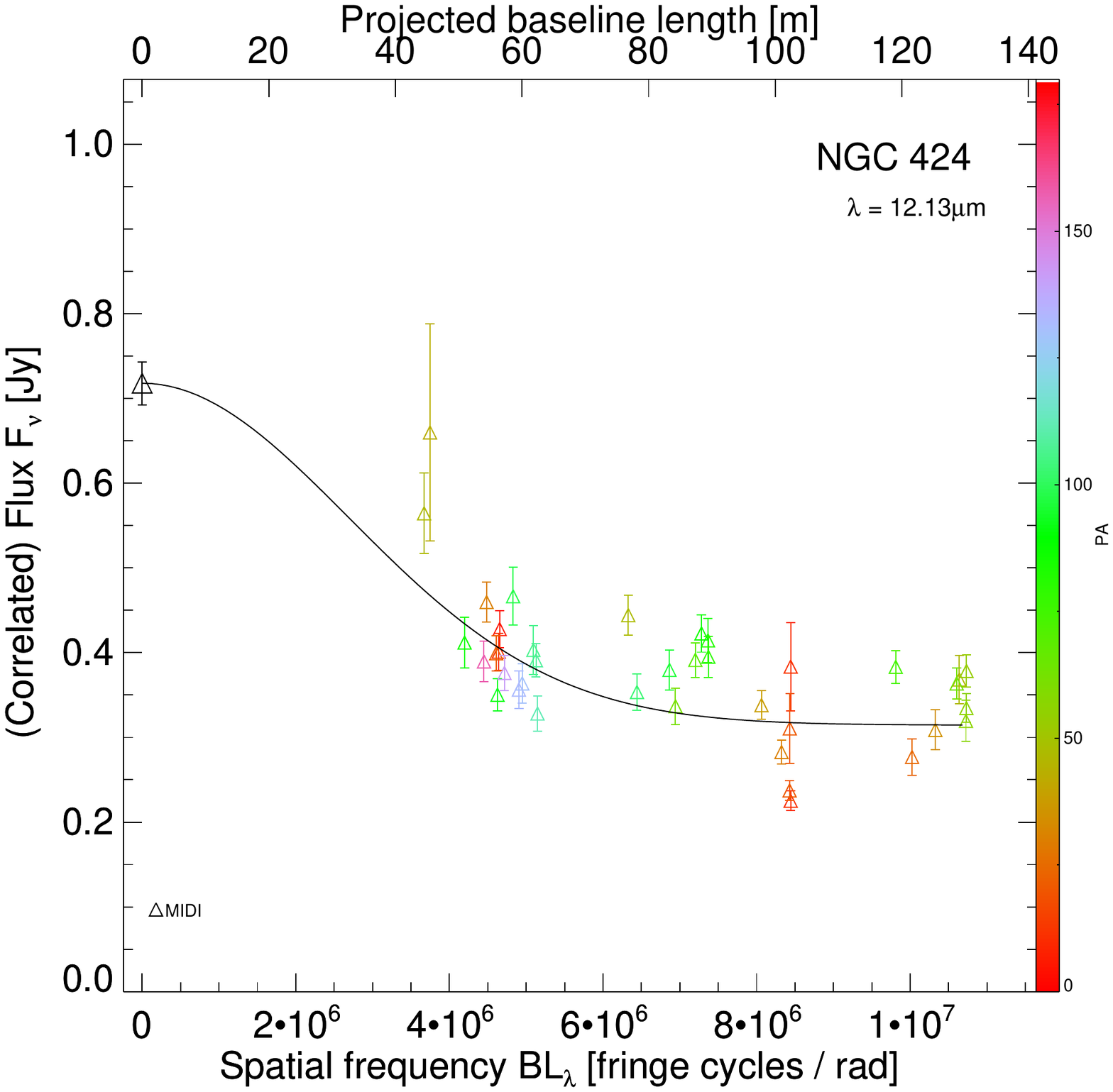}}
	~~~~~~~~~
	\subfloat{\includegraphics[trim=3cm 0cm 3cm 0cm, width=0.5\hsize]{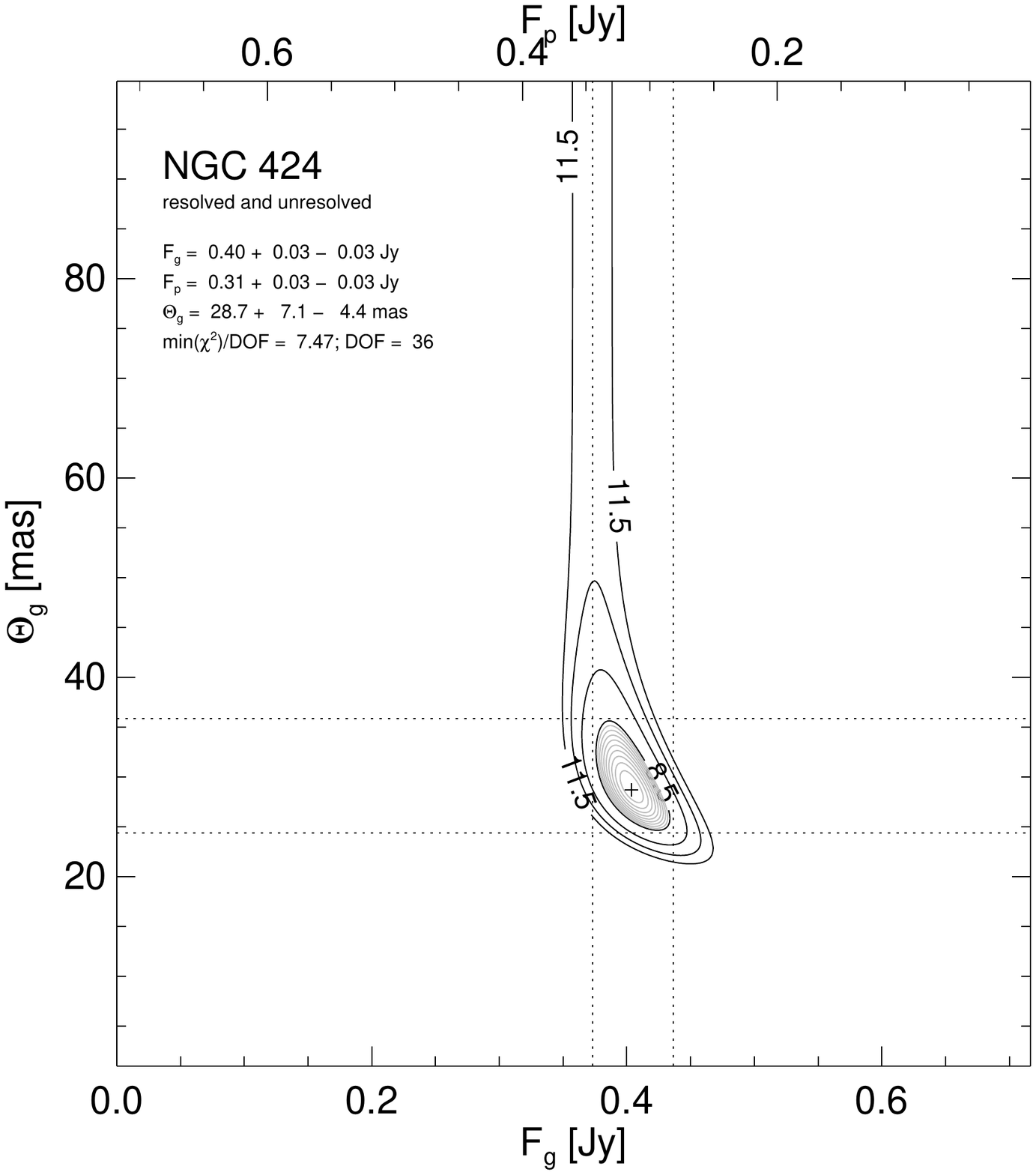}}
	\caption{\label{fig:rad:NGC424}The same as Fig. \ref{fig:rad:IZwicky1} but for NGC~424. The large value for the reduced $\chi^2$ is due to the elongation of the extended component \citep{hoenig2012} that was not fitted here.}
\end{figure*}

\begin{figure*}
	\centering
	\subfloat{\includegraphics[trim=7cm 4cm 7cm 4cm, width=0.5\hsize]{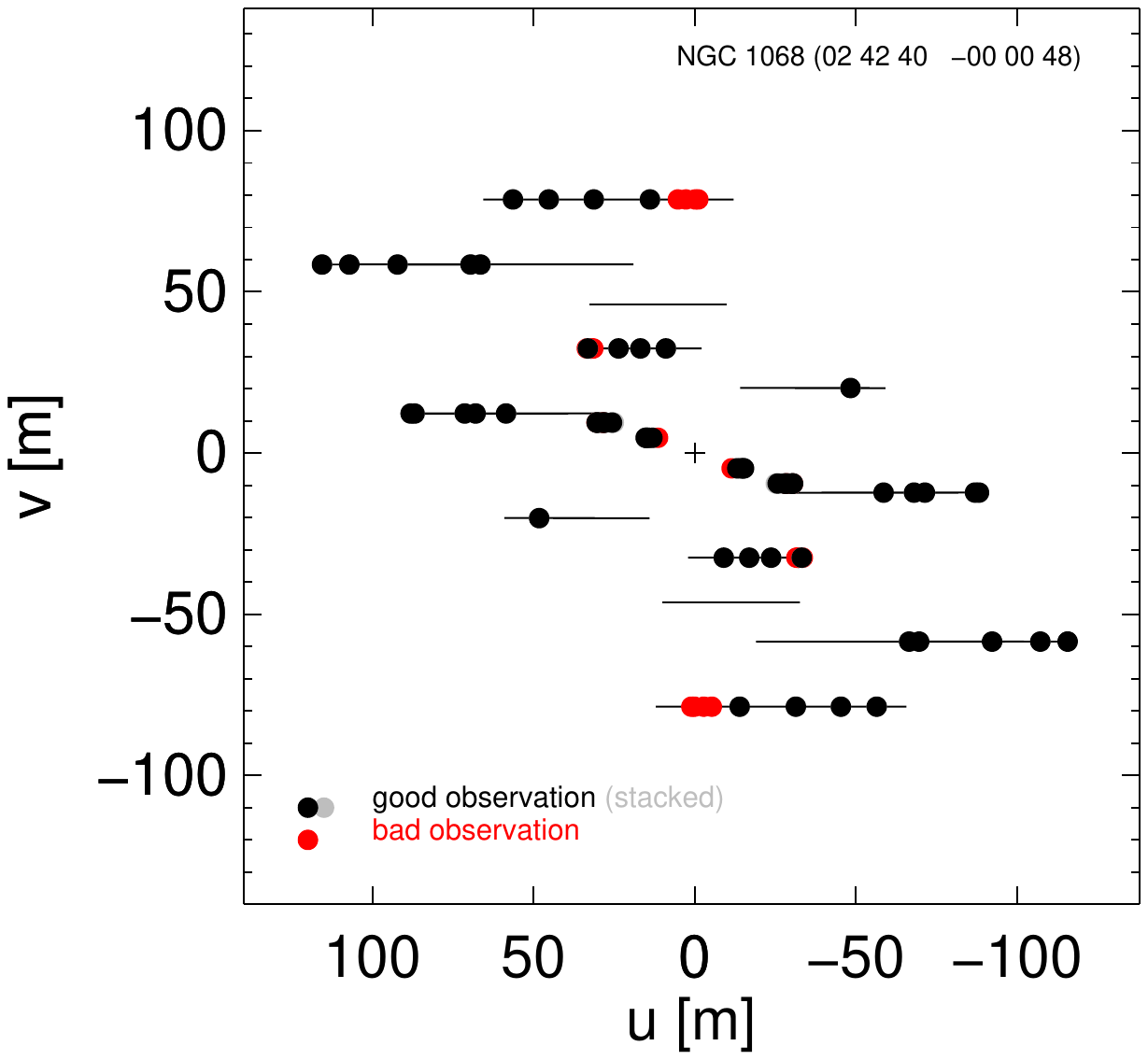}}
	~~~~~~~~~
	\subfloat{\includegraphics[trim=7cm 4cm 7cm 4cm, width=0.5\hsize]{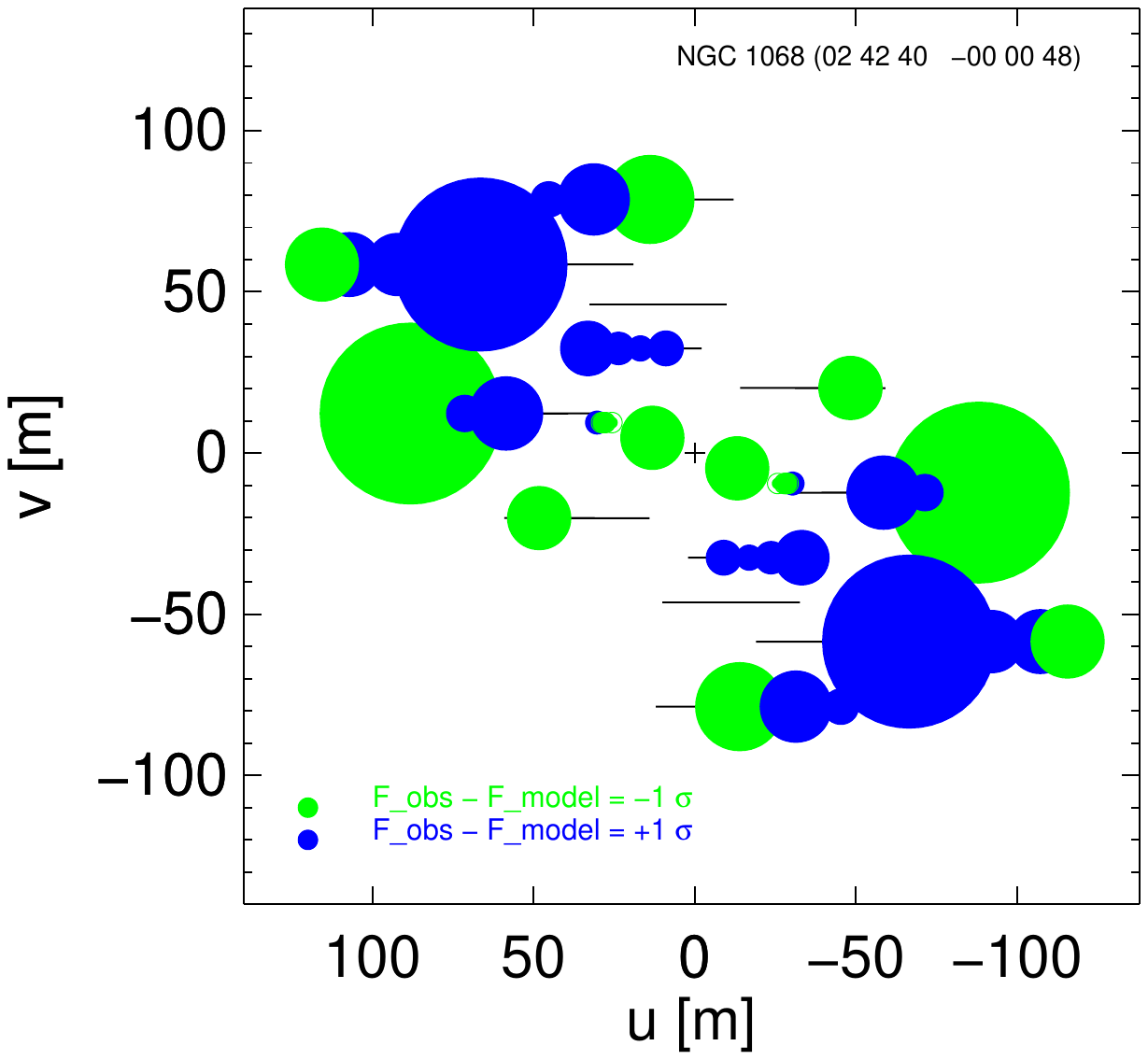}}\\
	\subfloat{\includegraphics[trim=3cm 0cm 3cm 0cm, width=0.5\hsize]{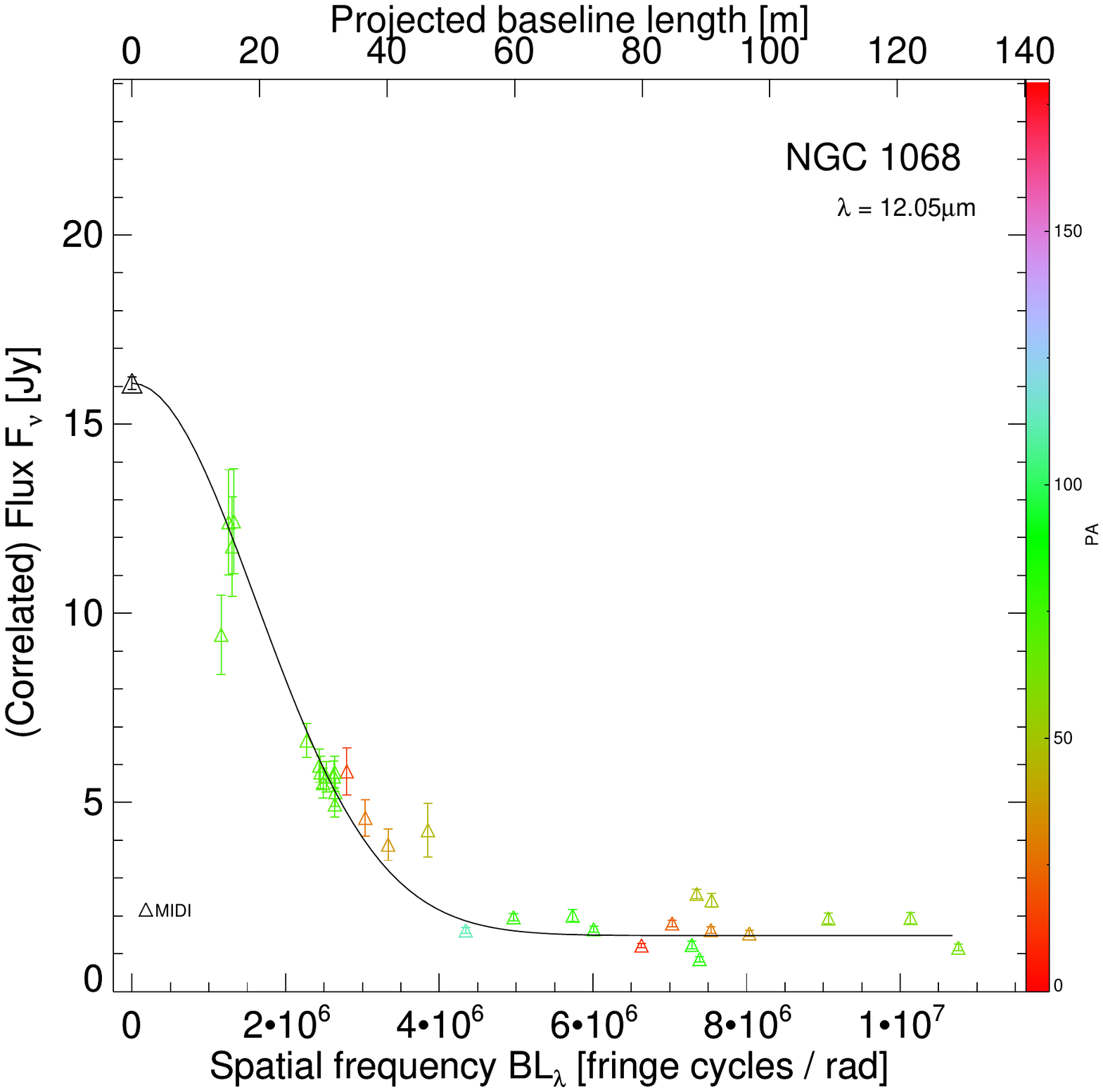}}
	~~~~~~~~~
	\subfloat{\includegraphics[trim=3cm 0cm 3cm 0cm, width=0.5\hsize]{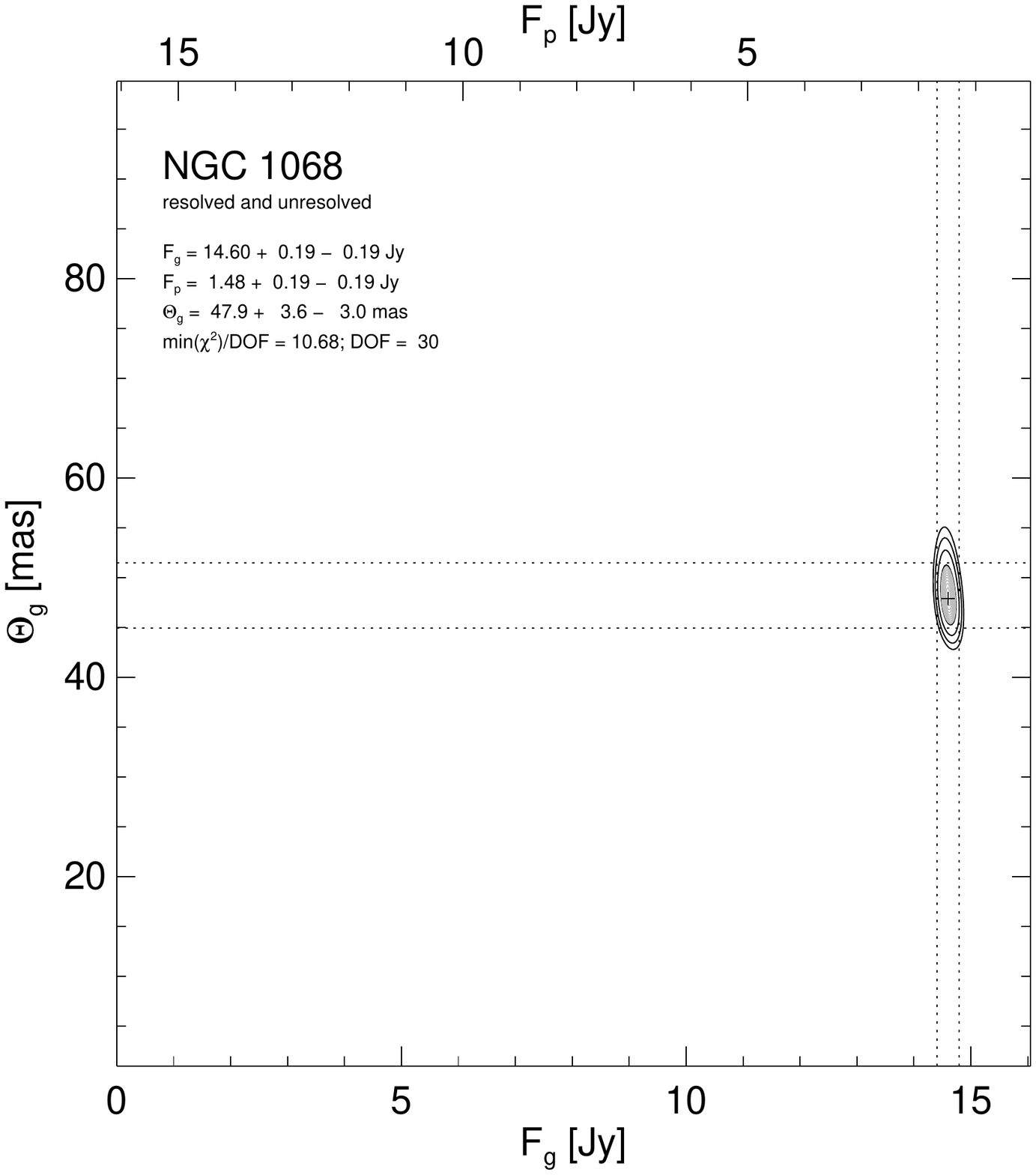}}
	\caption{\label{fig:rad:NGC1068}The same as Fig. \ref{fig:rad:IZwicky1} but for NGC~1068. The very large value for the reduced $\chi^2$ is due to the elongation of the compact disk component \citep{raban2009} that was not fitted here.}
\end{figure*}

\begin{figure*}
	\centering
	\subfloat{\includegraphics[trim=7cm 4cm 7cm 4cm, width=0.5\hsize]{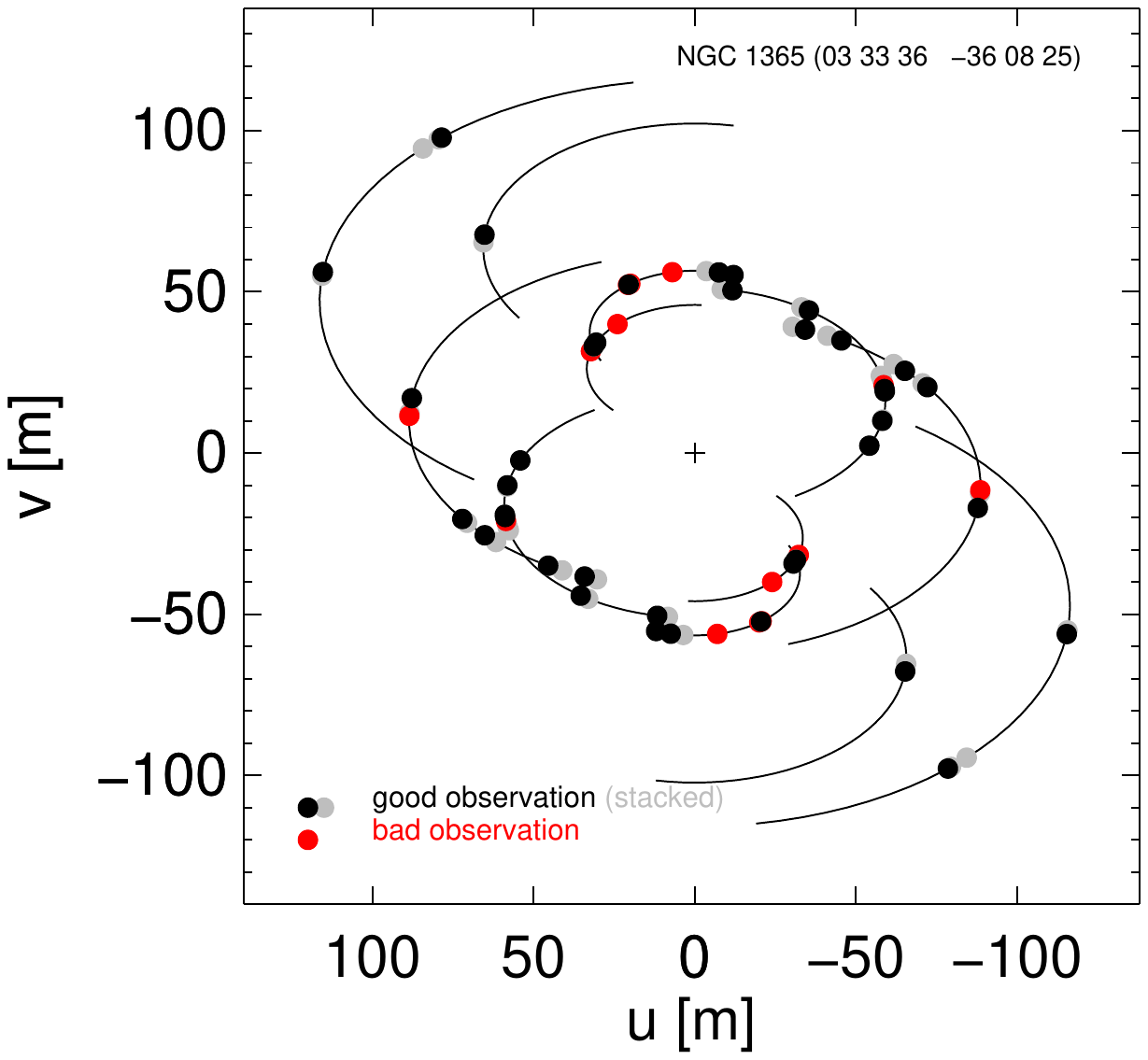}}
	~~~~~~~~~
	\subfloat{\includegraphics[trim=7cm 4cm 7cm 4cm, width=0.5\hsize]{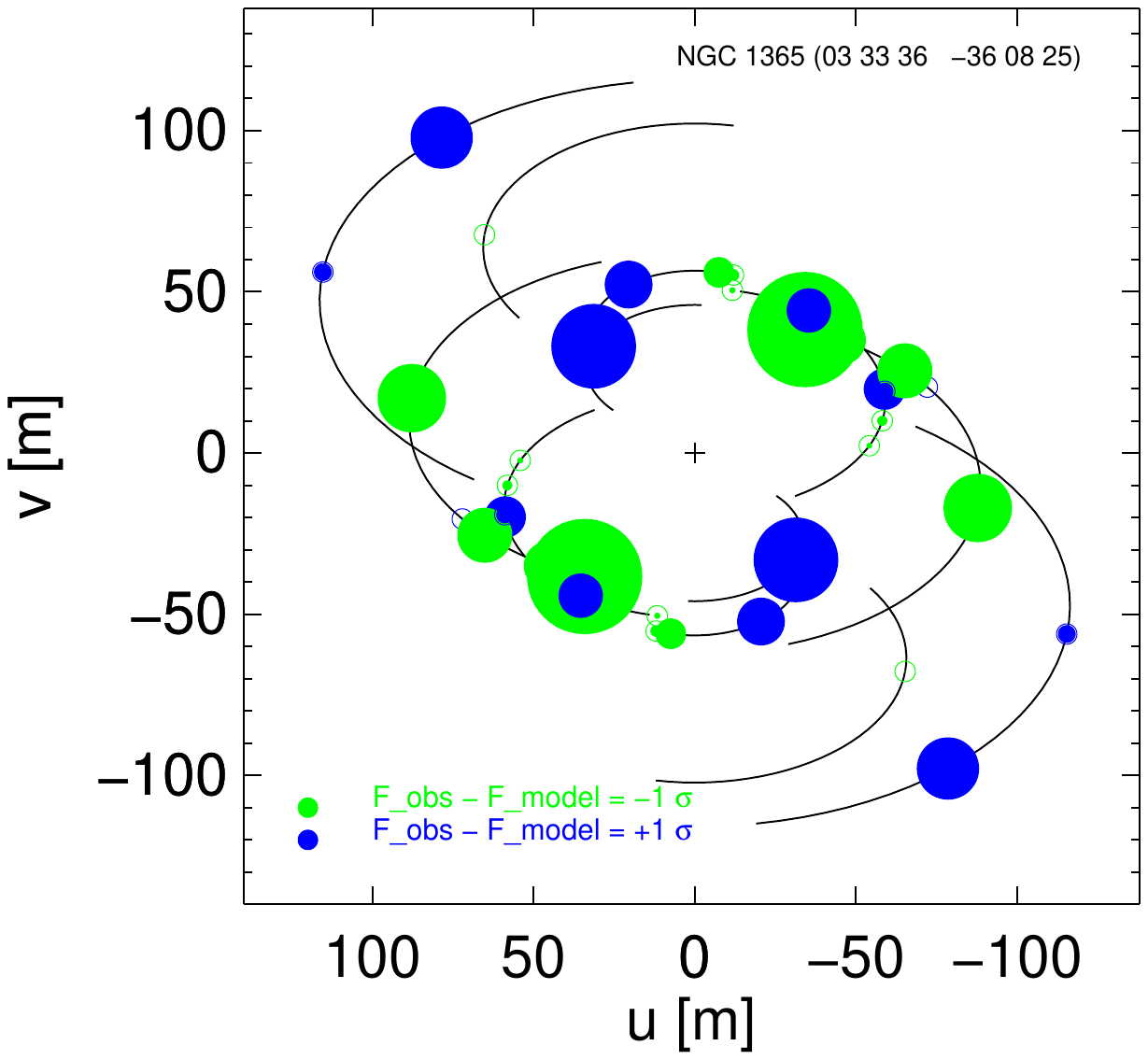}}\\
	\subfloat{\includegraphics[trim=3cm 0cm 3cm 0cm, width=0.5\hsize]{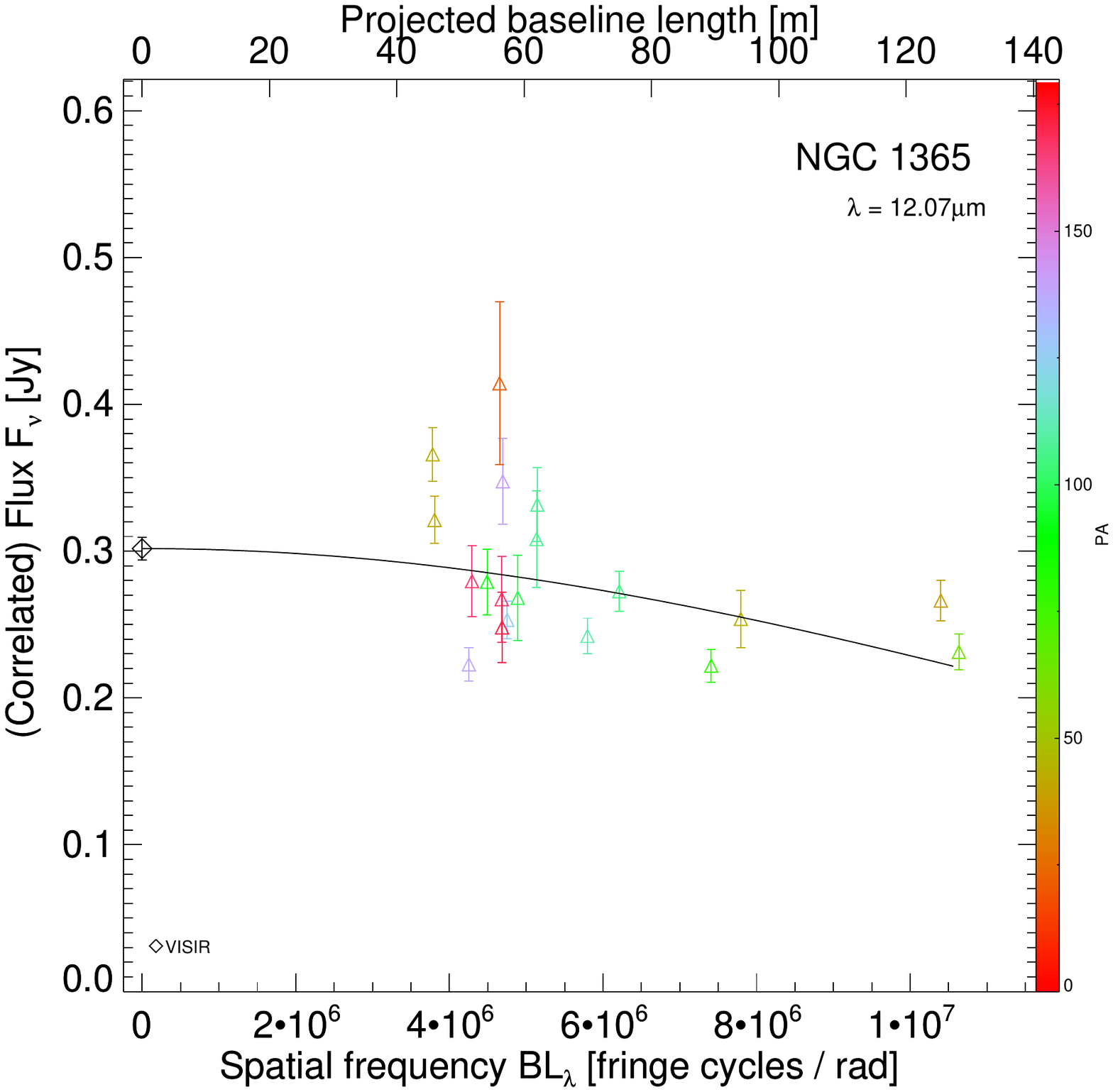}}
	~~~~~~~~~
	\subfloat{\includegraphics[trim=3cm 0cm 3cm 0cm, width=0.5\hsize]{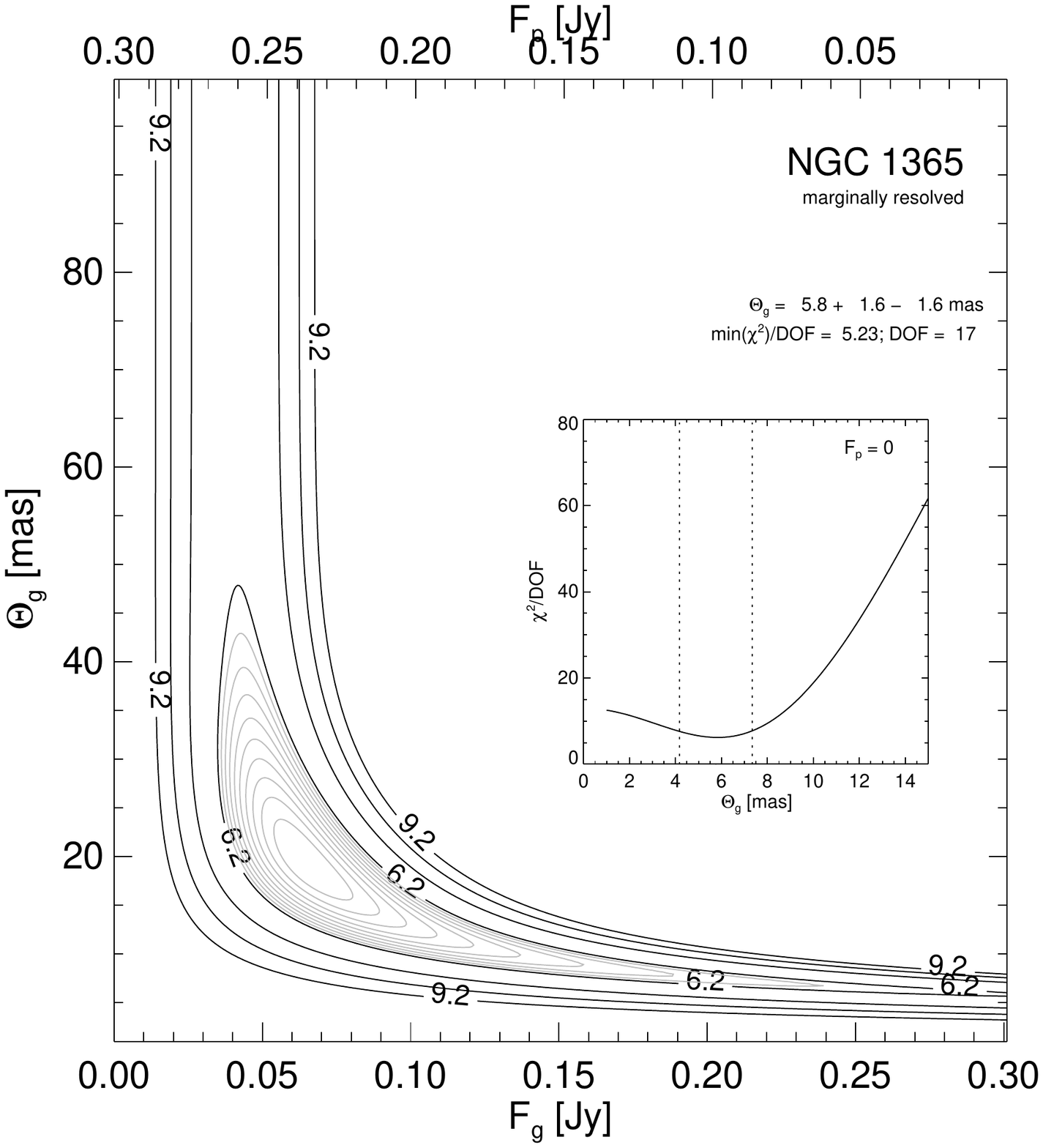}}
	\caption{\label{fig:rad:NGC1365}The same as Fig. \ref{fig:rad:IZwicky1} but for NGC~1365}
\end{figure*}

\begin{figure*}
	\centering
	\subfloat{\includegraphics[trim=7cm 4cm 7cm 4cm, width=0.5\hsize]{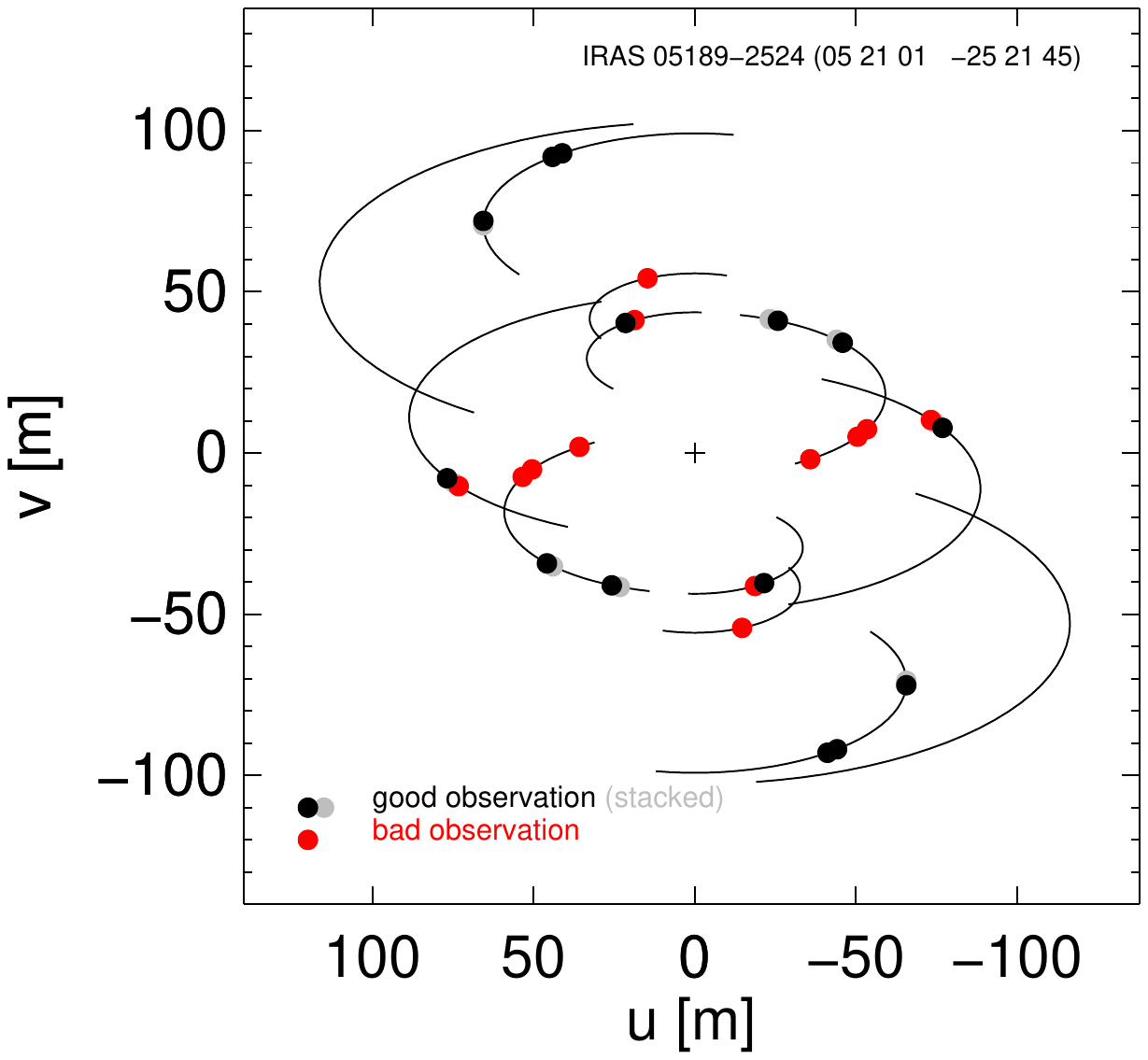}}
	~~~~~~~~~
	\subfloat{\includegraphics[trim=7cm 4cm 7cm 4cm, width=0.5\hsize]{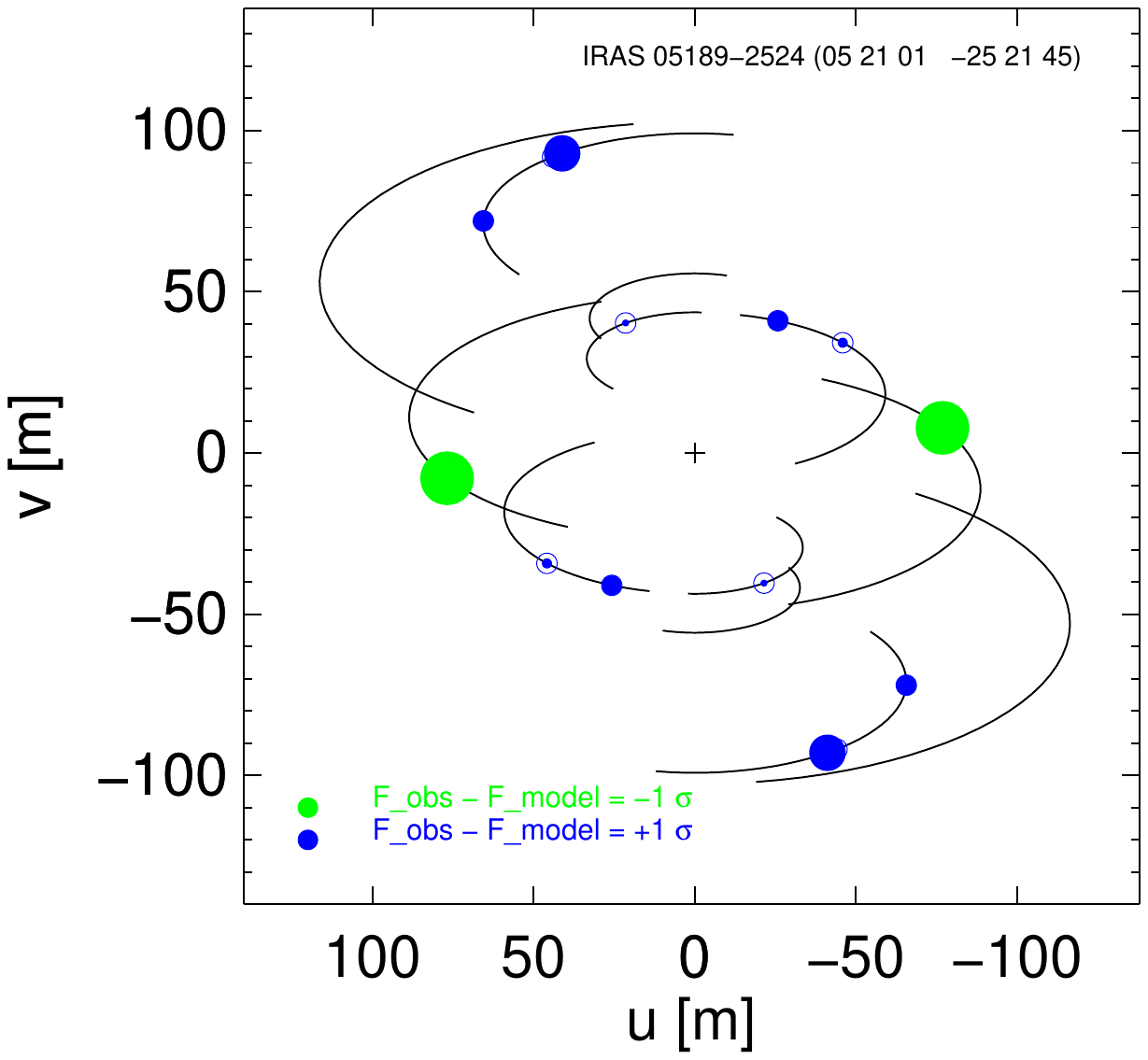}}\\
	\subfloat{\includegraphics[trim=3cm 0cm 3cm 0cm, width=0.5\hsize]{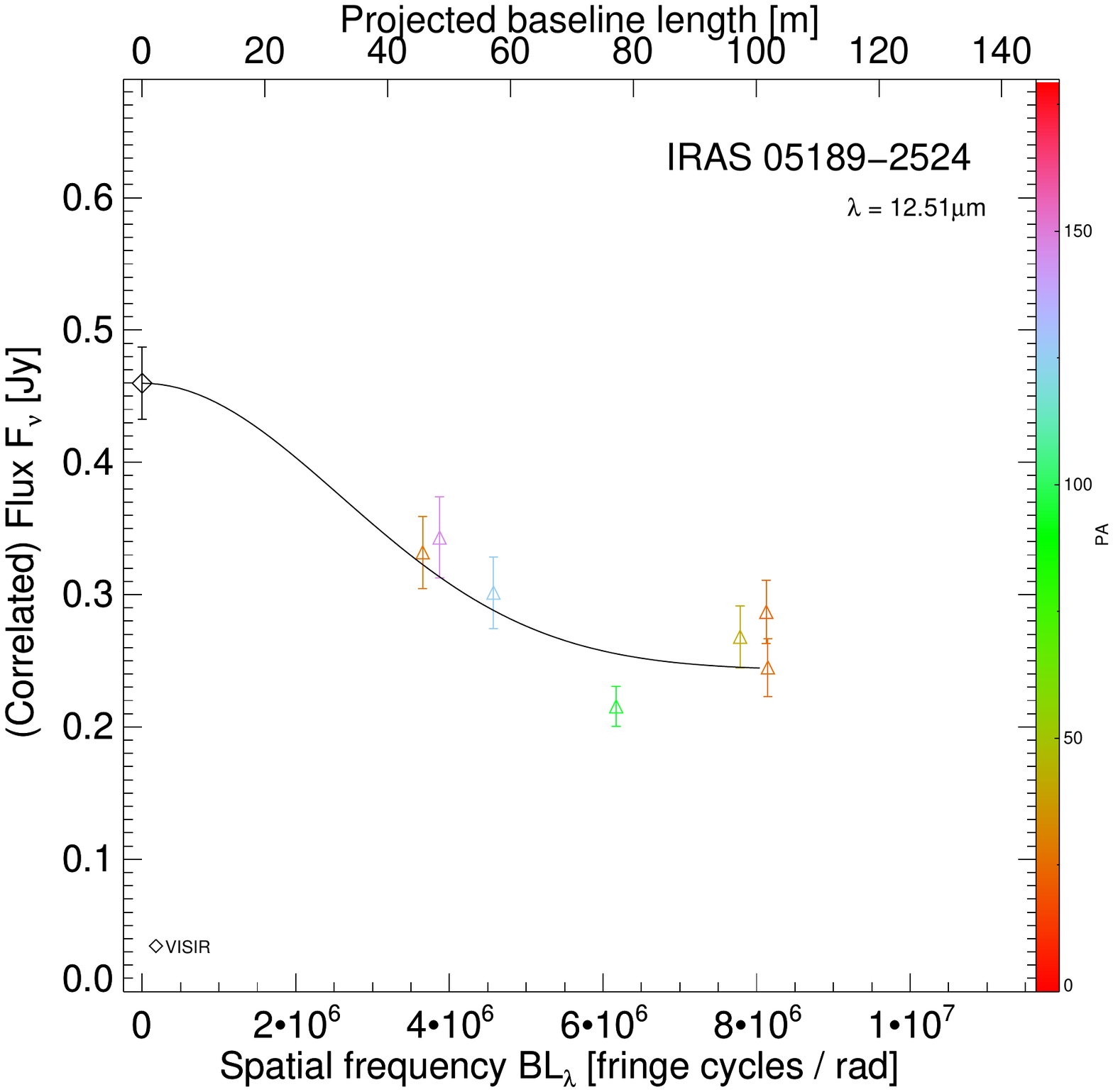}}
	~~~~~~~~~
	\subfloat{\includegraphics[trim=3cm 0cm 3cm 0cm, width=0.5\hsize]{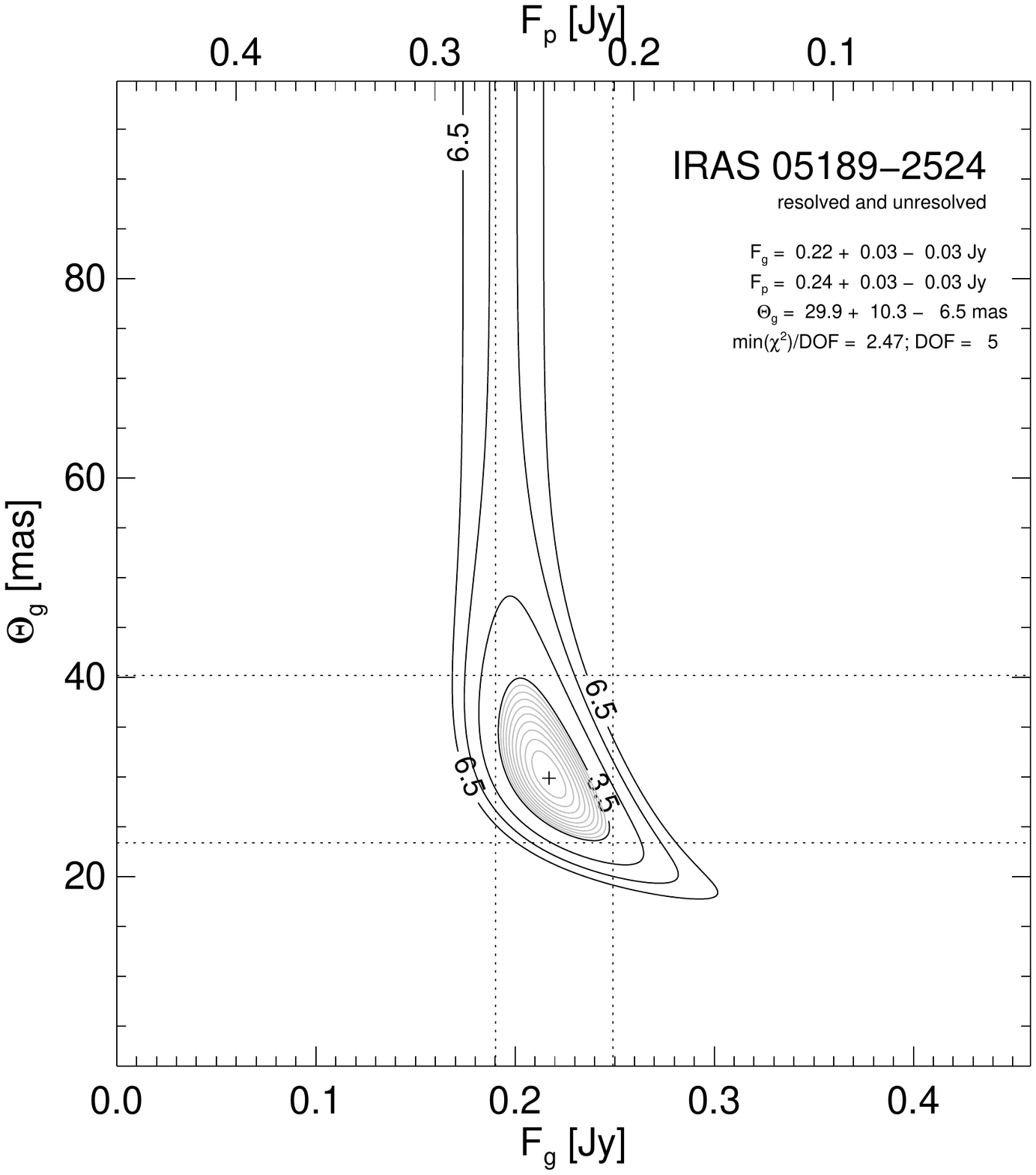}}
	\caption{\label{fig:rad:LEDA17155}The same as Fig. \ref{fig:rad:IZwicky1} but for IRAS~05189-2524}
\end{figure*}

\begin{figure*}
	\centering
	\subfloat{\includegraphics[trim=7cm 4cm 7cm 4cm, width=0.5\hsize]{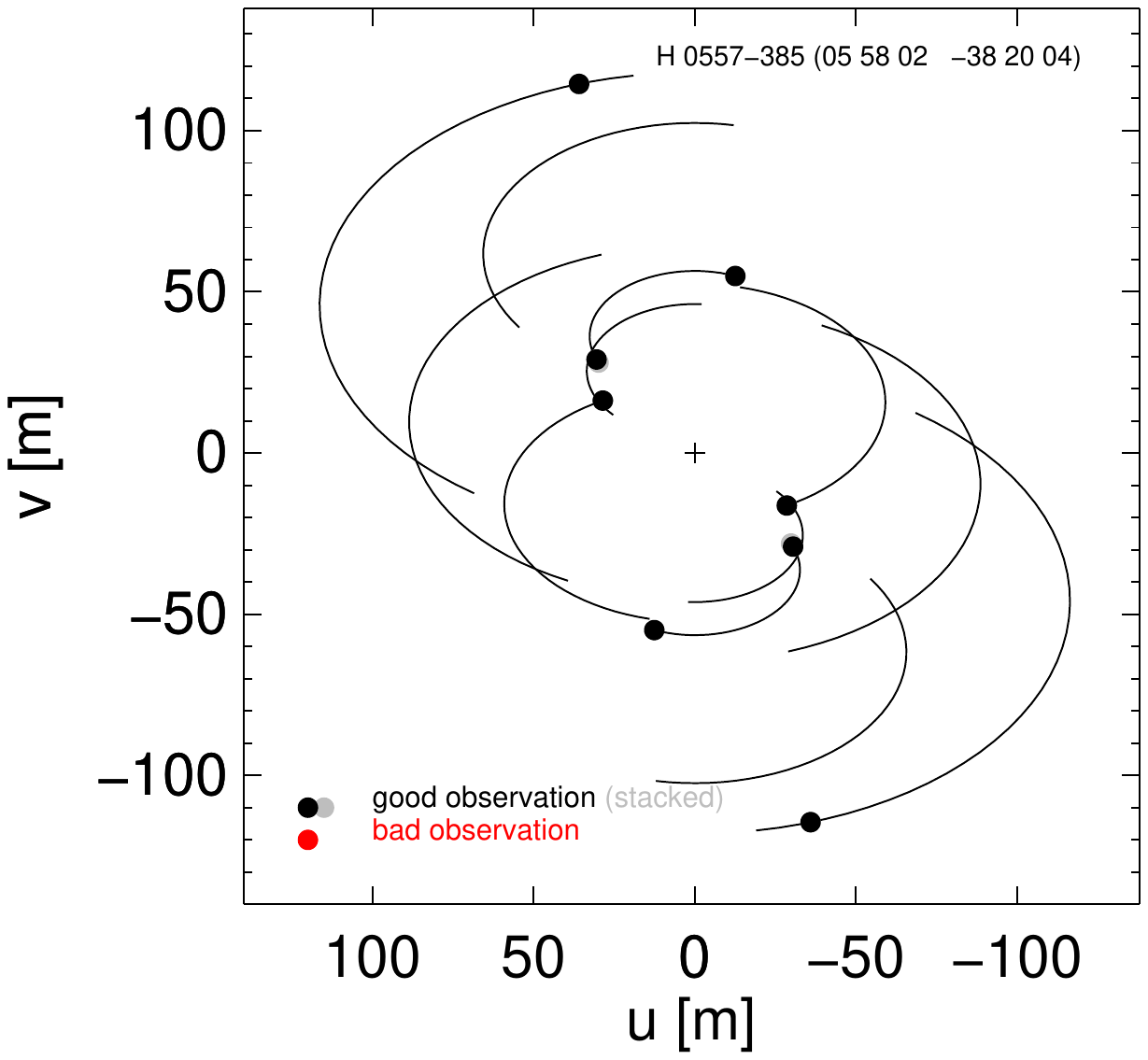}}
	~~~~~~~~~
	\subfloat{\includegraphics[trim=7cm 4cm 7cm 4cm, width=0.5\hsize]{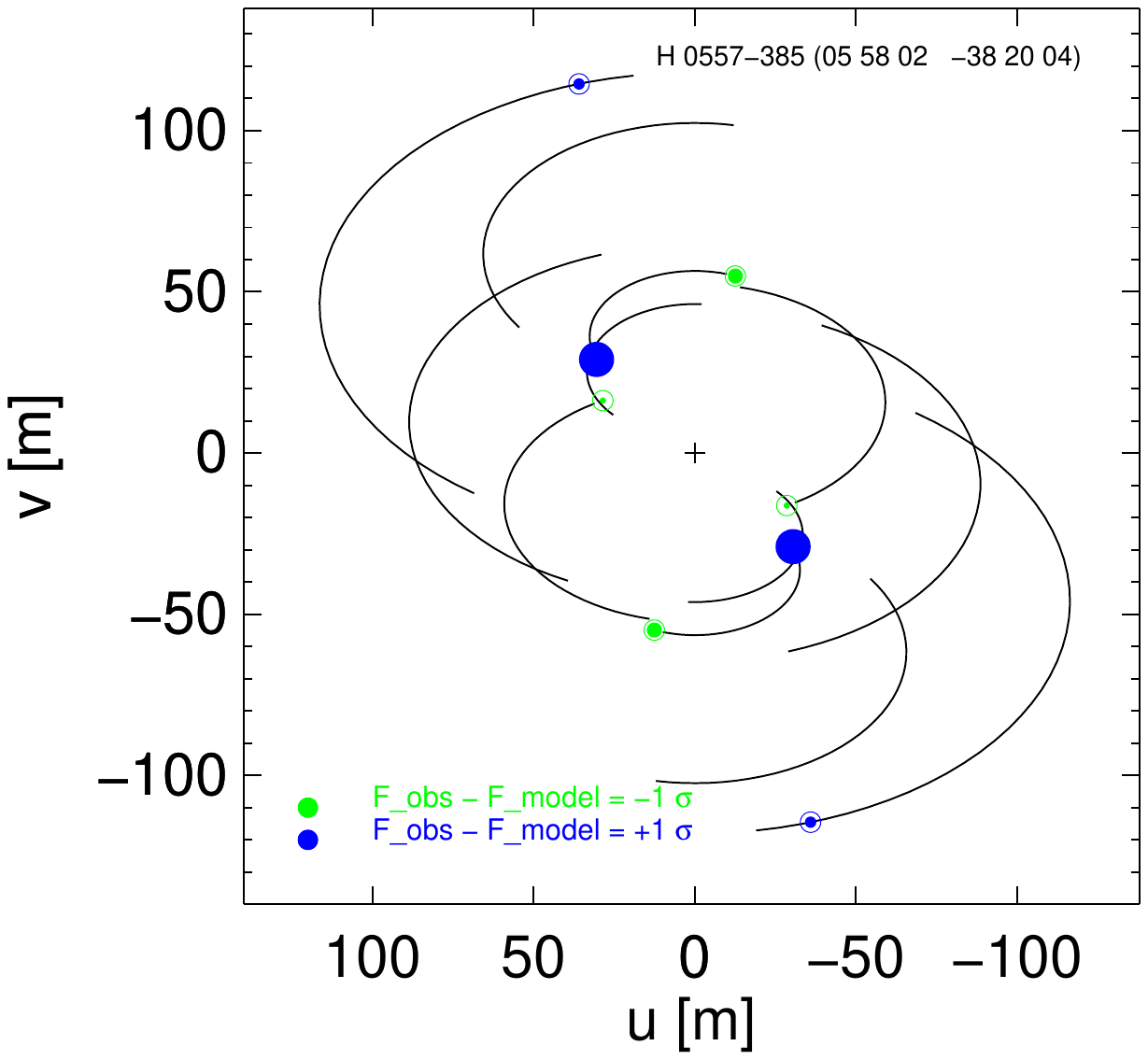}}\\
	\subfloat{\includegraphics[trim=3cm 0cm 3cm 0cm, width=0.5\hsize]{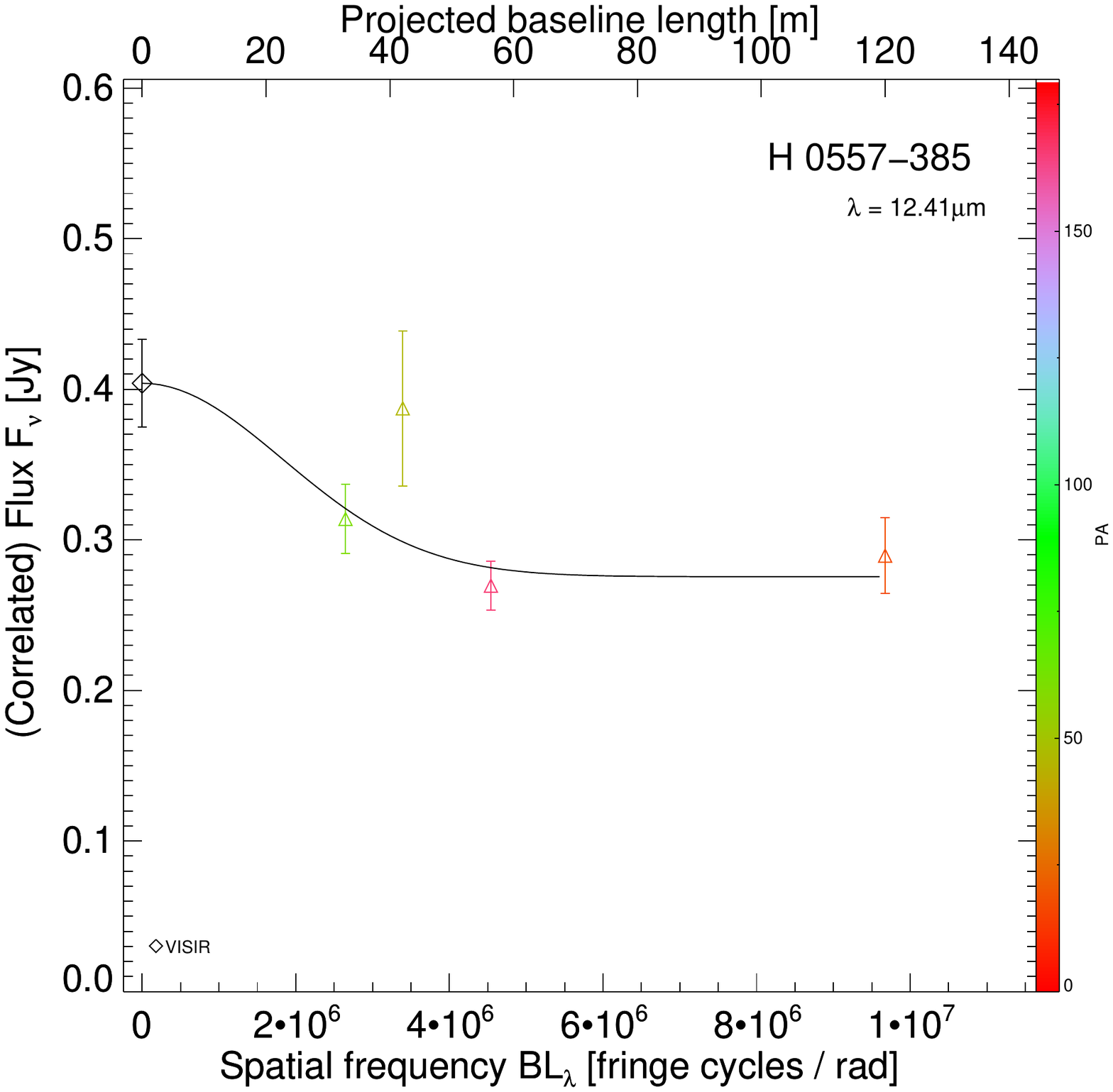}}
	~~~~~~~~~
	\subfloat{\includegraphics[trim=3cm 0cm 3cm 0cm, width=0.5\hsize]{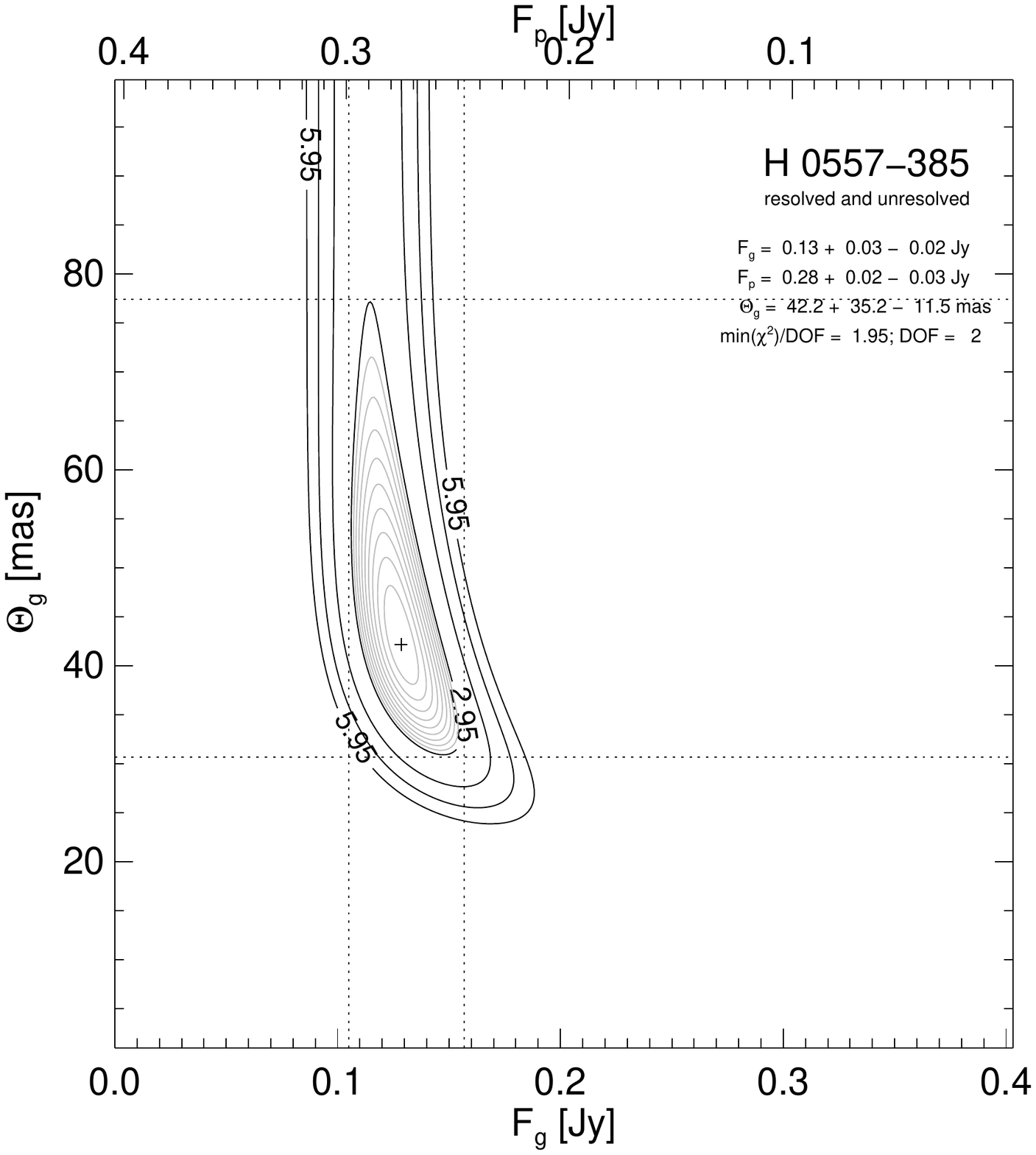}}
	\caption{\label{fig:rad:H0557-385}The same as Fig. \ref{fig:rad:IZwicky1} but for H0557-385}
\end{figure*}

\begin{figure*}
	\centering
	\subfloat{\includegraphics[trim=7cm 4cm 7cm 4cm, width=0.5\hsize]{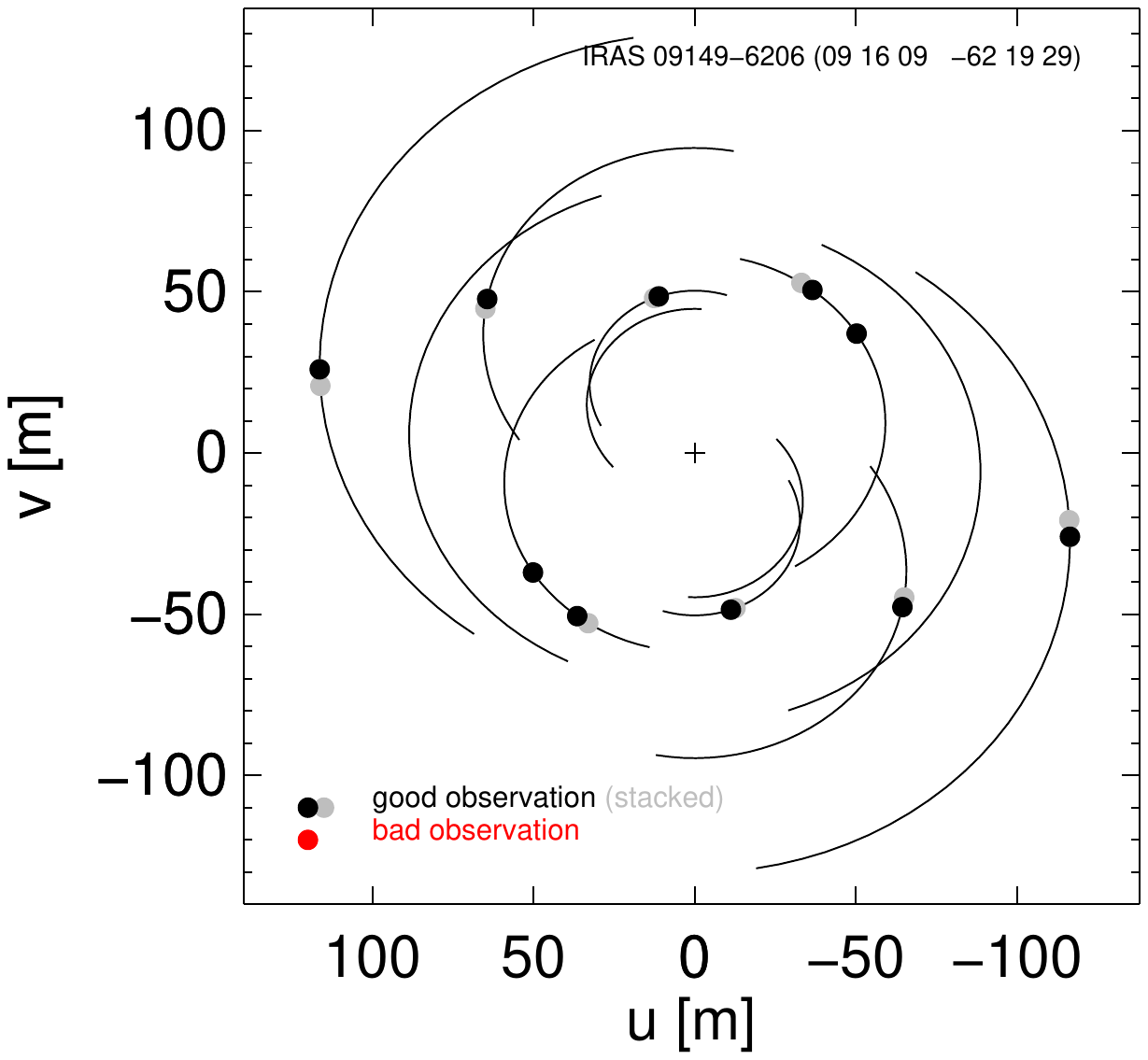}}
	~~~~~~~~~
	\subfloat{\includegraphics[trim=7cm 4cm 7cm 4cm, width=0.5\hsize]{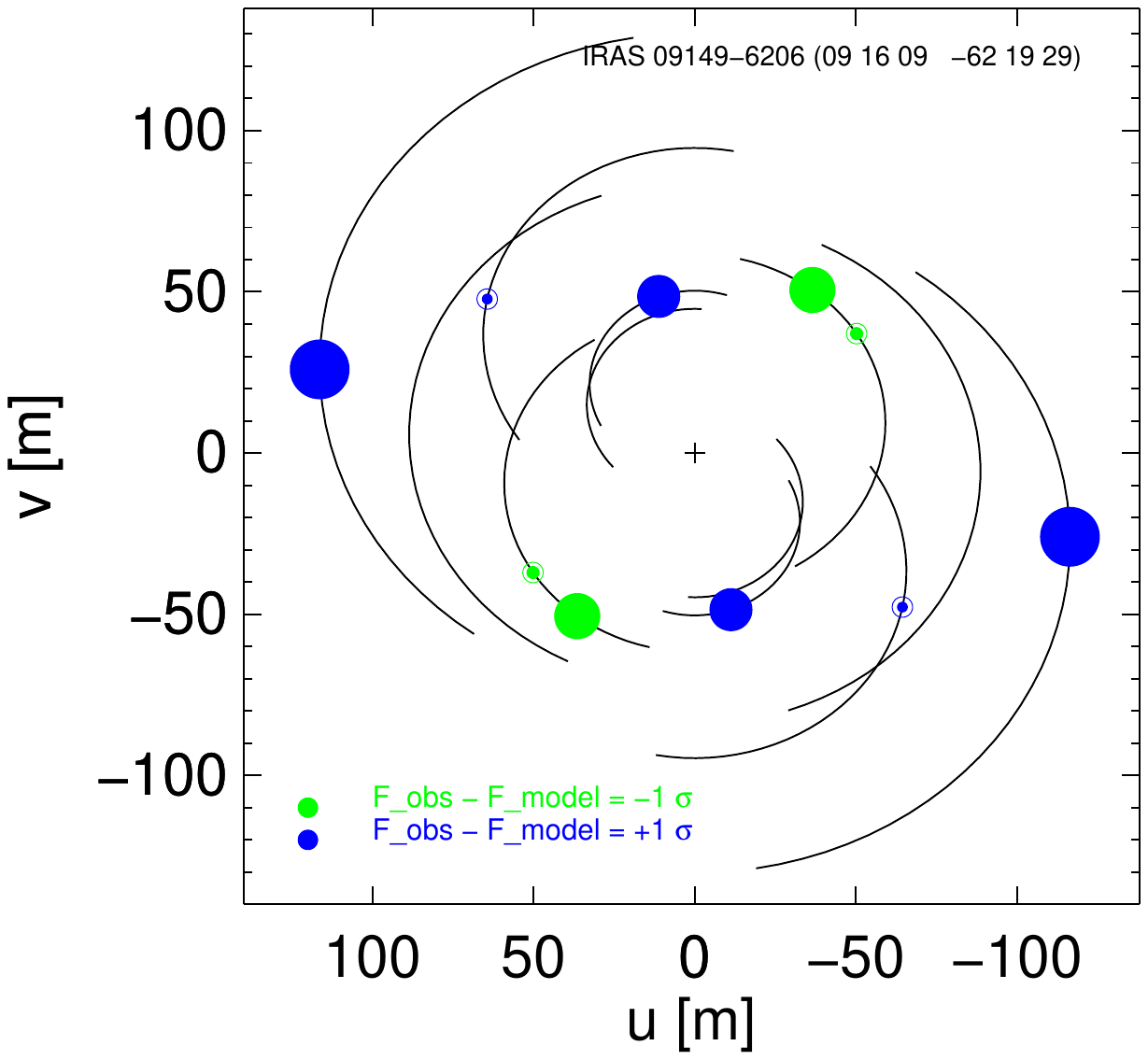}}\\
	\subfloat{\includegraphics[trim=3cm 0cm 3cm 0cm, width=0.5\hsize]{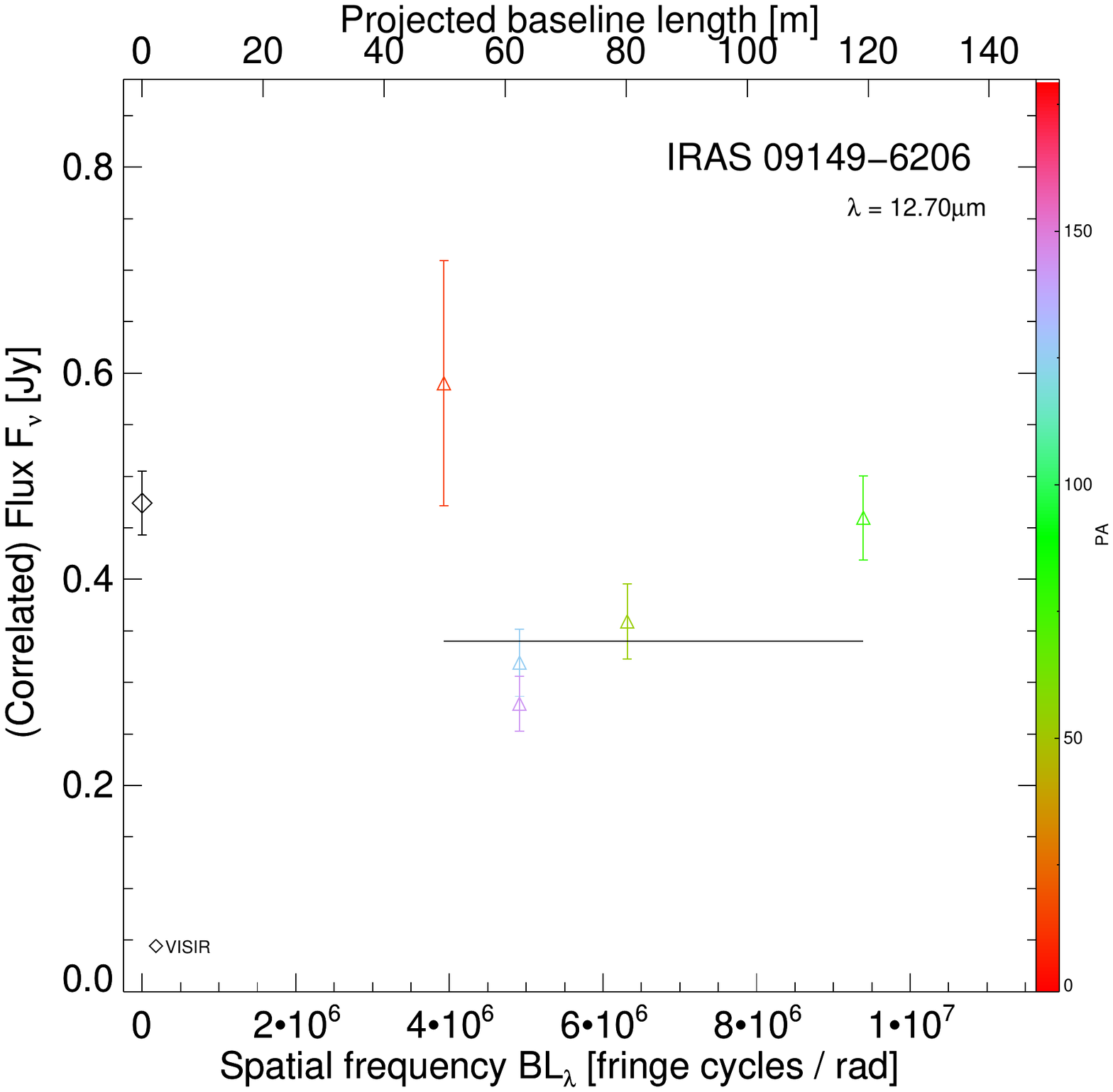}}
	~~~~~~~~~
	\subfloat{\includegraphics[trim=3cm 0cm 3cm 0cm, width=0.5\hsize]{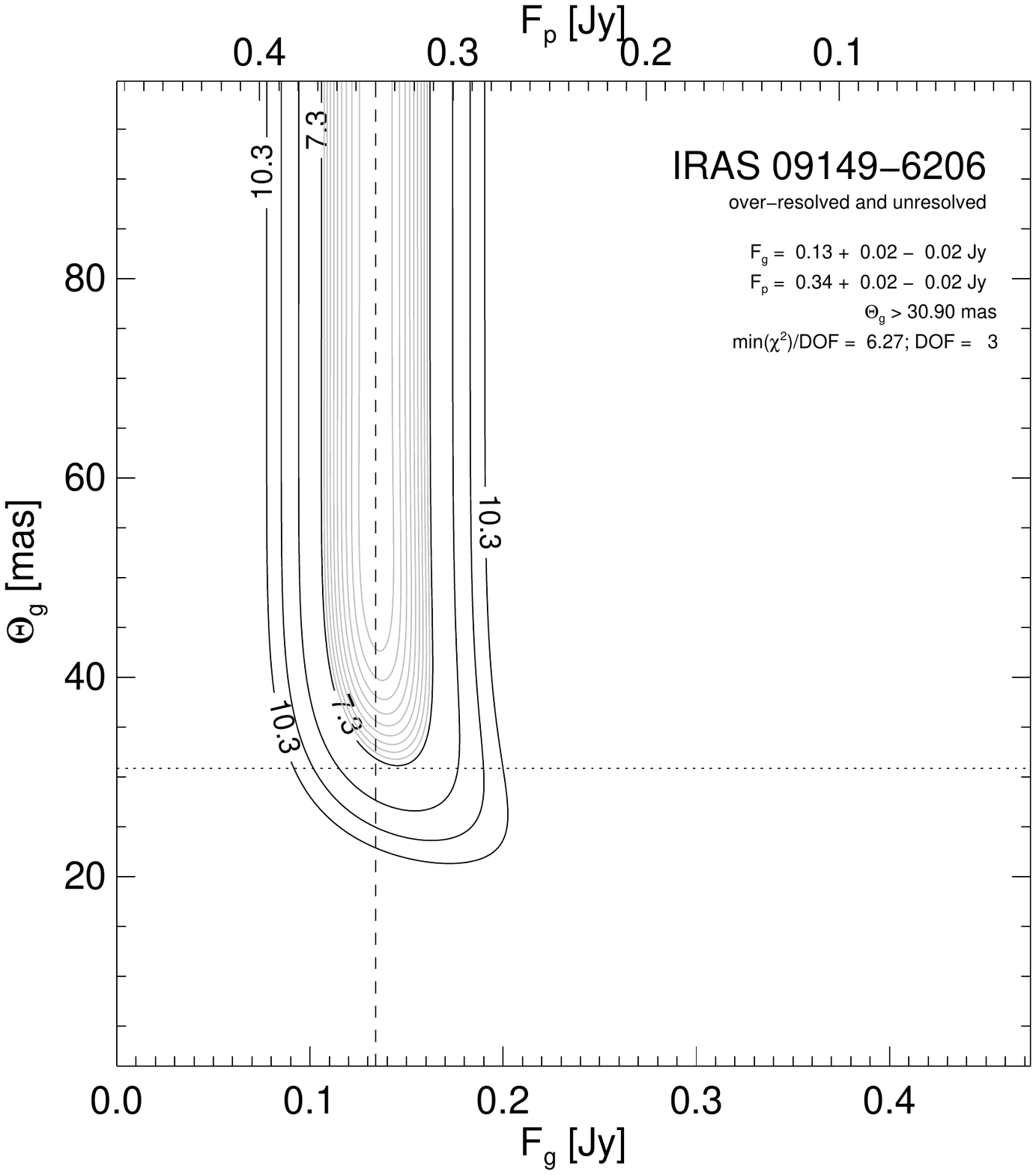}}
	\caption{\label{fig:rad:IRAS09149-6206}The same as Fig. \ref{fig:rad:IZwicky1} but for IRAS~09149-6206. Since the $\min(\chi_r^2)+1$ contour is open to large values of $\Theta_g$, we consider this source to consist of an over-resolved and an unresolved source and determine the model fit parameter $f_p$ (and its derived values $F_g$ and $F_p$) for the limit $\Theta_g \rightarrow \infty$. This is equivalent to computing $F_p$ from the weighted average of all correlated fluxes and $F_g$ from $F_g = F_{\nu, \rm tot} - F_p$. The flux level of $F_p$ is indicated in the radial plot (bottom left) as a straight line. Note that in this radial plots two $(u,v)$ points at nearly the same spatial frequency of $\approx 5 \cdot 10^8$ and similar position angle (color) are on top of each other.}
\end{figure*}

\begin{figure*}
	\centering
	\subfloat{\includegraphics[trim=7cm 4cm 7cm 4cm, width=0.5\hsize]{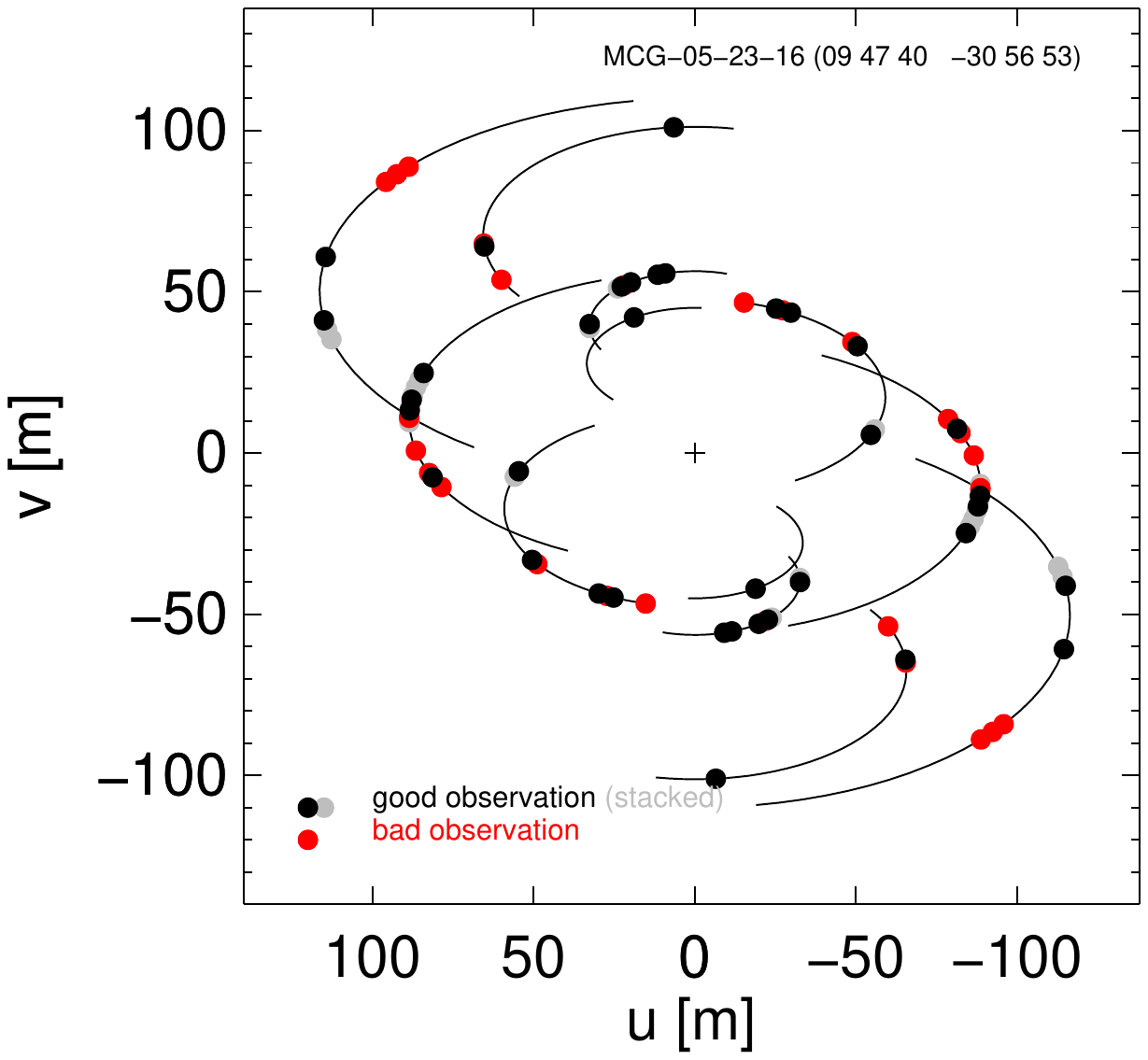}}
	~~~~~~~~~
	\subfloat{\includegraphics[trim=7cm 4cm 7cm 4cm, width=0.5\hsize]{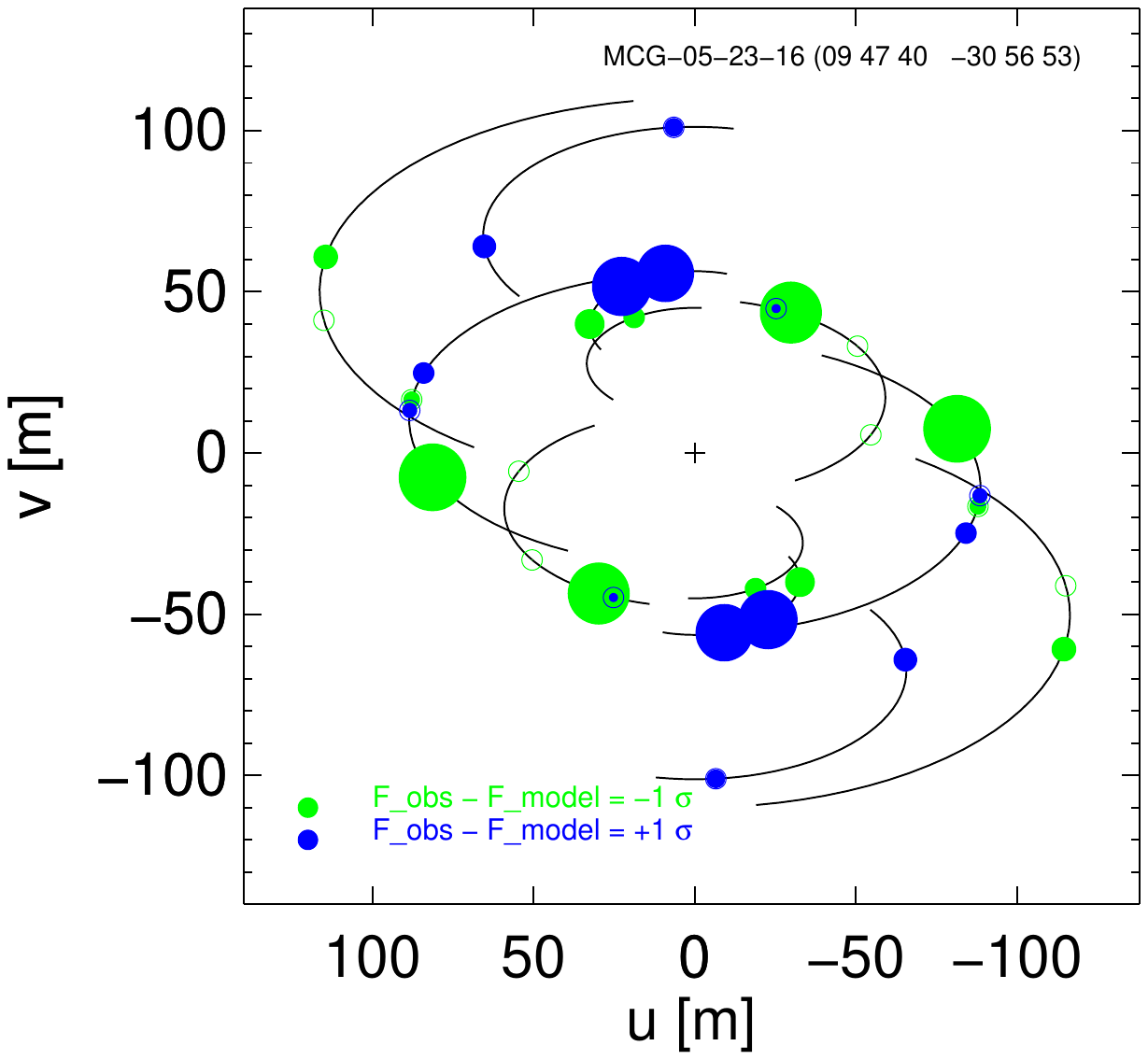}}\\
	\subfloat{\includegraphics[trim=3cm 0cm 3cm 0cm, width=0.5\hsize]{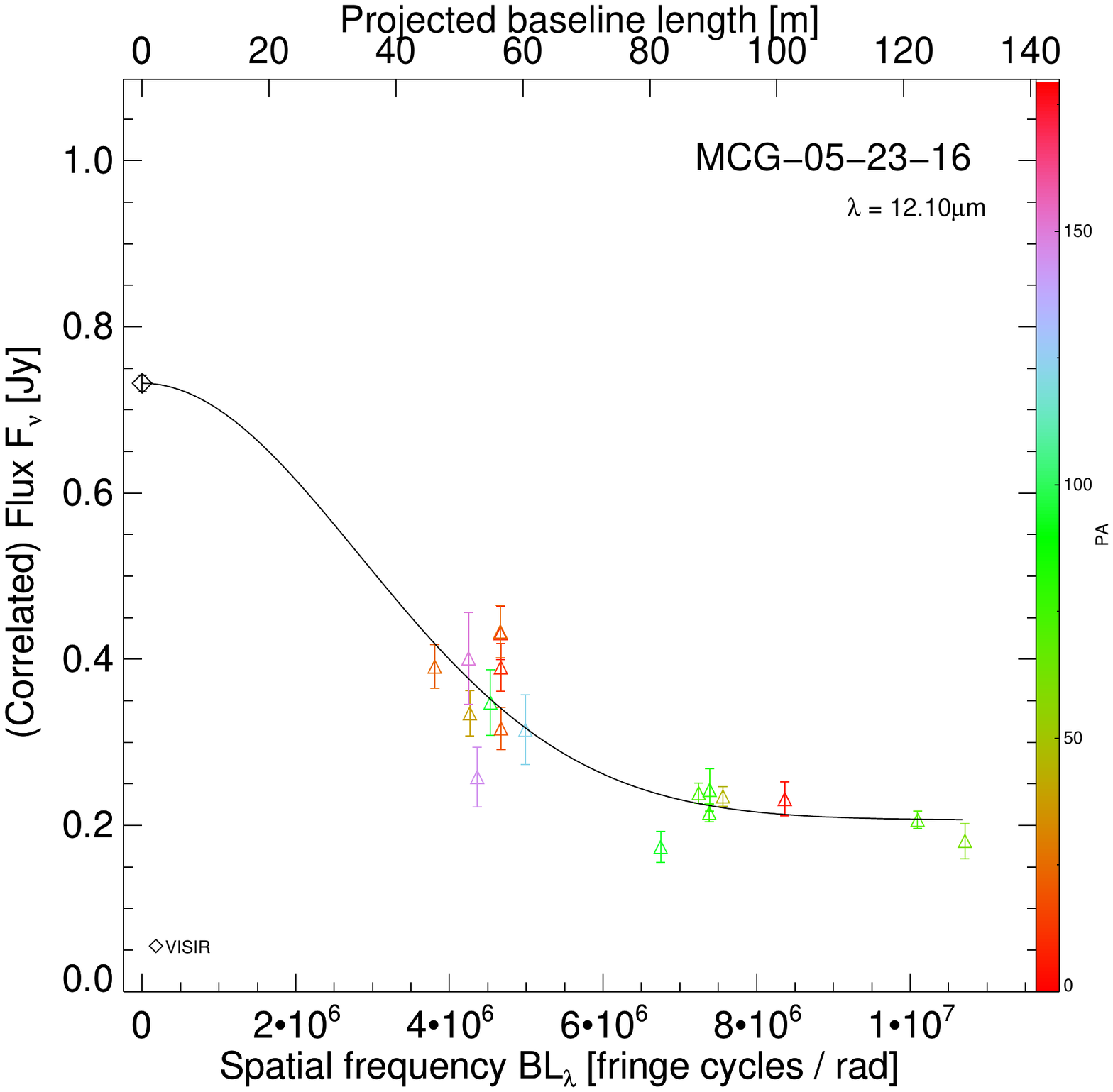}}
	~~~~~~~~~
	\subfloat{\includegraphics[trim=3cm 0cm 3cm 0cm, width=0.5\hsize]{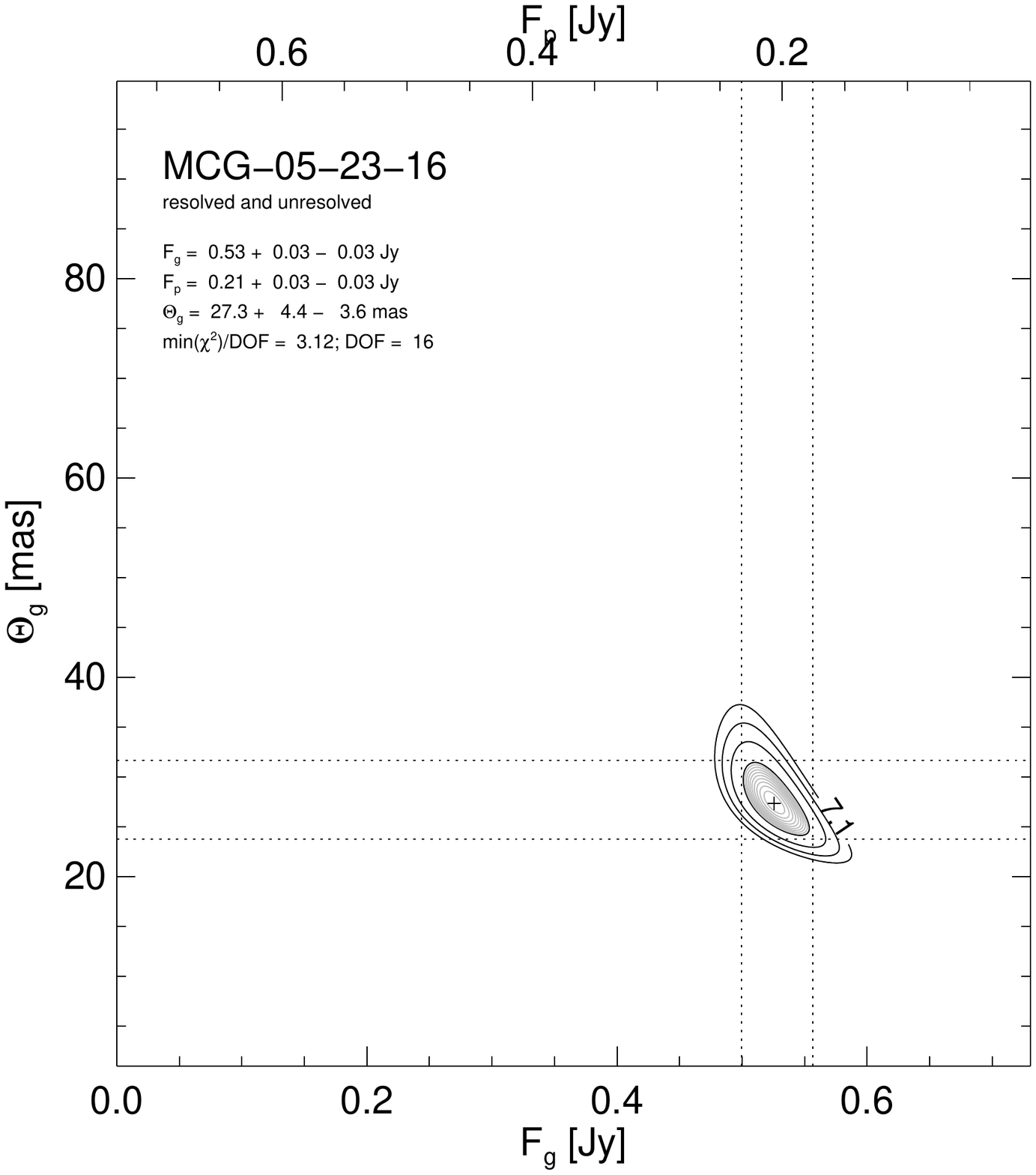}}
	\caption{\label{fig:rad:MCG-5-23-16}The same as Fig. \ref{fig:rad:IZwicky1} but for MCG-05-23-16}
\end{figure*}

\begin{figure*}
	\centering
	\subfloat{\includegraphics[trim=7cm 4cm 7cm 4cm, width=0.5\hsize]{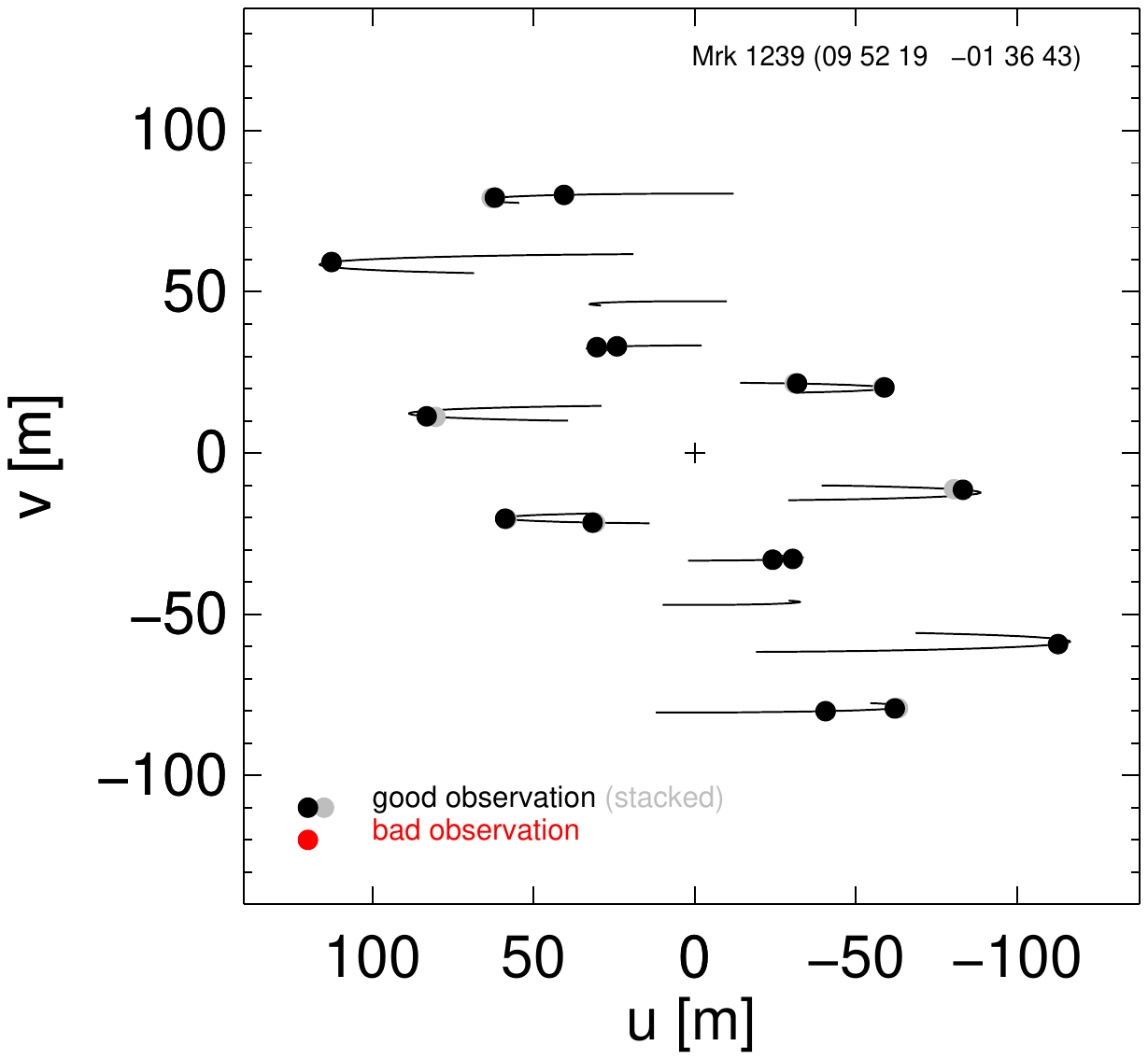}}
	~~~~~~~~~
	\subfloat{\includegraphics[trim=7cm 4cm 7cm 4cm, width=0.5\hsize]{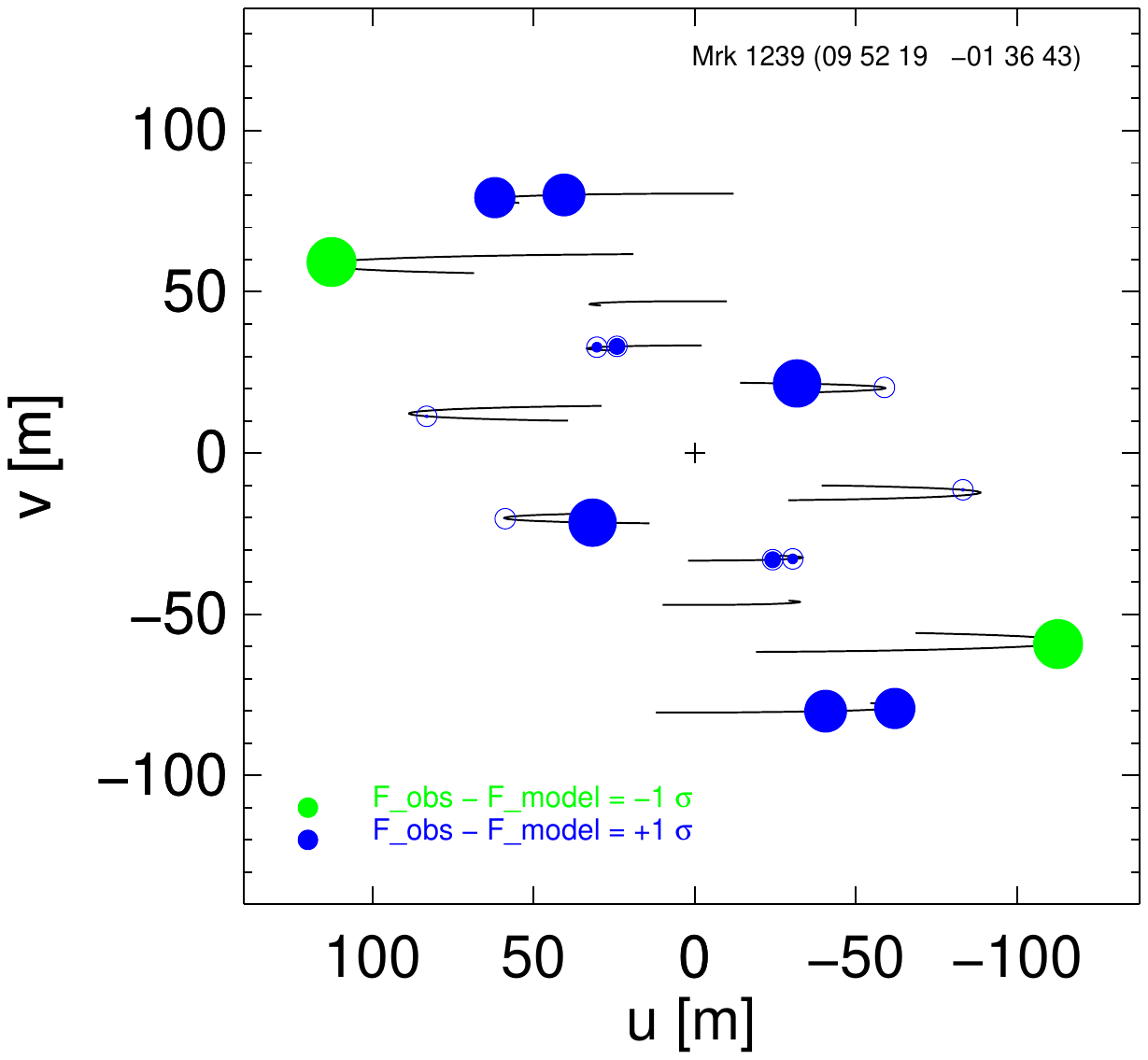}}\\
	\subfloat{\includegraphics[trim=3cm 0cm 3cm 0cm, width=0.5\hsize]{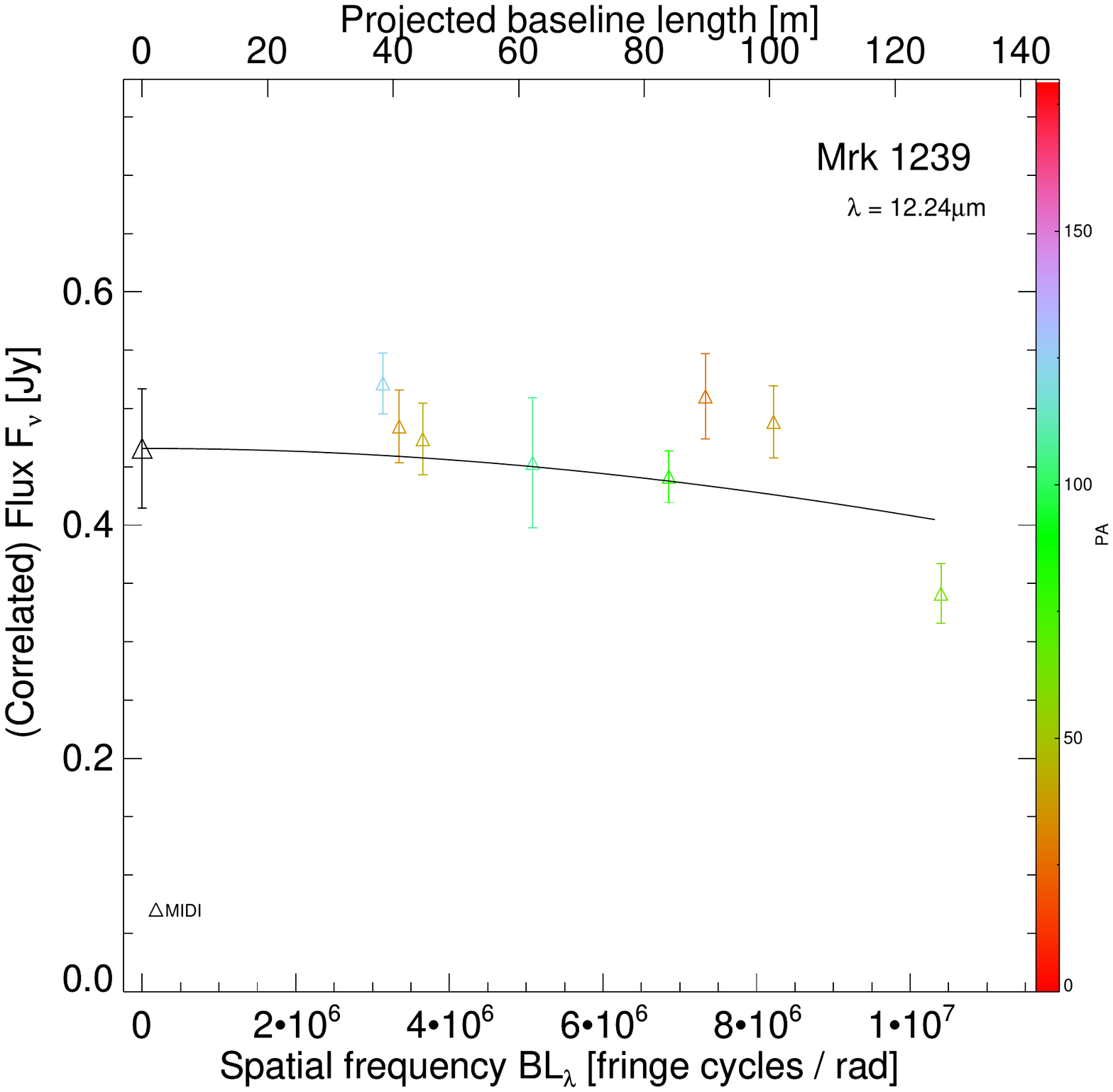}}
	~~~~~~~~~
	\subfloat{\includegraphics[trim=3cm 0cm 3cm 0cm, width=0.5\hsize]{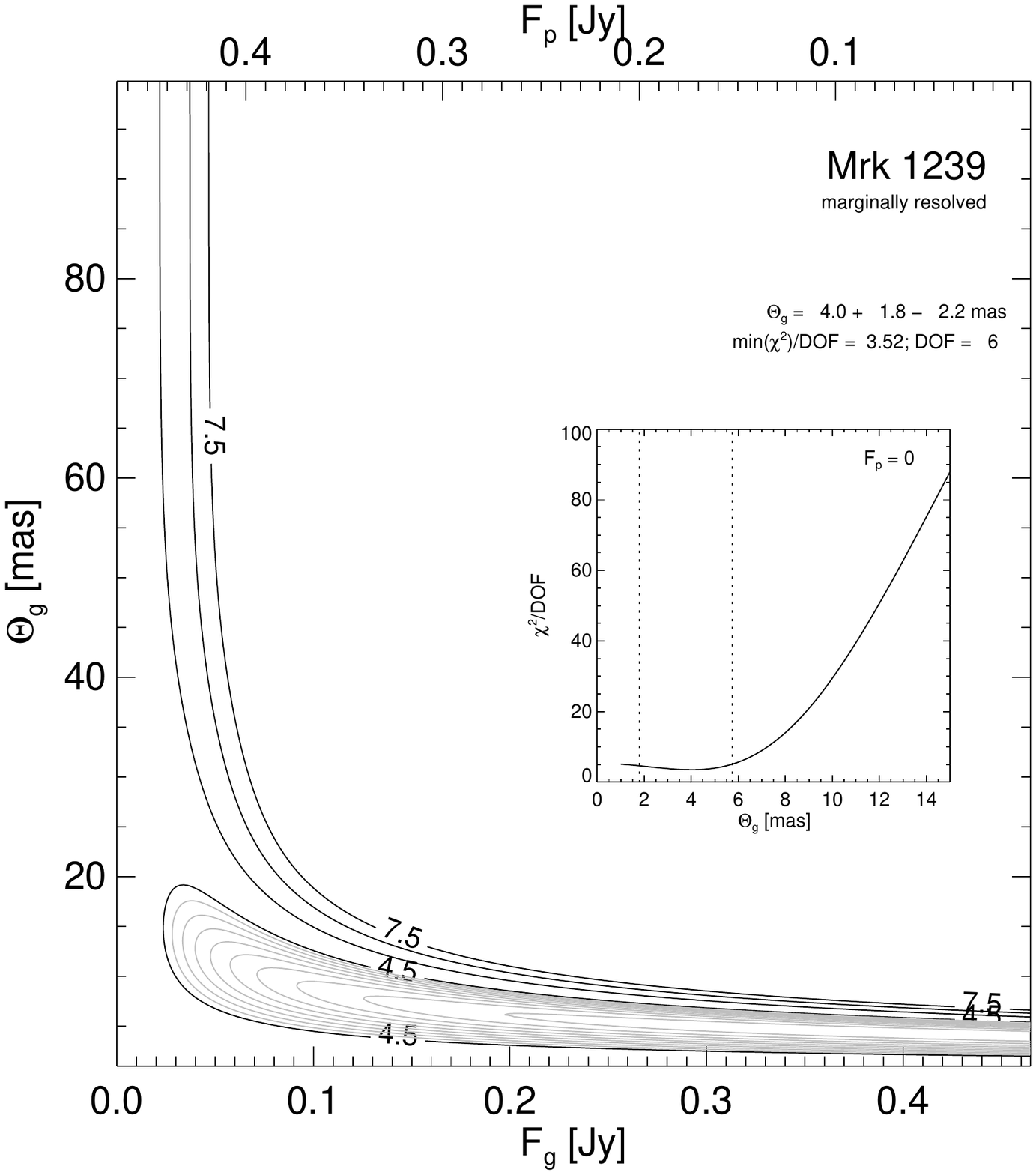}}
	\caption{\label{fig:rad:Mrk1239}The same as Fig. \ref{fig:rad:IZwicky1} but for Mrk~1239}
\end{figure*}

\begin{figure*}
	\centering
	\subfloat{\includegraphics[trim=7cm 4cm 7cm 4cm, width=0.5\hsize]{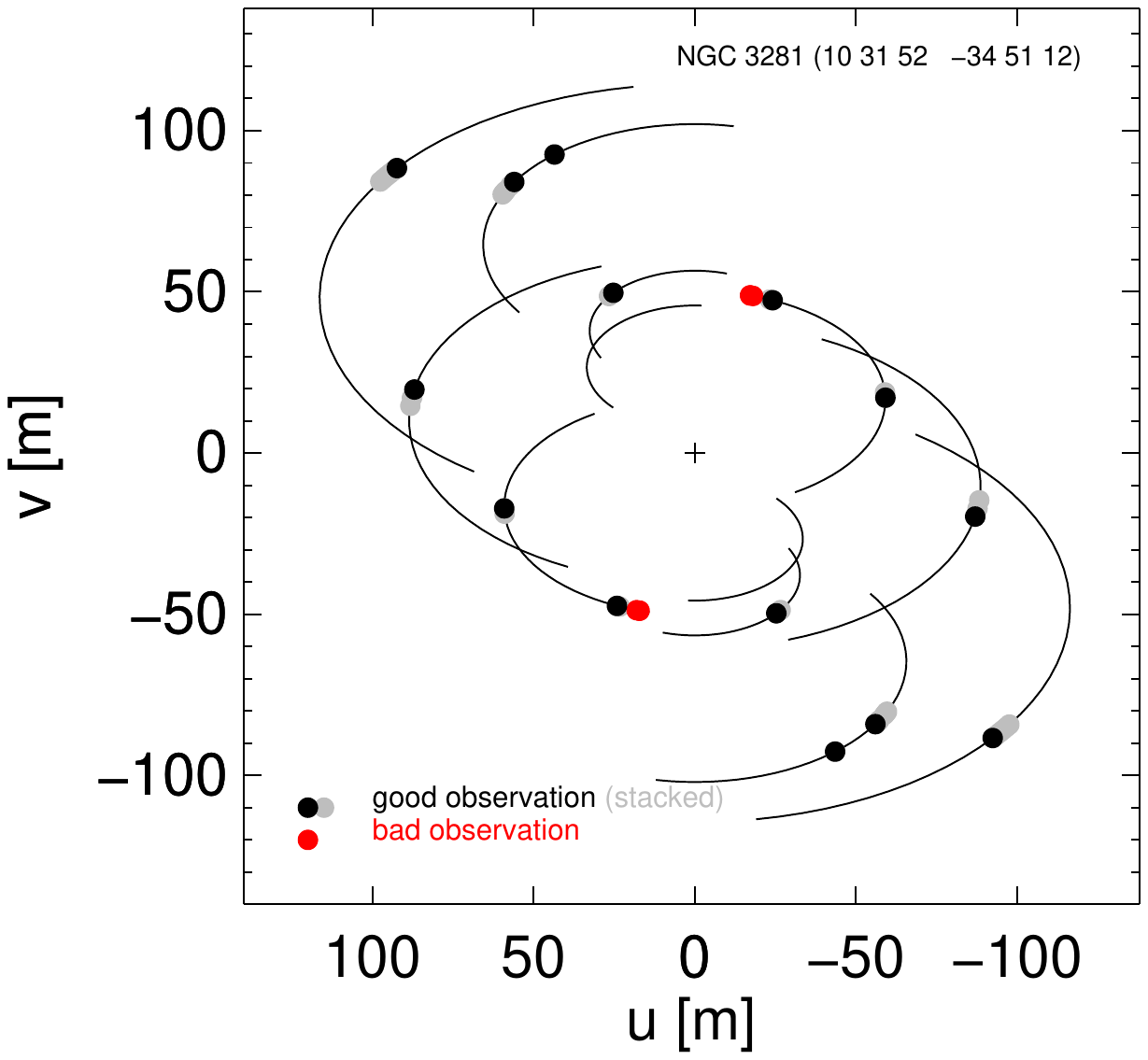}}
	~~~~~~~~~
	\subfloat{\includegraphics[trim=7cm 4cm 7cm 4cm, width=0.5\hsize]{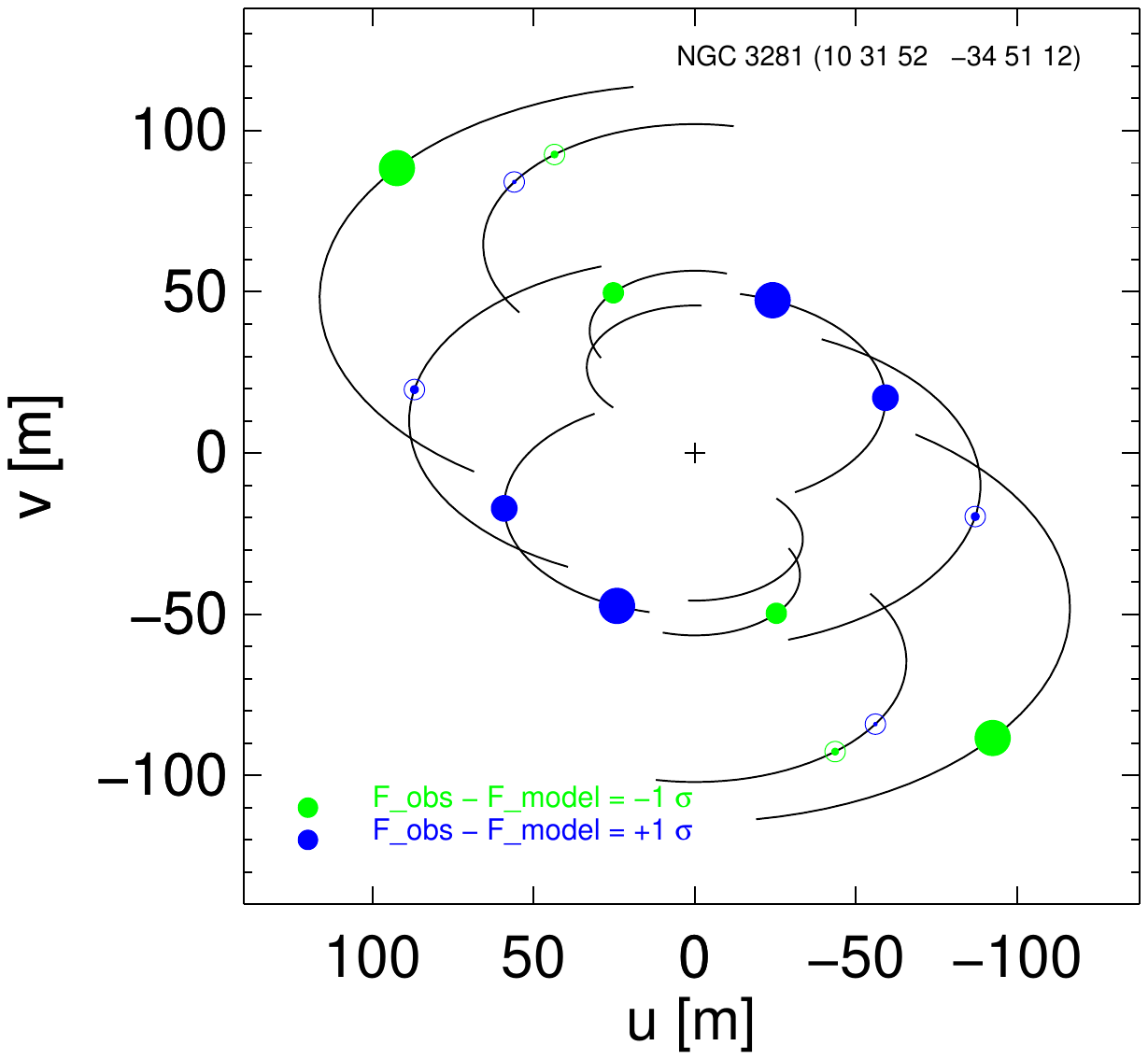}}\\
	\subfloat{\includegraphics[trim=3cm 0cm 3cm 0cm, width=0.5\hsize]{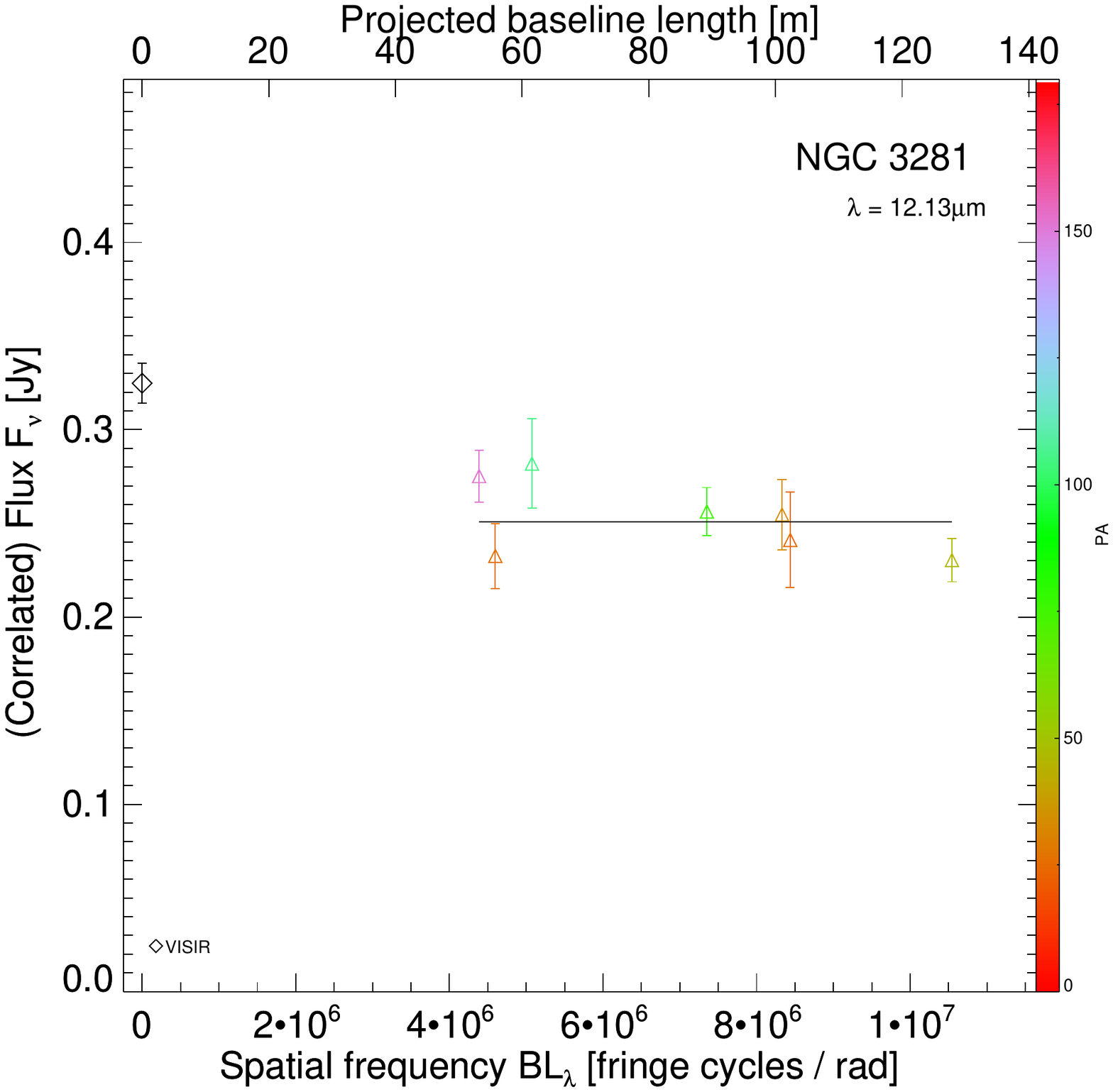}}
	~~~~~~~~~
	\subfloat{\includegraphics[trim=3cm 0cm 3cm 0cm, width=0.5\hsize]{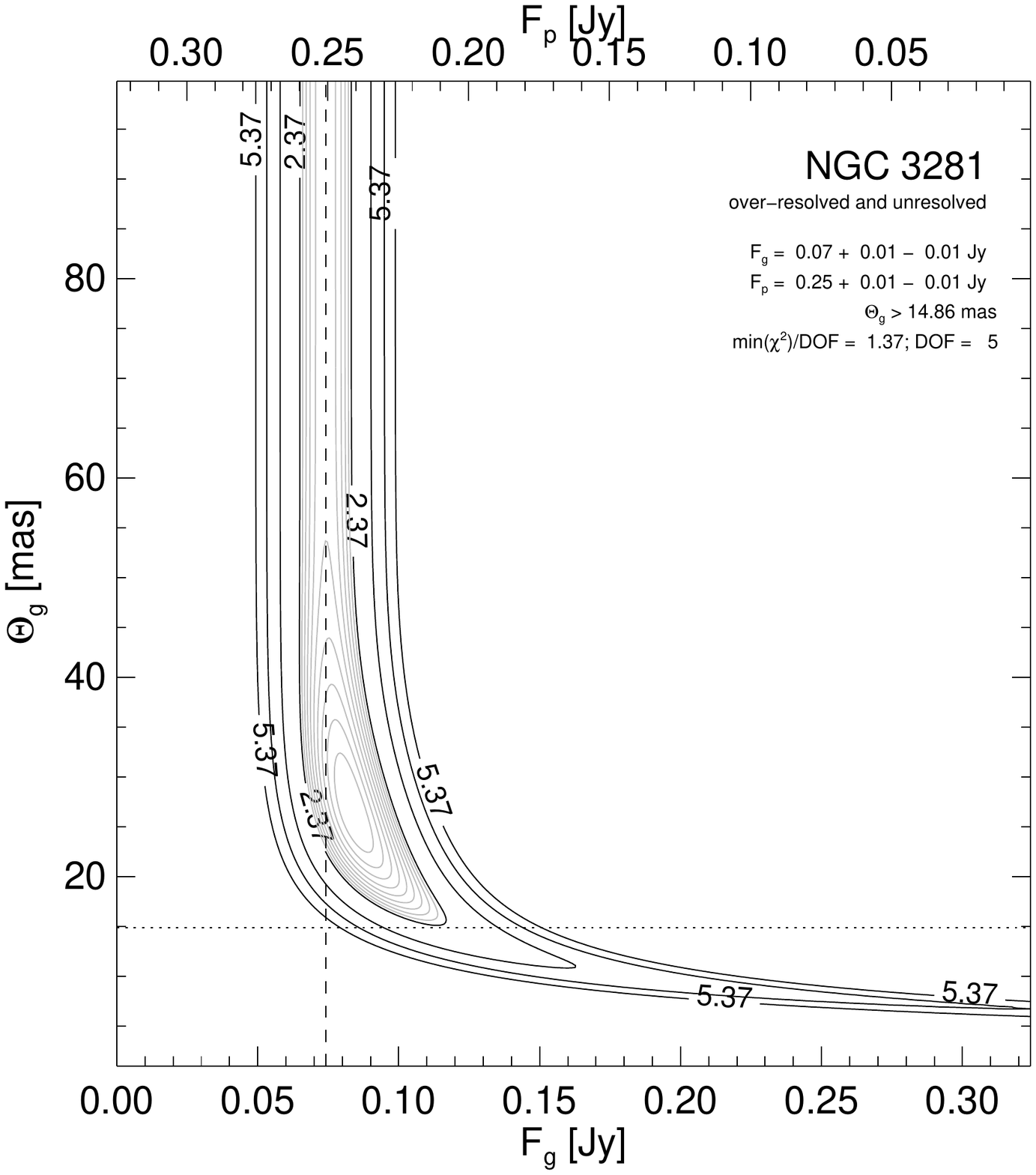}}
	\caption{\label{fig:rad:NGC3281}The same as Fig. \ref{fig:rad:IZwicky1} but for NGC~3281}
\end{figure*}

\begin{figure*}
	\centering
	\subfloat{\includegraphics[trim=7cm 4cm 7cm 4cm, width=0.5\hsize]{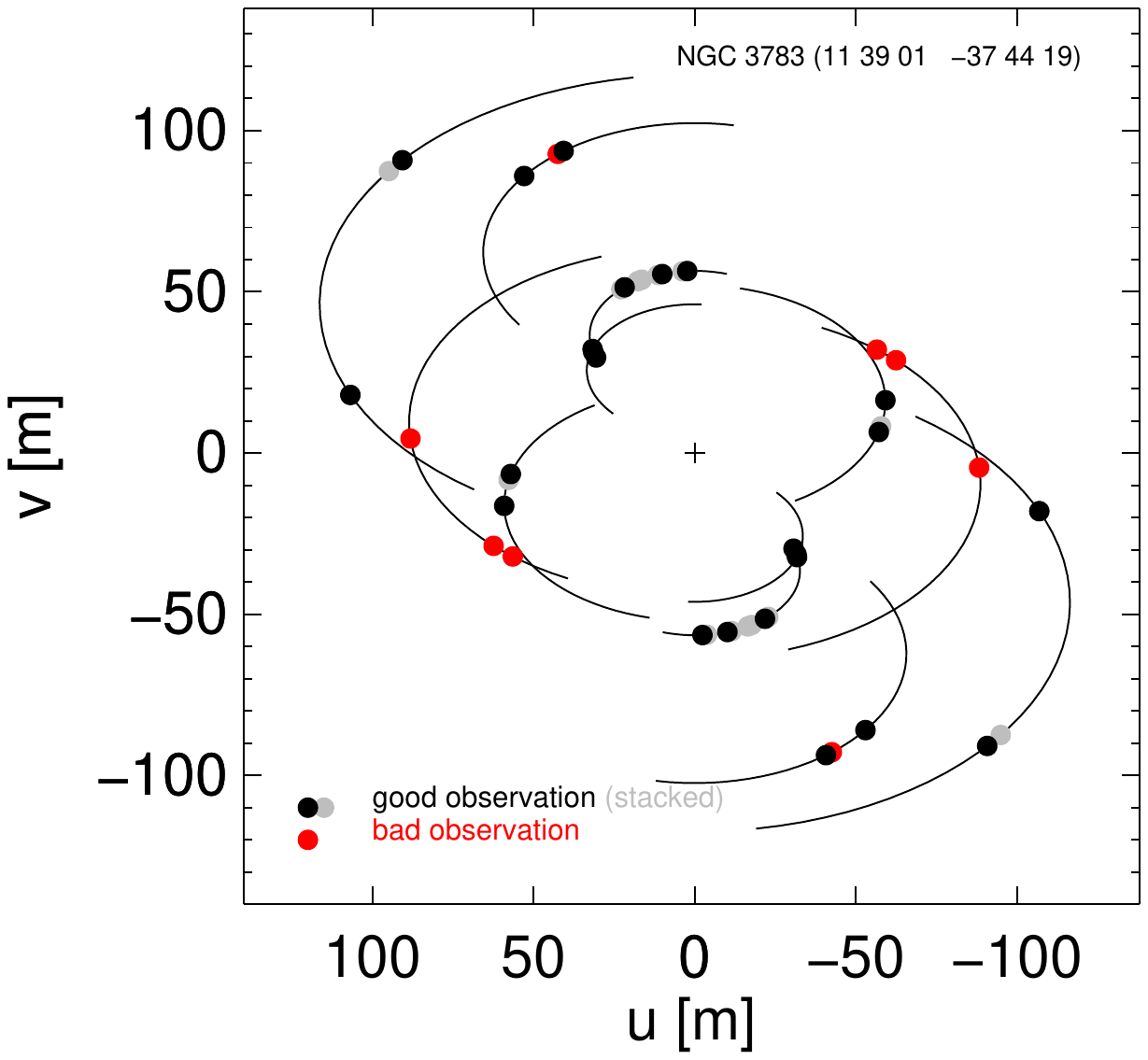}}
	~~~~~~~~~
	\subfloat{\includegraphics[trim=7cm 4cm 7cm 4cm, width=0.5\hsize]{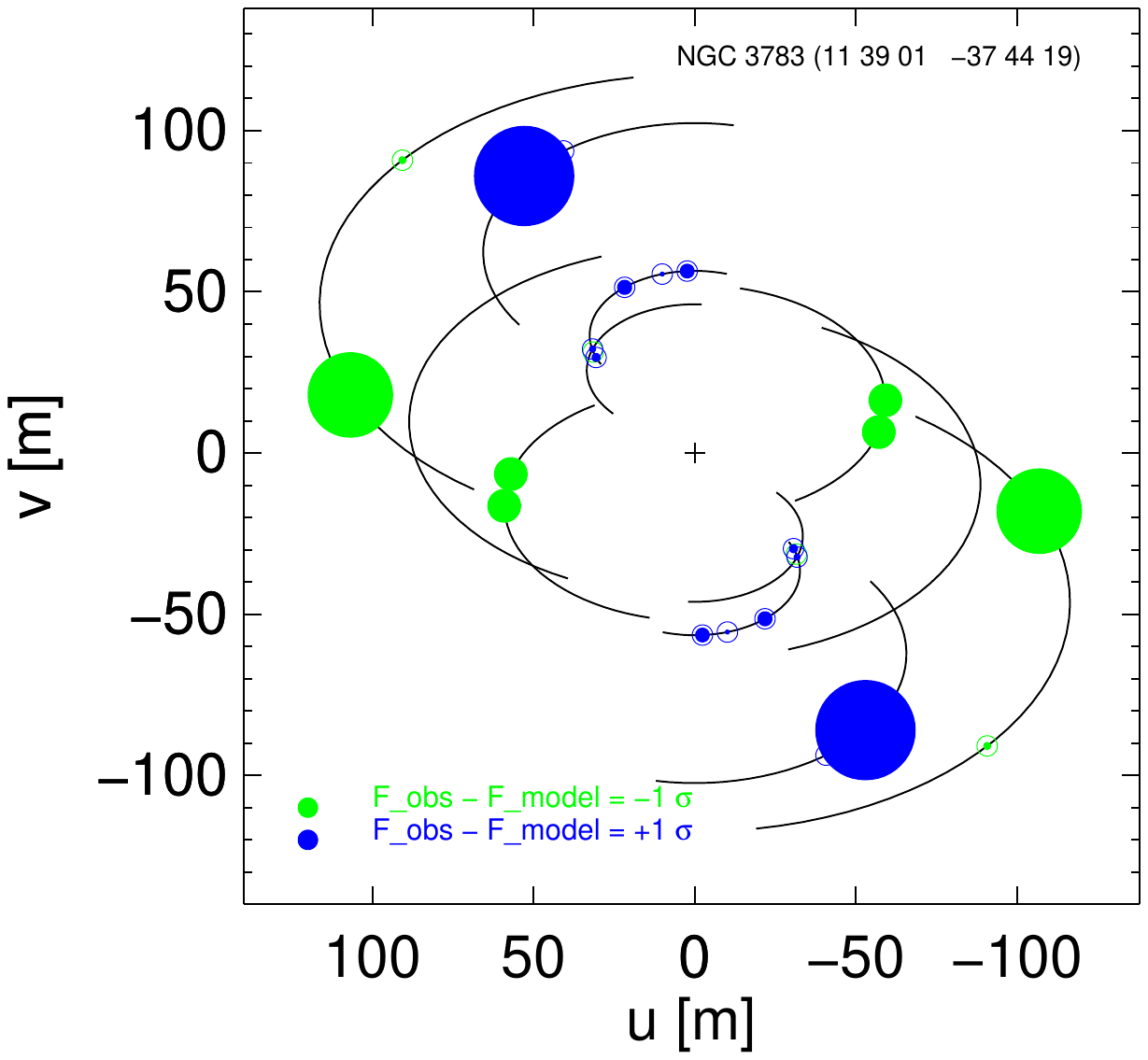}}\\
	\subfloat{\includegraphics[trim=3cm 0cm 3cm 0cm, width=0.5\hsize]{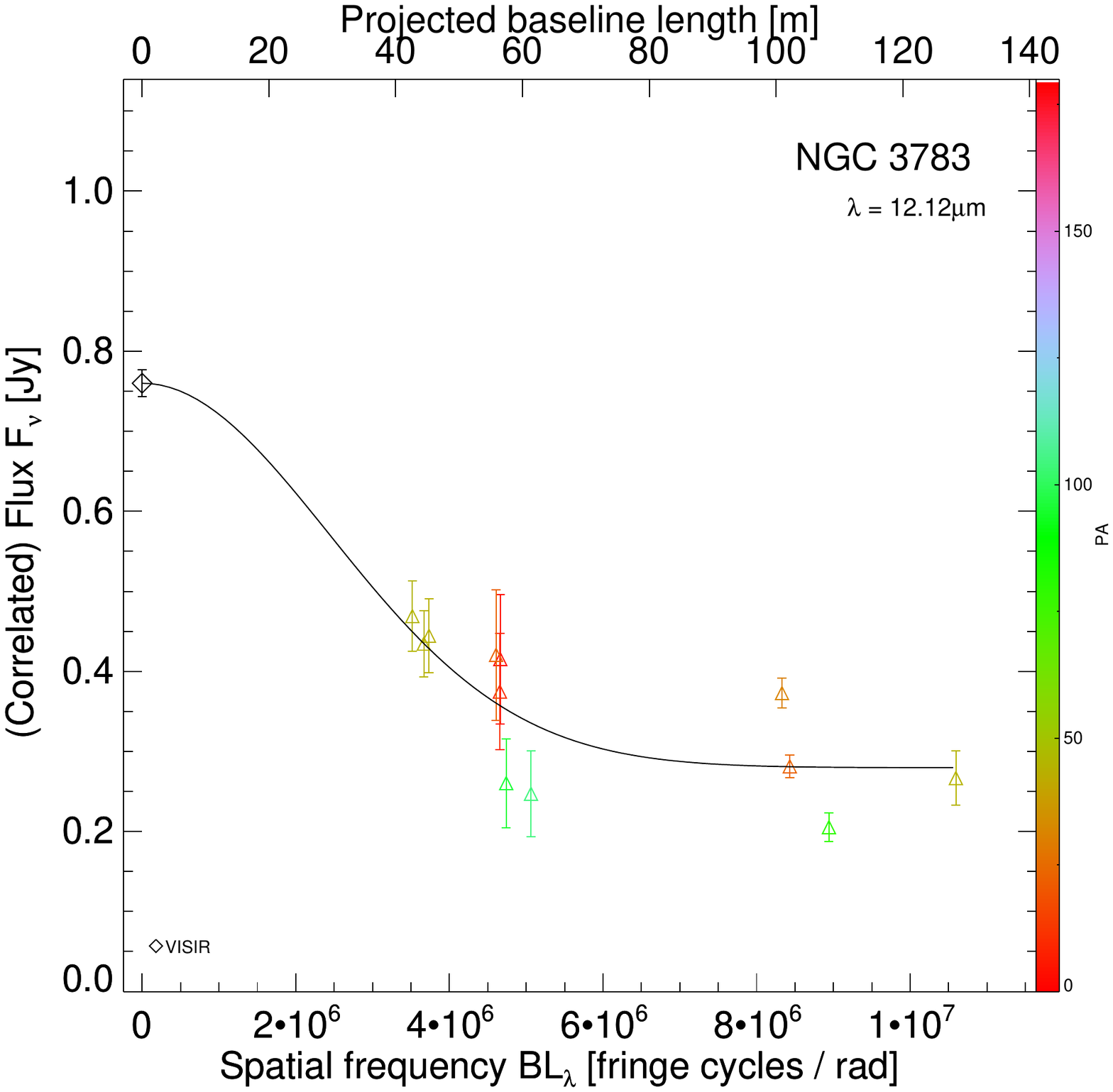}}
	~~~~~~~~~
	\subfloat{\includegraphics[trim=3cm 0cm 3cm 0cm, width=0.5\hsize]{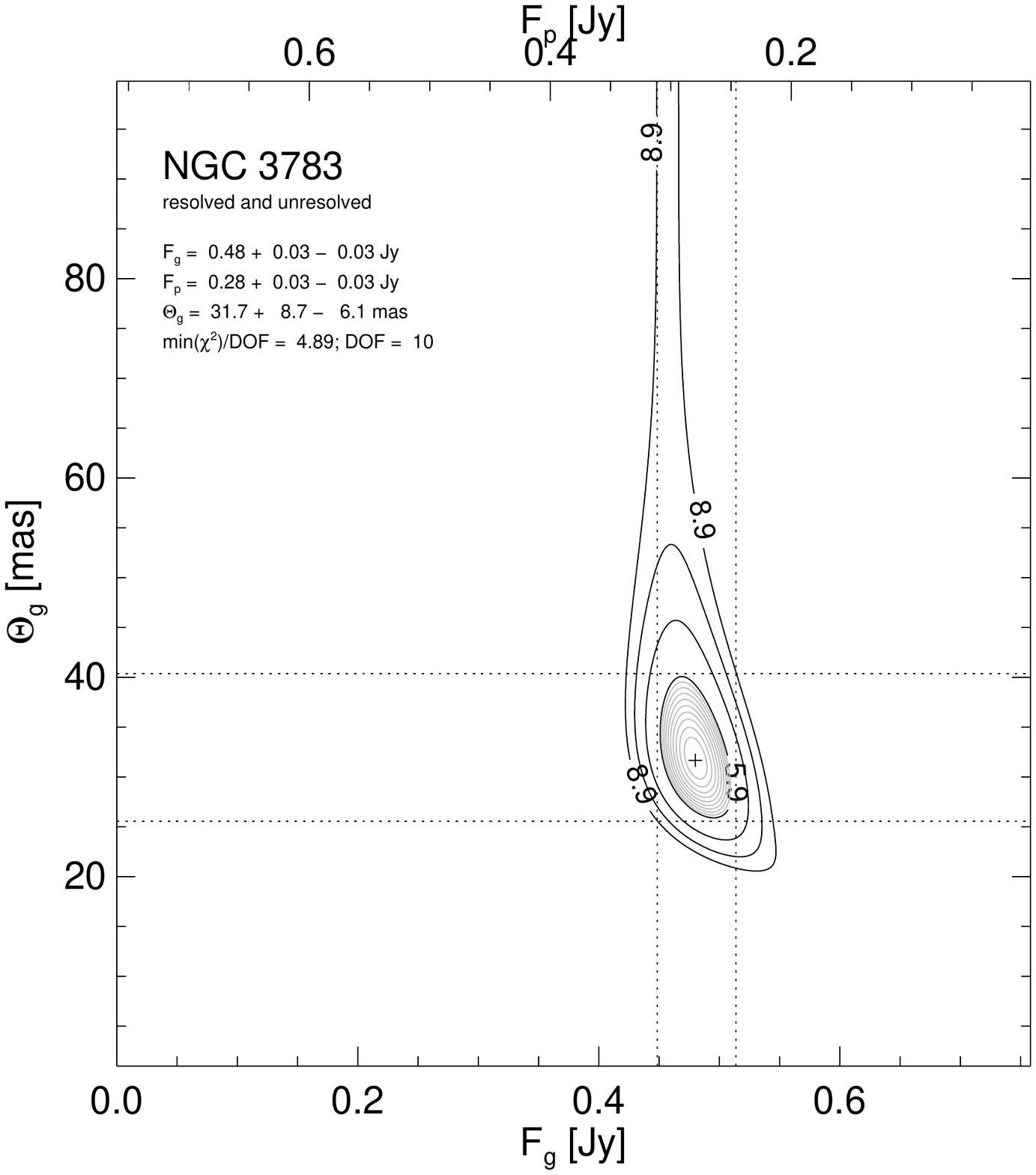}}
	\caption{\label{fig:rad:NGC3783}The same as Fig. \ref{fig:rad:IZwicky1} but for NGC~3783}
\end{figure*}

\begin{figure*}
	\centering
	\subfloat{\includegraphics[trim=7cm 4cm 7cm 4cm, width=0.5\hsize]{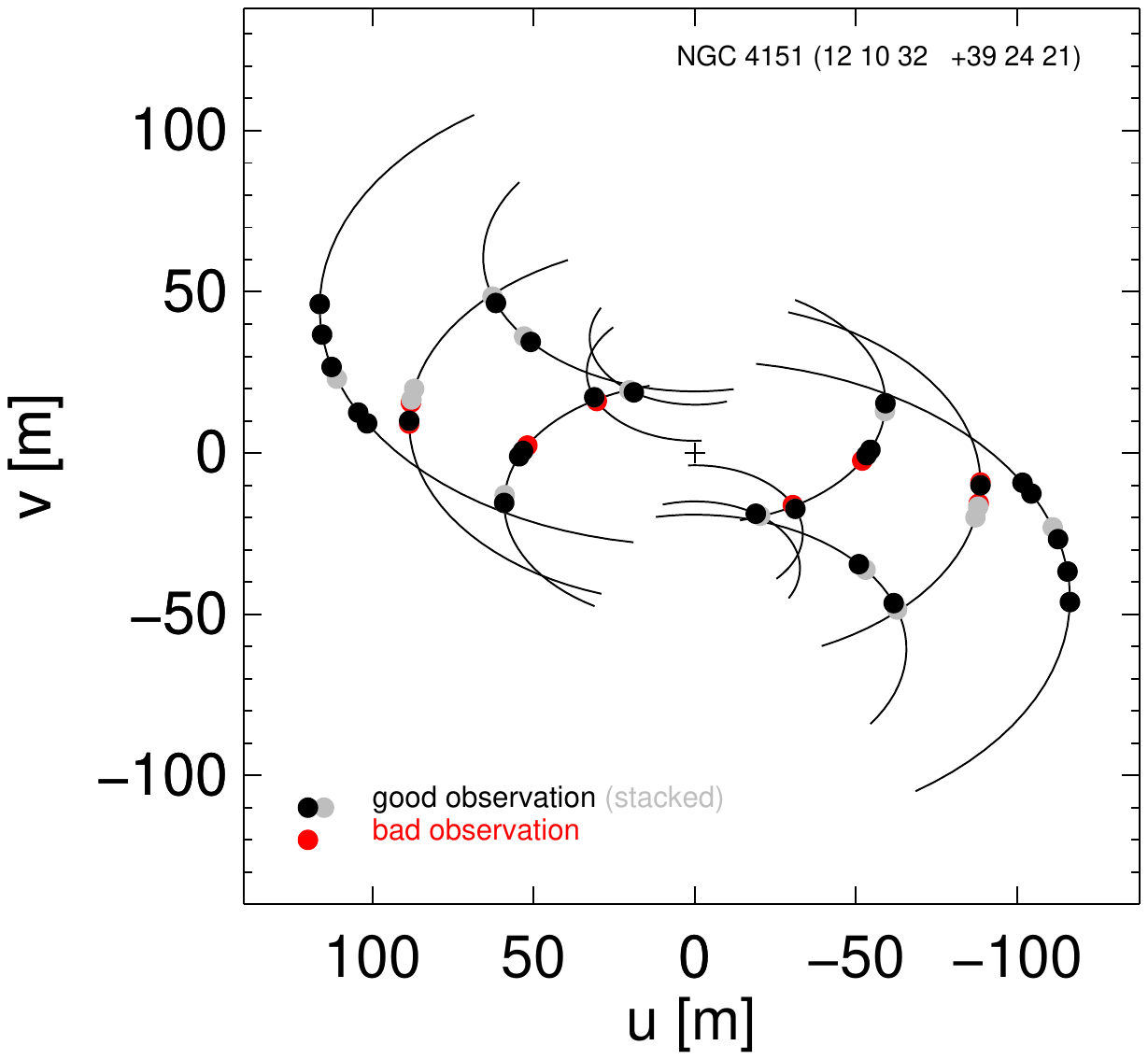}}
	~~~~~~~~~
	\subfloat{\includegraphics[trim=7cm 4cm 7cm 4cm, width=0.5\hsize]{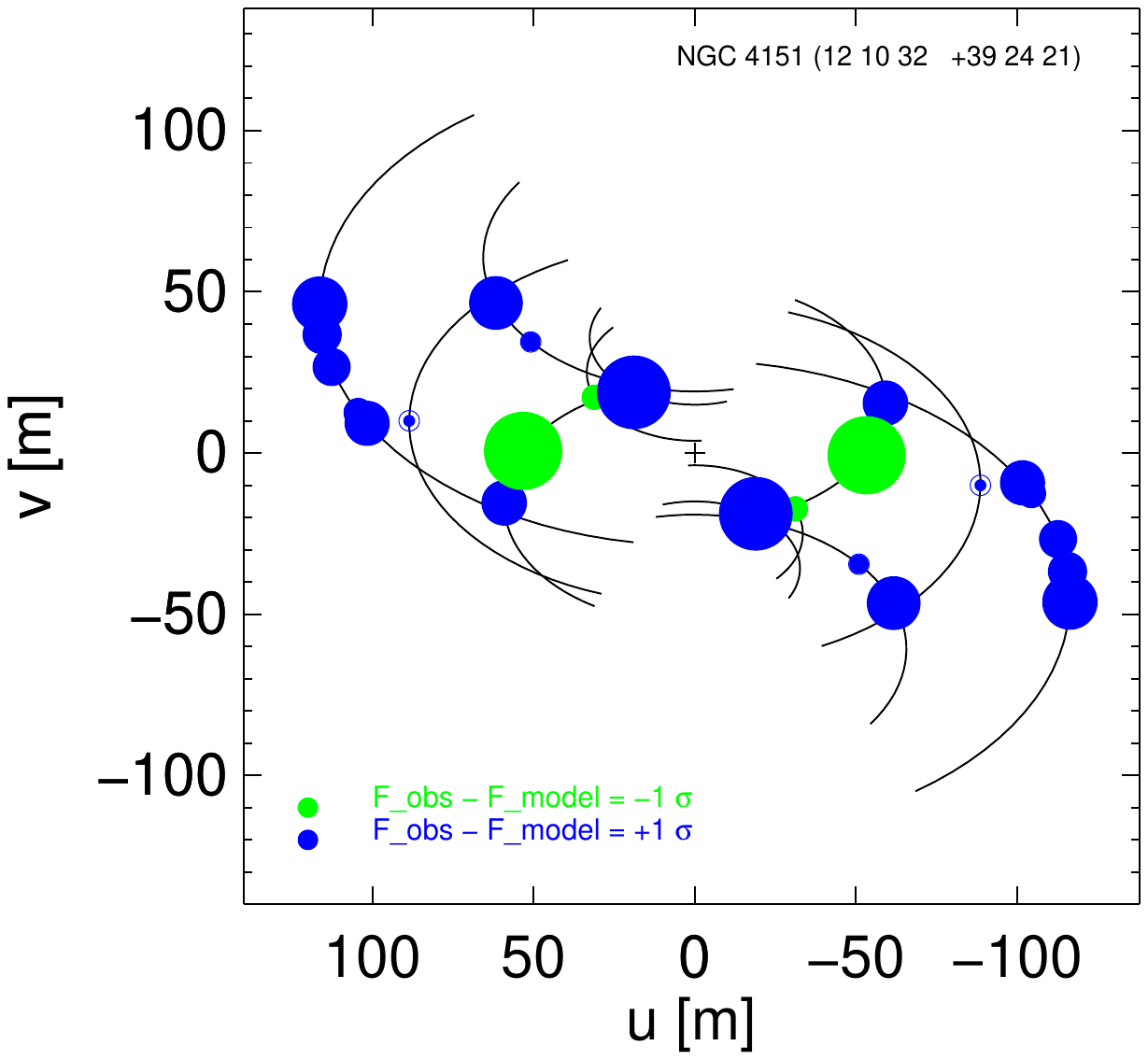}}\\
	\subfloat{\includegraphics[trim=3cm 0cm 3cm 0cm, width=0.5\hsize]{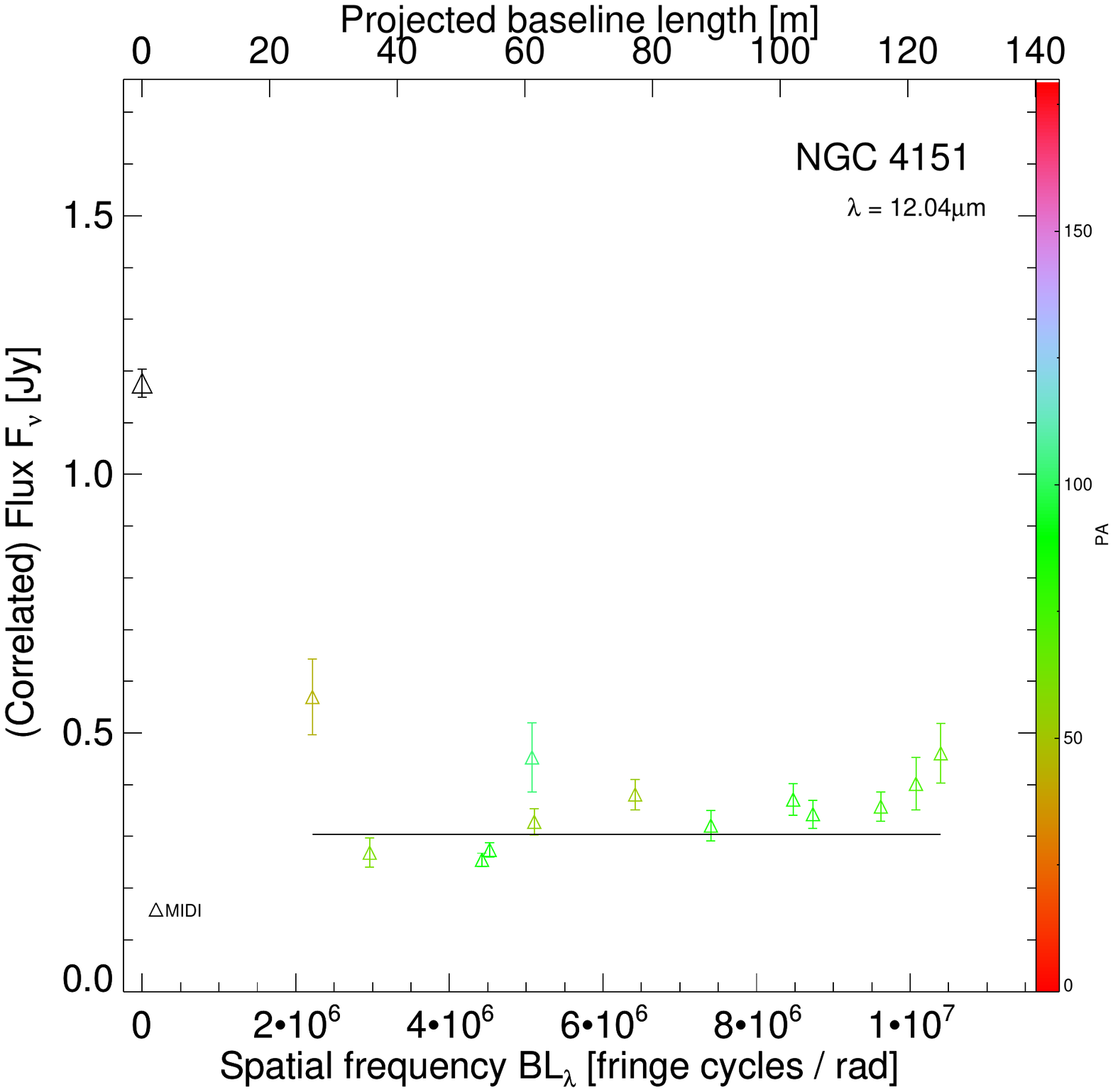}}
	~~~~~~~~~
	\subfloat{\includegraphics[trim=3cm 0cm 3cm 0cm, width=0.5\hsize]{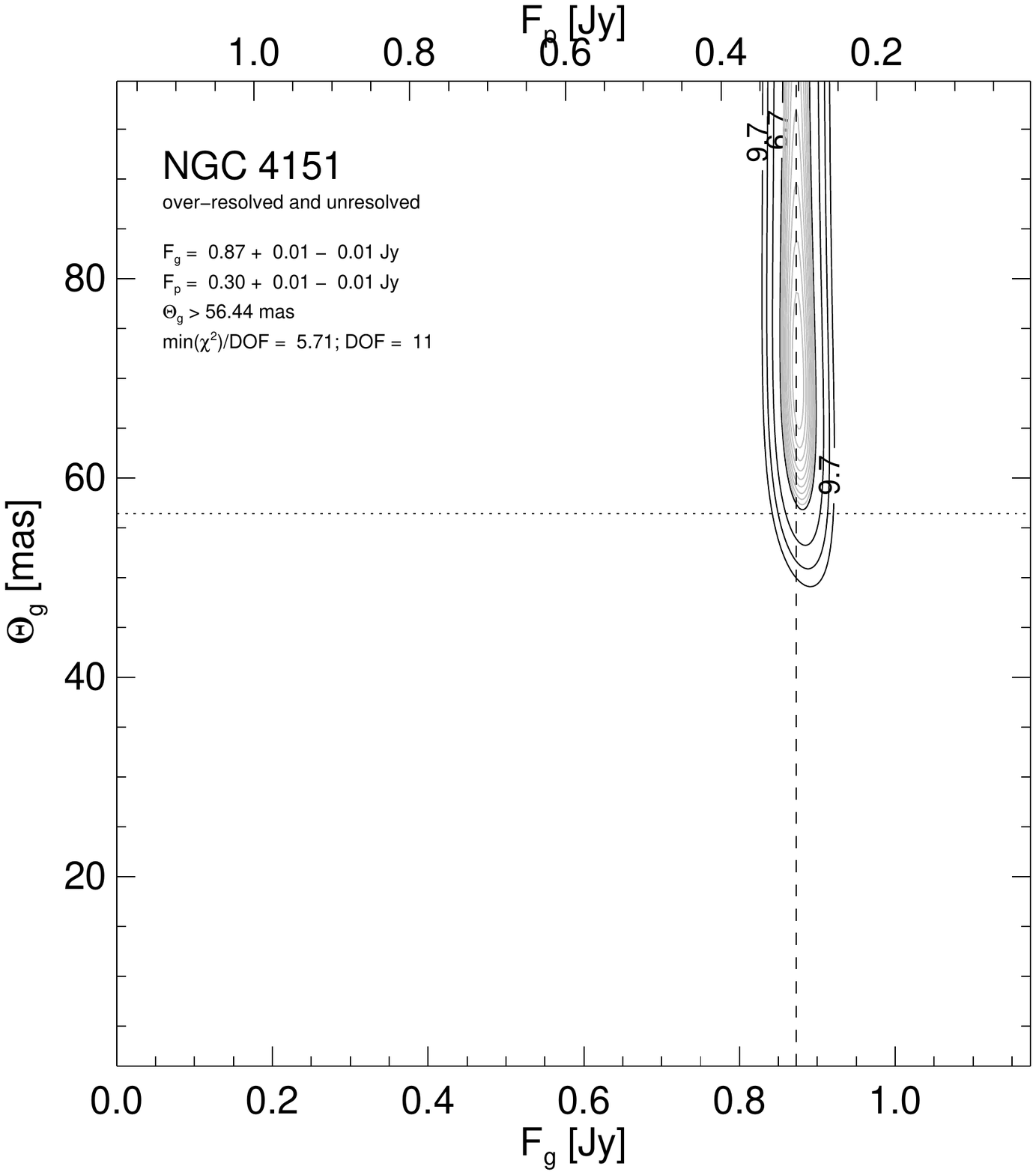}}
	\caption{\label{fig:rad:NGC4151}The same as Fig. \ref{fig:rad:IZwicky1} but for NGC~4151.}
\end{figure*}

\begin{figure*}
	\centering
	\subfloat{\includegraphics[trim=7cm 4cm 7cm 4cm, width=0.5\hsize]{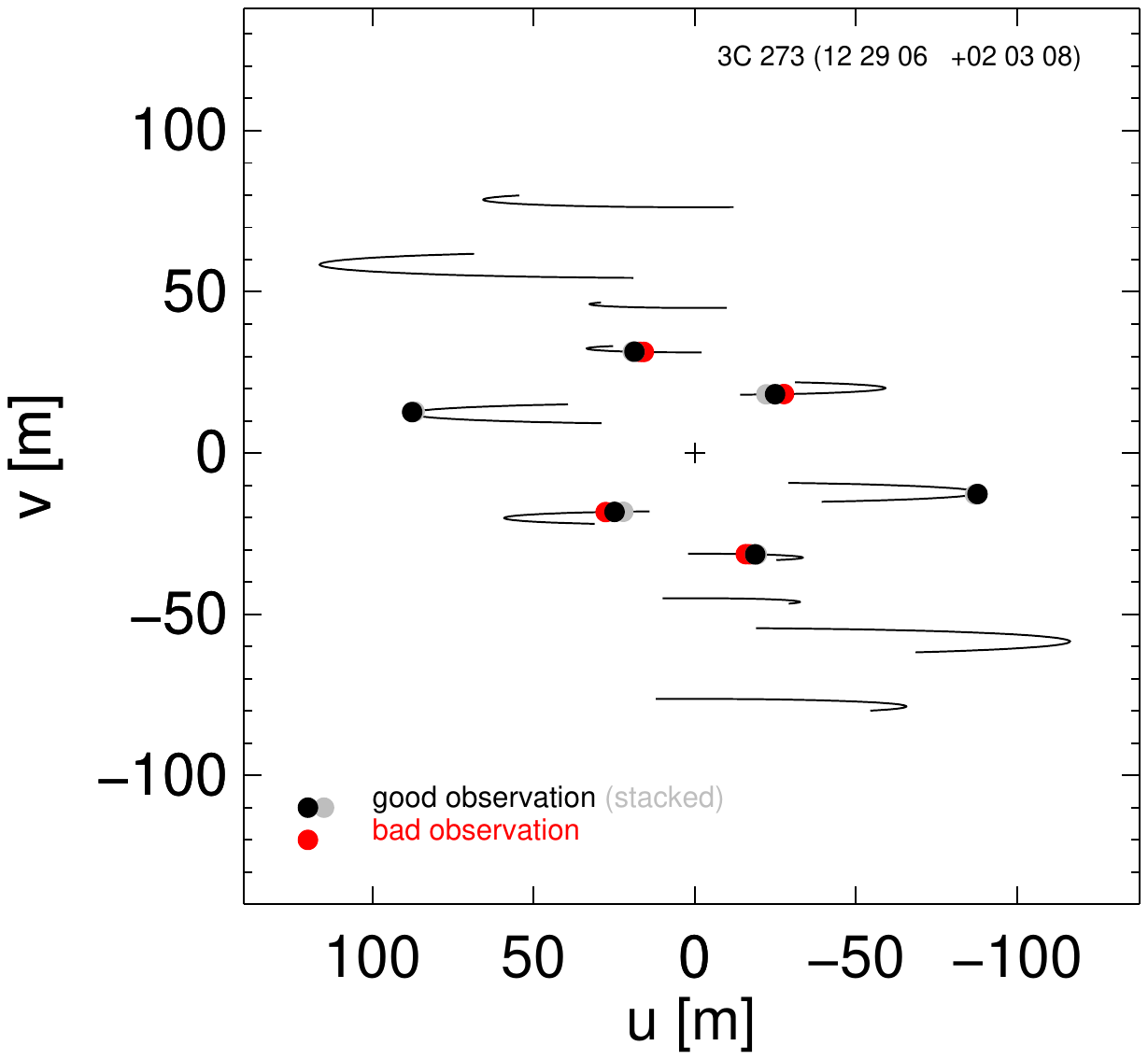}}
	~~~~~~~~~
	\subfloat{\includegraphics[trim=7cm 4cm 7cm 4cm, width=0.5\hsize]{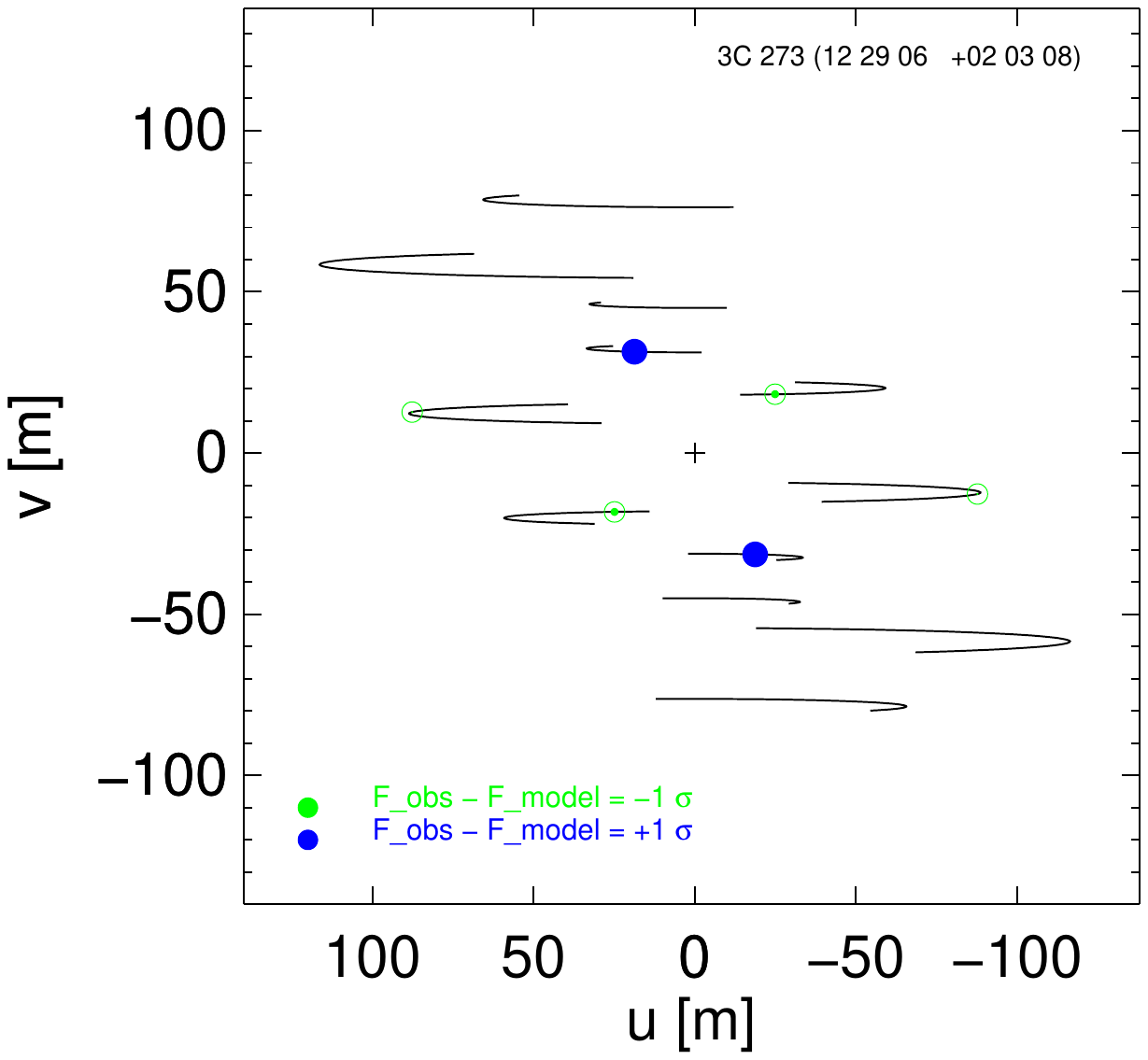}}\\
	\subfloat{\includegraphics[trim=3cm 0cm 3cm 0cm, width=0.5\hsize]{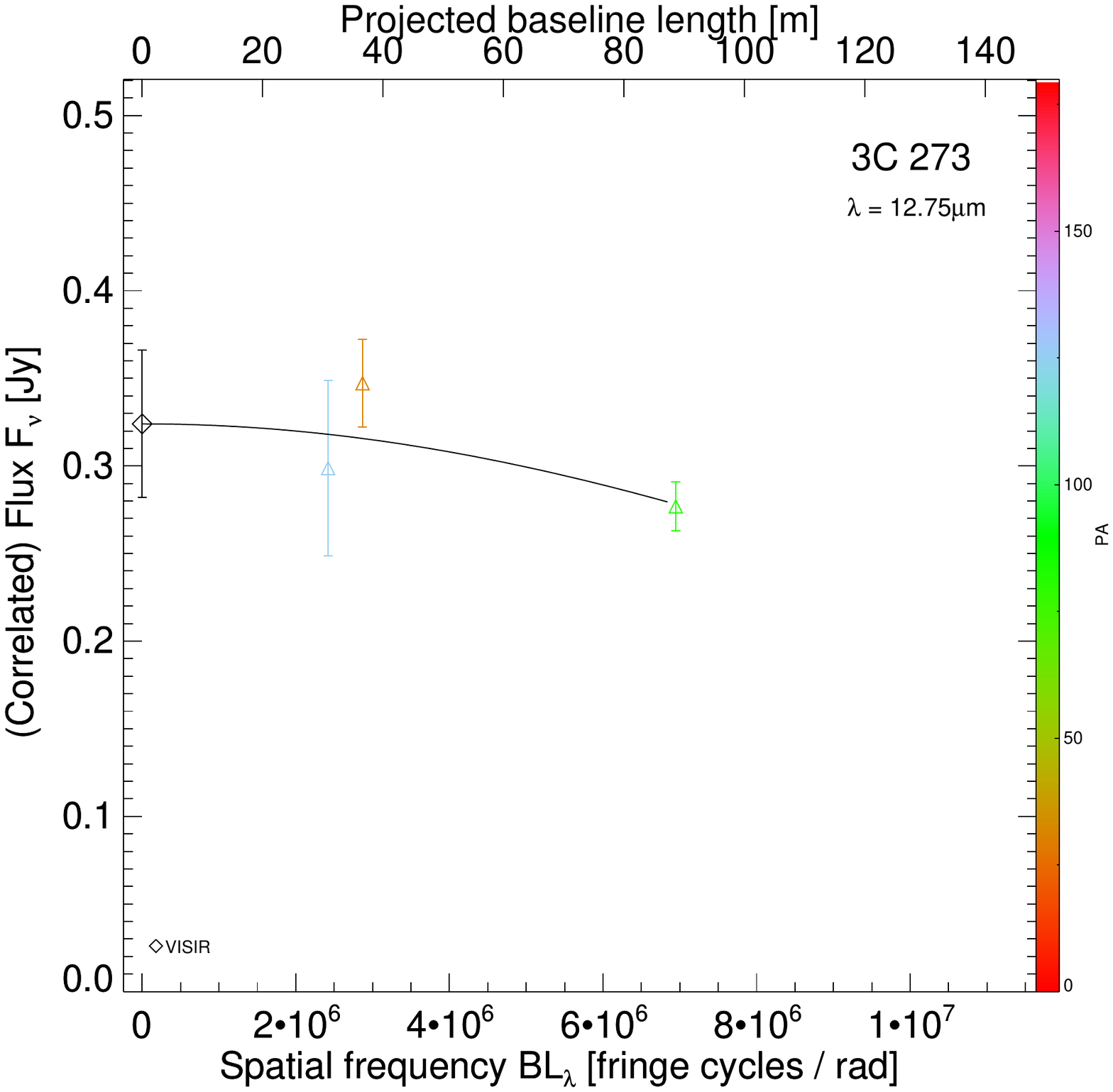}}
	~~~~~~~~~
	\subfloat{\includegraphics[trim=3cm 0cm 3cm 0cm, width=0.5\hsize]{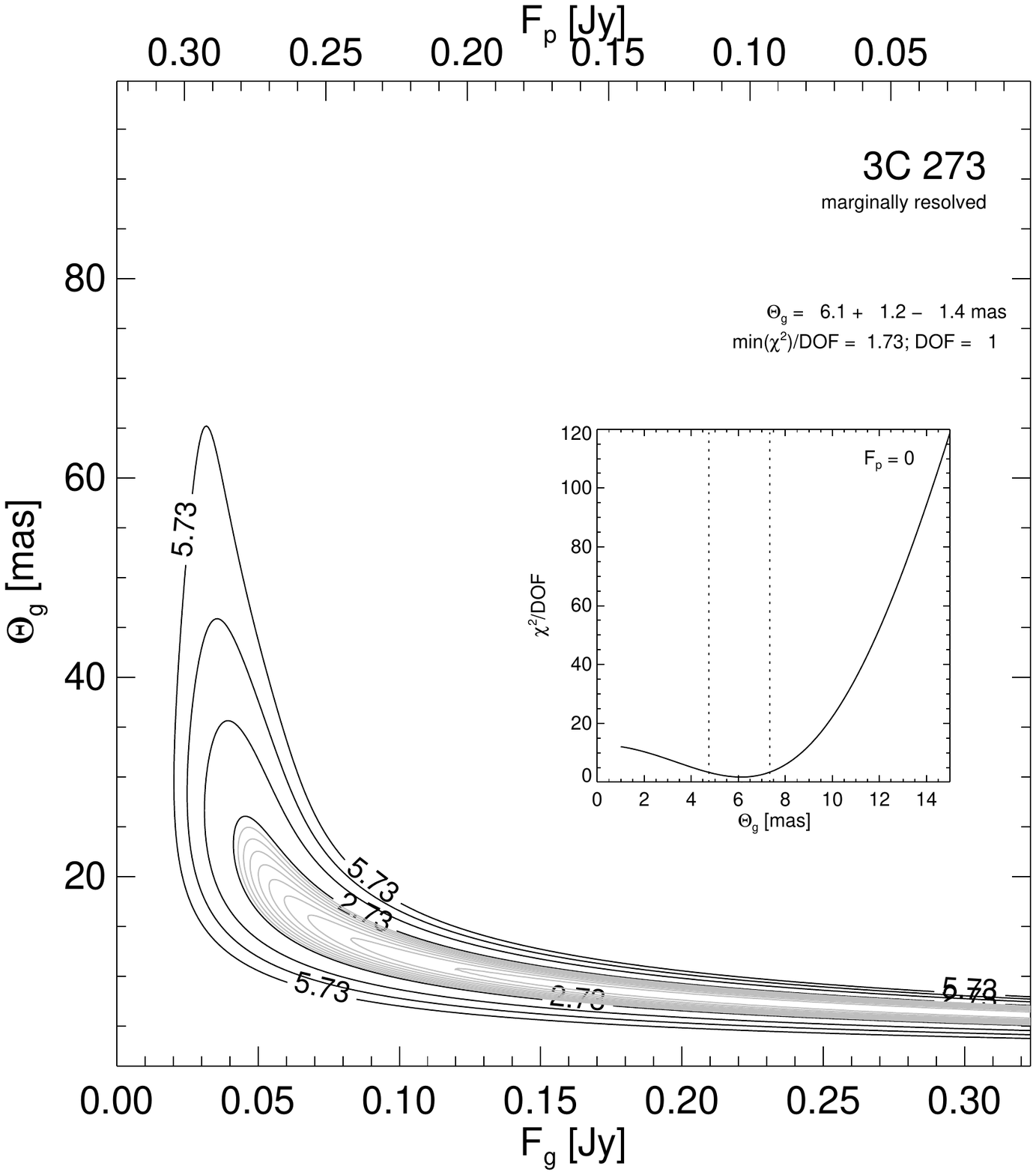}}
	\caption{\label{fig:rad:3C273}The same as Fig. \ref{fig:rad:IZwicky1} but for 3C~273 -- While the data shown here are certainly only marginal evidence for a resolved source, more recent data seem to support this view (K.R.W. Tristram in prep.)}
\end{figure*}

\begin{figure*}
	\centering
	\subfloat{\includegraphics[trim=7cm 4cm 7cm 4cm, width=0.5\hsize]{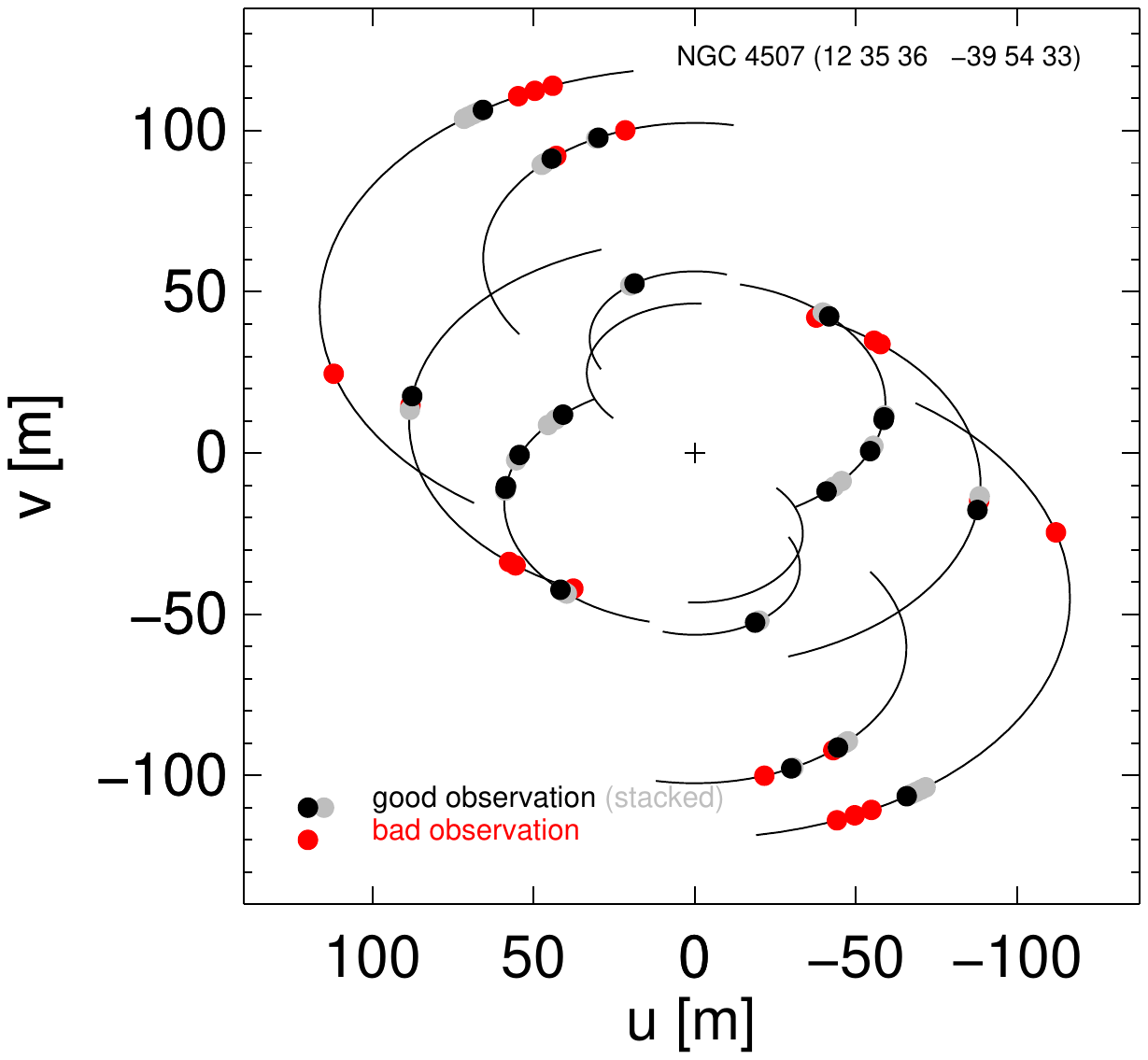}}
	~~~~~~~~~
	\subfloat{\includegraphics[trim=7cm 4cm 7cm 4cm, width=0.5\hsize]{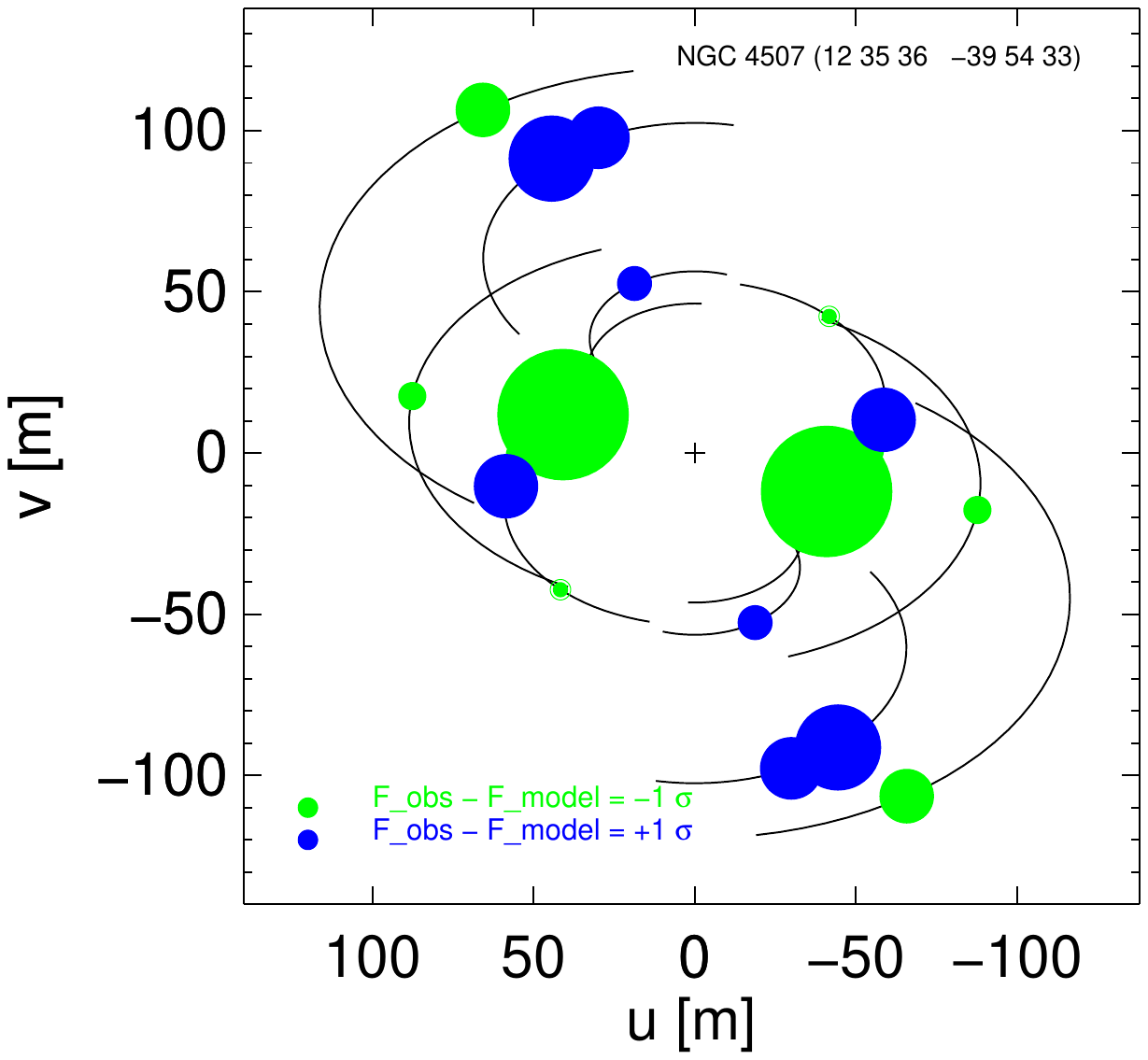}}\\
	\subfloat{\includegraphics[trim=3cm 0cm 3cm 0cm, width=0.5\hsize]{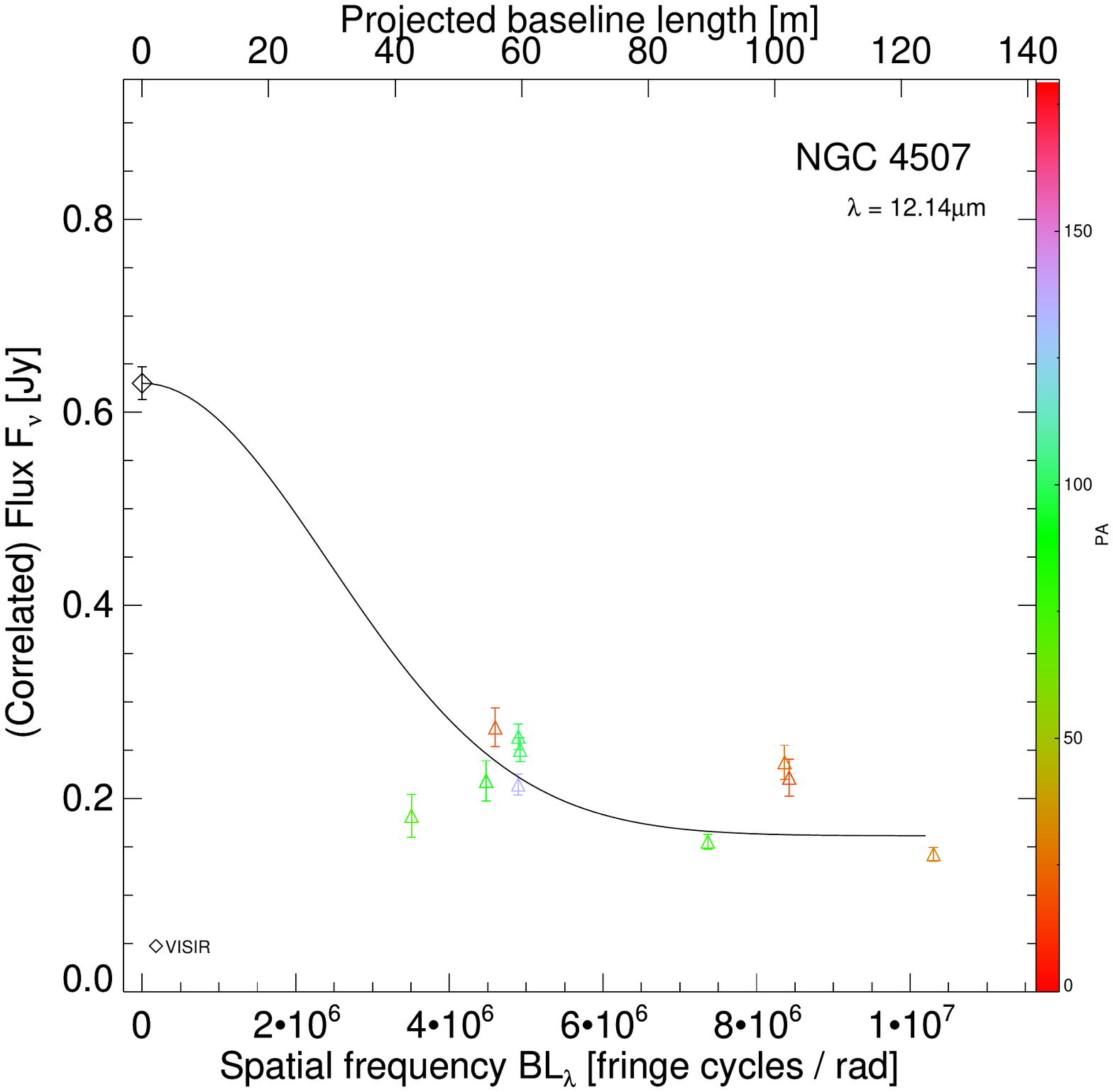}}
	~~~~~~~~~
	\subfloat{\includegraphics[trim=3cm 0cm 3cm 0cm, width=0.5\hsize]{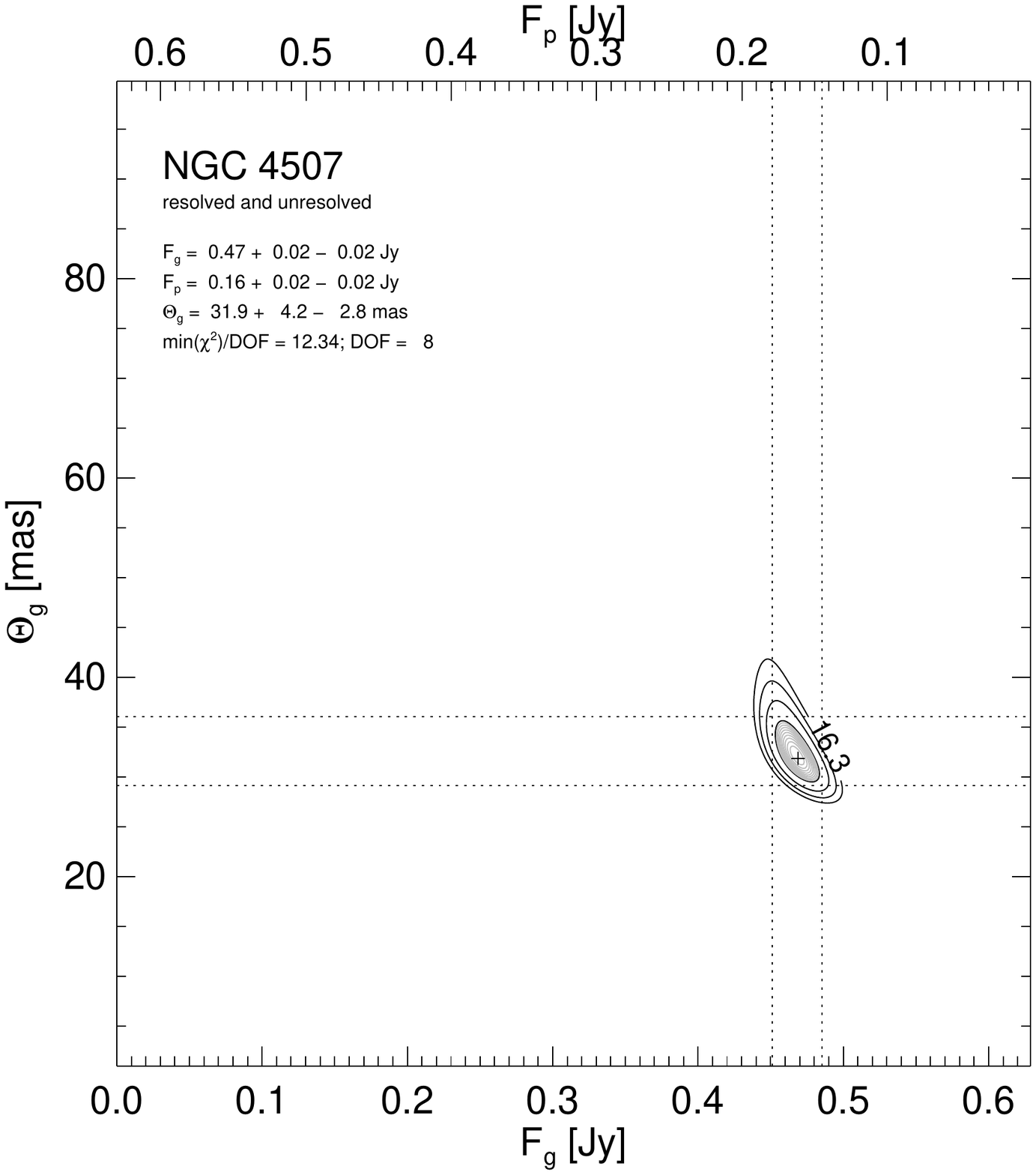}}
	\caption{\label{fig:rad:NGC4507}The same as Fig. \ref{fig:rad:IZwicky1} but for NGC~4507}
\end{figure*}

\begin{figure*}
	\centering
	\subfloat{\includegraphics[trim=7cm 4cm 7cm 4cm, width=0.5\hsize]{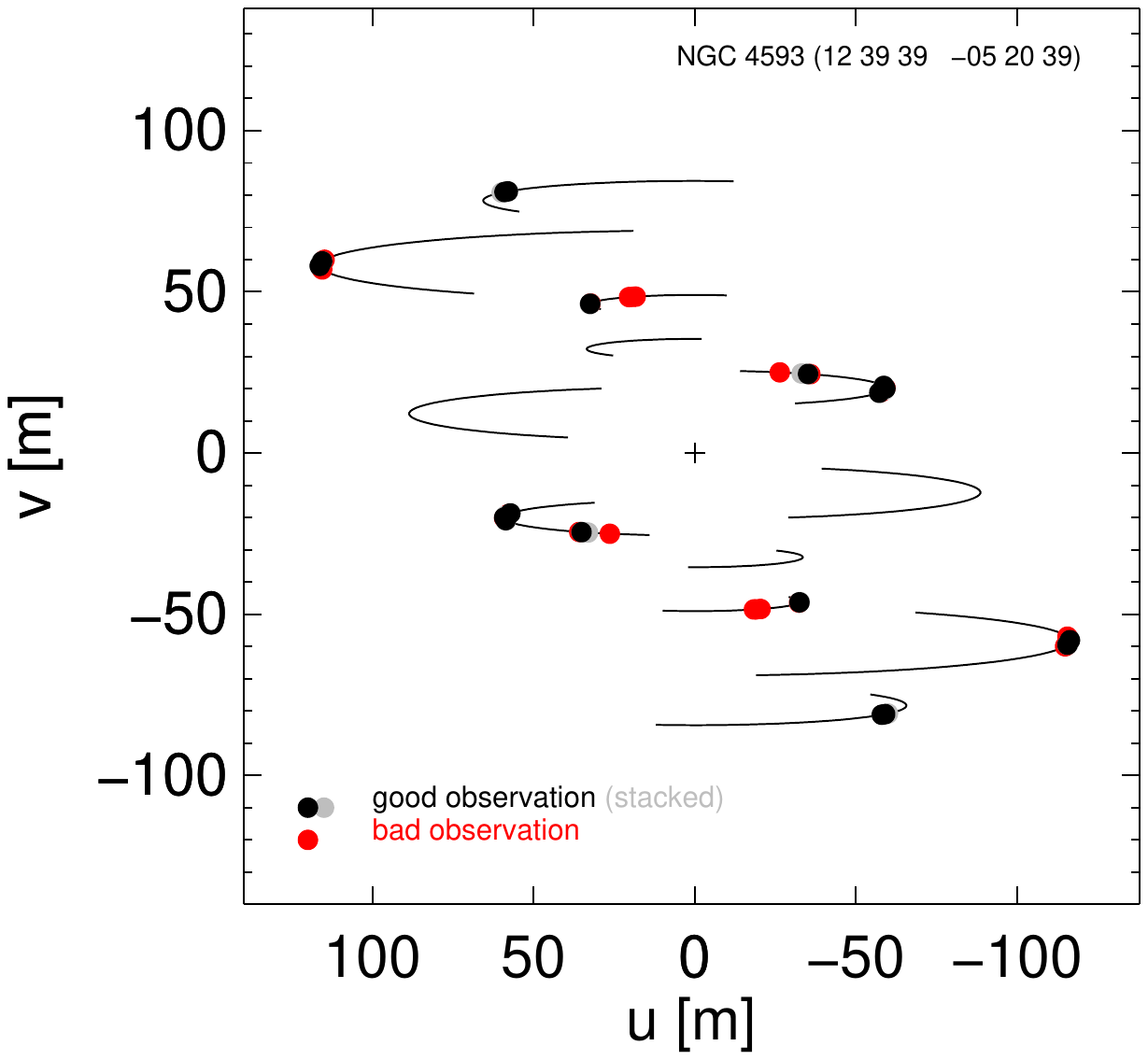}}
	~~~~~~~~~
	\subfloat{\includegraphics[trim=7cm 4cm 7cm 4cm, width=0.5\hsize]{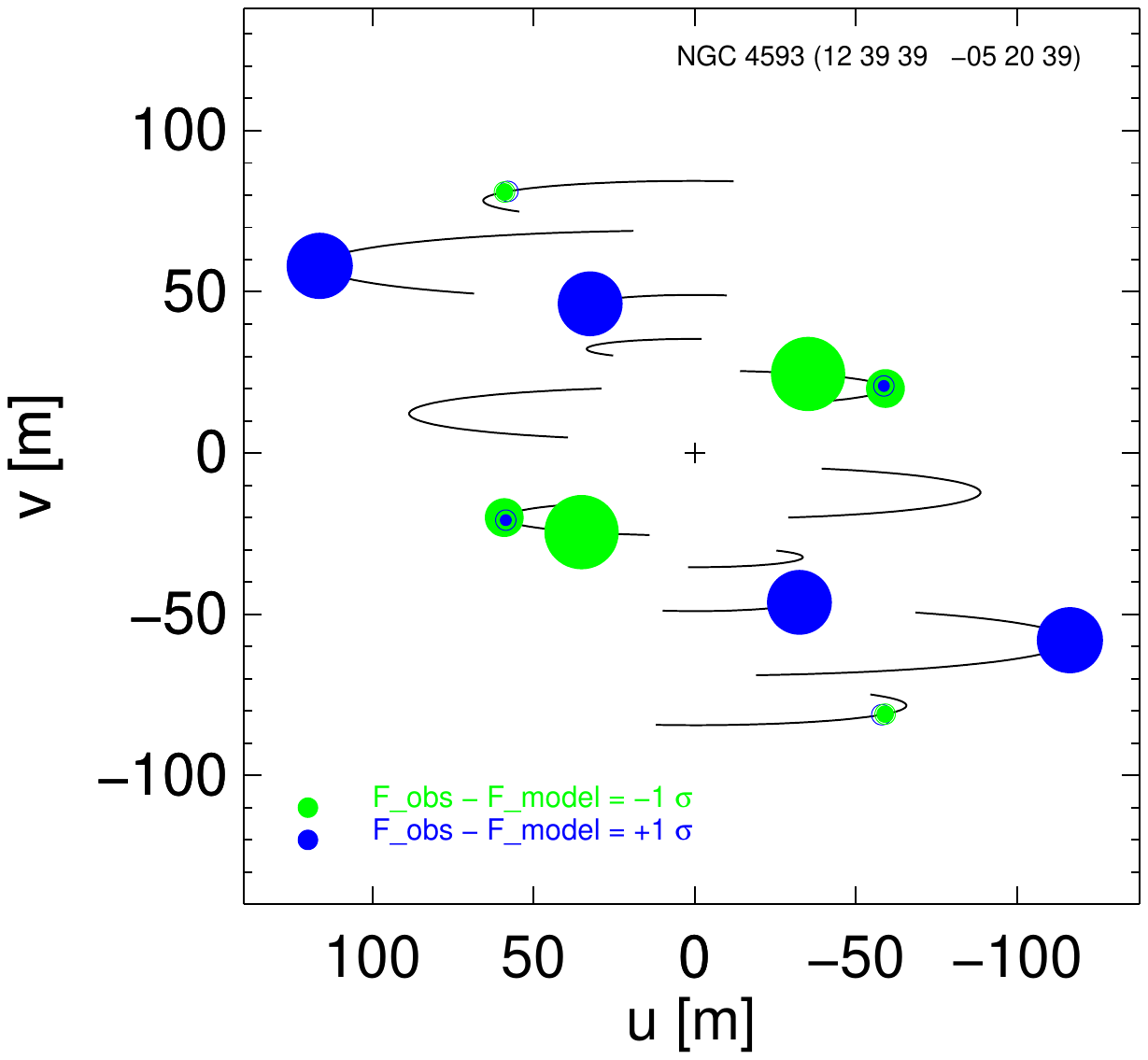}}\\
	\subfloat{\includegraphics[trim=3cm 0cm 3cm 0cm, width=0.5\hsize]{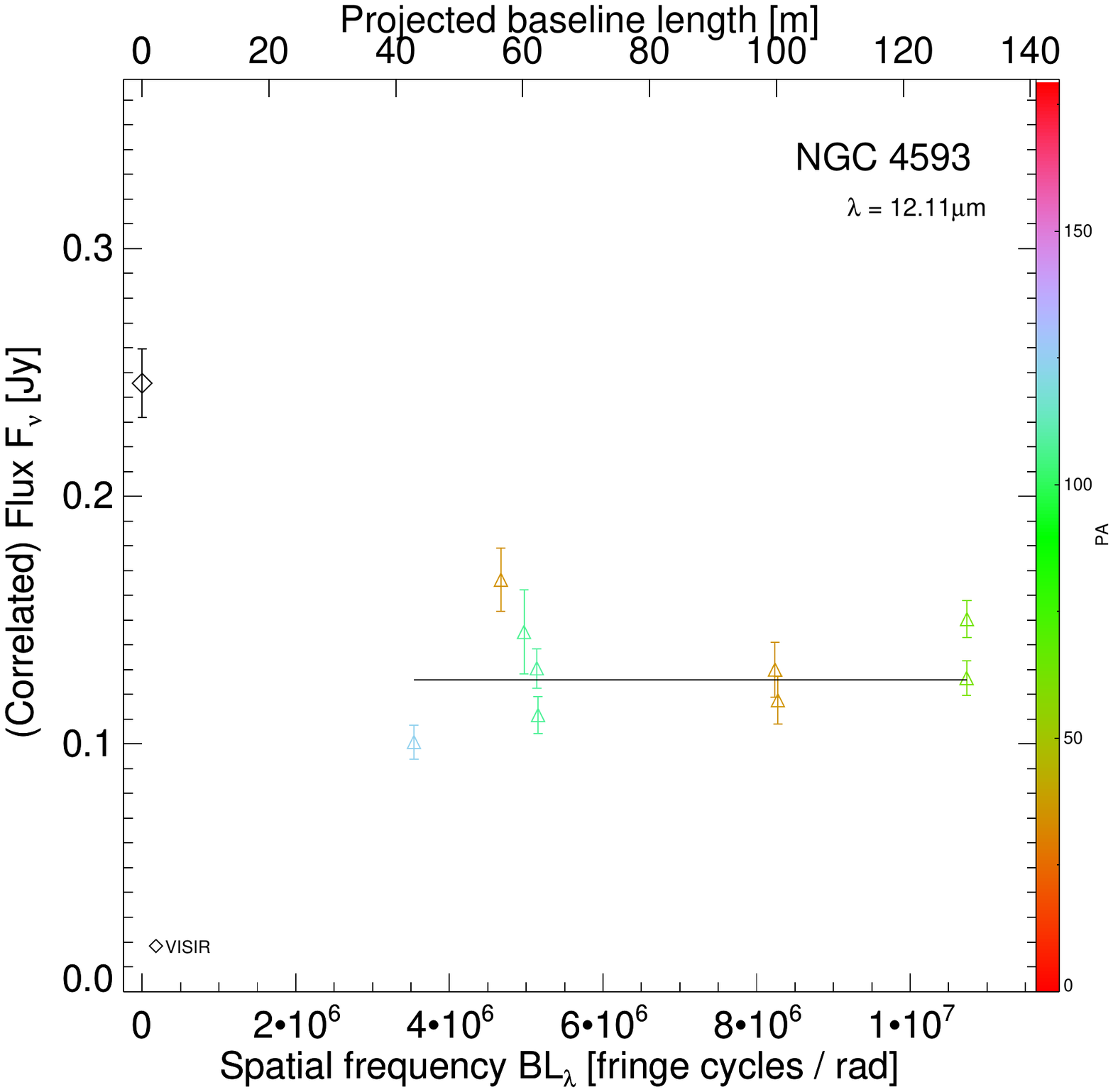}}
	~~~~~~~~~
	\subfloat{\includegraphics[trim=3cm 0cm 3cm 0cm, width=0.5\hsize]{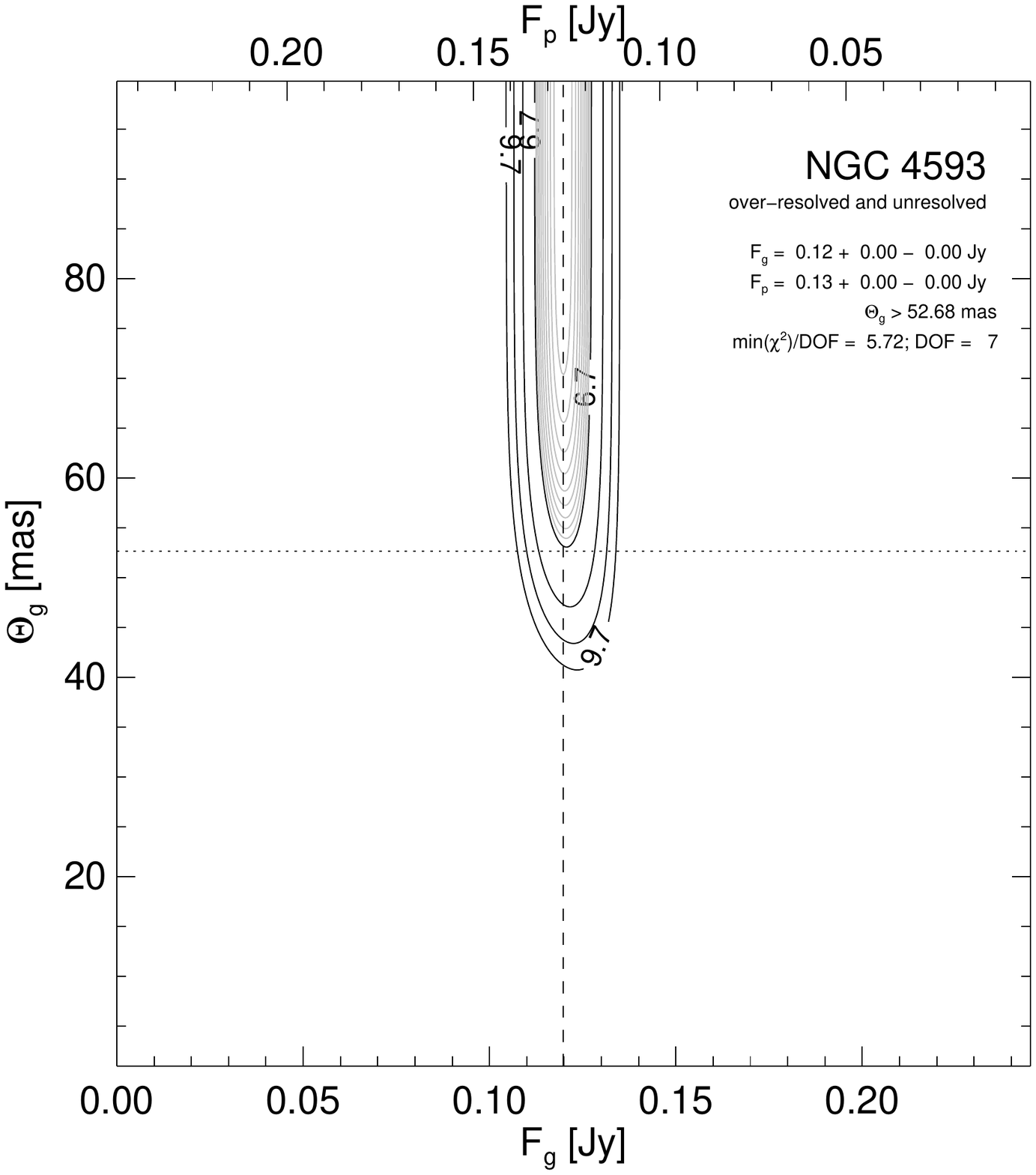}}
	\caption{\label{fig:rad:NGC4593}The same as Fig. \ref{fig:rad:IZwicky1} but for NGC4593}
\end{figure*}

\begin{figure*}
	\centering
	\subfloat{\includegraphics[trim=7cm 4cm 7cm 4cm, width=0.5\hsize]{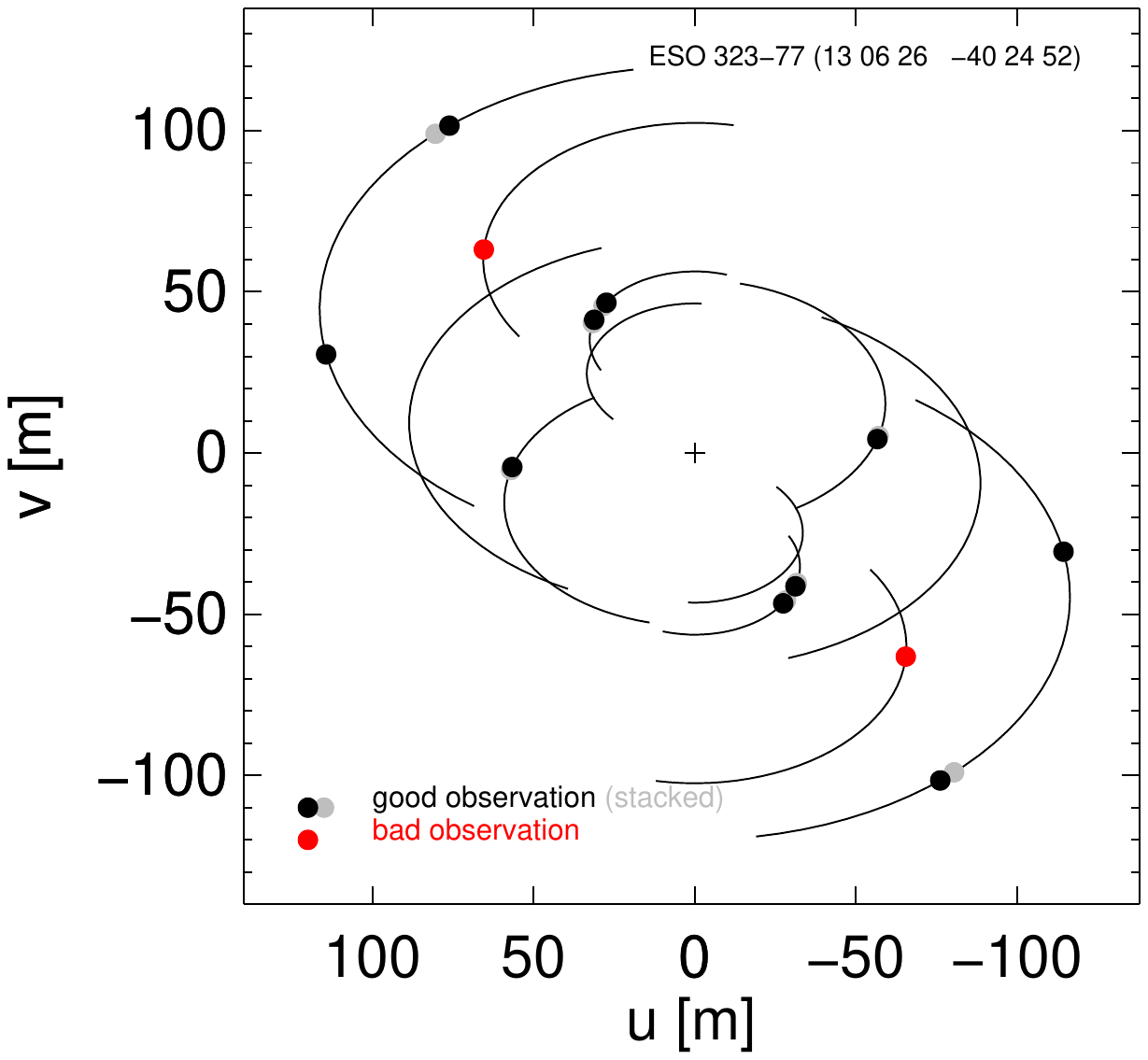}}
	~~~~~~~~~
	\subfloat{\includegraphics[trim=7cm 4cm 7cm 4cm, width=0.5\hsize]{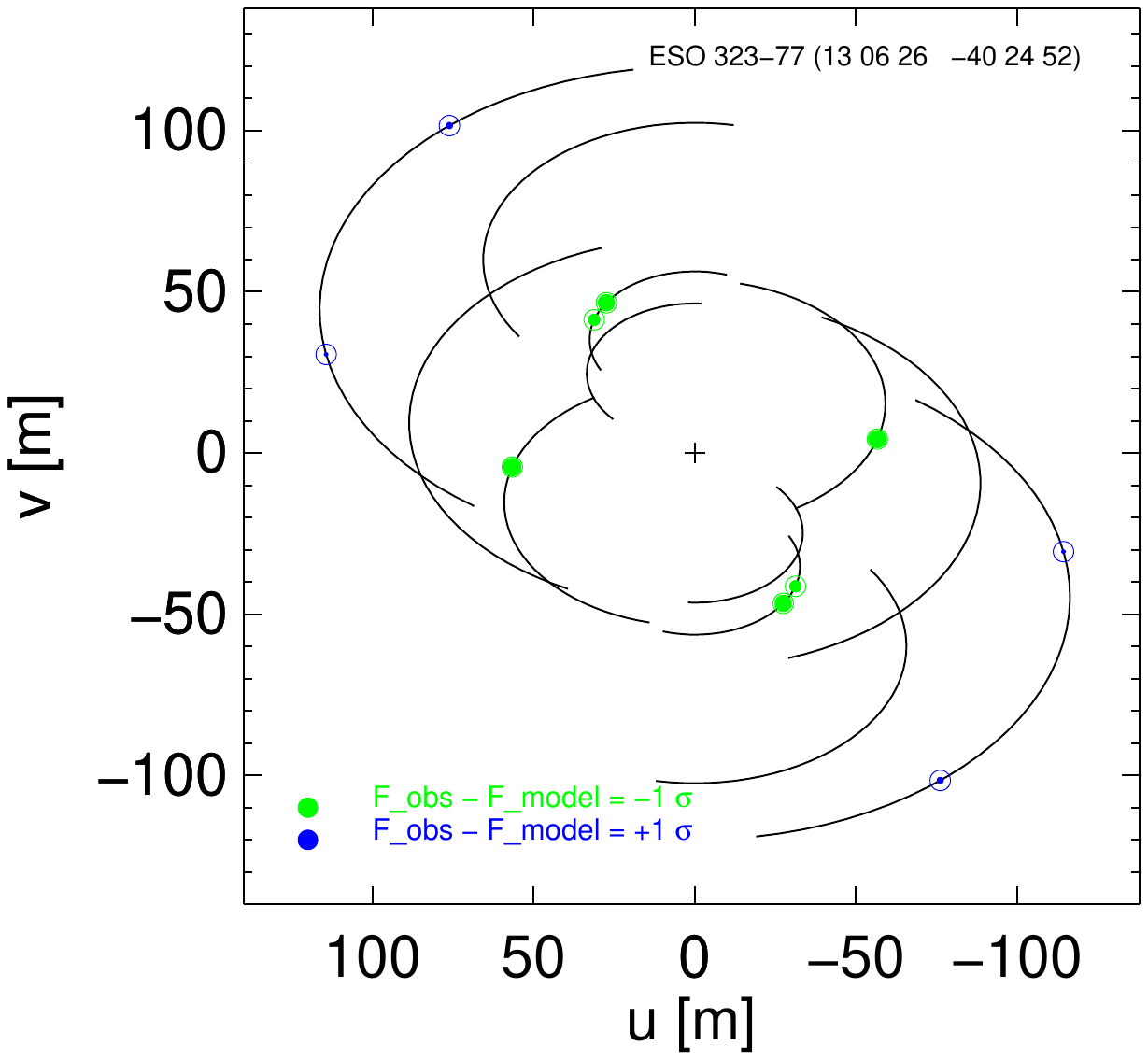}}\\
	\subfloat{\includegraphics[trim=3cm 0cm 3cm 0cm, width=0.5\hsize]{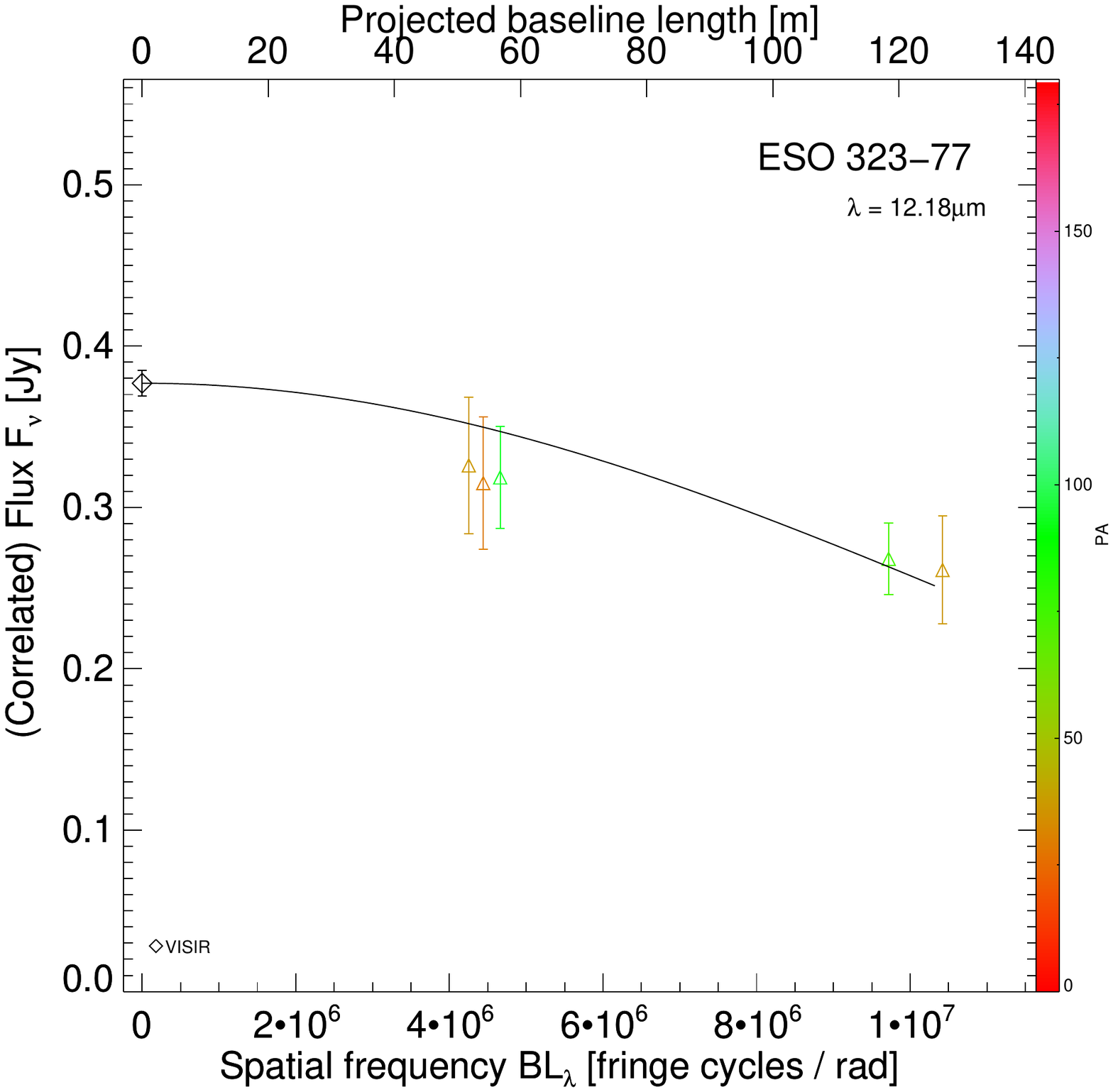}}
	~~~~~~~~~
	\subfloat{\includegraphics[trim=3cm 0cm 3cm 0cm, width=0.5\hsize]{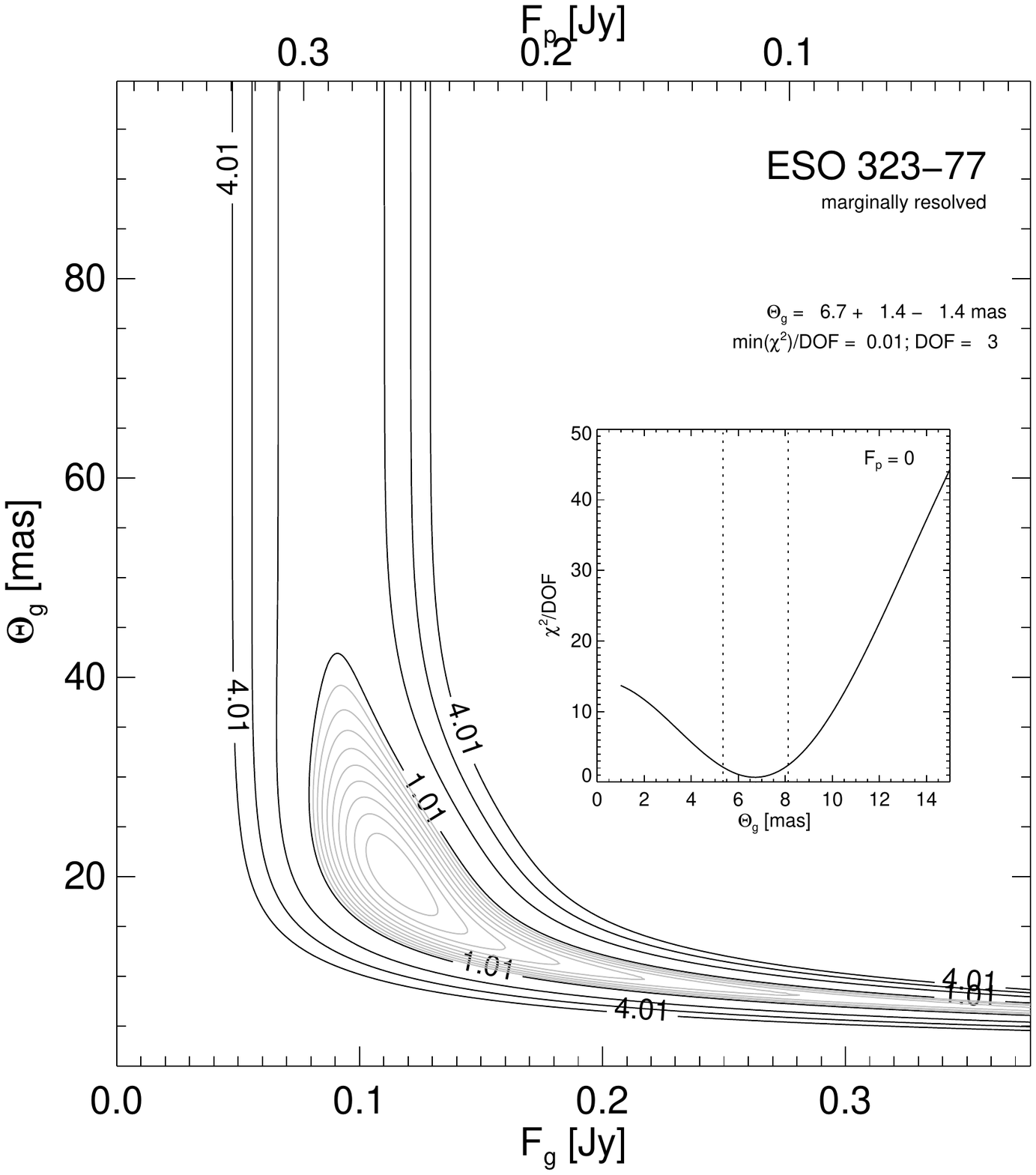}}
	\caption{\label{fig:rad:ESO323-77}The same as Fig. \ref{fig:rad:IZwicky1} but for ESO~323-77}
\end{figure*}

\begin{figure*}
	\centering
	\subfloat{\includegraphics[trim=7cm 4cm 7cm 4cm, width=0.5\hsize]{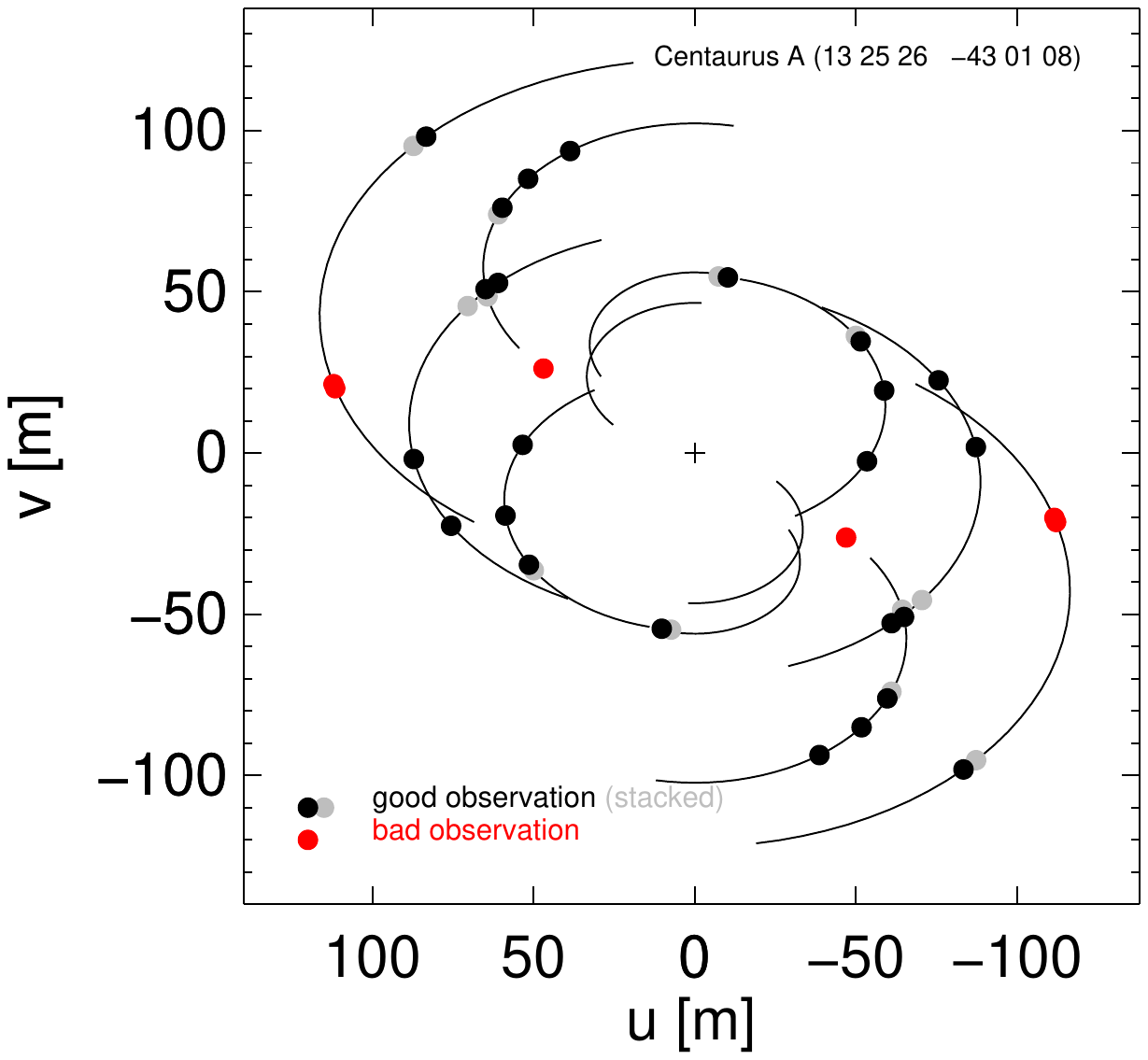}}
	~~~~~~~~~
	\subfloat{\includegraphics[trim=7cm 4cm 7cm 4cm, width=0.5\hsize]{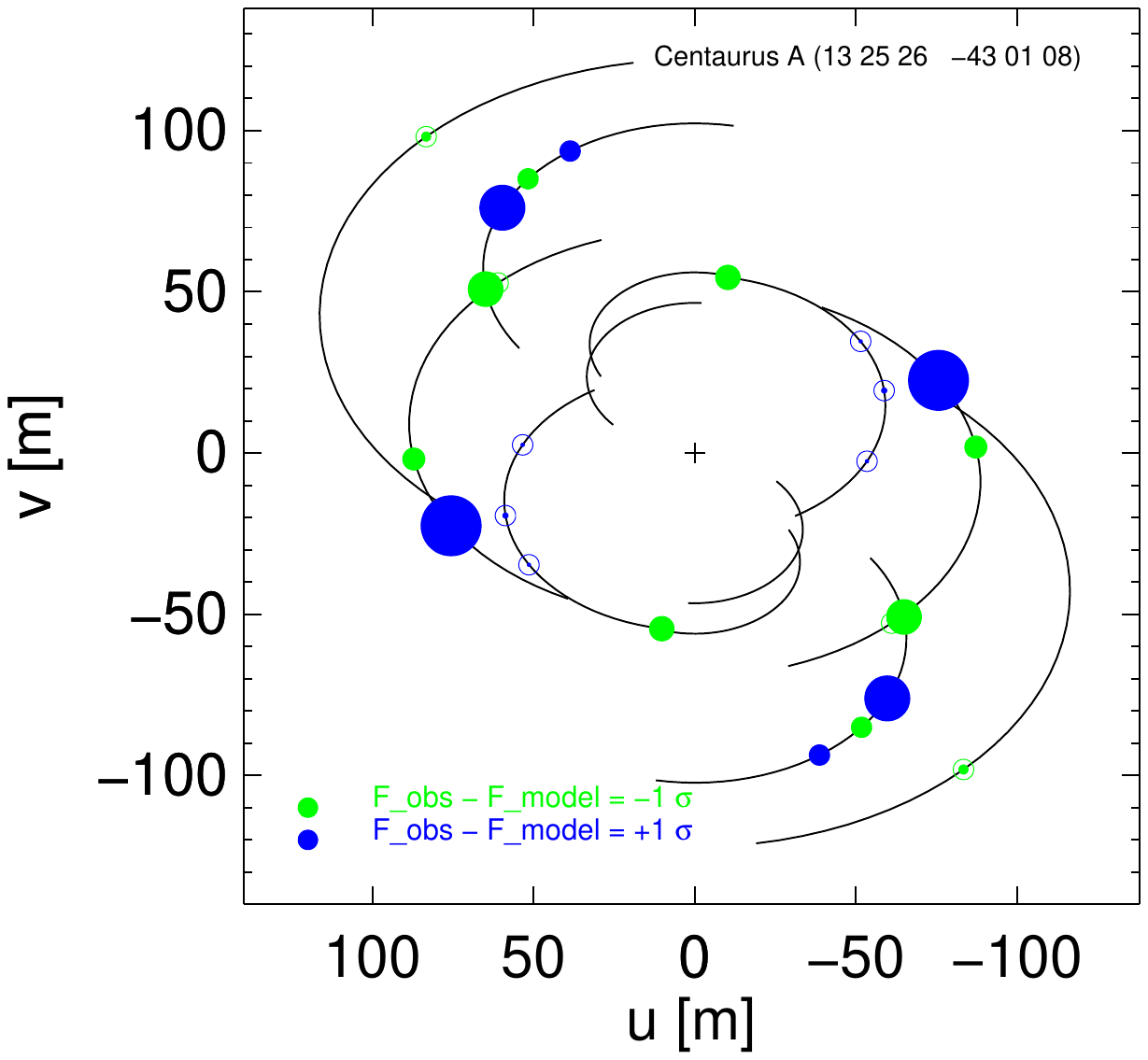}}\\
	\subfloat{\includegraphics[trim=3cm 0cm 3cm 0cm, width=0.5\hsize]{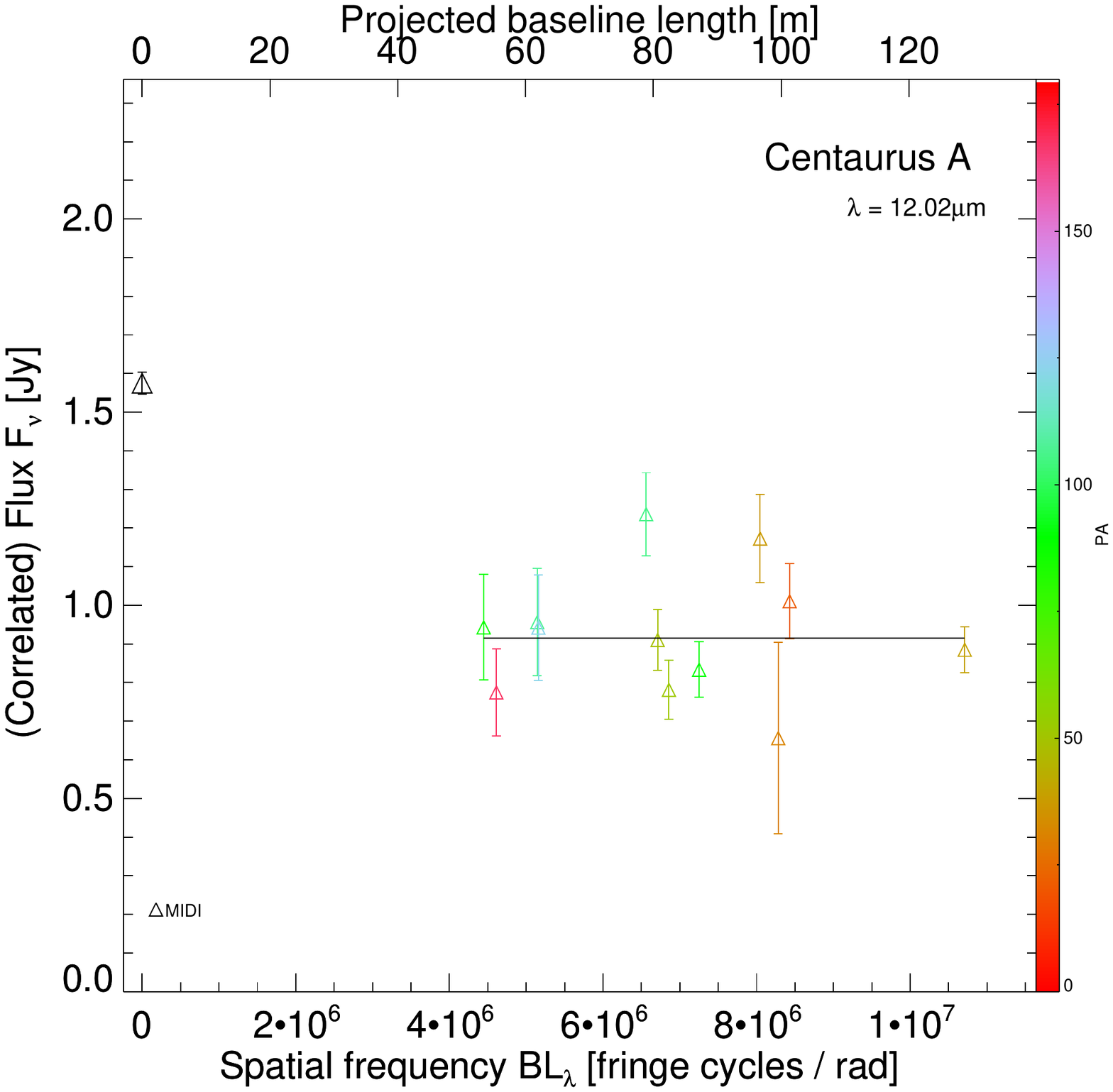}}
	~~~~~~~~~
	\subfloat{\includegraphics[trim=3cm 0cm 3cm 0cm, width=0.5\hsize]{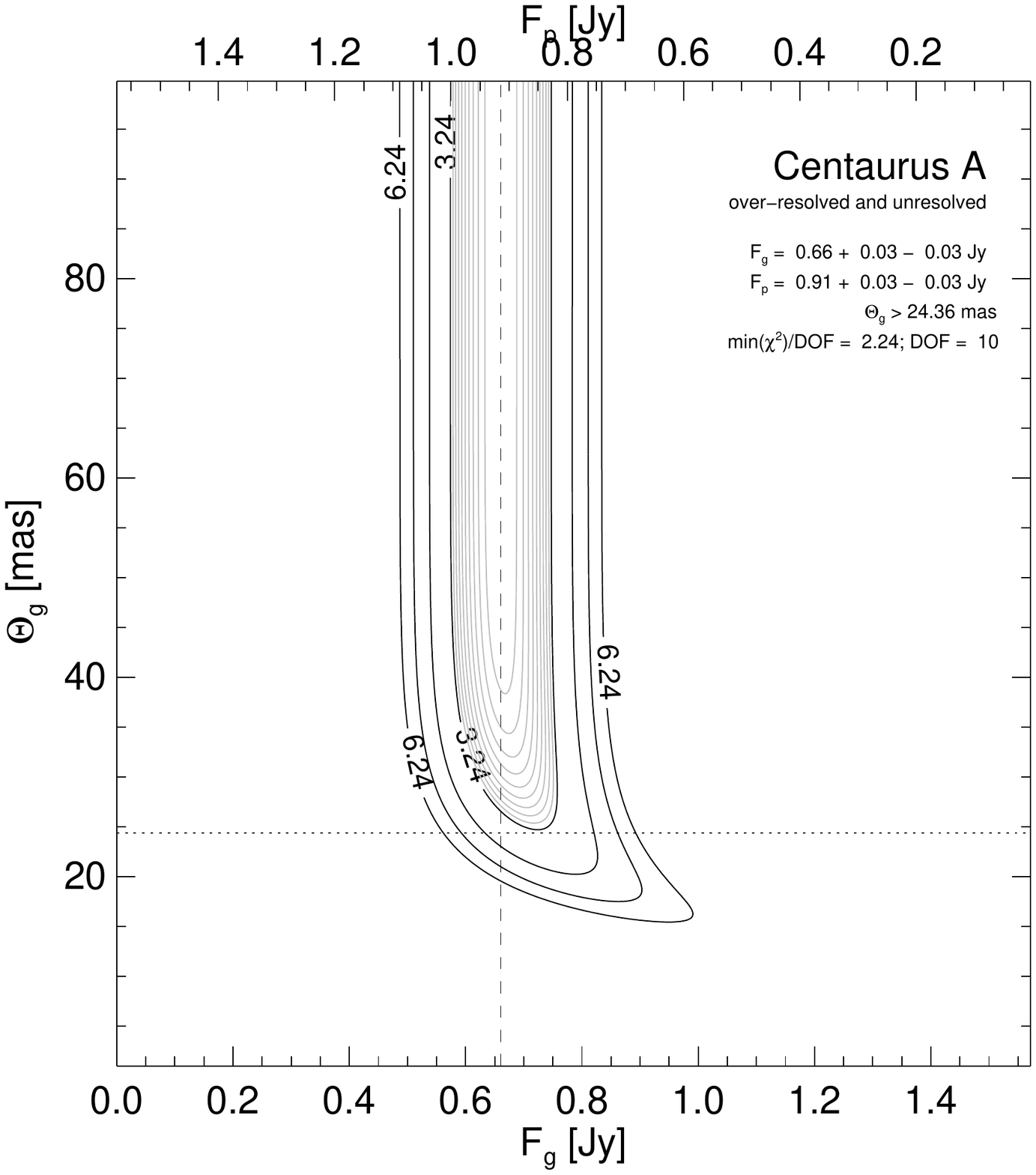}}
	\caption{\label{fig:rad:NGC5128}The same as Fig. \ref{fig:rad:IZwicky1} but for Centaurus~A. The source is variable in the mid-infrared \citep[][and in prep.]{burtscher2010}. Here we only show the data from the 2008 epoch.}
\end{figure*}

\begin{figure*}
	\centering
	\subfloat{\includegraphics[trim=7cm 4cm 7cm 4cm, width=0.5\hsize]{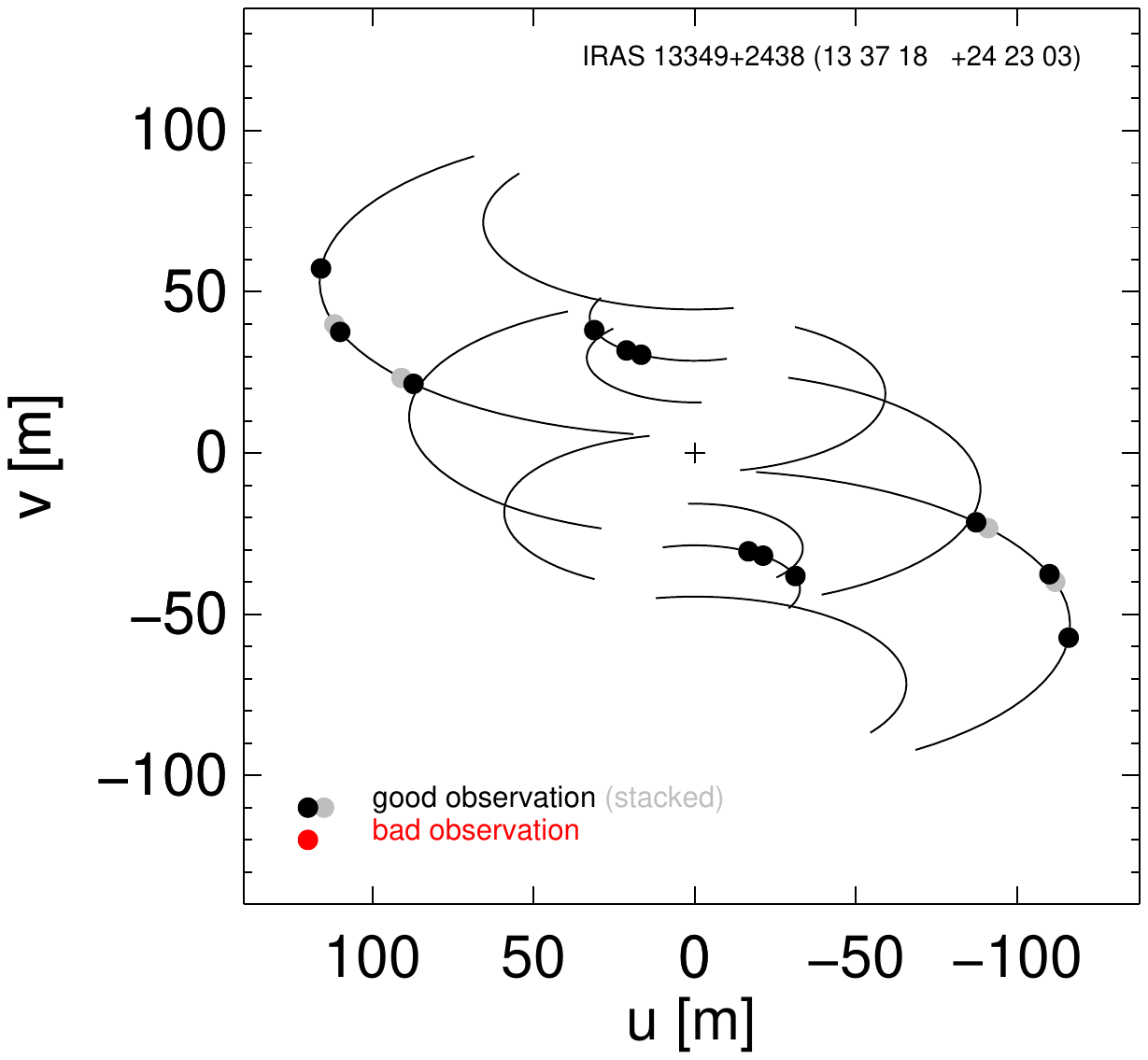}}
	~~~~~~~~~
	\subfloat{\includegraphics[trim=7cm 4cm 7cm 4cm, width=0.5\hsize]{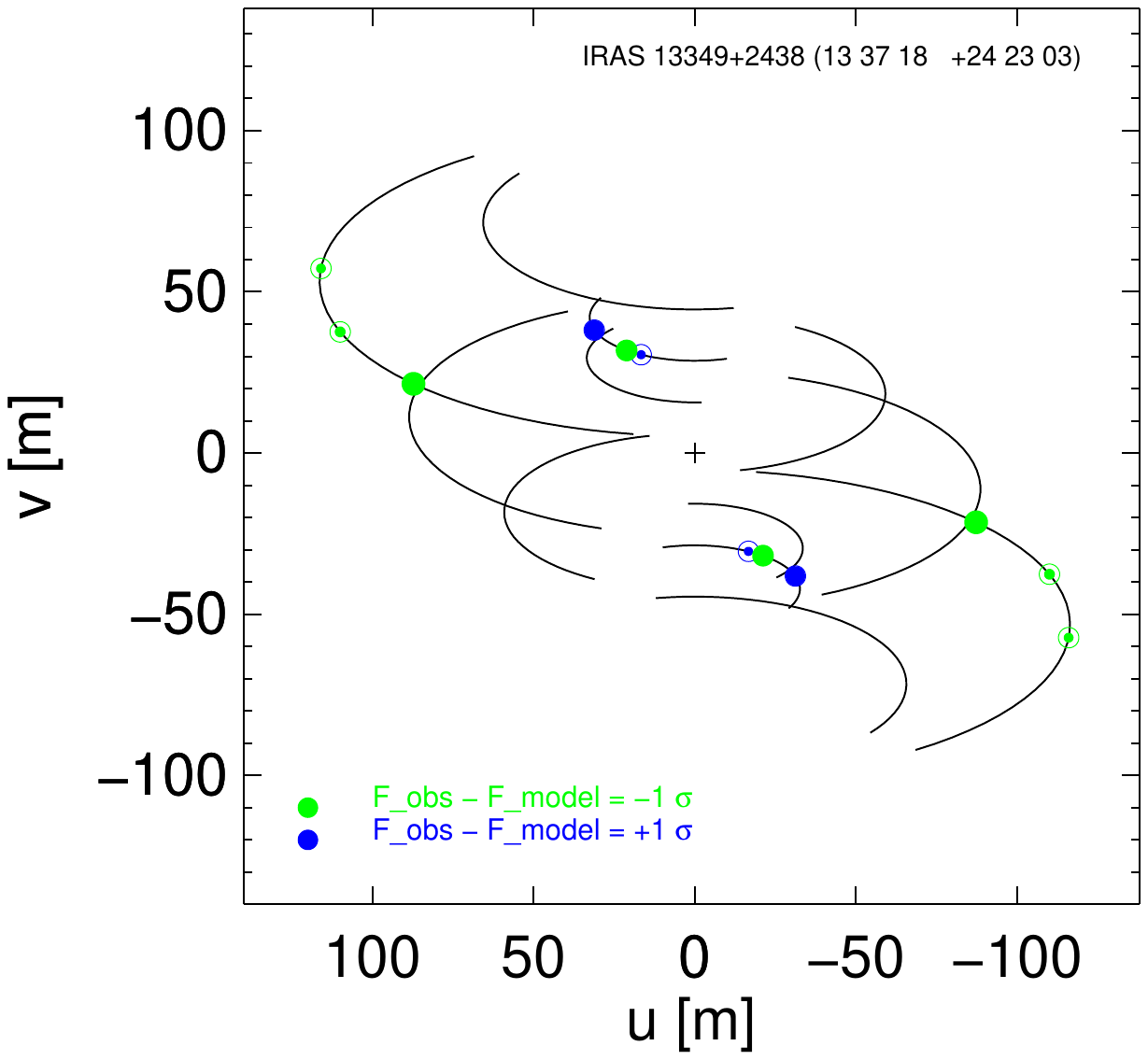}}\\
	\subfloat{\includegraphics[trim=3cm 0cm 3cm 0cm, width=0.5\hsize]{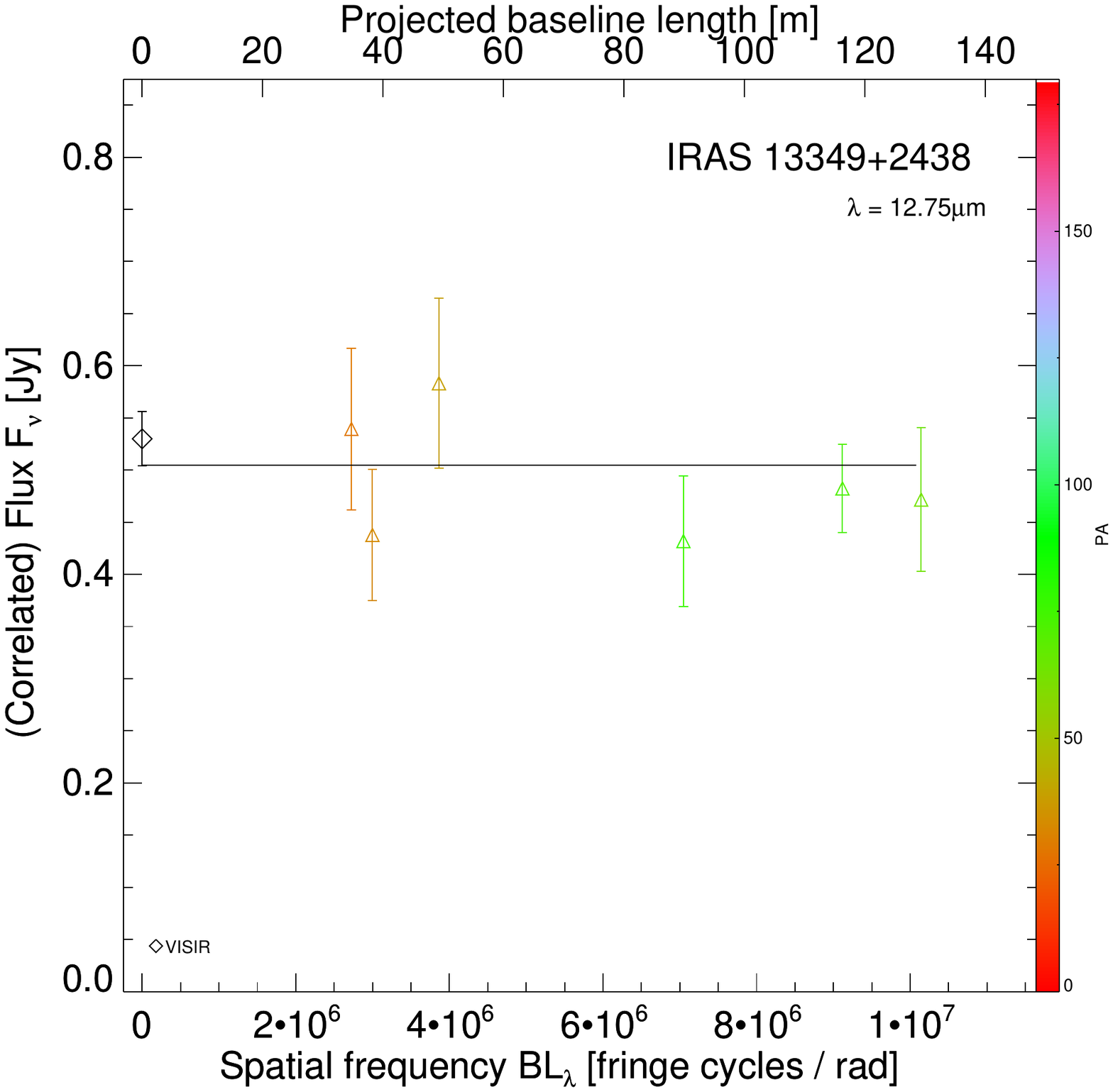}}
	~~~~~~~~~
	\subfloat{\includegraphics[trim=3cm 0cm 3cm 0cm, width=0.5\hsize]{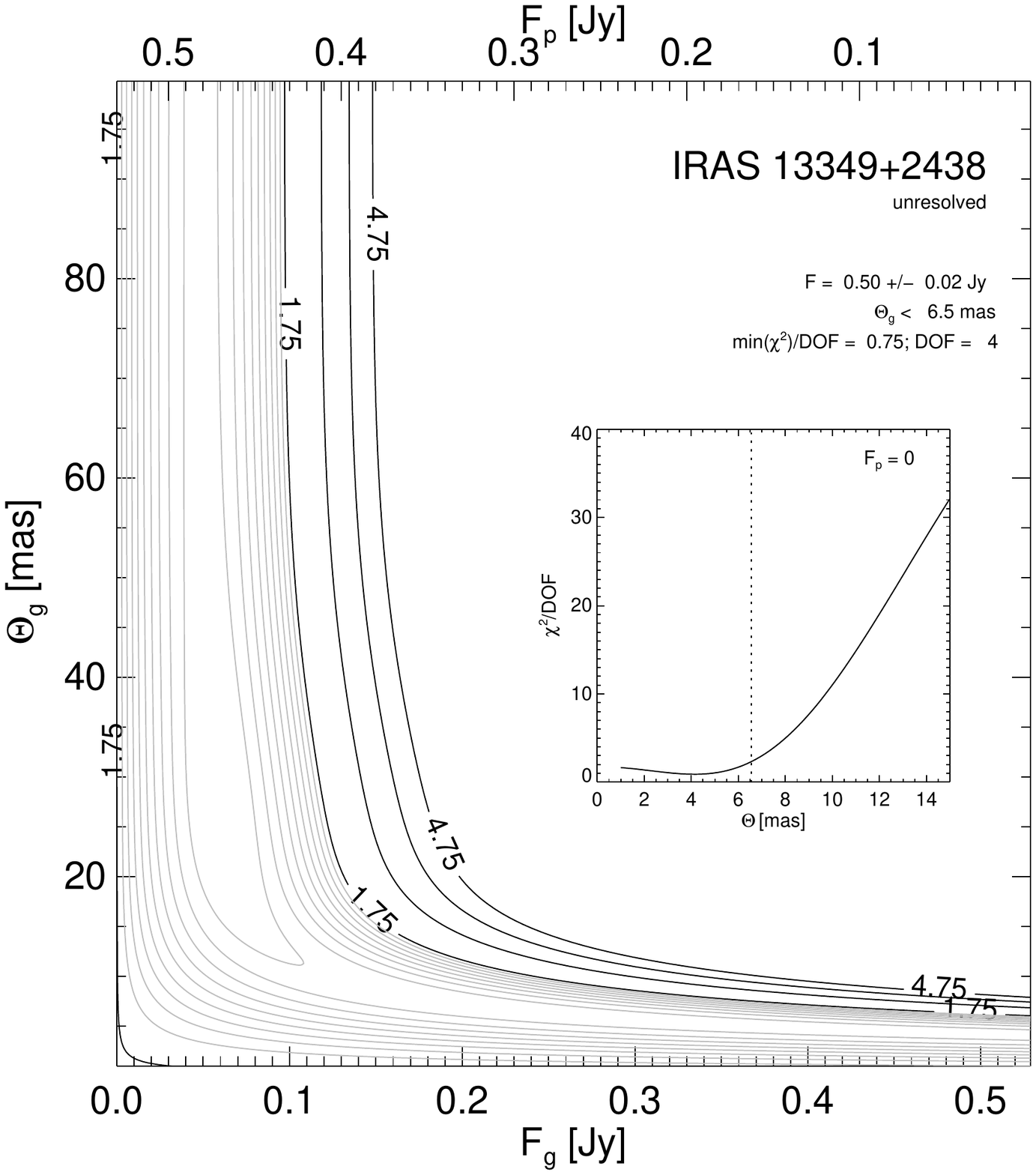}}
	\caption{\label{fig:rad:IRAS13349+2438}The same as Fig. \ref{fig:rad:IZwicky1} but for IRAS~13349+2438}
\end{figure*}

\begin{figure*}
	\centering
	\subfloat{\includegraphics[trim=7cm 4cm 7cm 4cm, width=0.5\hsize]{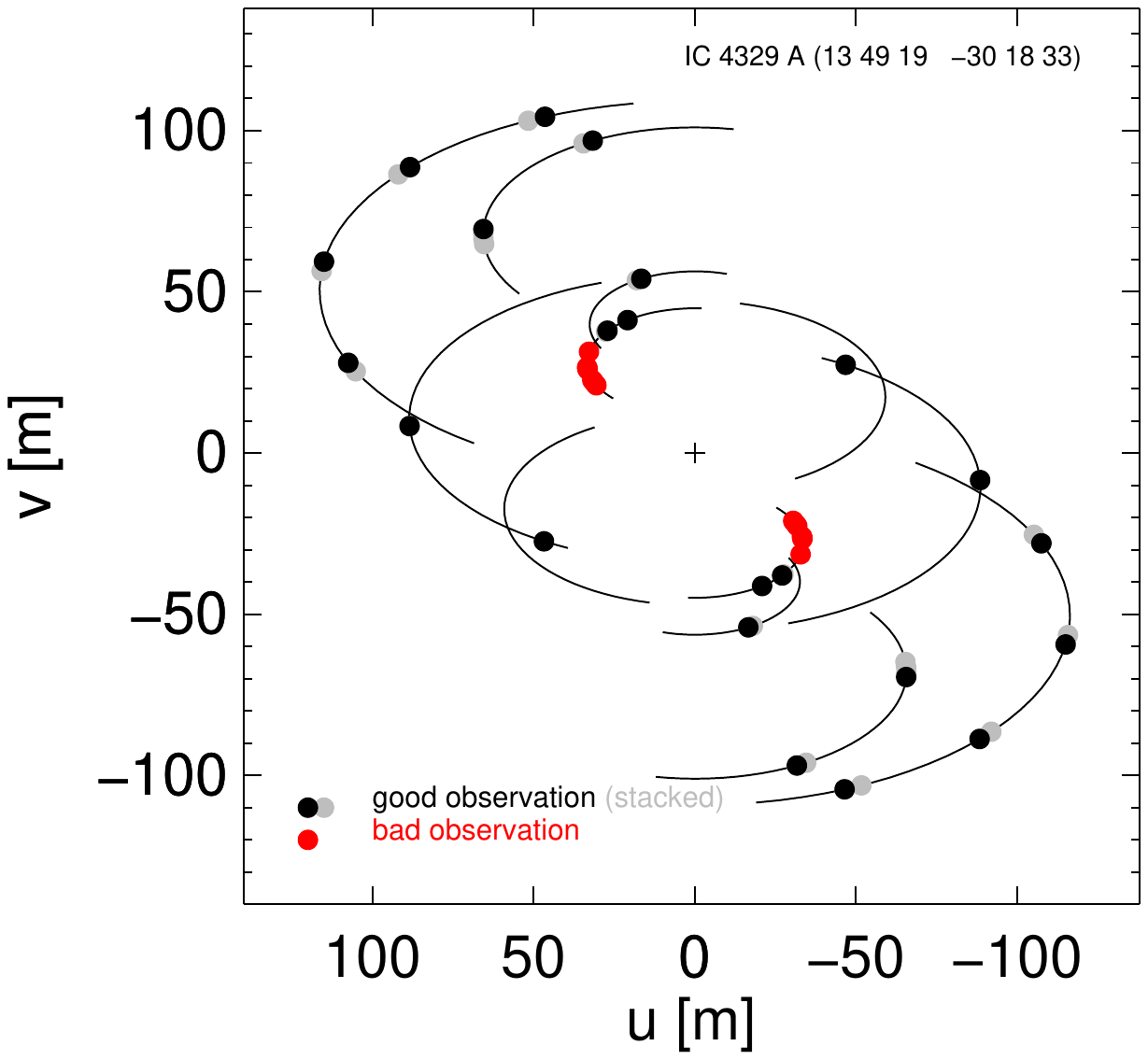}}
	~~~~~~~~~
	\subfloat{\includegraphics[trim=7cm 4cm 7cm 4cm, width=0.5\hsize]{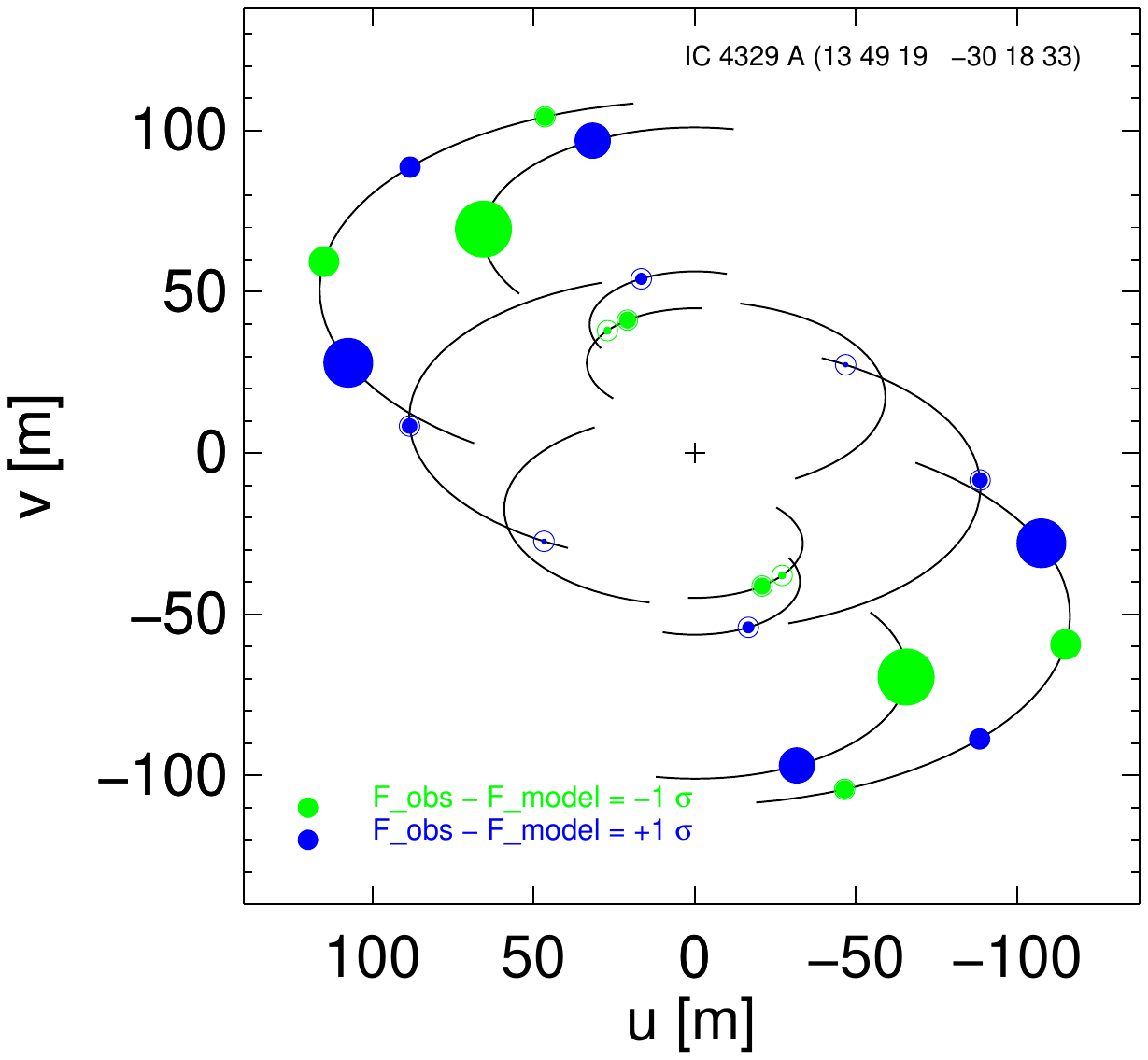}}\\
	\subfloat{\includegraphics[trim=3cm 0cm 3cm 0cm, width=0.5\hsize]{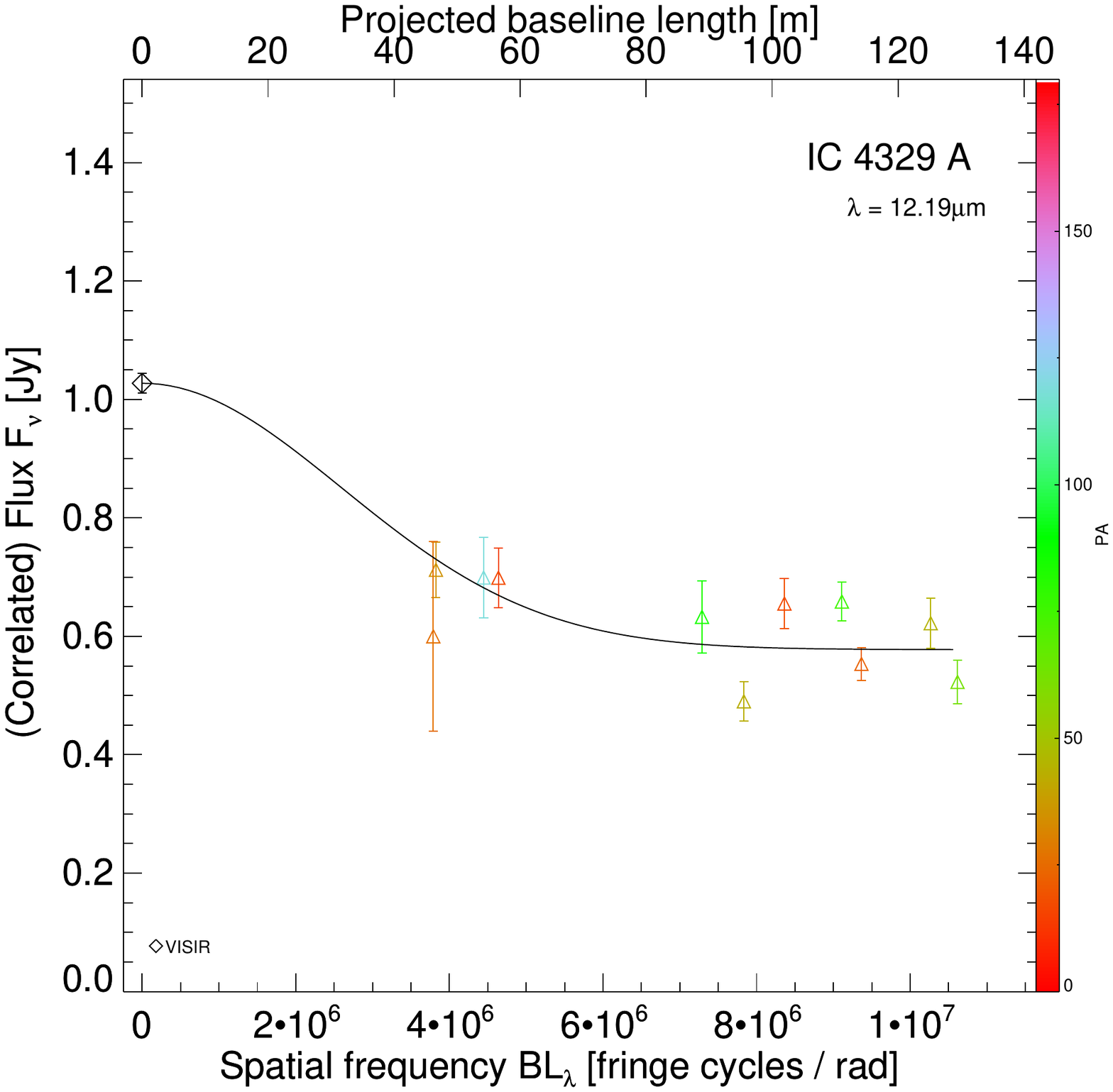}}
	~~~~~~~~~
	\subfloat{\includegraphics[trim=3cm 0cm 3cm 0cm, width=0.5\hsize]{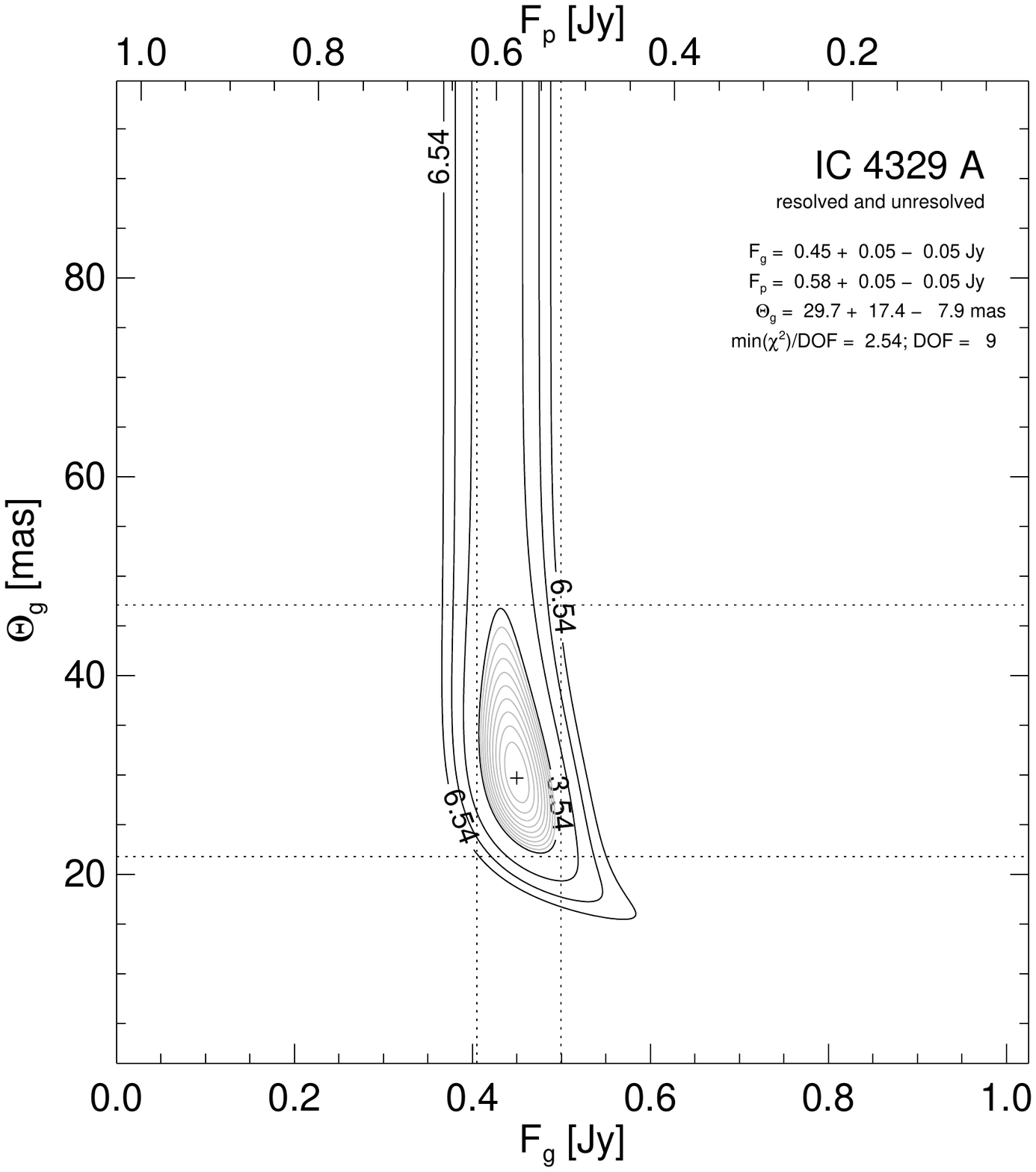}}
	\caption{\label{fig:rad:IC4329A}The same as Fig. \ref{fig:rad:IZwicky1} but for IC~4329~A}
\end{figure*}

\begin{figure*}
	\centering
	\subfloat{\includegraphics[trim=7cm 4cm 7cm 4cm, width=0.5\hsize]{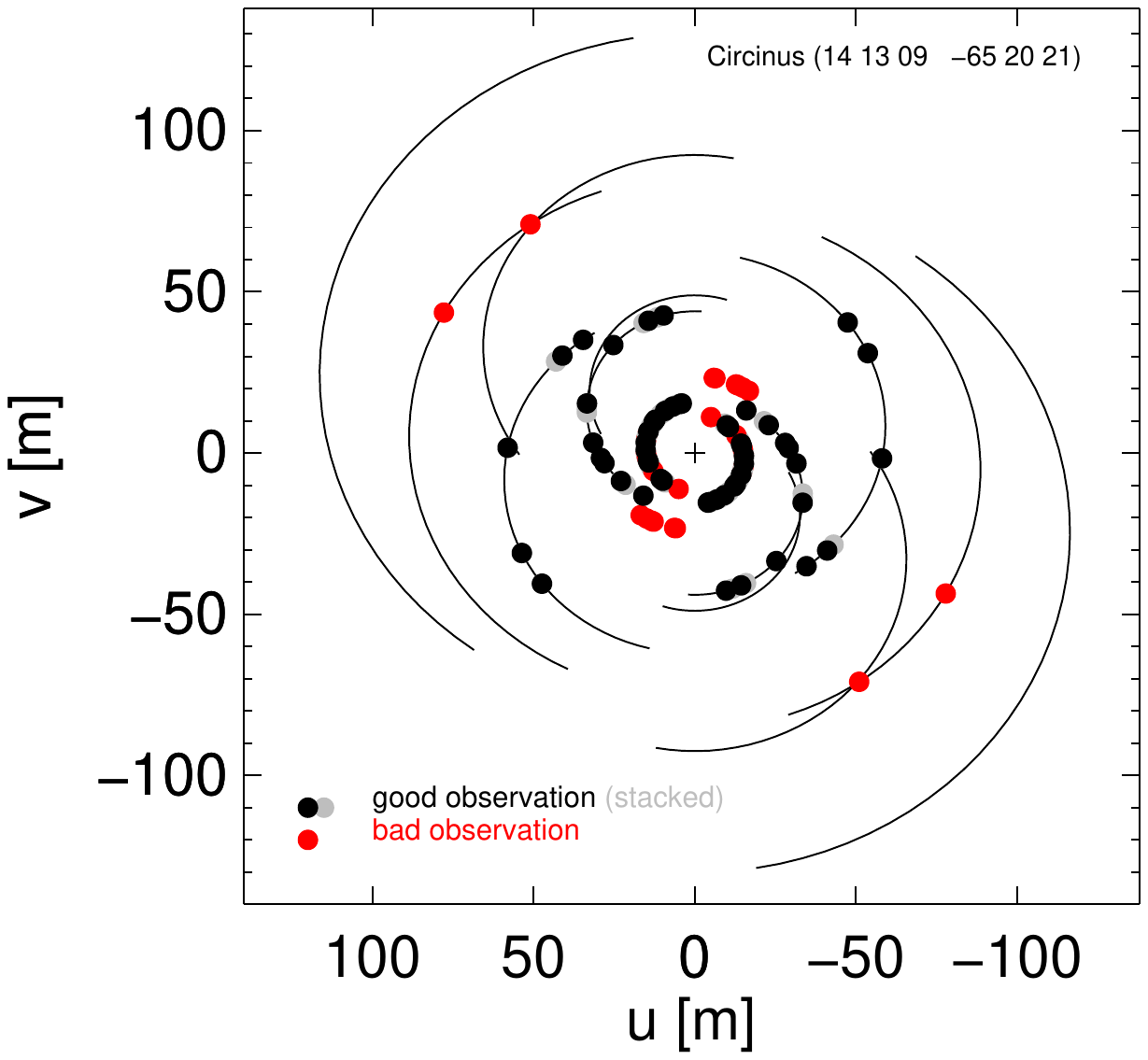}}
	~~~~~~~~~
	\subfloat{\includegraphics[trim=7cm 4cm 7cm 4cm, width=0.5\hsize]{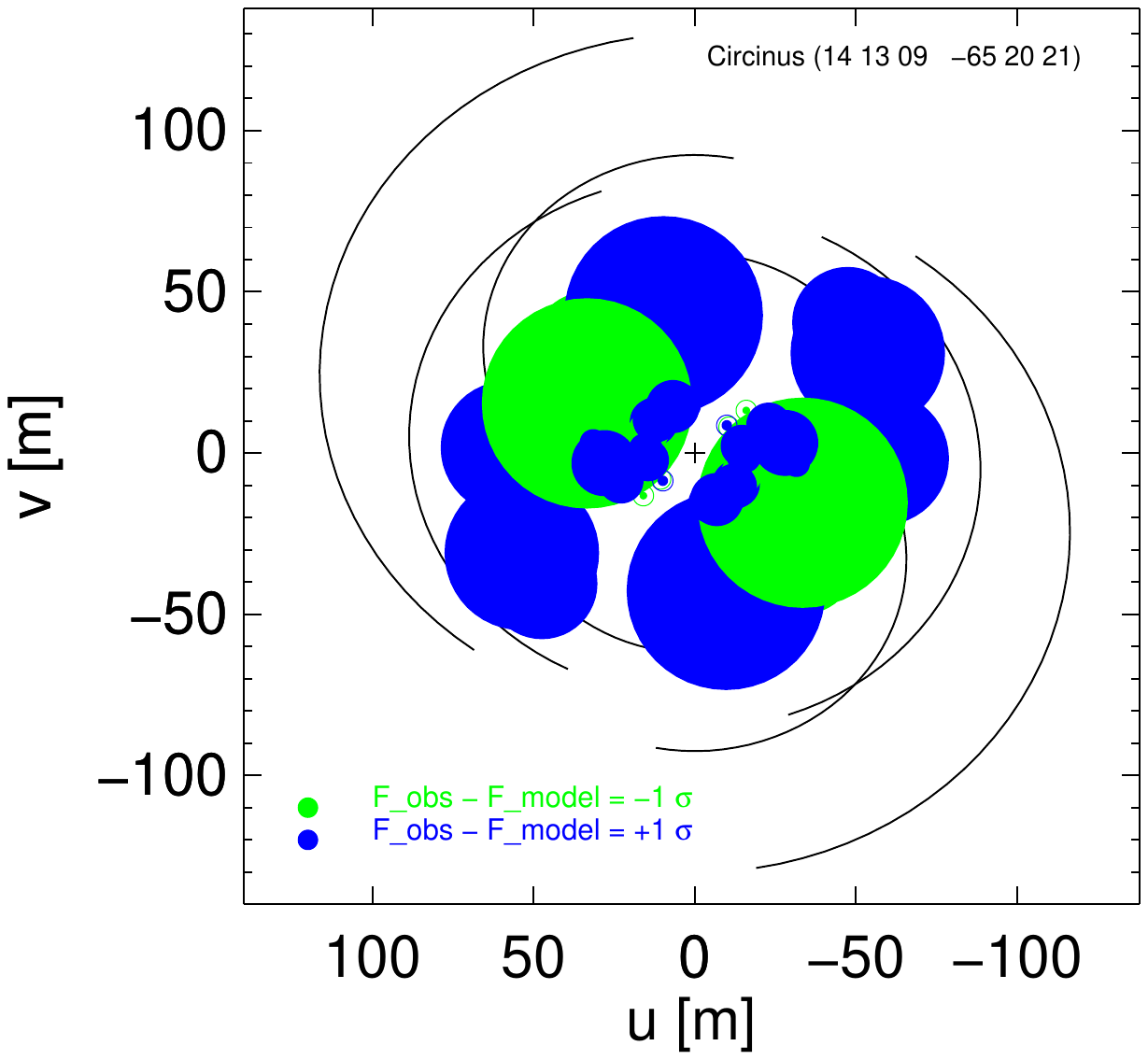}}\\
	\subfloat{\includegraphics[trim=3cm 0cm 3cm 0cm, width=0.5\hsize]{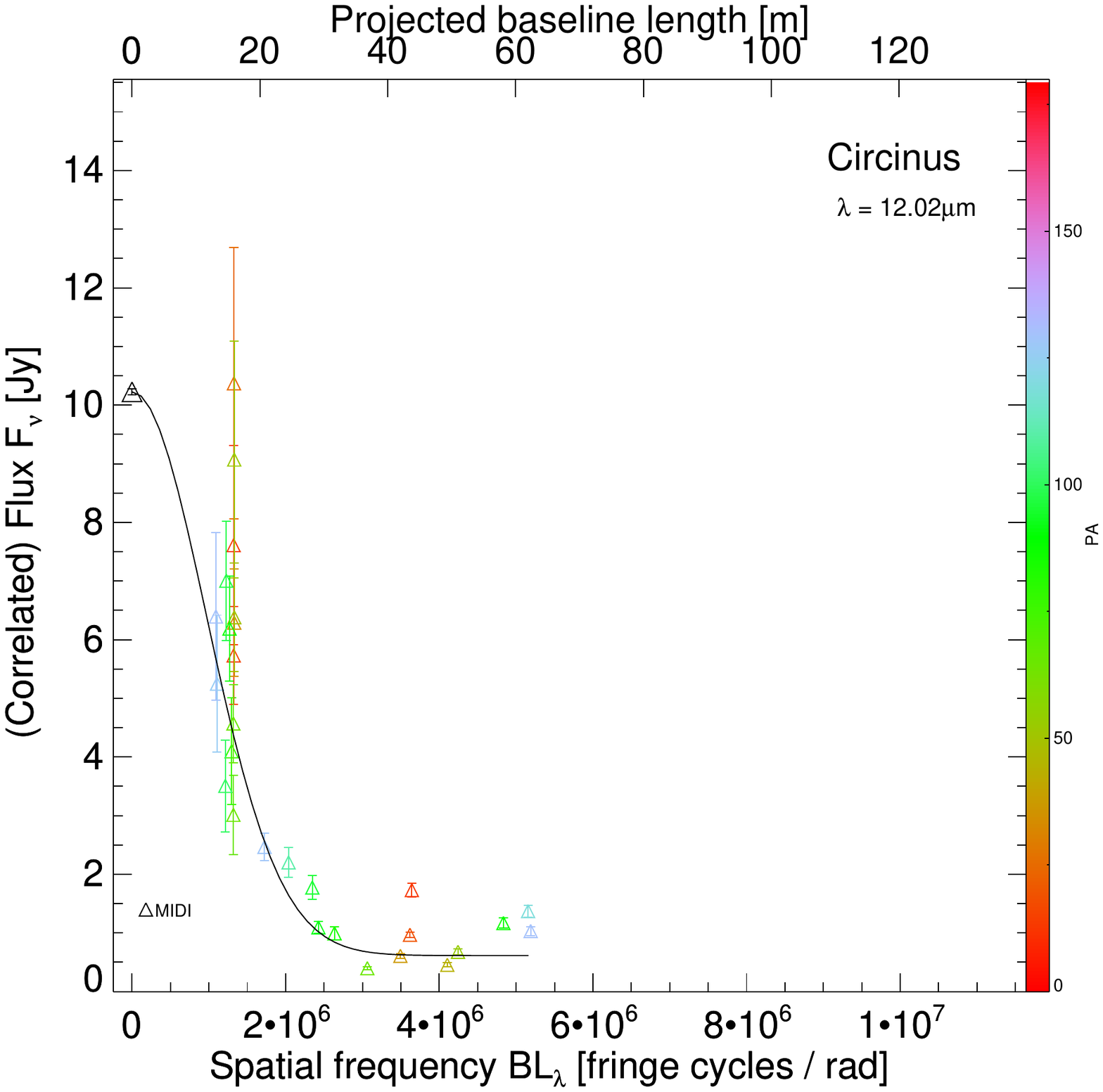}}
	~~~~~~~~~
	\subfloat{\includegraphics[trim=3cm 0cm 3cm 0cm, width=0.5\hsize]{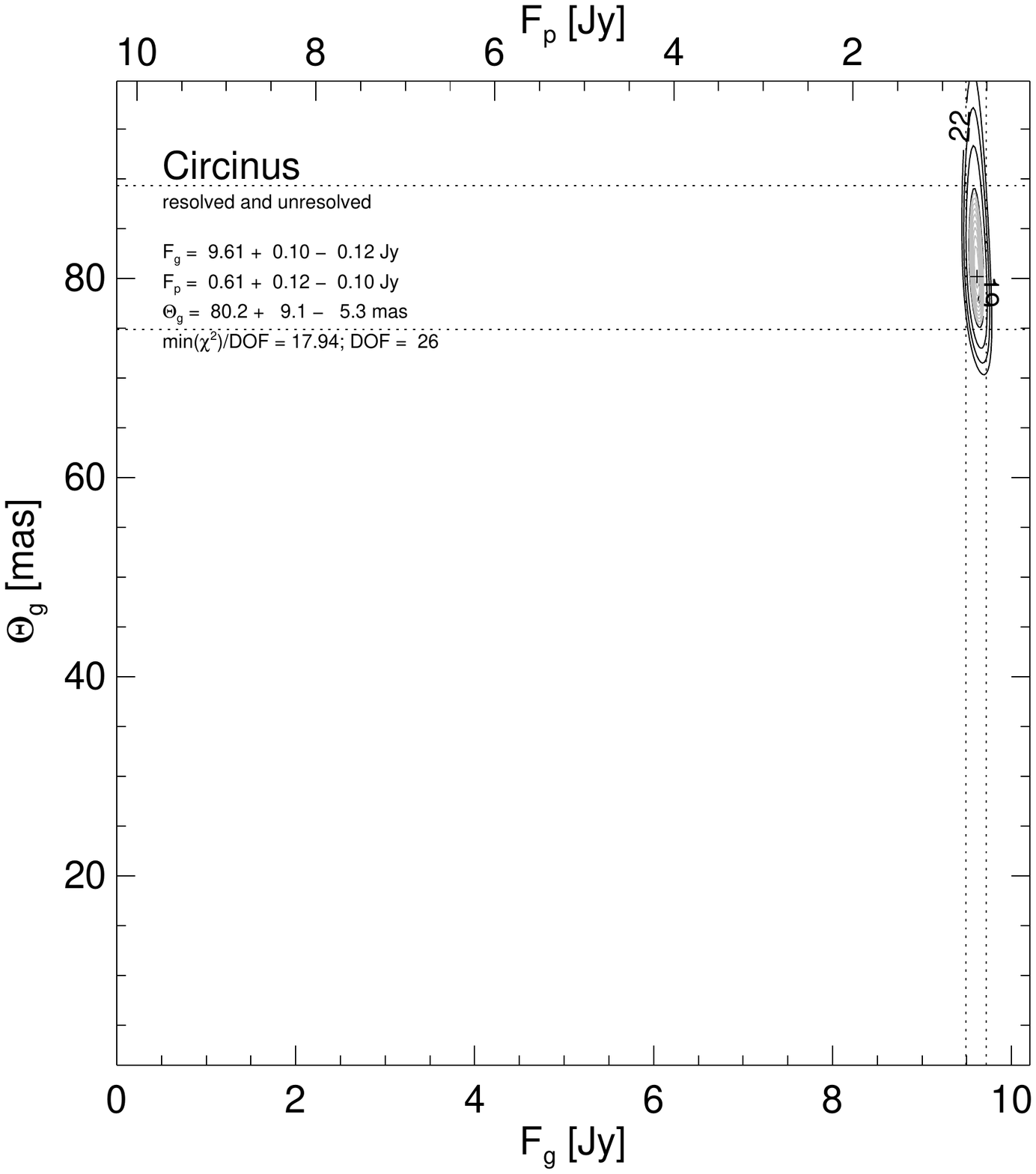}}
	\caption{\label{fig:rad:Circinus}The same as Fig. \ref{fig:rad:IZwicky1} but for Circinus. The very large value for the reduced $\chi^2$ is due to the elongation of both the extended and the compact component \citep{tristram2012} that was not fitted here.}
\end{figure*}

\begin{figure*}
	\centering
	\subfloat{\includegraphics[trim=7cm 4cm 7cm 4cm, width=0.5\hsize]{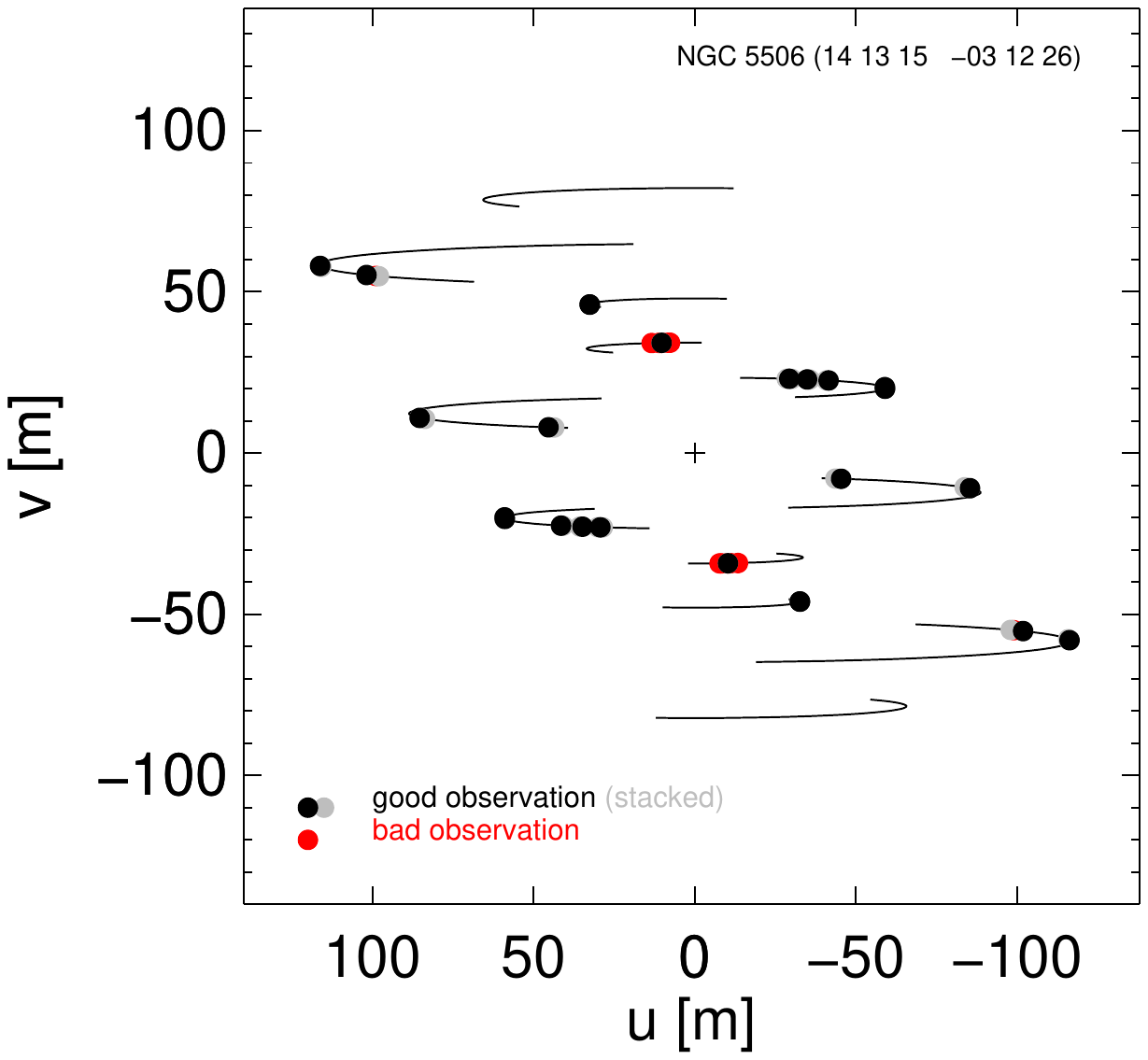}}
	~~~~~~~~~
	\subfloat{\includegraphics[trim=7cm 4cm 7cm 4cm, width=0.5\hsize]{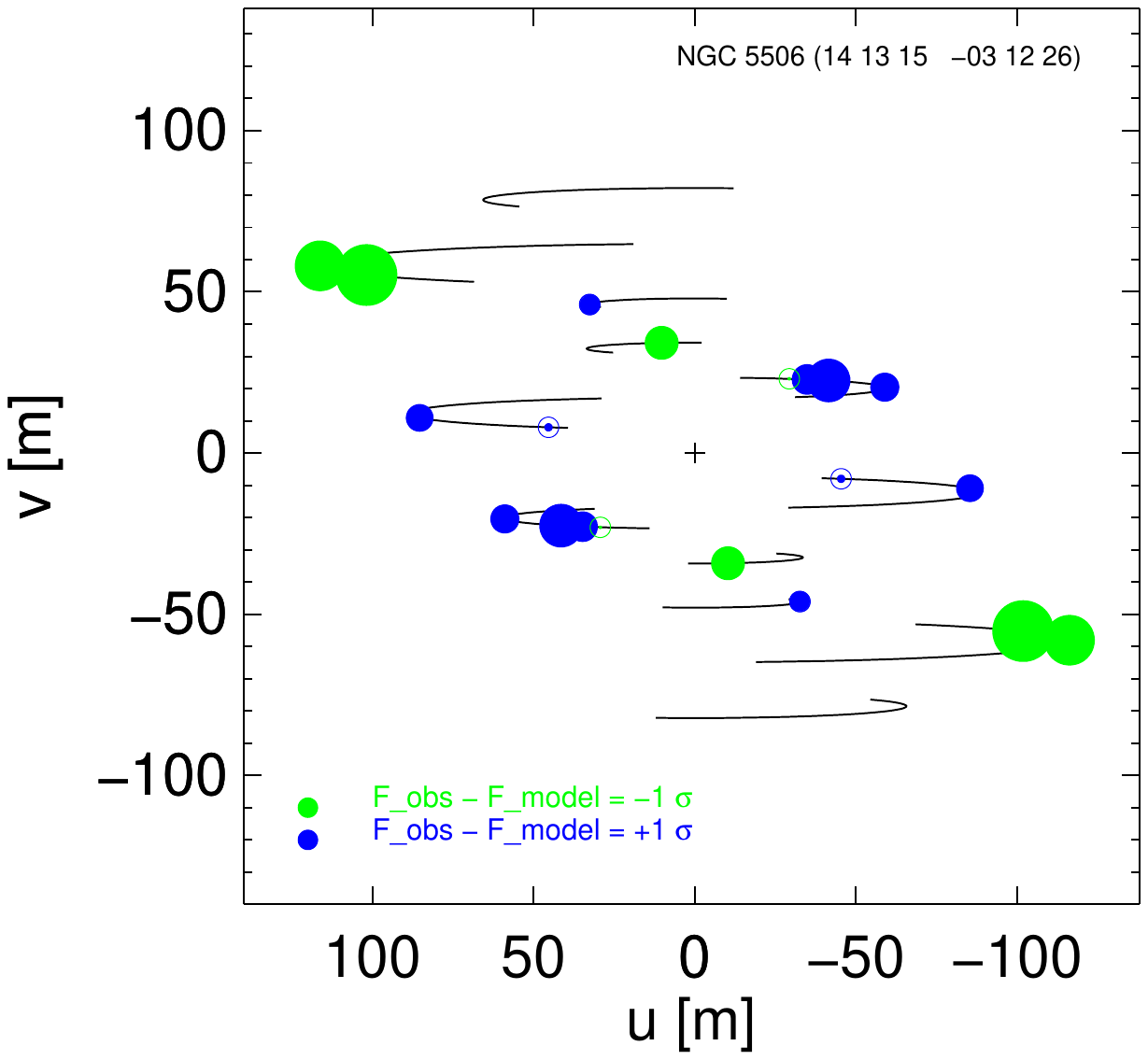}}\\
	\subfloat{\includegraphics[trim=3cm 0cm 3cm 0cm, width=0.5\hsize]{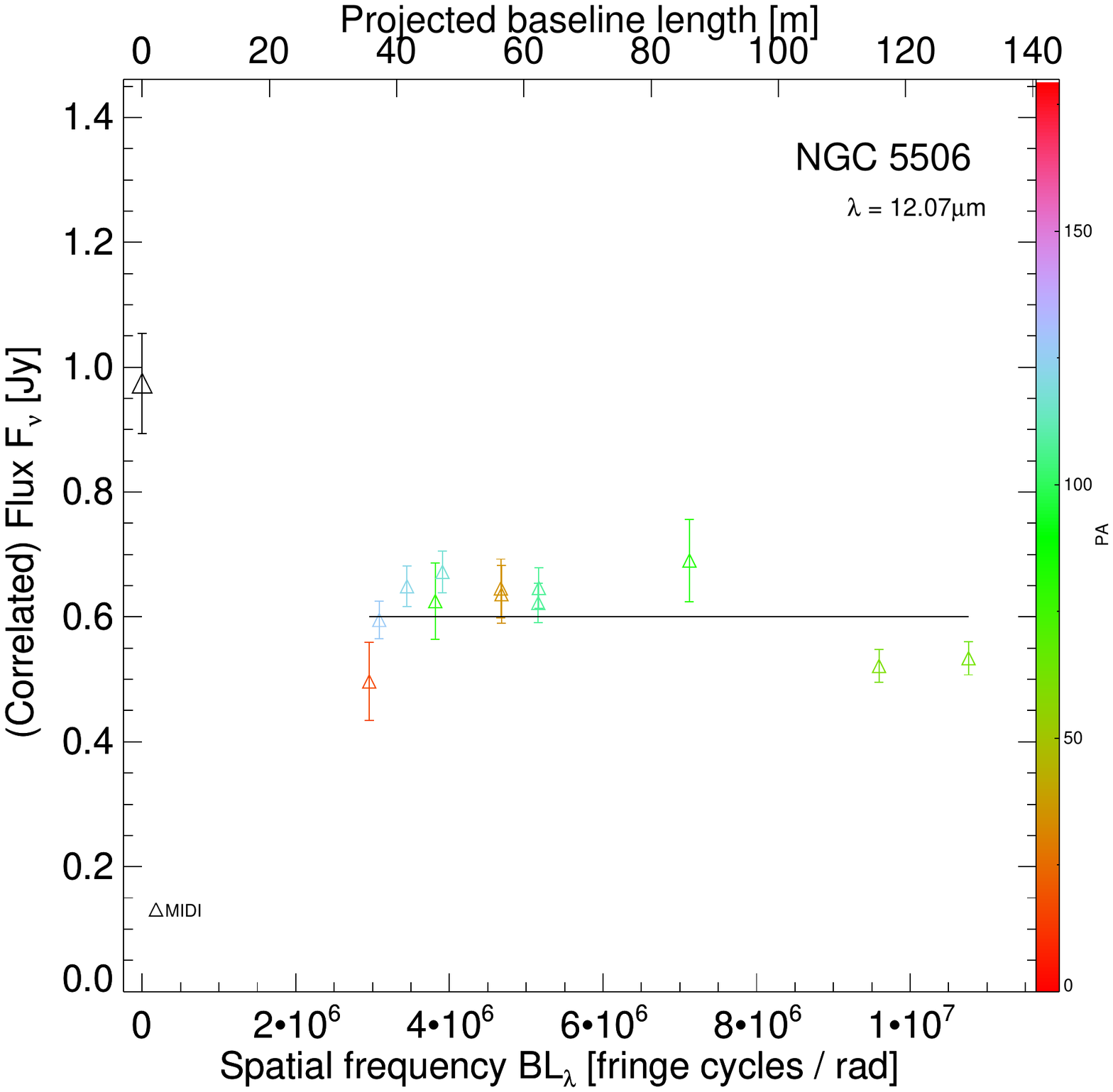}}
	~~~~~~~~~
	\subfloat{\includegraphics[trim=3cm 0cm 3cm 0cm, width=0.5\hsize]{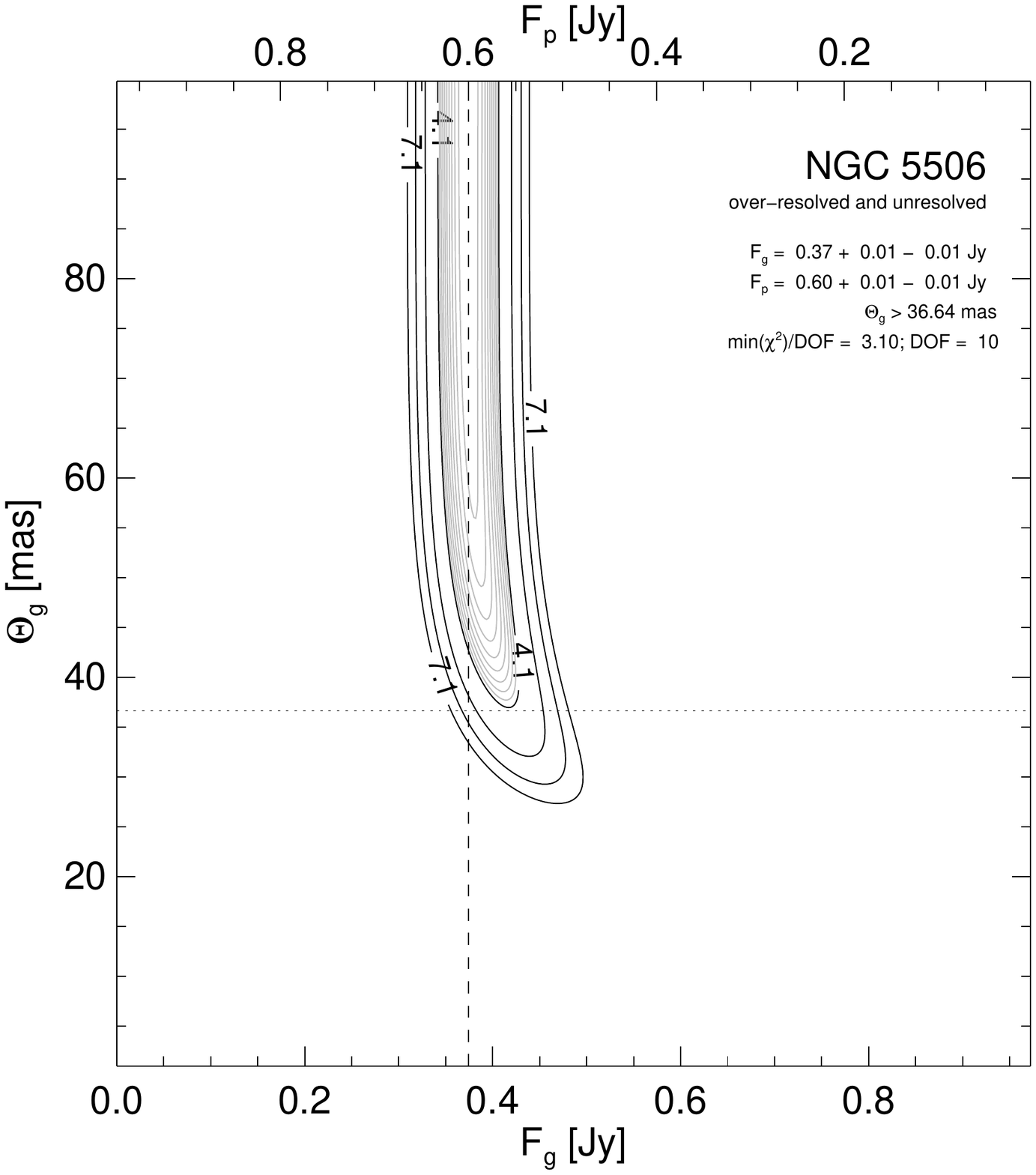}}
	\caption{\label{fig:rad:NGC5506}The same as Fig. \ref{fig:rad:IZwicky1} but for NGC~5506}
\end{figure*}

\begin{figure*}
	\centering
	\subfloat{\includegraphics[trim=7cm 4cm 7cm 4cm, width=0.5\hsize]{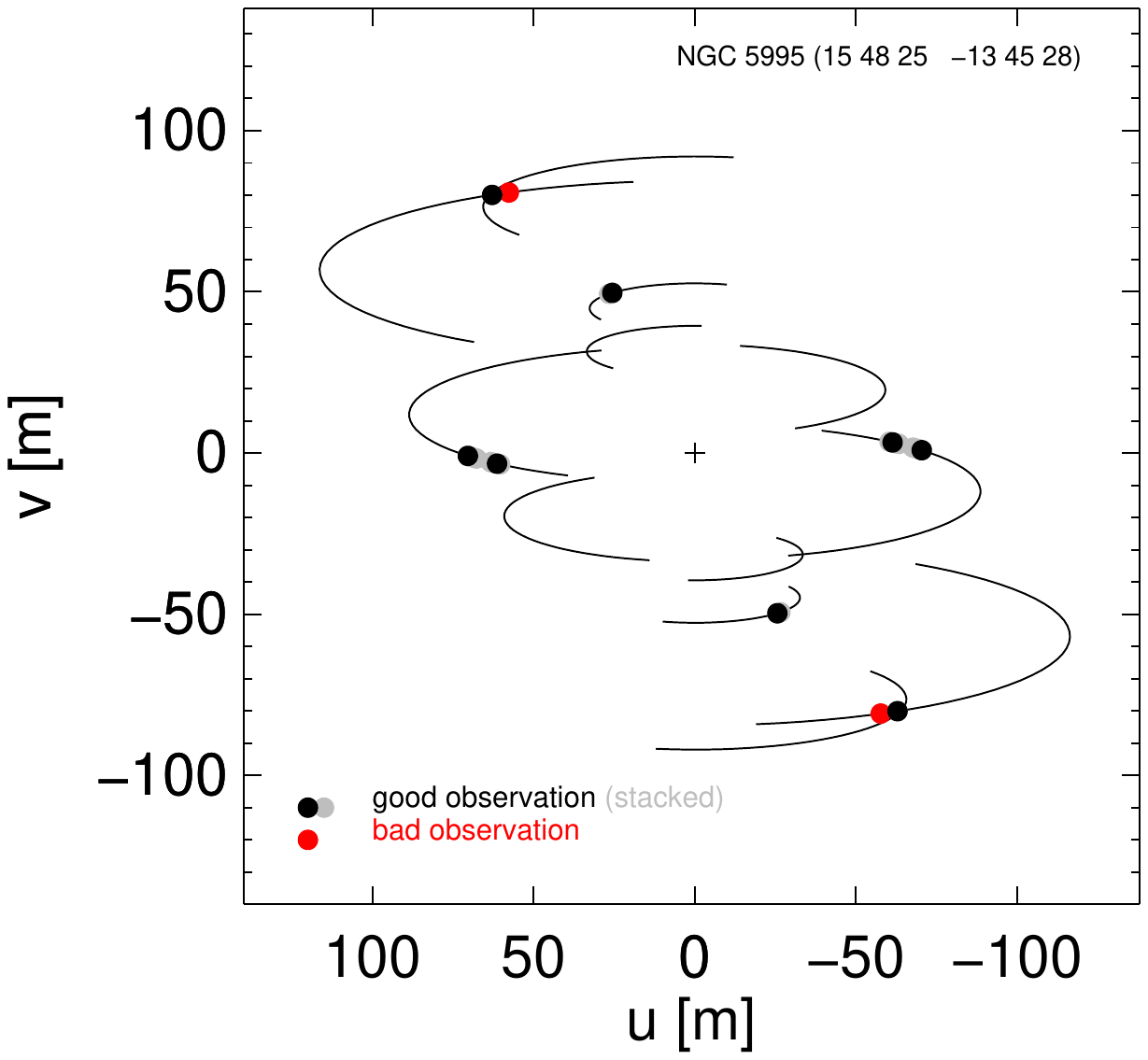}}
	~~~~~~~~~
	\subfloat{\includegraphics[trim=7cm 4cm 7cm 4cm, width=0.5\hsize]{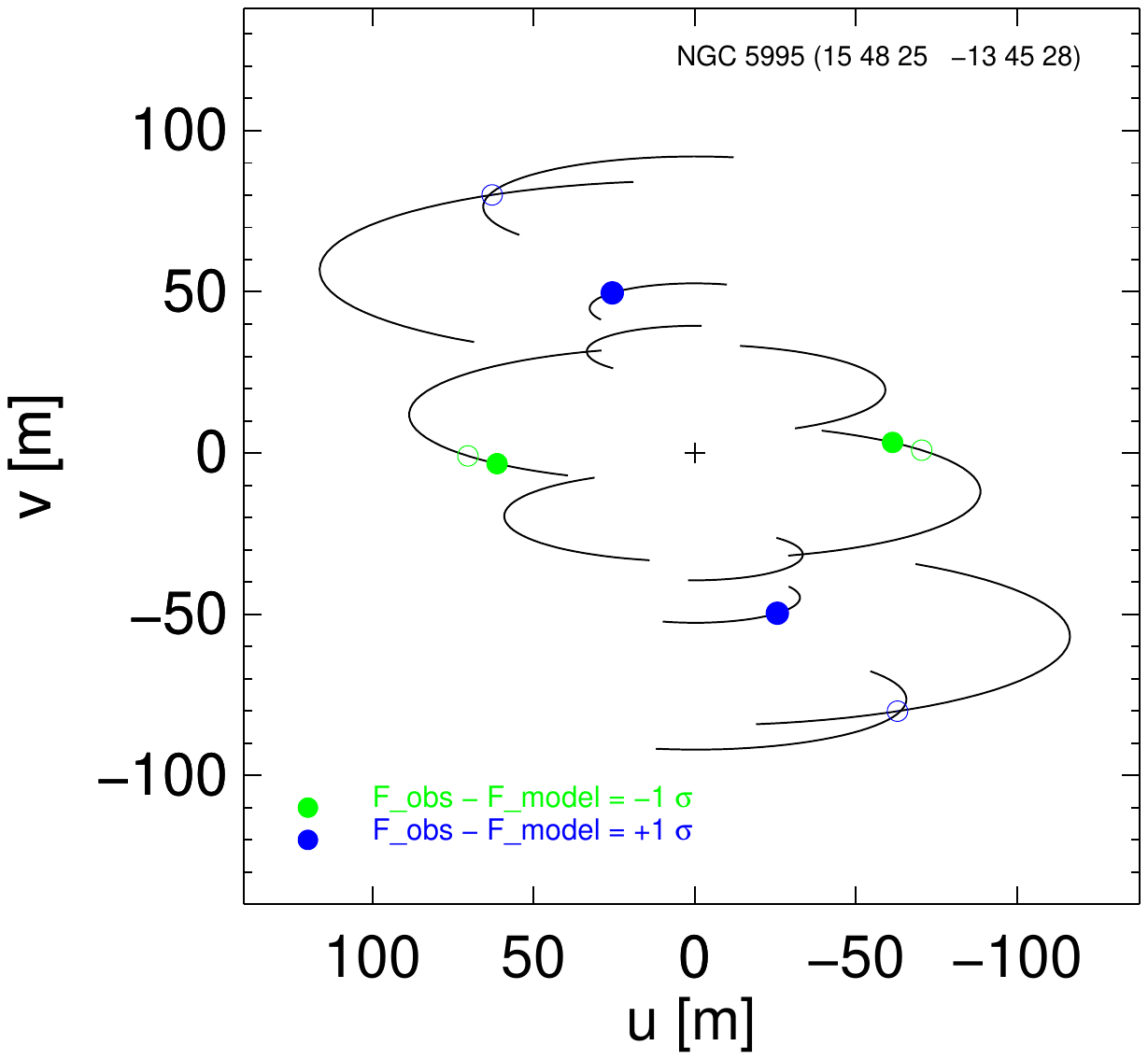}}\\
	\subfloat{\includegraphics[trim=3cm 0cm 3cm 0cm, width=0.5\hsize]{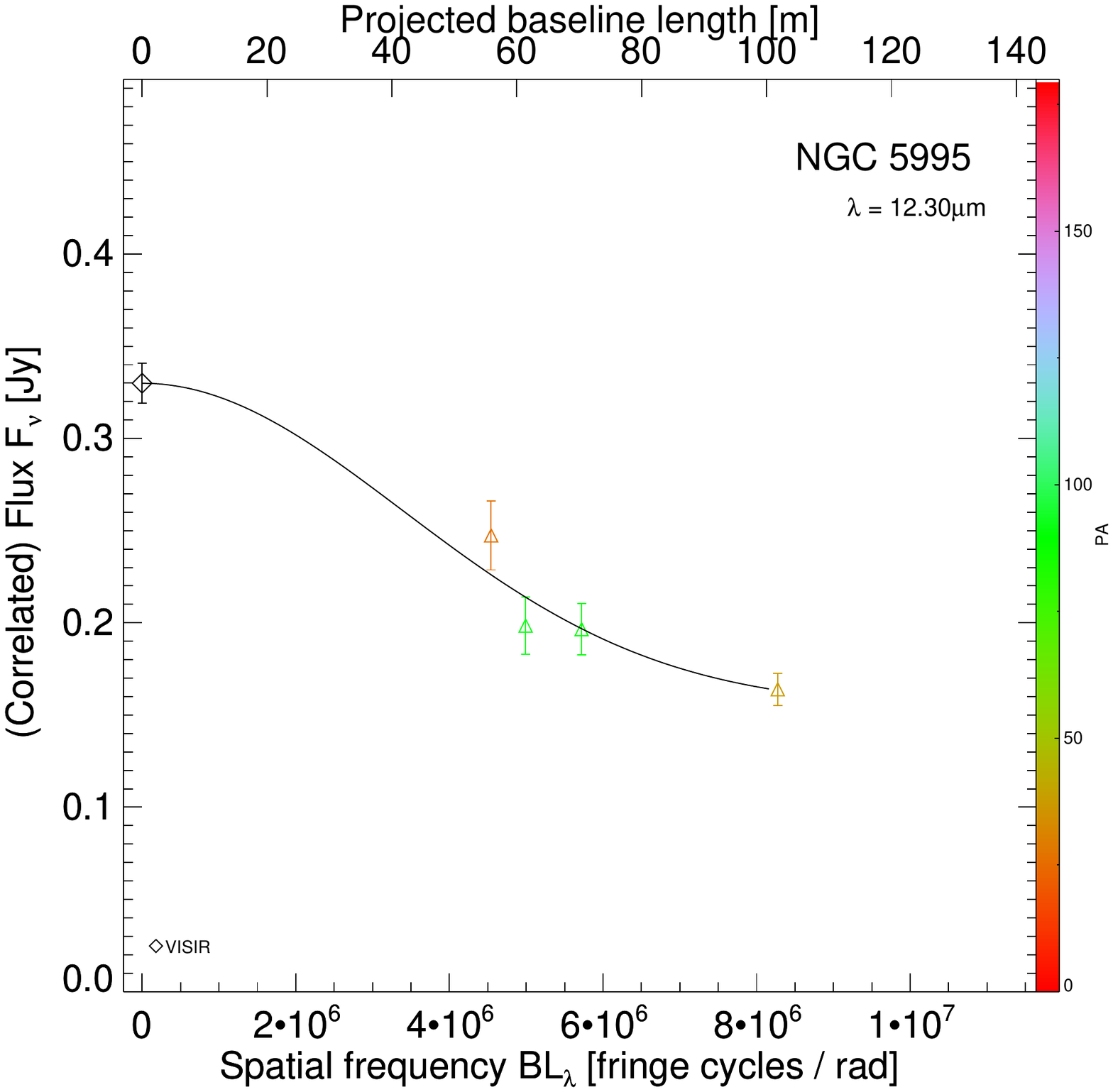}}
	~~~~~~~~~
	\subfloat{\includegraphics[trim=3cm 0cm 3cm 0cm, width=0.5\hsize]{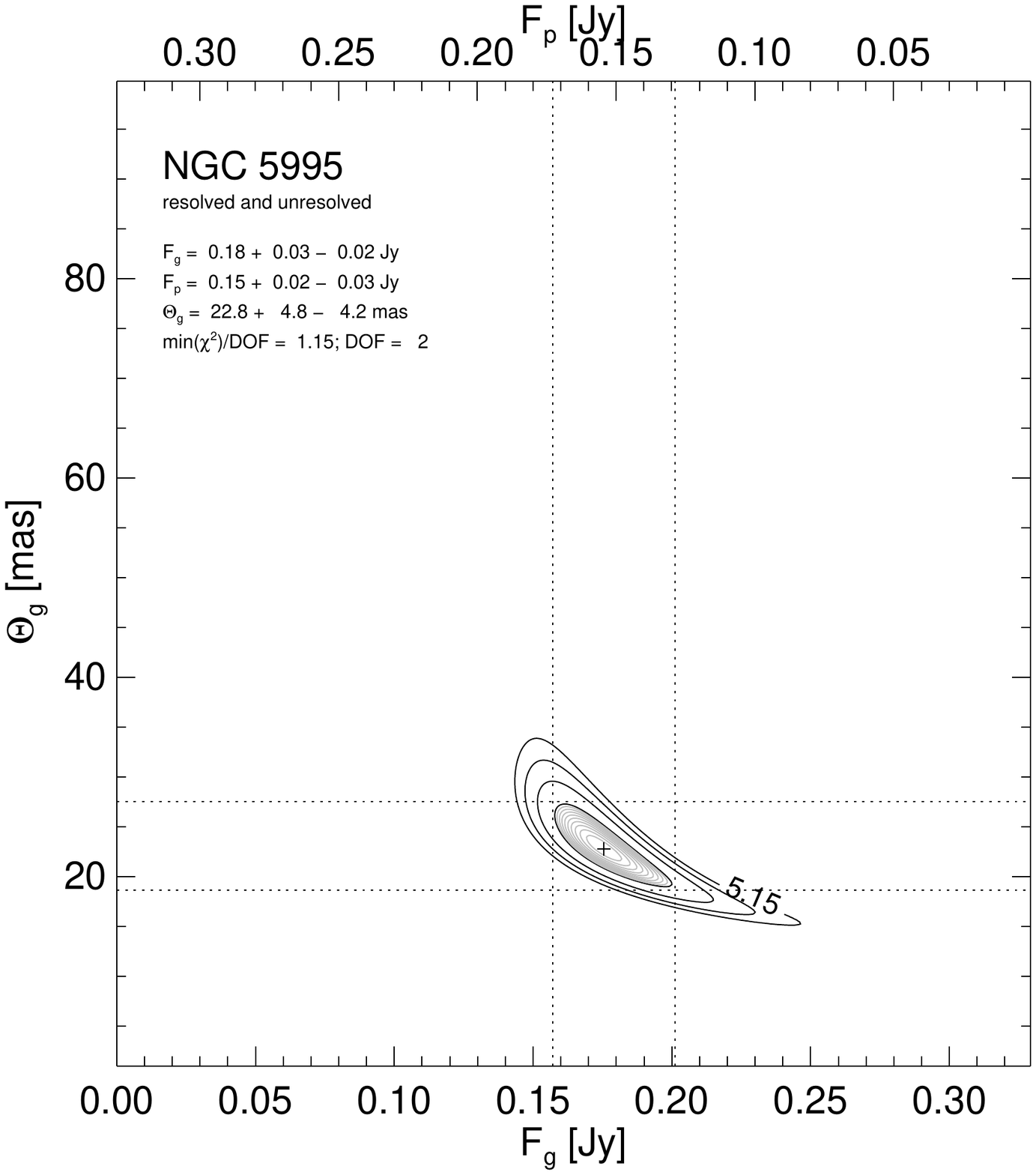}}
	\caption{\label{fig:rad:NGC5995}The same as Fig. \ref{fig:rad:IZwicky1} but for NGC~5995}
\end{figure*}

\begin{figure*}
	\centering
	\subfloat{\includegraphics[trim=7cm 4cm 7cm 4cm, width=0.5\hsize]{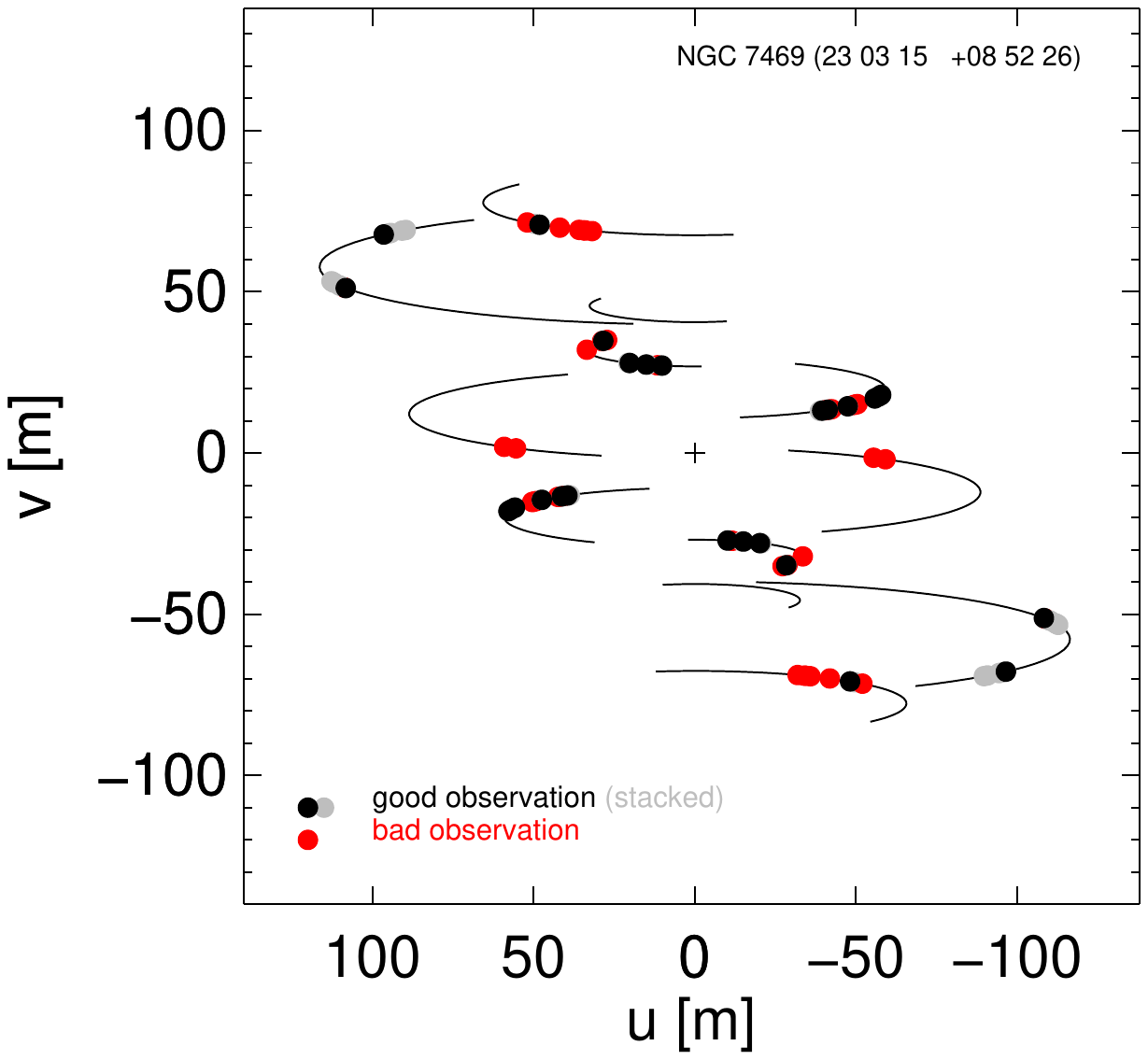}}
	~~~~~~~~~
	\subfloat{\includegraphics[trim=7cm 4cm 7cm 4cm, width=0.5\hsize]{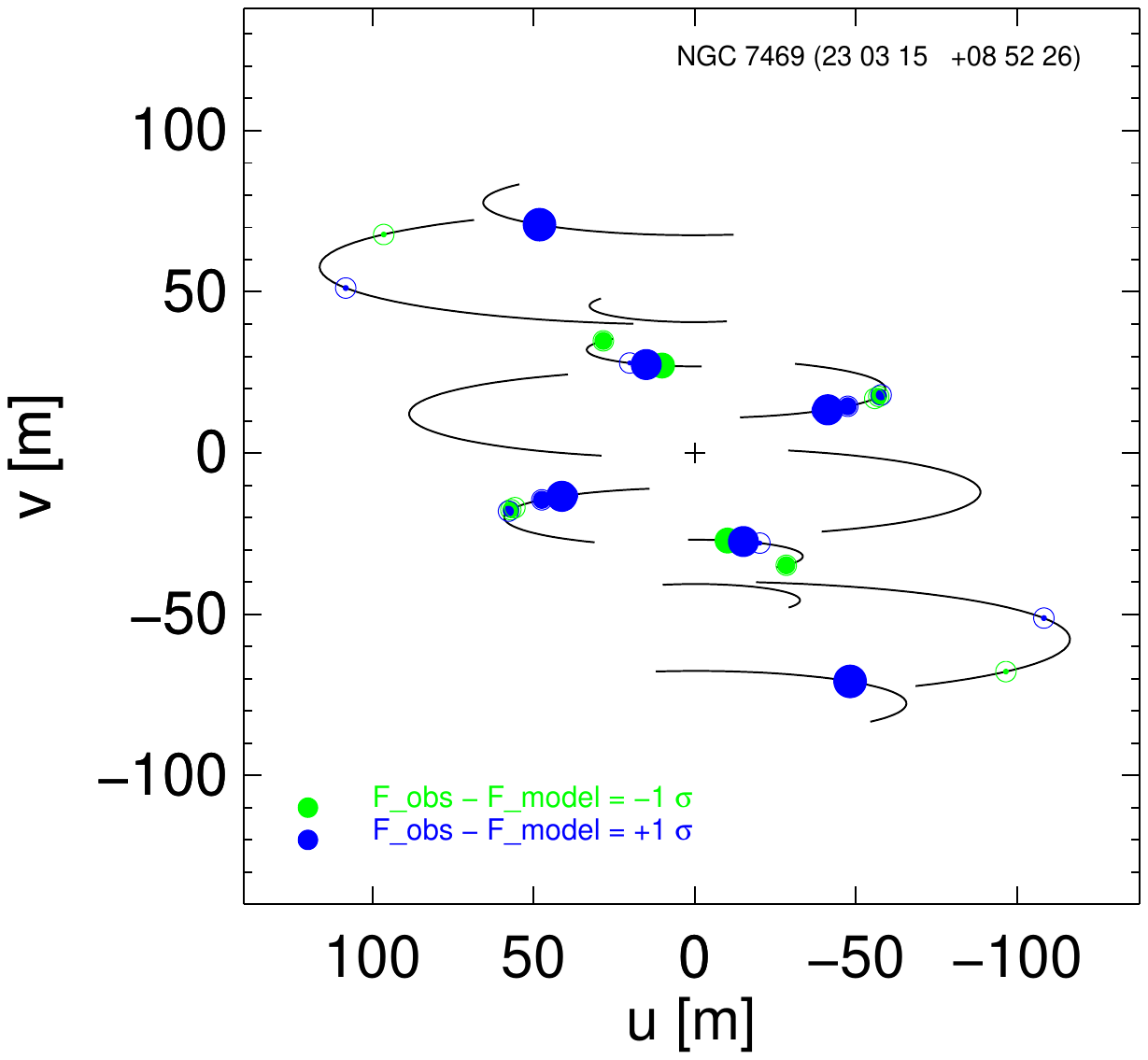}}\\
	\subfloat{\includegraphics[trim=3cm 0cm 3cm 0cm, width=0.5\hsize]{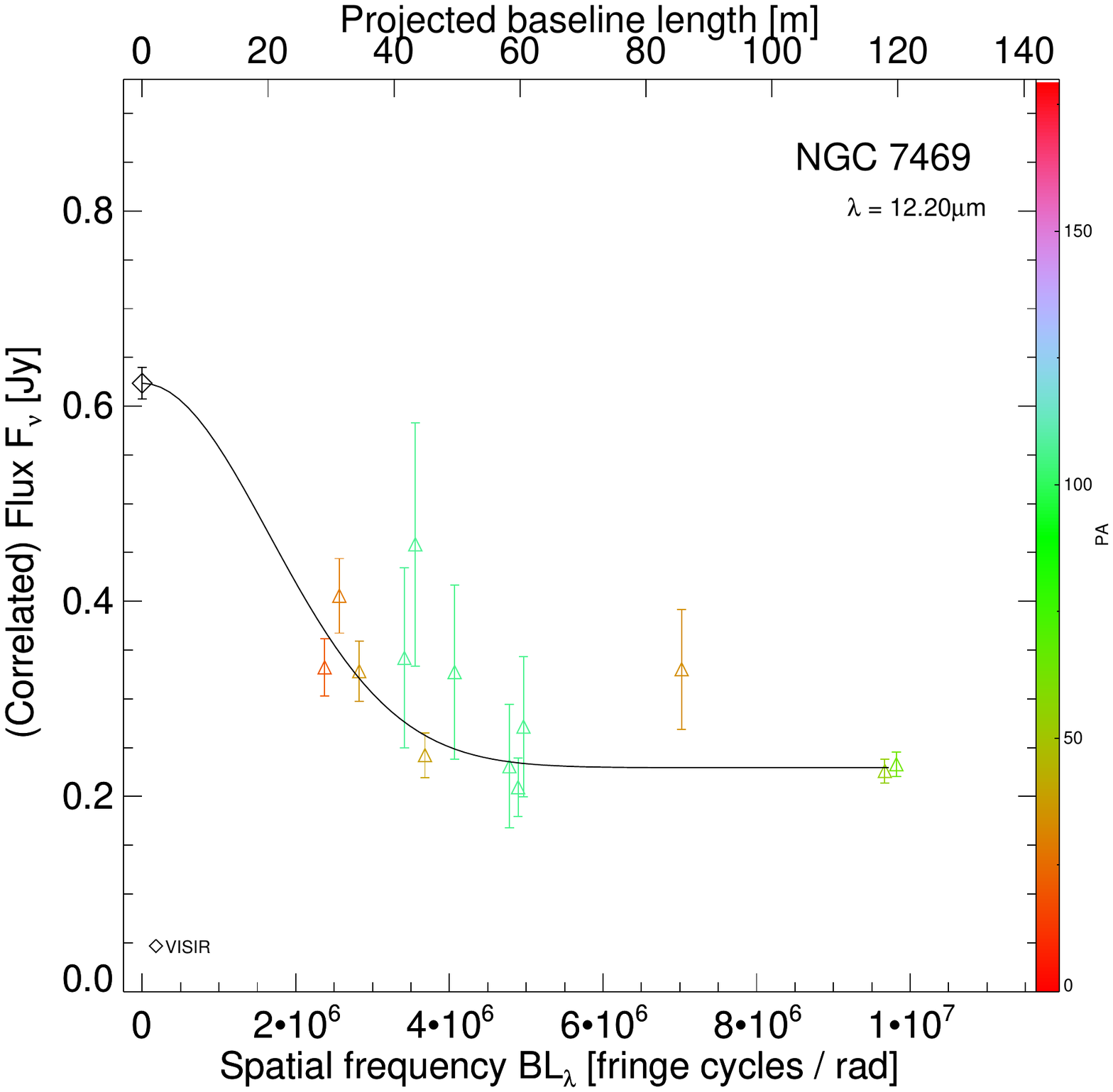}}
	~~~~~~~~~
	\subfloat{\includegraphics[trim=3cm 0cm 3cm 0cm, width=0.5\hsize]{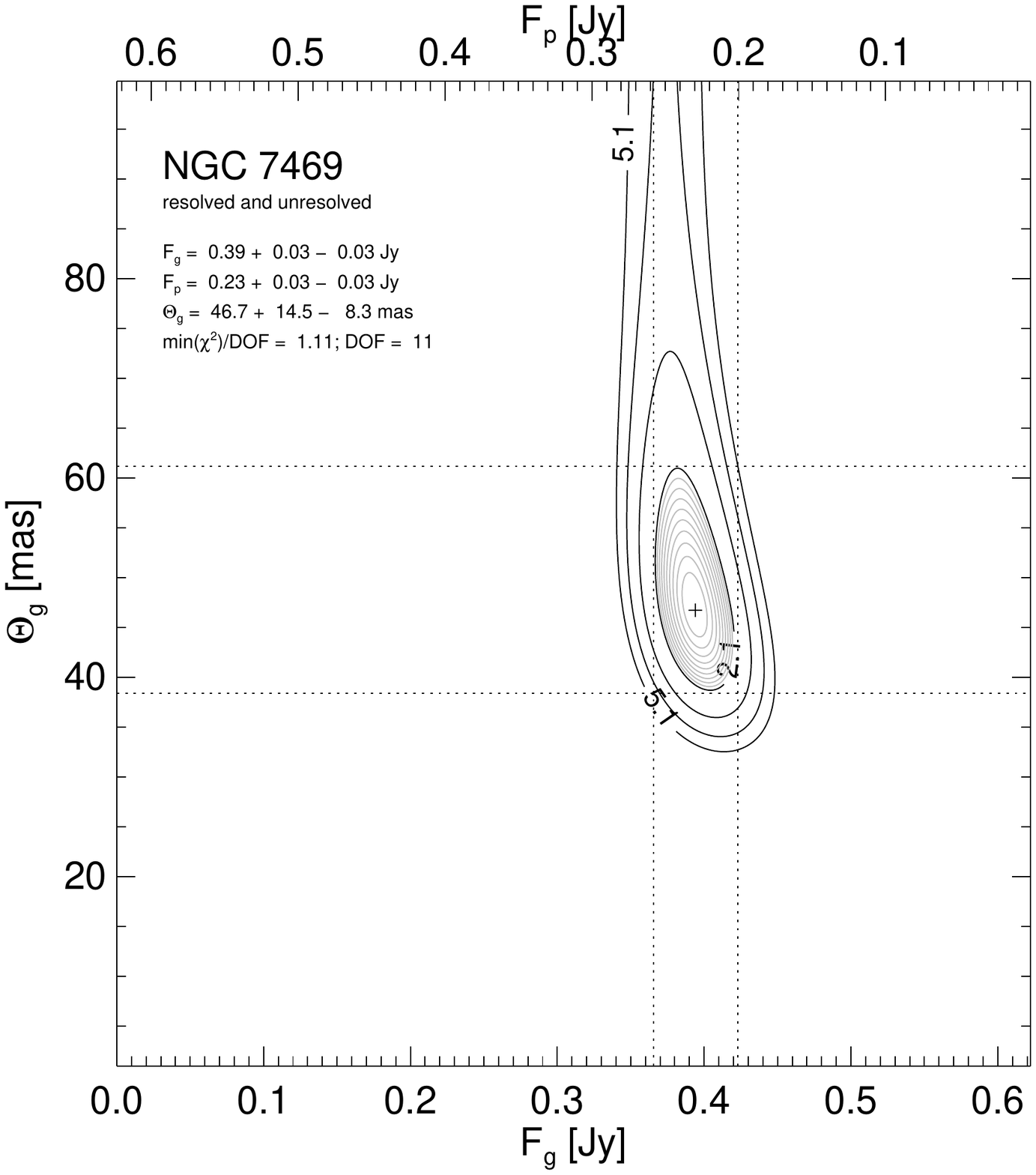}}
	\caption{\label{fig:rad:NGC7469}The same as Fig. \ref{fig:rad:IZwicky1} but for NGC~7469}
\end{figure*}

\begin{figure*}
\subfloat{\includegraphics[trim=5cm 0cm 5cm 0cm, width=0.17\hsize]{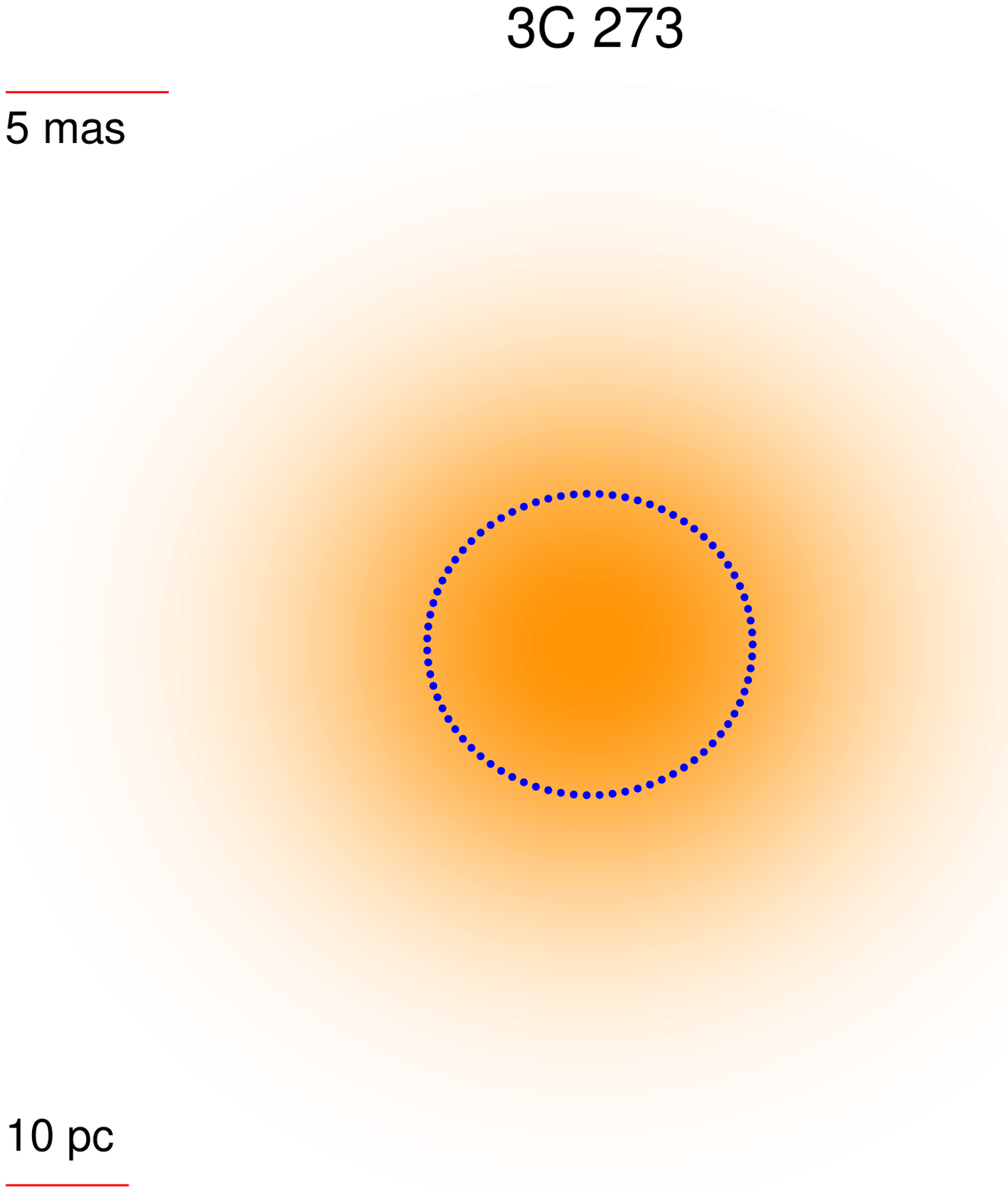}}
~~~~~~~~
\subfloat{\includegraphics[trim=5cm 0cm 5cm 0cm, width=0.17\hsize]{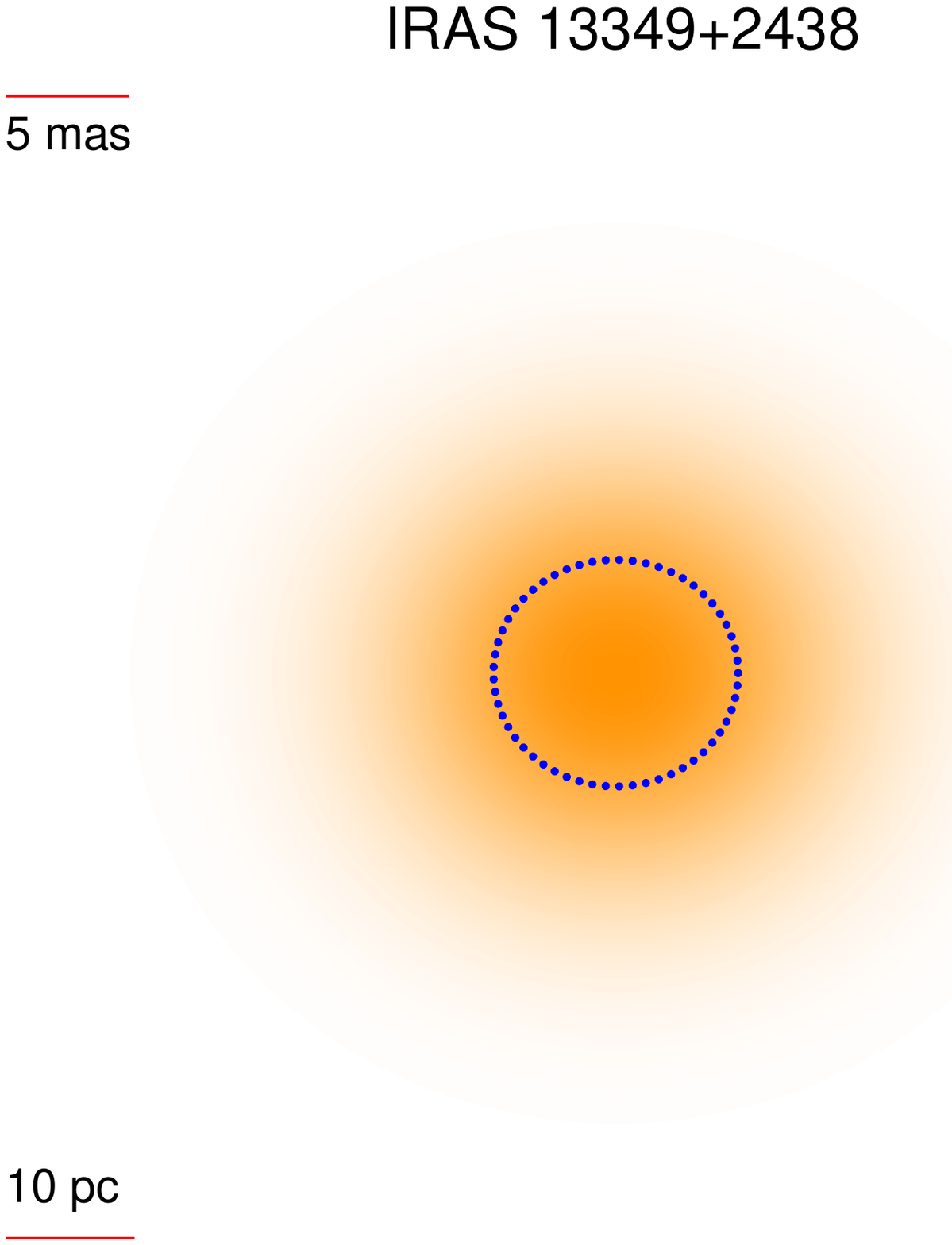}}
~~~~~~~~
\subfloat{\includegraphics[trim=5cm 0cm 5cm 0cm, width=0.17\hsize]{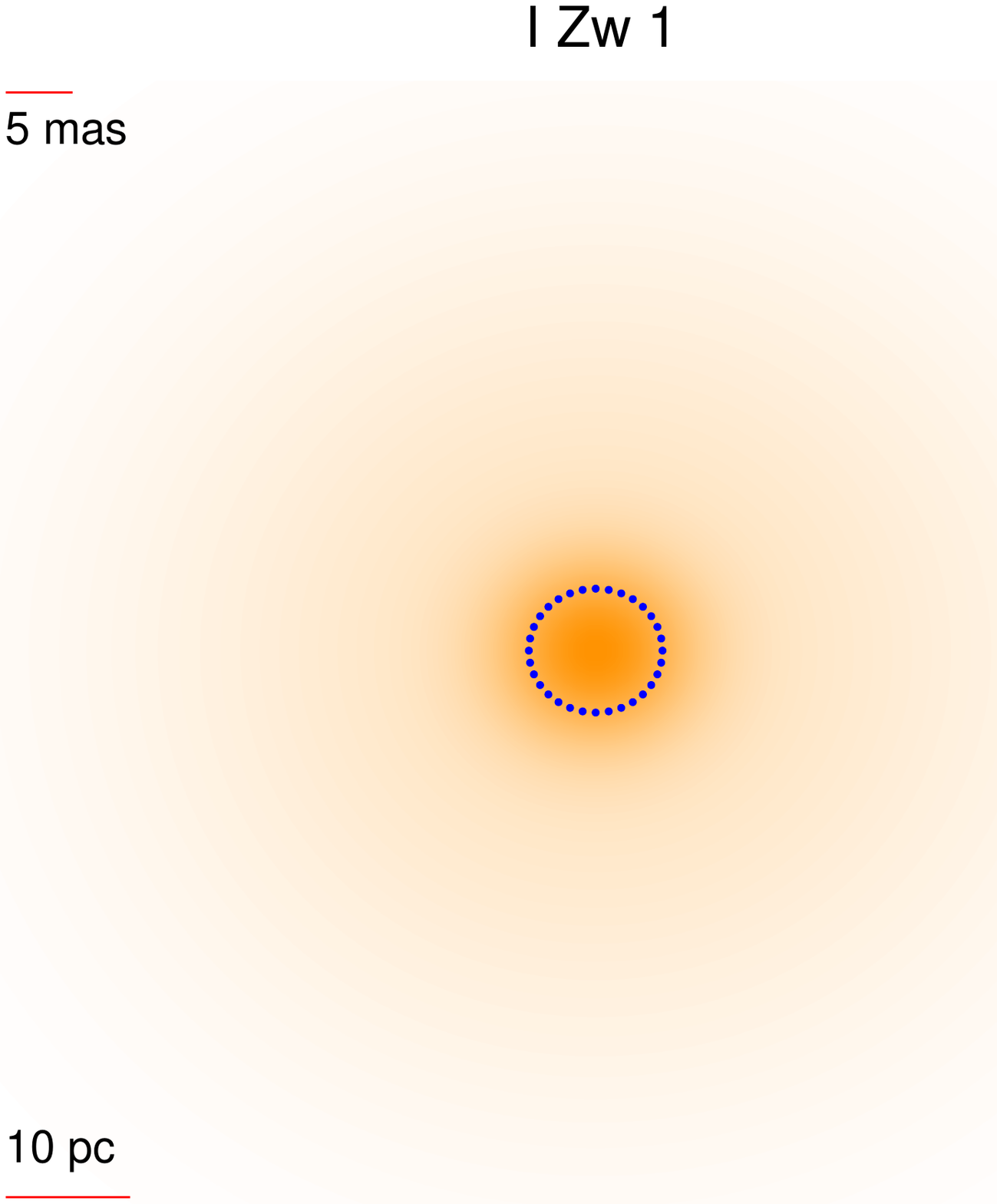}}
~~~~~~~~
\subfloat{\includegraphics[trim=5cm 0cm 5cm 0cm, width=0.17\hsize]{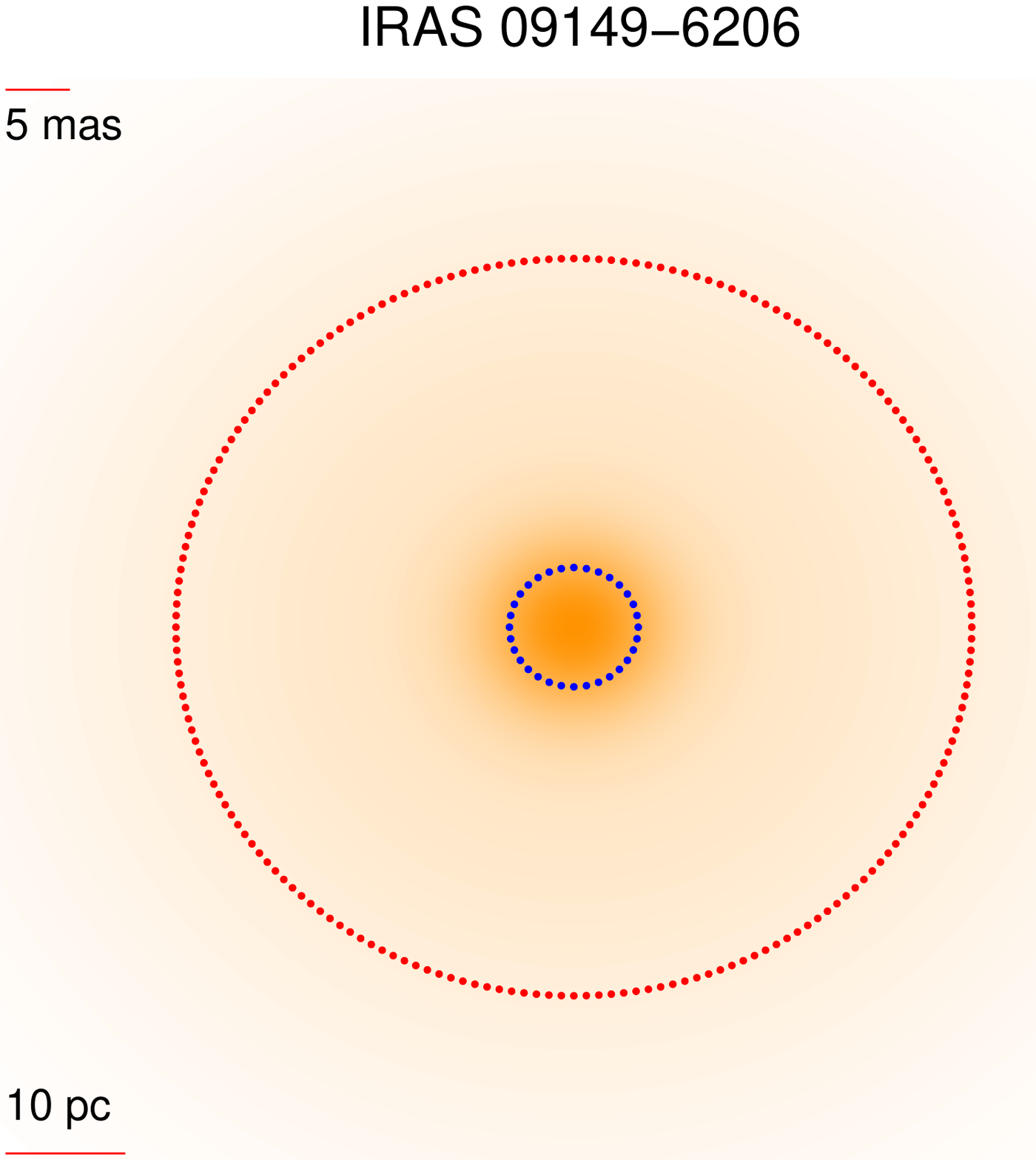}}
~~~~~~~~
\subfloat{\includegraphics[trim=5cm 0cm 5cm 0cm, width=0.17\hsize]{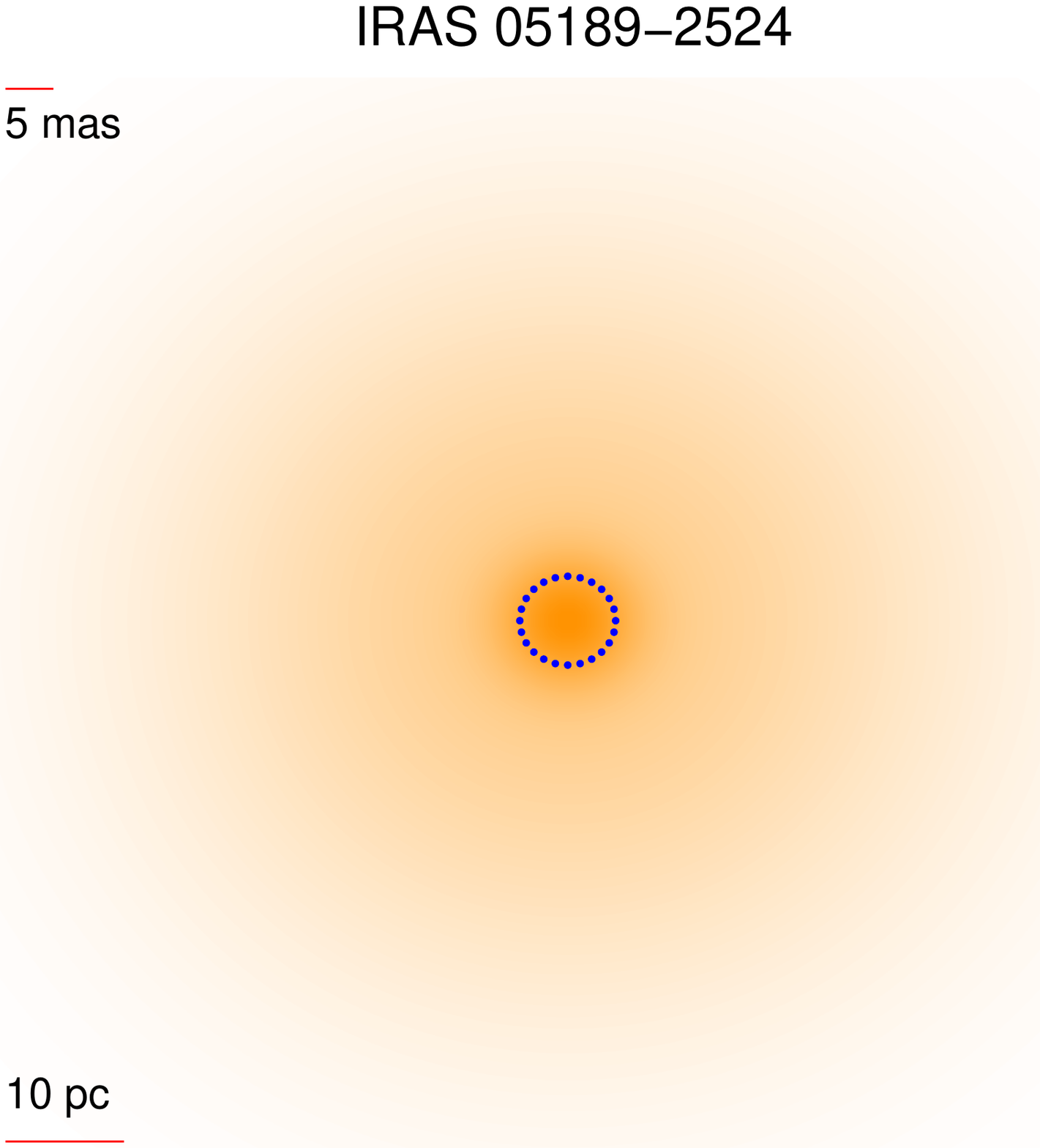}}
\\\vspace{-0,5cm}
\subfloat{\includegraphics[trim=5cm 0cm 5cm 0cm, width=0.17\hsize]{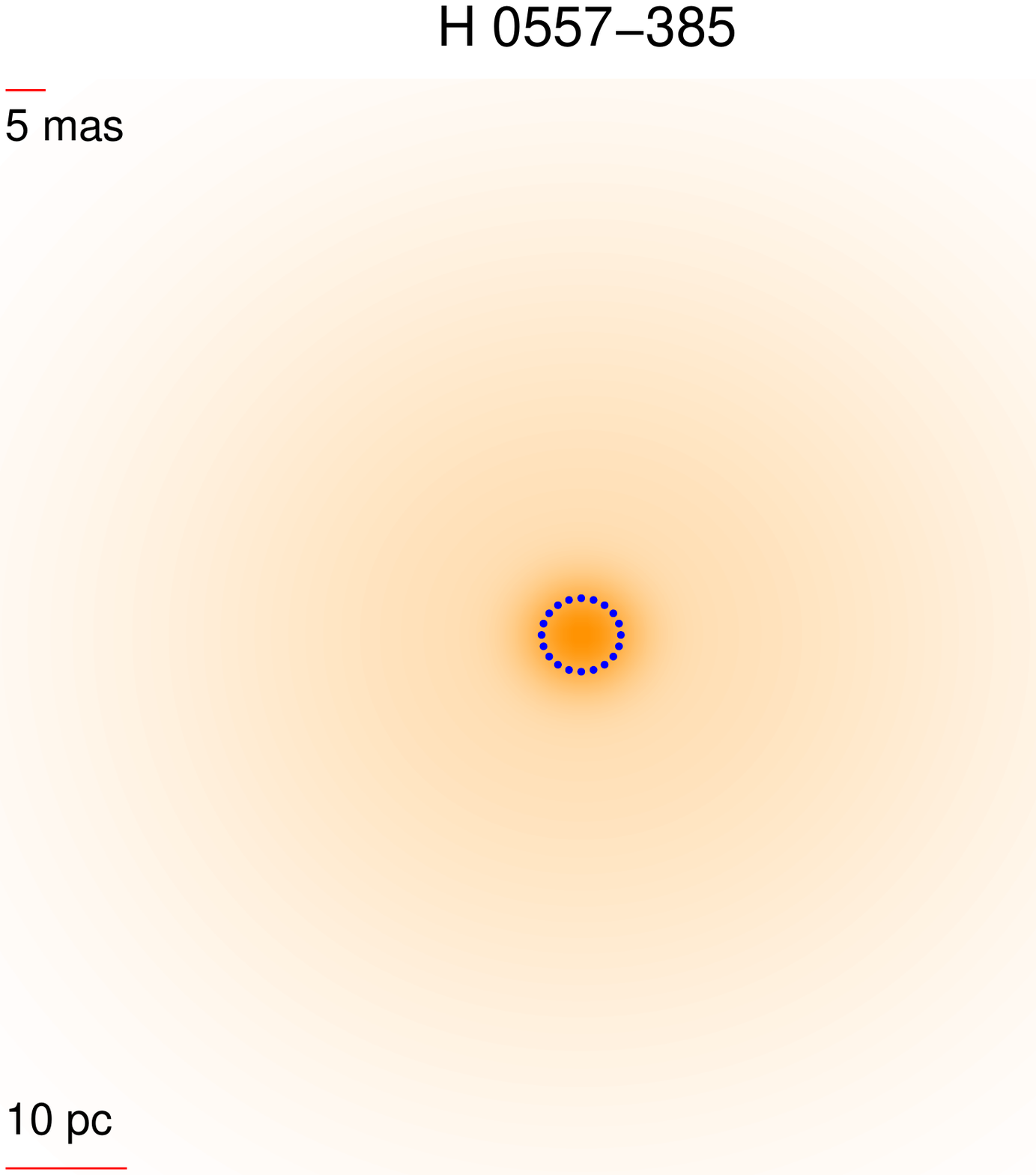}}
~~~~~~~~~
\subfloat{\includegraphics[trim=5cm 0cm 5cm 0cm, width=0.17\hsize]{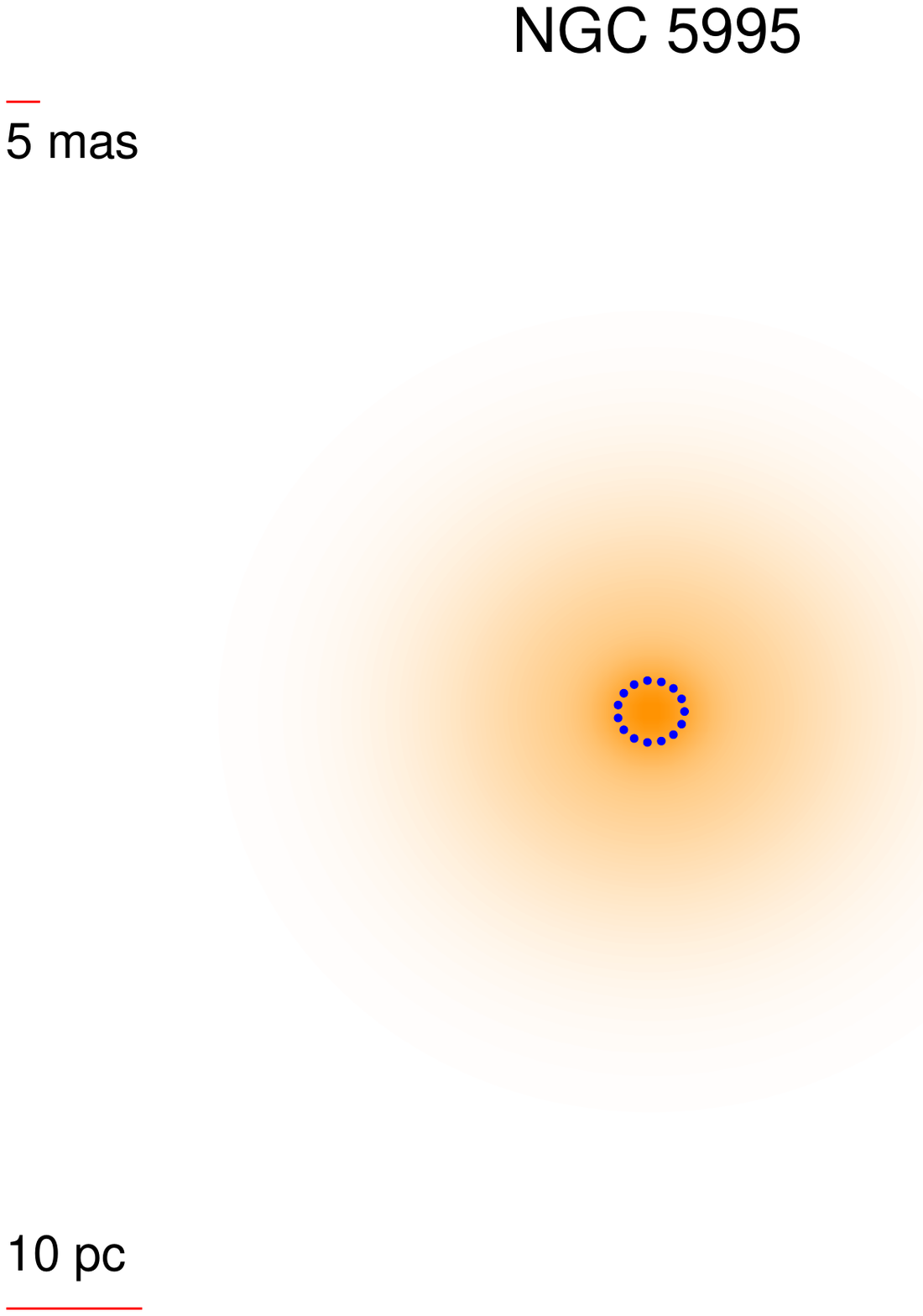}}
~~~~~~~~~
\subfloat{\includegraphics[trim=5cm 0cm 5cm 0cm, width=0.17\hsize]{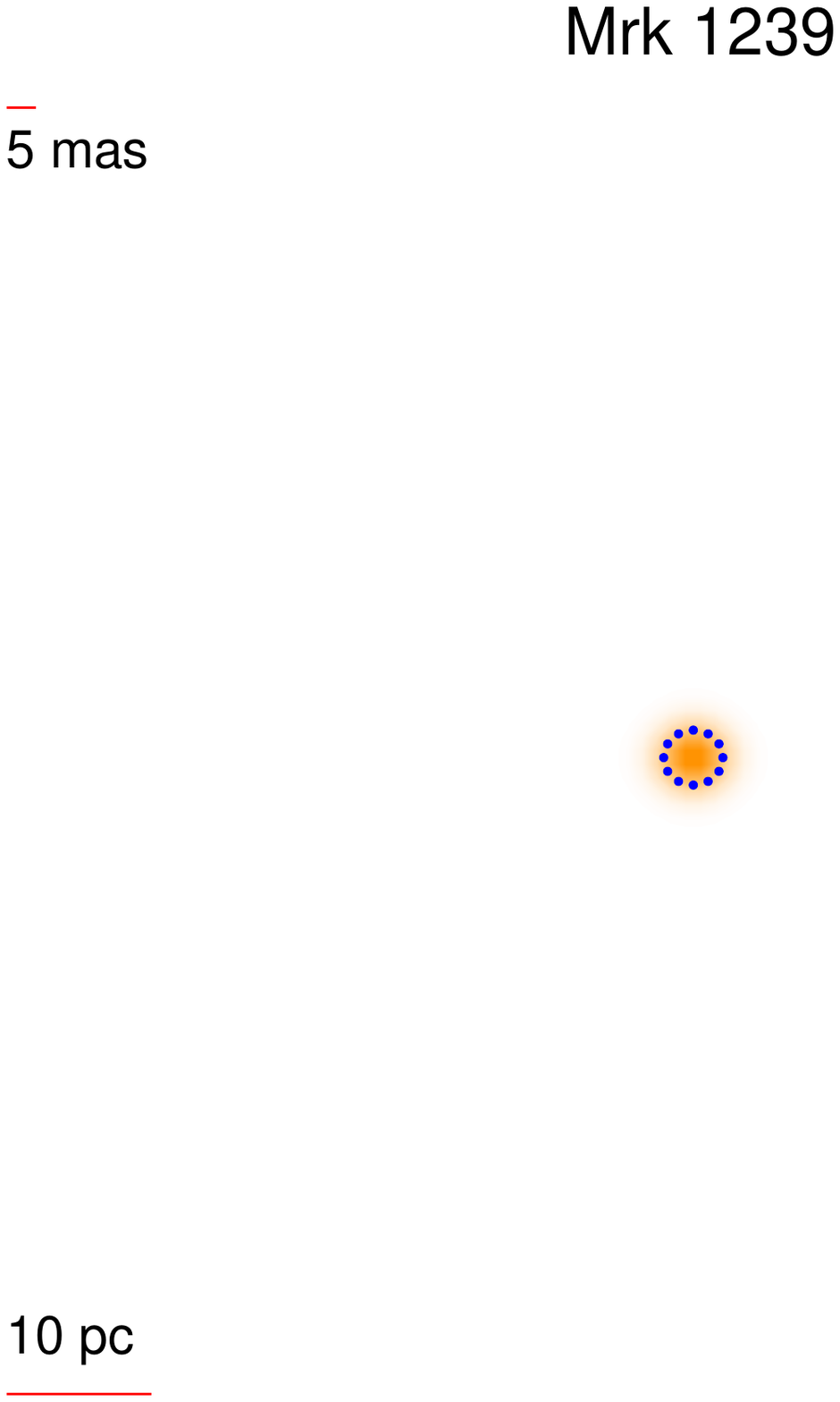}}
~~~~~~~~~
\subfloat{\includegraphics[trim=5cm 0cm 5cm 0cm, width=0.17\hsize]{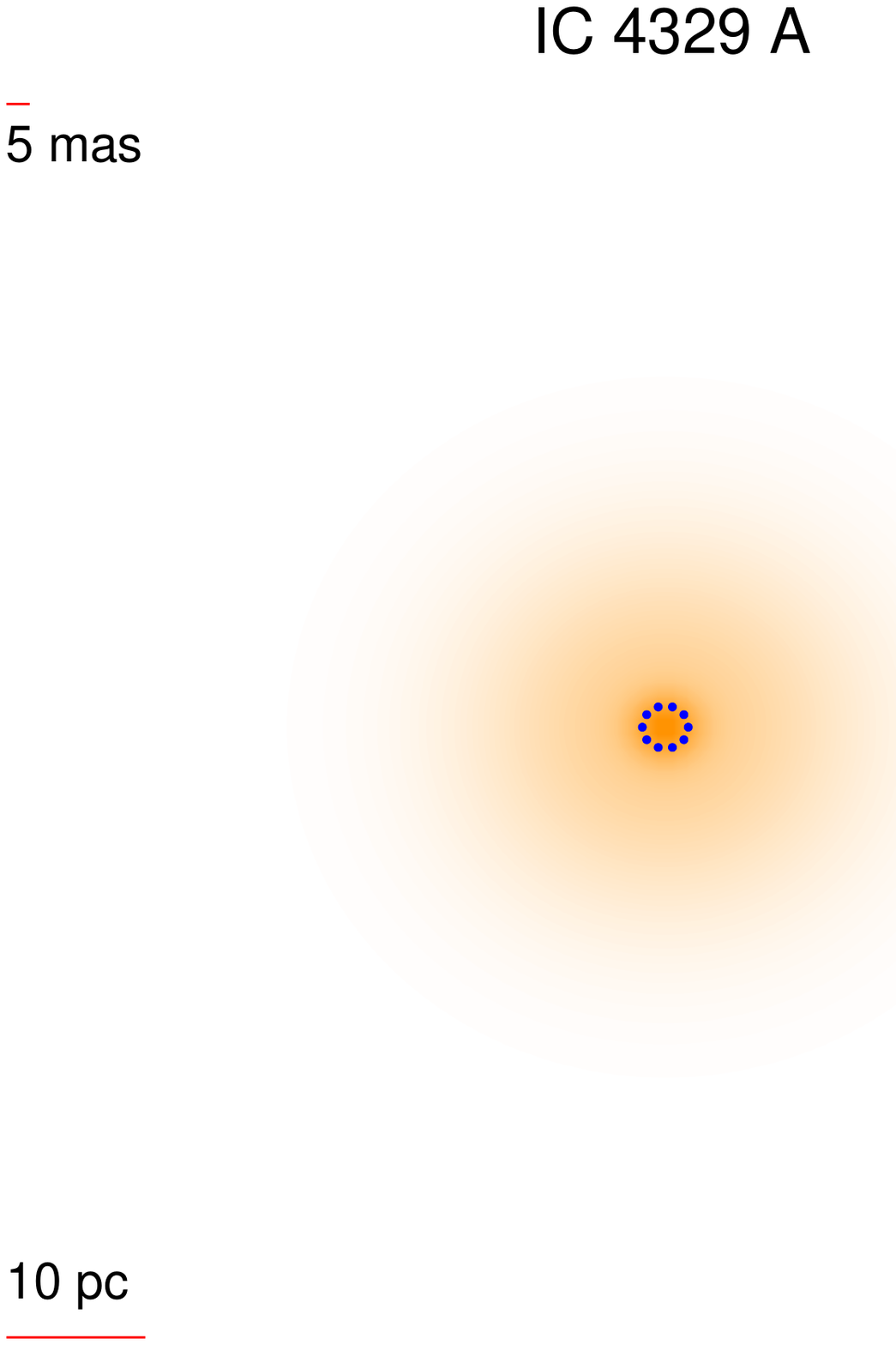}}
~~~~~~~~~
\subfloat{\includegraphics[trim=5cm 0cm 5cm 0cm, width=0.17\hsize]{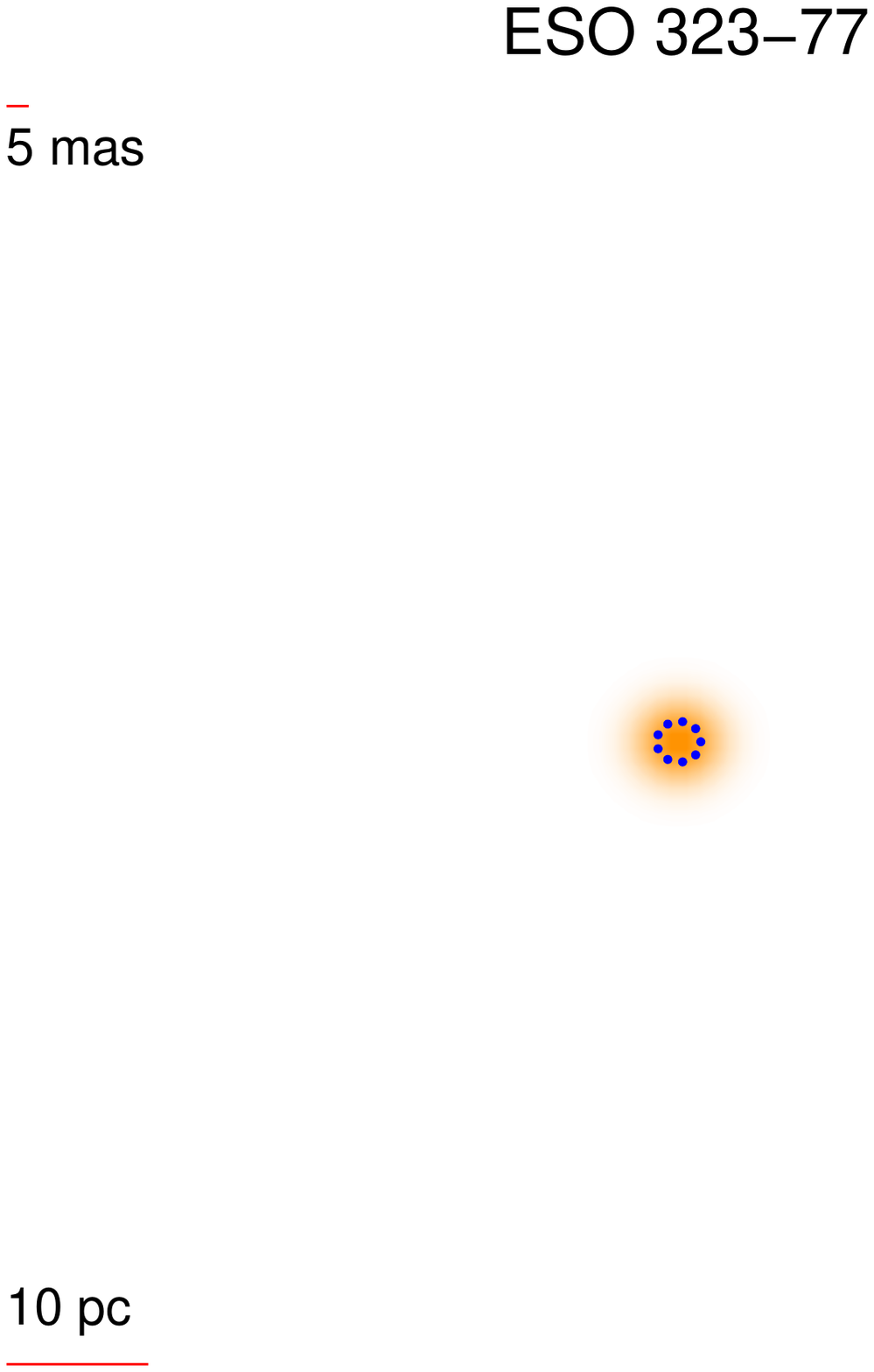}}
\\\vspace{-0,5cm}
\subfloat{\includegraphics[trim=5cm 0cm 5cm 0cm, width=0.17\hsize]{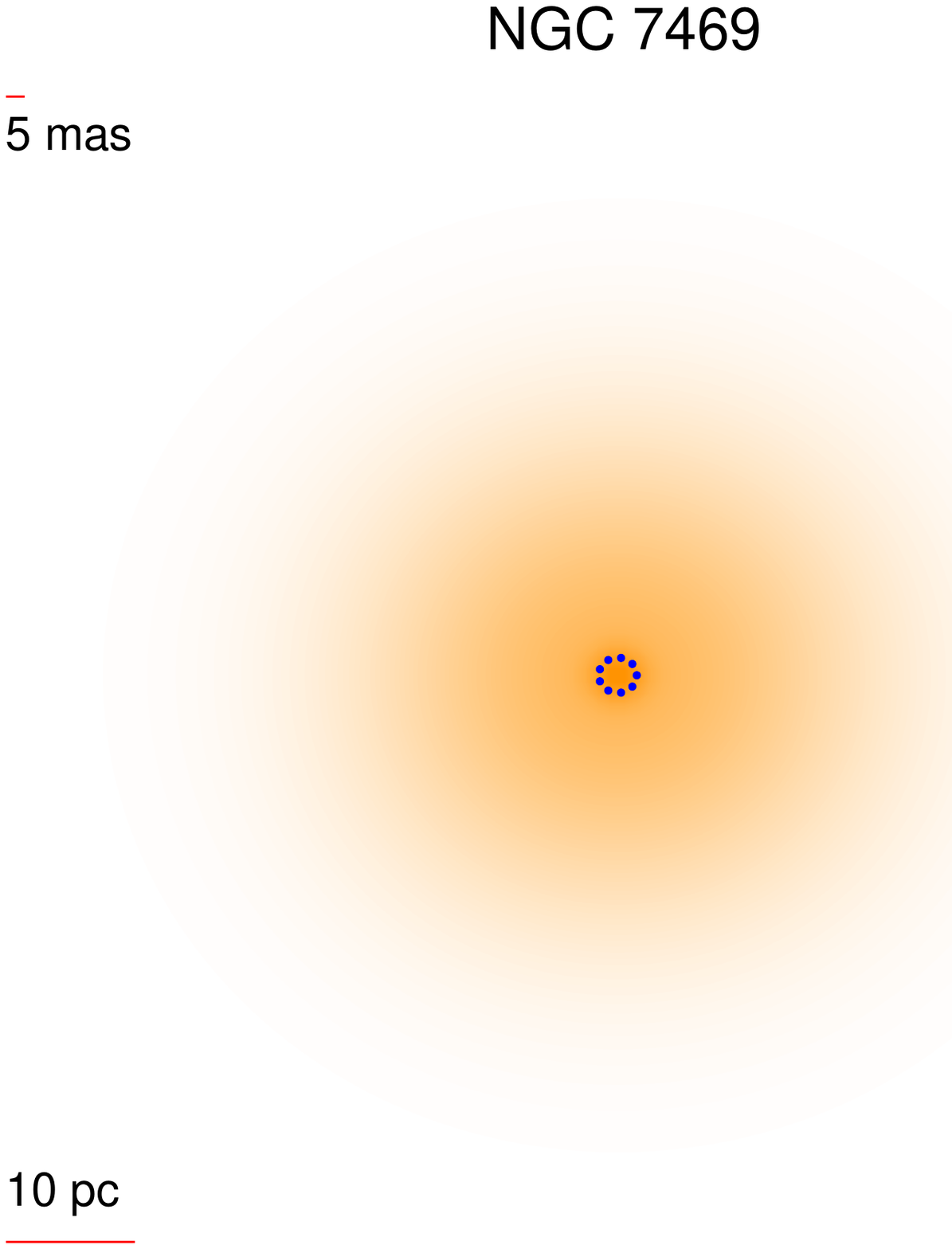}}
~~~~~~~~~
\subfloat{\includegraphics[trim=5cm 0cm 5cm 0cm, width=0.17\hsize]{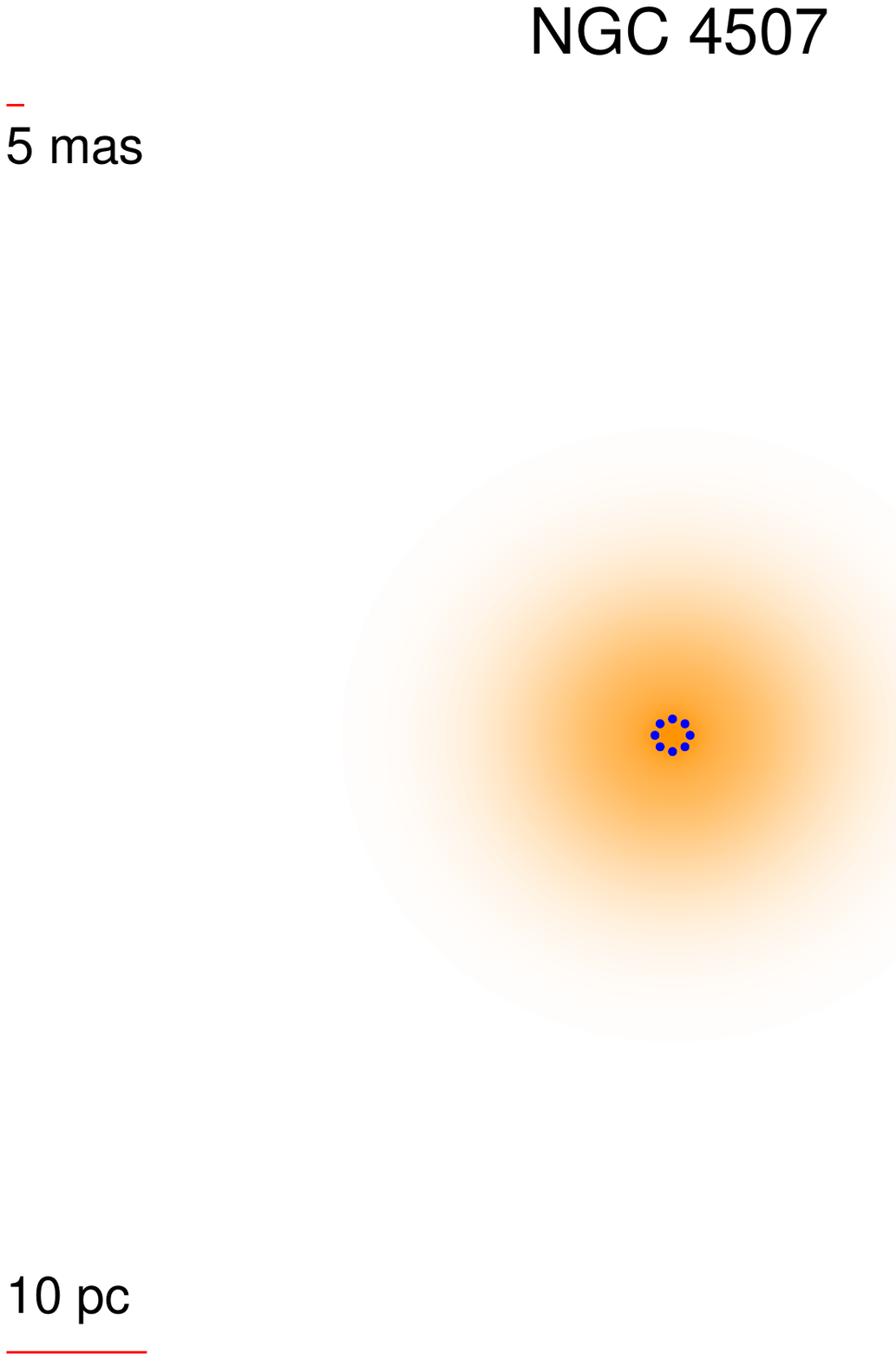}}
~~~~~~~~~
\subfloat{\includegraphics[trim=5cm 0cm 5cm 0cm, width=0.17\hsize]{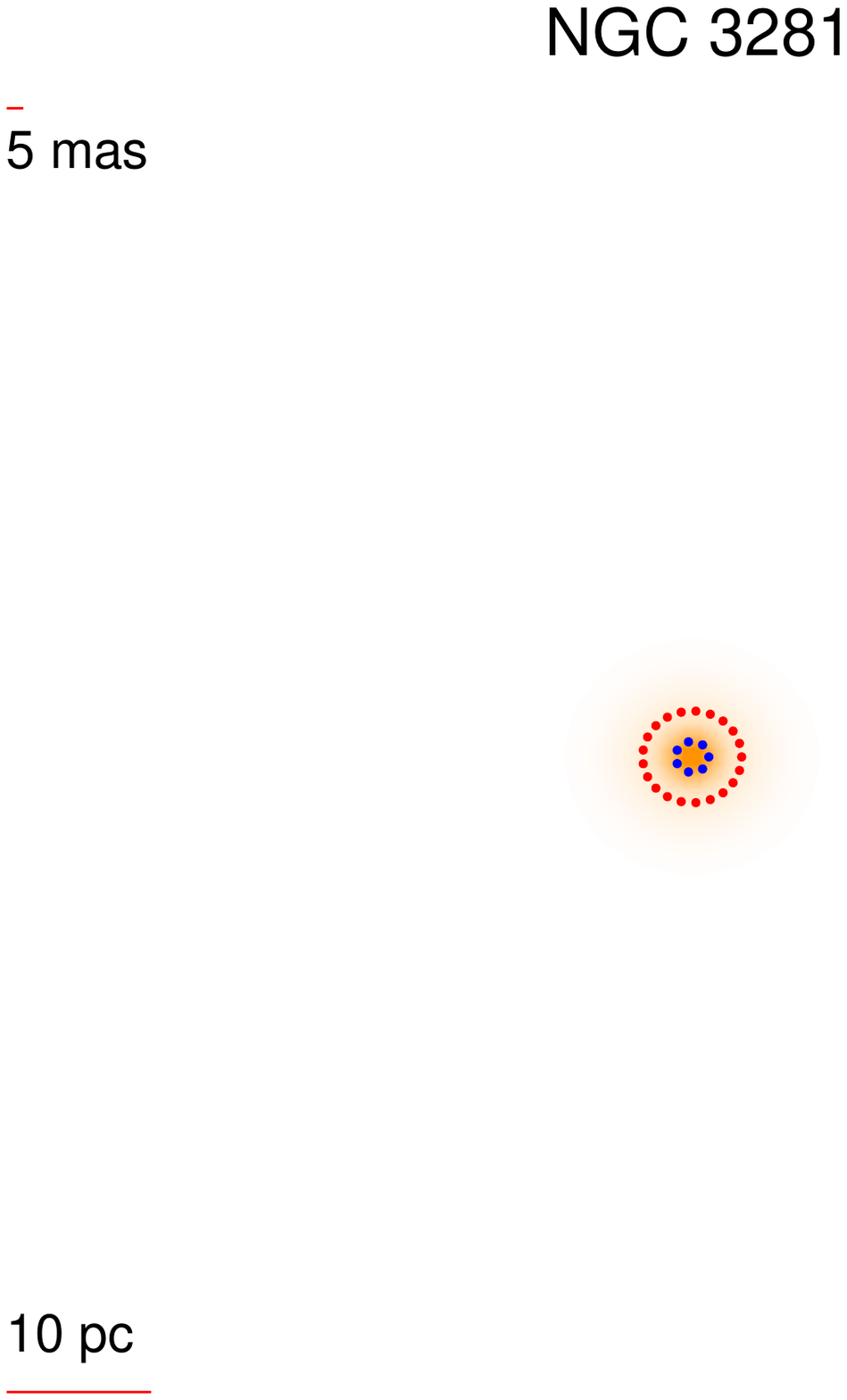}}
~~~~~~~~~
\subfloat{\includegraphics[trim=5cm 0cm 5cm 0cm, width=0.17\hsize]{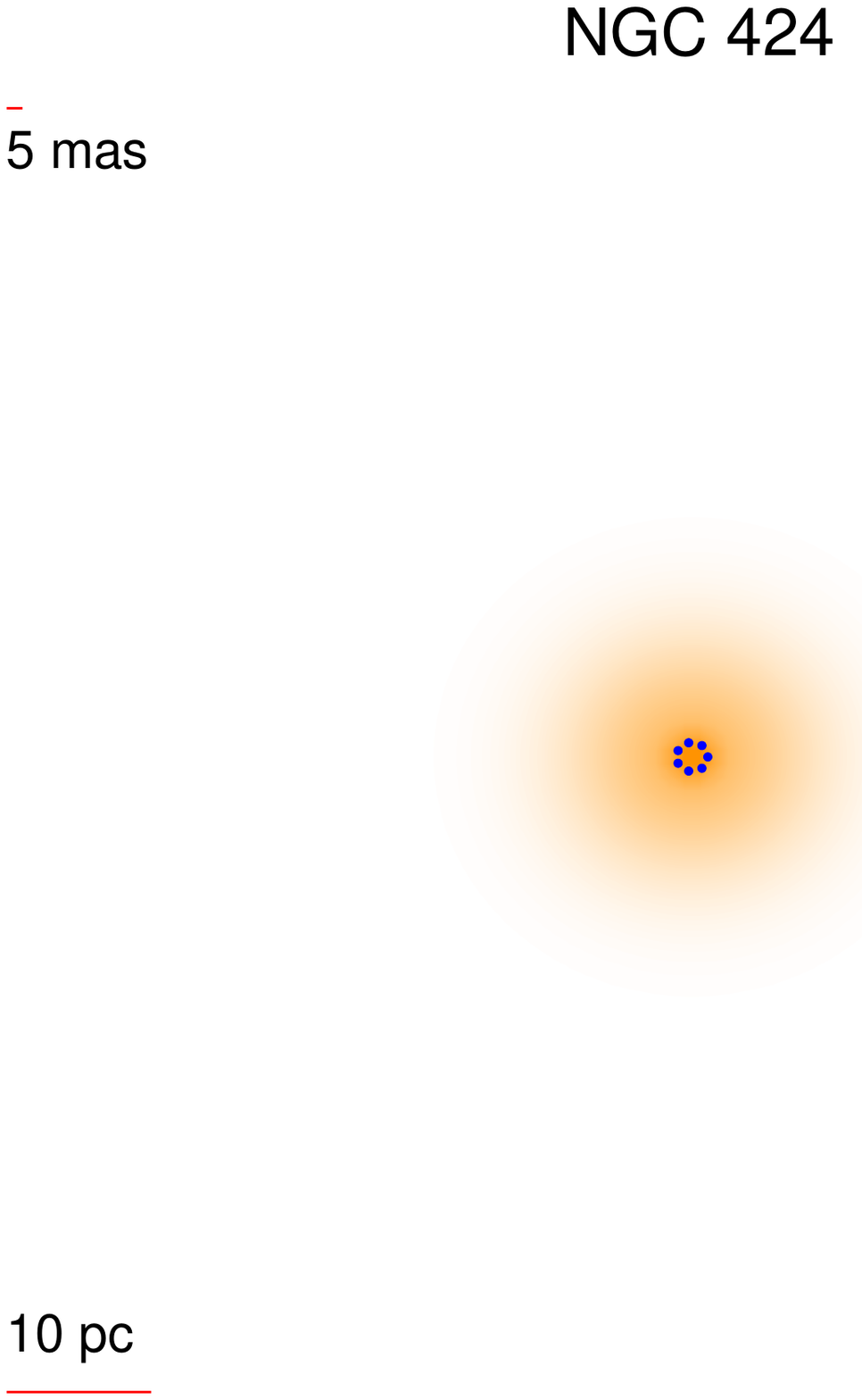}}
~~~~~~~~~
\subfloat{\includegraphics[trim=5cm 0cm 5cm 0cm, width=0.17\hsize]{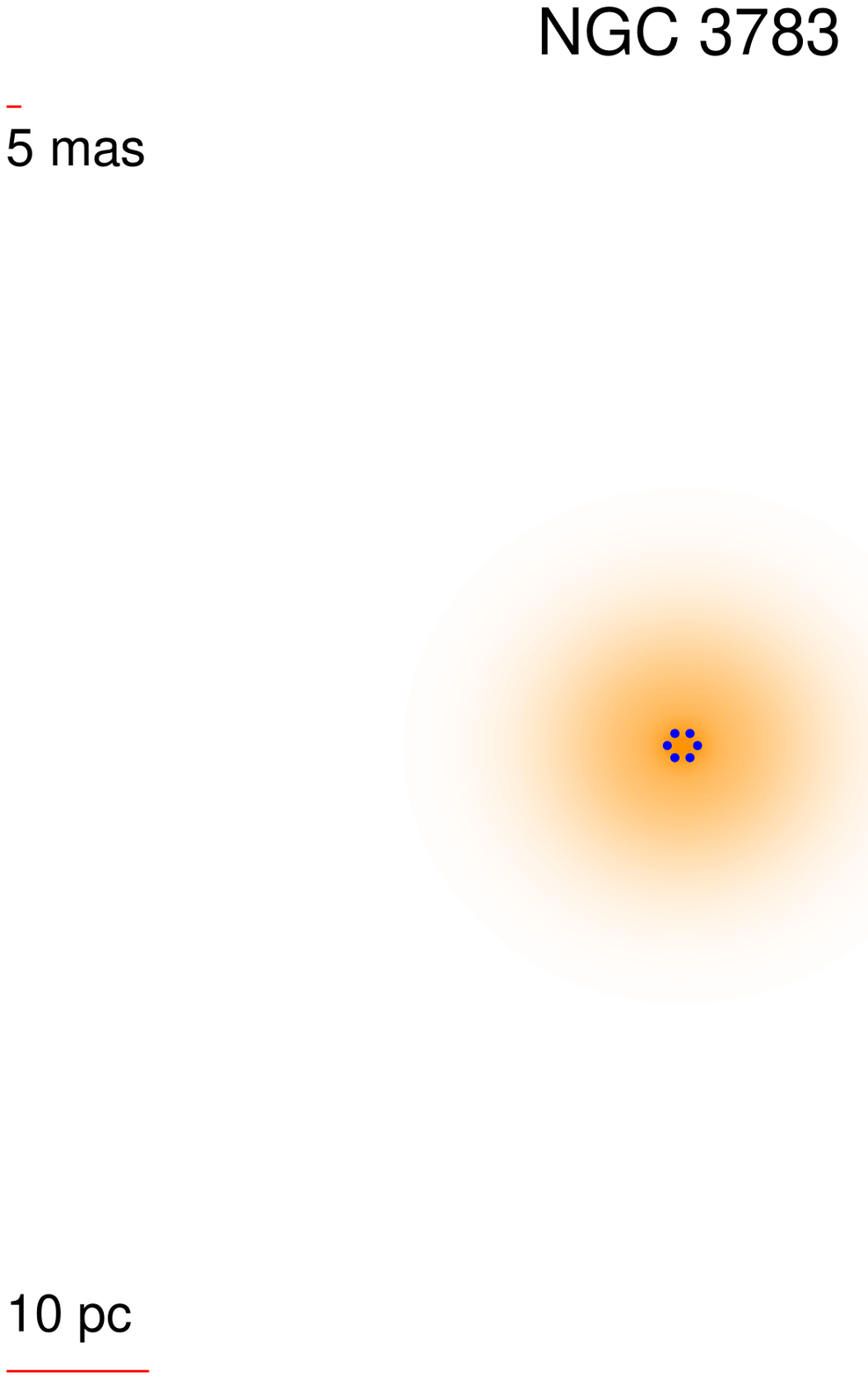}}
\\\vspace{-0,5cm}
\subfloat{\includegraphics[trim=5cm 0cm 5cm 0cm, width=0.17\hsize]{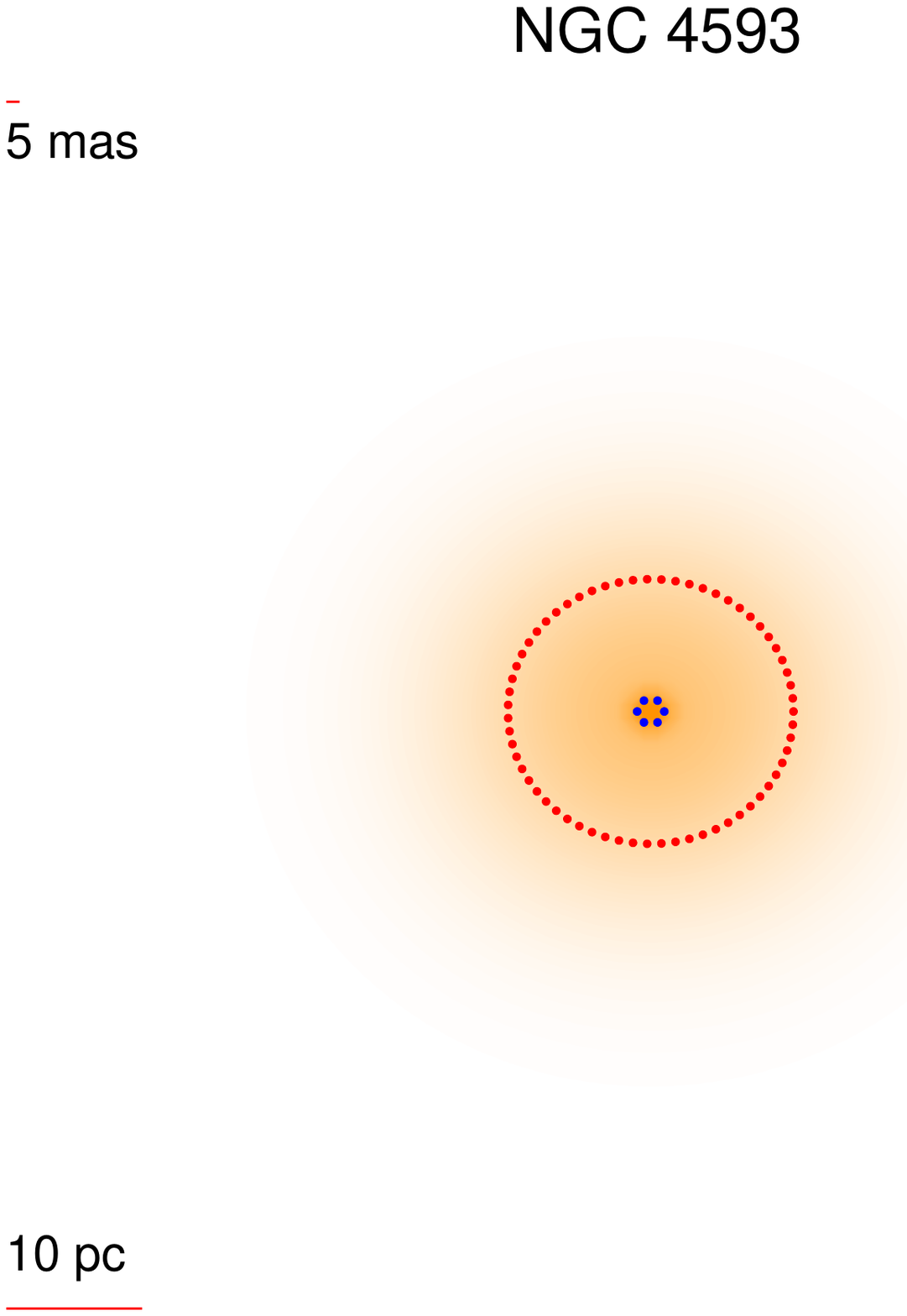}}
~~~~~~~~~
\subfloat{\includegraphics[trim=5cm 0cm 5cm 0cm, width=0.17\hsize]{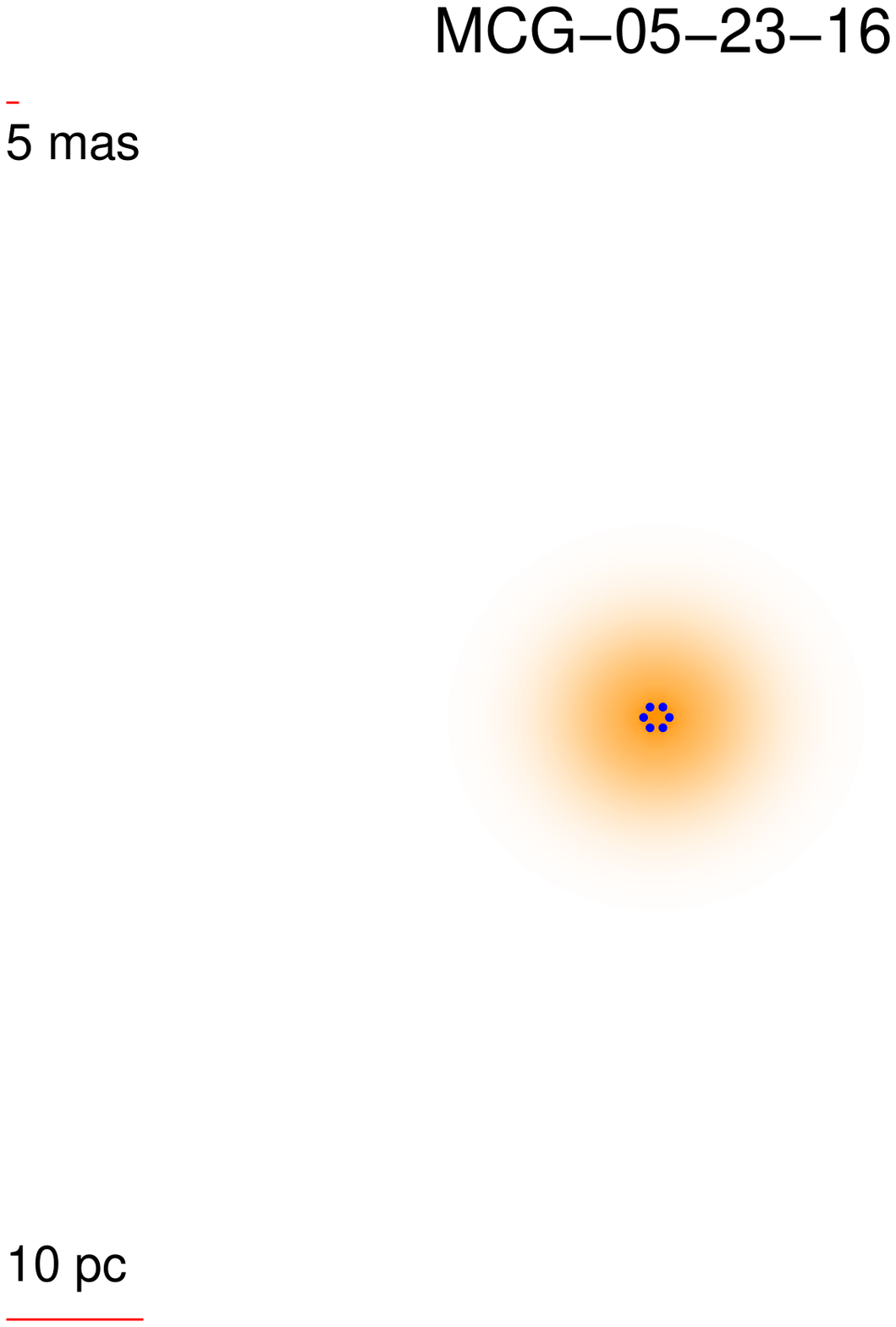}}
~~~~~~~~~
\subfloat{\includegraphics[trim=5cm 0cm 5cm 0cm, width=0.17\hsize]{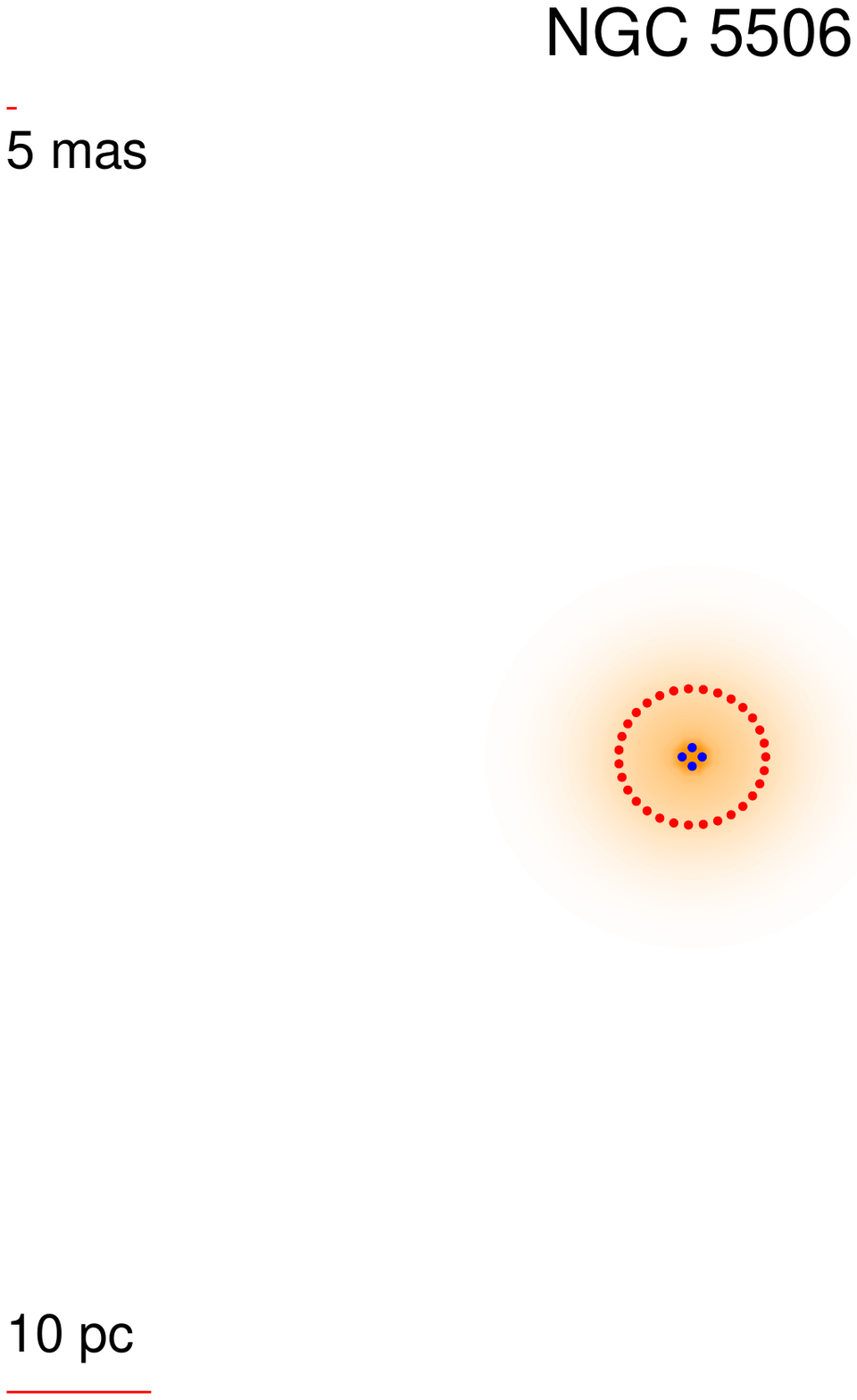}}
~~~~~~~~~
\subfloat{\includegraphics[trim=5cm 0cm 5cm 0cm, width=0.17\hsize]{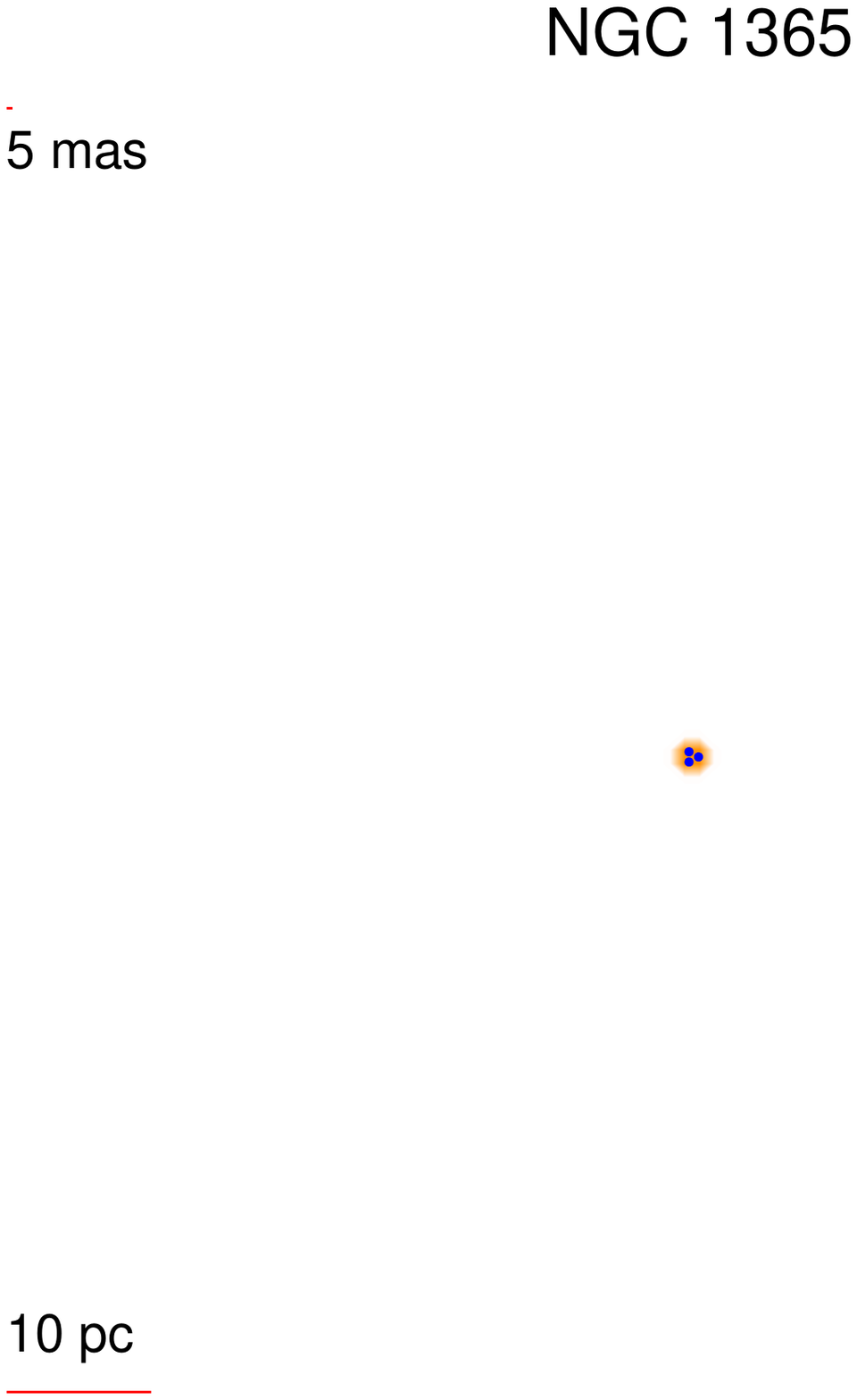}}
~~~~~~~~~
\subfloat{\includegraphics[trim=5cm 0cm 5cm 0cm, width=0.17\hsize]{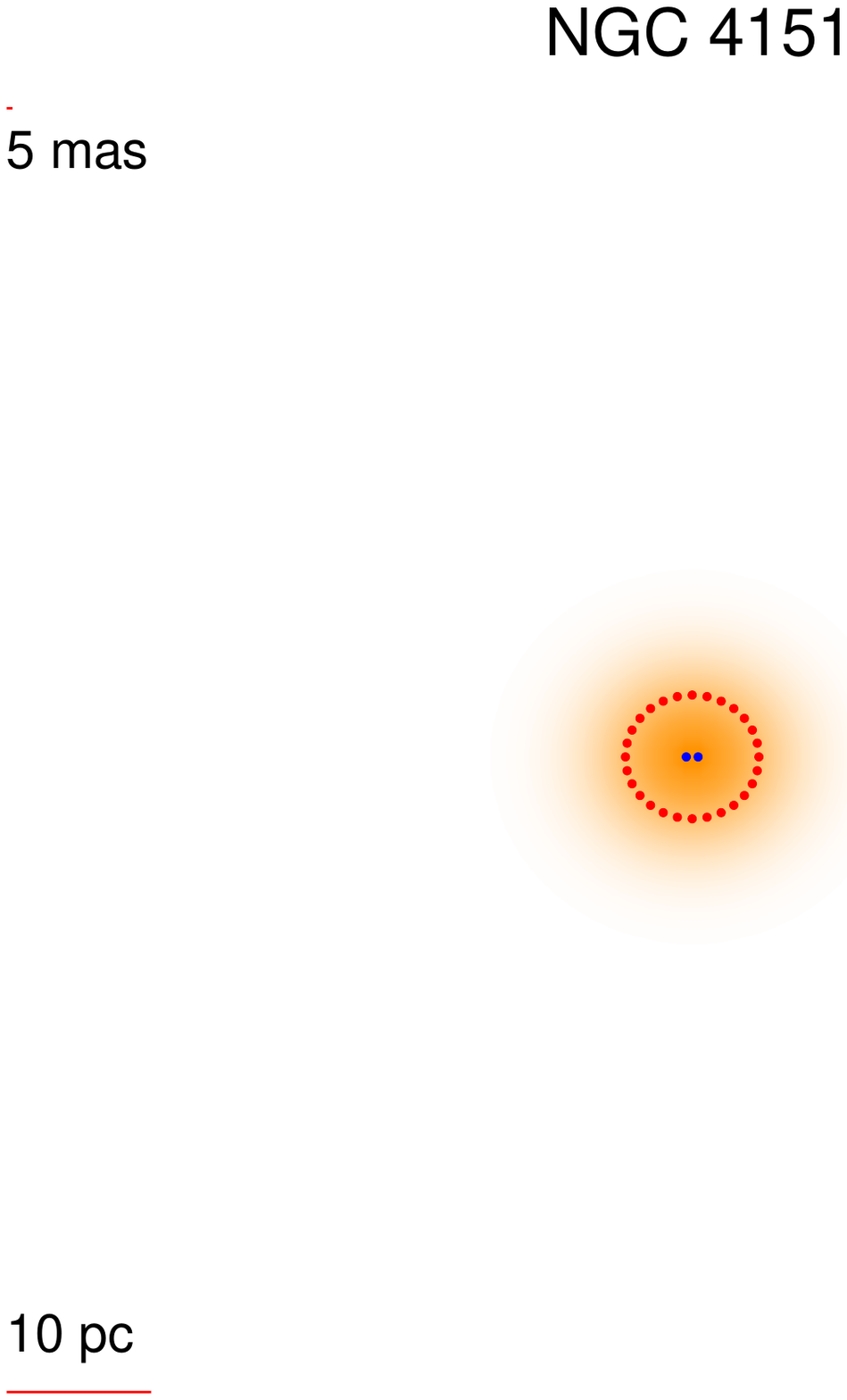}}
\\\vspace{-0,5cm}
\subfloat{\includegraphics[trim=5cm 0cm 5cm 0cm, width=0.17\hsize]{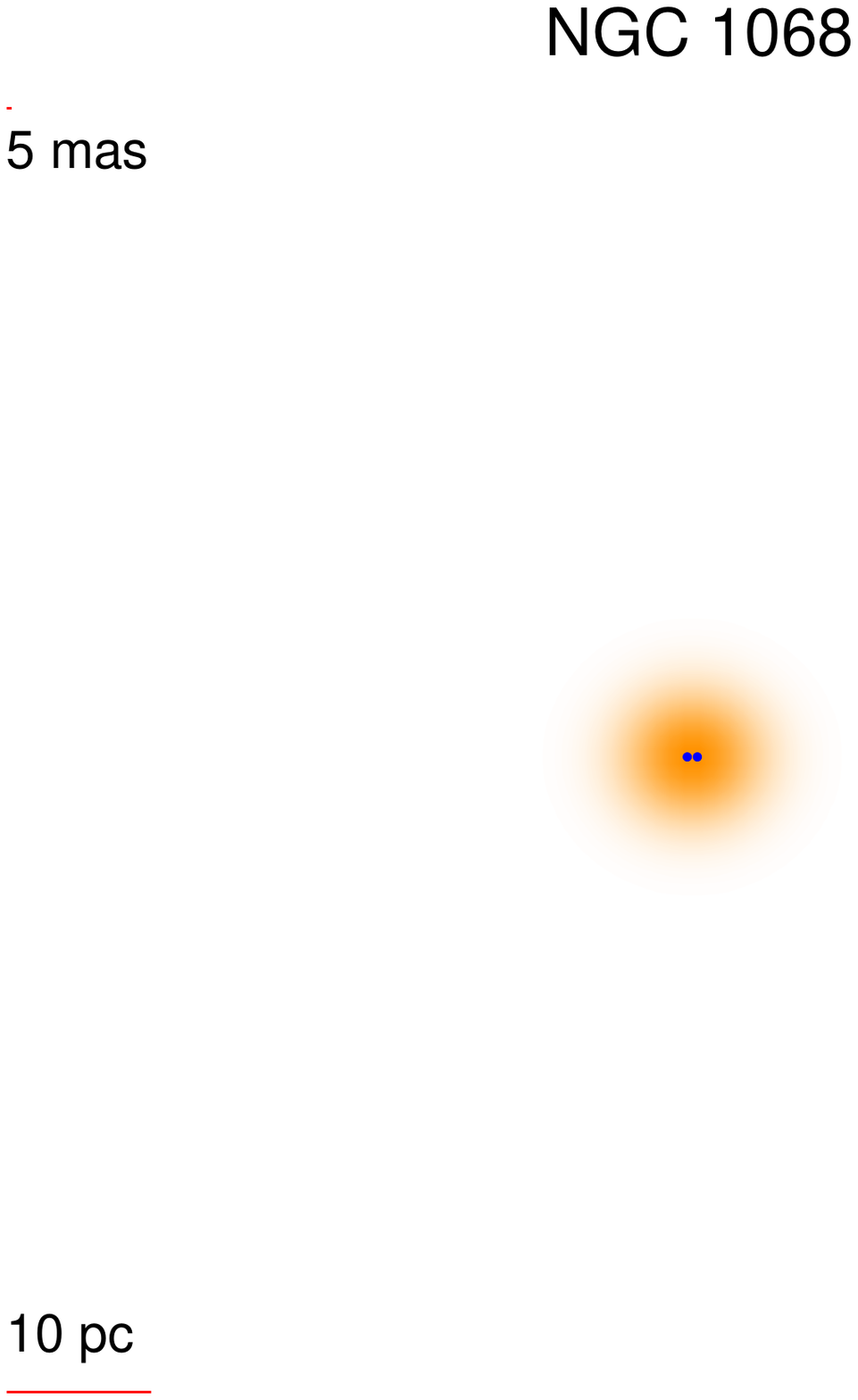}}
~~~~~~~~~
\subfloat{\includegraphics[trim=5cm 0cm 5cm 0cm, width=0.17\hsize]{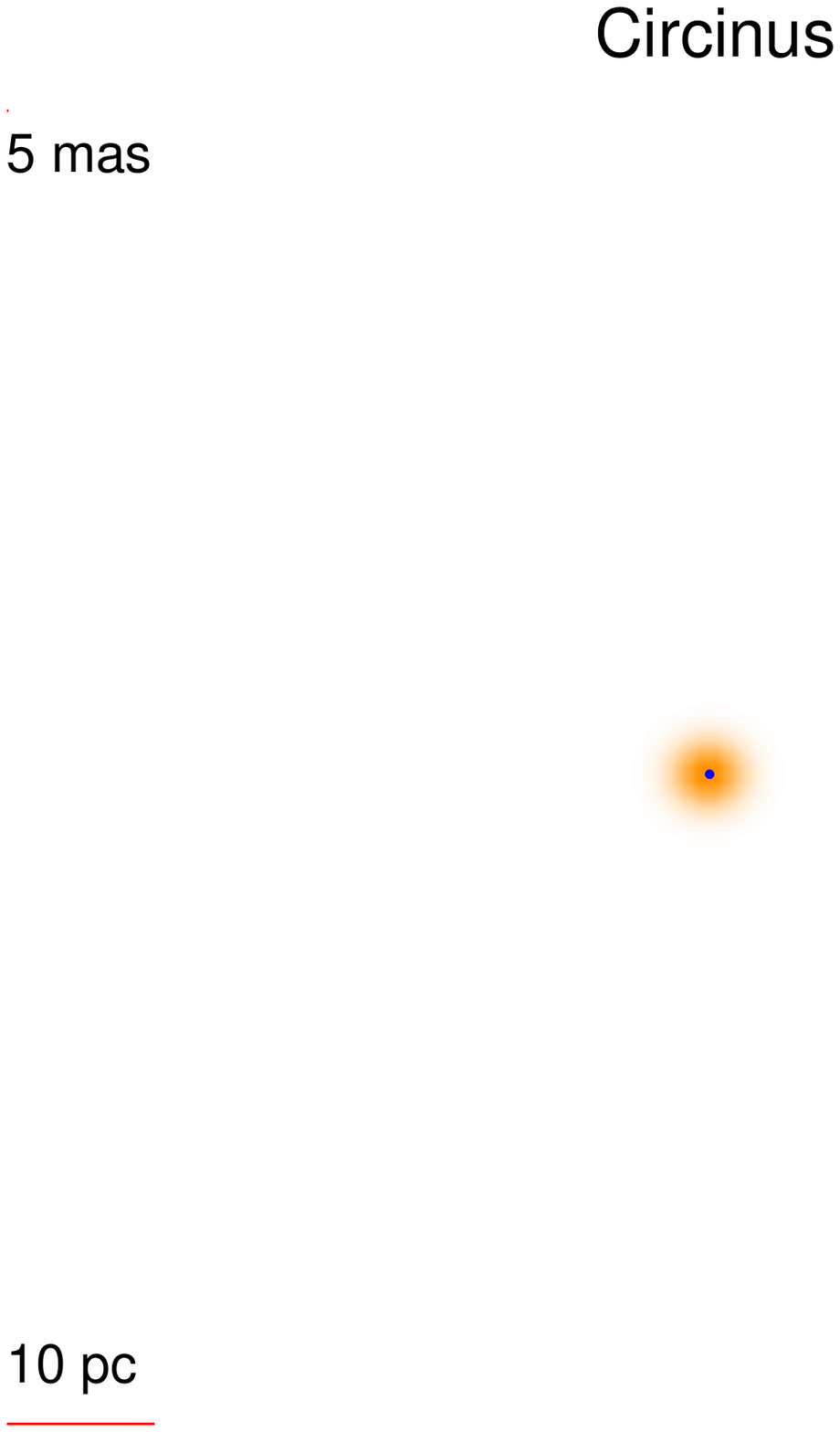}}
~~~~~~~~~
\subfloat{\includegraphics[trim=5cm 0cm 5cm 0cm, width=0.17\hsize]{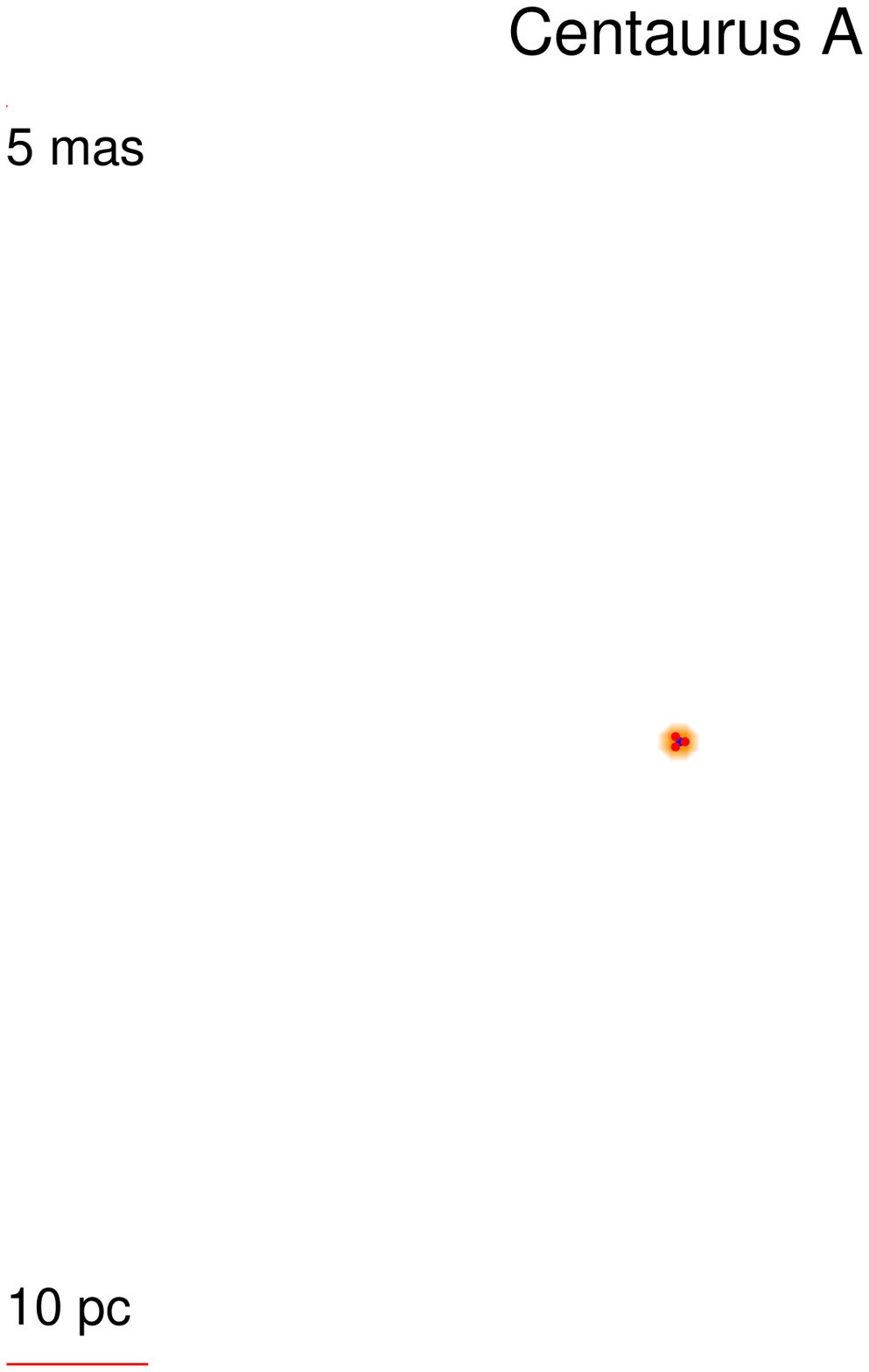}}	
\caption{\label{fig:modelimages}Model images for all \nsources sources, sorted by scale (most distant sources first). The color scale is normalized to the total flux of the target. The dotted blue line indicates the adopted resolution limit (5 mas), inside which is the point source. The dotted red line, when plotted, indicates sources for which we only derived a limit on the resolved structure (see Tab. \ref{tab:radialfit}). The pixel scale in parsec is constant over all images; the scale in angular units is also given. While more distant sources are less resolved, it can also be seen that the morphology of the source does not only depend on resolution.}
\end{figure*}

In Figs.~\ref{fig:rad:IZwicky1} -- \ref{fig:rad:NGC7469} (upper right panel) we show the residuals of the data with respect to our one dimensional models on the $(u,v)$ plane. These plots can be appreciated by looking how ``clustered'' and how large equal-colored dots are: If, as in NGC~3783, the blue dots (positive residuals) are along one direction and the green ones along the perpendicular direction, it is a clear sign for elongation; in the case of NGC~3783, one would deduce a PA for the major axis of the ellipse of $\approx -40 \deg$, in good agreement with the detailed modeling of \citet{hoenig2013}. 

For the other sources with known asymmetry in the nuclear dust component, a strong signal is also evident in these plots, i.e. NGC~1068 \citep{raban2009}, the Circinus~galaxy \citep{tristram2012} and NGC~424 \citep{hoenig2012}. The published axis ratios for these sources are about 2 in the Circinus~galaxy, 1.5 in NGC~3783, 1.3 in NGC~1068, and 1.3 in NGC~424. 

We further detect elongations of the nuclear mid-IR brightness distribution in NGC~3281, NGC~4507, NGC~4593 and in MCG-5-23-16 and can exclude strong elongations only in I~Zw~1 and IC~4329A. For the other sources either the signal/noise and / or the $(u,v)$ coverage are not sufficient to detect or exclude elongations. We plan to analyze the elongation signal for all sources in a quantitative way in a future publication.

For the most elongated source, the Circinus~galaxy, the derived half-light radius should therefore only be considered a rough scale and not a precise size determination. For all other sources, however, the PA-averaged half-light radii are probably a fair description of the source's extension.

%
\clearpage
\begin{landscape}
\begin{table}
\caption{\label{tab:radialfit}Fit results for all sources in this sample. For the model case selection, see Fig.~\ref{fig:rad:models} and text. Case '0' refers to those sources for which we only have upper limits on the total flux, no model was fitted here.}
\begin{tabular}{l l l l l l l l l p{8cm}}
\hline
Name               & case        & $F_{\rm tot}$ & $f_p$ & $F_p$ & $F_g$      &$\Theta_g$  & $r_{1/2}$  & $r_{1/2}$      &            comment\\
                   &             & [Jy]          &       & [Jy]  & [Jy]       & [mas]    & [mas]       &   [pc]\\
\hline\hline
         I Zw 1&1& 0.53$\pm$ 0.02&0.73$^{+0.08}_{-0.31}$&0.31$^{+0.03}_{-0.13}$& 0.11$^{+0.13}_{-0.03}$  &24$^{+20}_{-15}$          &$<$ 2.5                   &$<$ 2.7& resolved and unresolved\\
        NGC 424&1& 0.72$\pm$ 0.03&0.54$\pm 0.05$        &0.31$^{+0.03}_{-0.03}$& 0.50$^{+0.03}_{-0.03}$  &28.7$^{+ 7.1}_{- 4.4}$    & 5.9$^{+ 1.5}_{- 0.9}$    & 1.28$^{+ 0.32}_{- 0.19}$& resolved and unresolved\\
       NGC 1068&1&16.08$\pm$ 0.16&0.09$\pm 0.01$        &1.48$^{+0.19}_{-0.19}$&14.60$^{+0.19}_{-0.19}$  &47.9$^{+ 3.6}_{- 3.0}$    &22.2$^{+ 1.6}_{- 1.2}$    & 1.55$^{+ 0.11}_{- 0.08}$& resolved and unresolved\\
       NGC 1365&3& 0.30$\pm$ 0.01&1.00                  &\ldots                &\ldots                   & 5.75$^{+ 1.58}_{- 1.58}$ & 2.88$^{+ 0.79}_{- 0.79}$ &0.25$^{+ 0.07}_{- 0.07}$  & marginally resolved\\
IRAS 05189-2524&1& 0.56$\pm$ 0.03&0.53$^{+0.07}_{-0.08}$&0.24$^{+0.03}_{-0.03}$& 0.22$^{+0.03}_{-0.03}$  &30$^{+10}_{- 7}$          &$<$ 4.3                   &$<$ 3.56 & resolved and unresolved\\
     H 0557-385&1& 0.50$\pm$ 0.03&0.58$^{+0.08}_{-0.09}$&0.28$^{+0.02}_{-0.03}$& 0.13$^{+0.03}_{-0.02}$  &42$^{+35}_{-11}$          &$<$ 2.5                   &$<$ 1.6& resolved and unresolved\\
IRAS 09149-6206&2& 0.57$\pm$ 0.03&0.72$\pm 0.06$        &0.34$^{+0.02}_{-0.02}$& 0.13$^{+0.02}_{-0.02}$  &$>$31                     &$<$ 2.5                   &$<$ 2.7& over-resolved and unresolved\\
   MCG-05-23-16&1& 0.73$\pm$ 0.01&0.28$\pm 0.04$        &0.21$^{+0.03}_{-0.03}$& 0.53$^{+0.03}_{-0.03}$  &27.3$^{+ 4.4}_{- 3.6}$    & 9.9$^{+ 1.6}_{- 1.3}$    & 1.86$^{+ 0.30}_{- 0.24}$& resolved and unresolved\\
       Mrk 1239&3& 0.57$\pm$ 0.05&1.00                  &\ldots                &\ldots                   & 3.97$^{+ 1.78}_{- 2.18}$ & 1.99$^{+ 0.89}_{- 1.09}$ & 0.81$^{+ 0.37}_{- 0.55}$ & marginally resolved\\
       NGC 3281&2& 0.32$\pm$ 0.01&0.77$\pm 0.03$        &0.25$^{+0.01}_{-0.01}$& 0.07$^{+0.01}_{-0.01}$  &$>$15                     &$<$ 2.5                   &$<$ 0.58& over-resolved and unresolved\\
       NGC 3783&1& 0.76$\pm$ 0.02&0.37$^{+0.04}_{-0.05}$&0.28$^{+0.03}_{-0.03}$& 0.58$^{+0.03}_{-0.03}$  &31.7$^{+ 8.7}_{- 6.1}$    & 9.2$^{+ 2.5}_{- 1.8}$    & 1.96$^{+ 0.54}_{- 0.38}$& resolved and unresolved\\
       NGC 4151&2& 1.18$\pm$ 0.03&0.26$\pm 0.01$        &0.30$^{+0.01}_{-0.01}$& 0.87$^{+0.01}_{-0.01}$  &$>$56                     &$>$21.3                   &$>$  1.7& over-resolved and unresolved\\
         3C 273&3& 0.32$\pm$ 0.04&1.00                  &\ldots                &\ldots                   & 6.15$^{+ 1.19}_{- 1.39}$ & 3.07$^{+ 0.59}_{- 0.59}$ & 8.1$^{+ 1.6}_{- 1.8}$& marginally resolved\\
       NGC 4507&1& 0.53$\pm$ 0.02&0.26$\pm 0.03$        &0.16$^{+0.02}_{-0.02}$& 0.57$^{+0.02}_{-0.02}$  &31.9$^{+ 4.2}_{- 2.8}$    &12.1$^{+ 1.6}_{- 1.1}$    & 3.03$^{+ 0.39}_{- 0.26}$& resolved and unresolved\\
       NGC 4593&2& 0.25$\pm$ 0.01&0.51$\pm 0.03$        &0.13$^{+0.00}_{-0.00}$& 0.12$^{+0.00}_{-0.00}$  &$>$52                     & \ldots                   & \ldots & over-resolved and unresolved\\
     ESO 323-77&3& 0.38$\pm$ 0.01&1.00                  &\ldots                &\ldots                   & 6.74$^{+ 1.39}_{- 1.39}$ & 3.37$^{+ 0.59}_{- 0.59}$ & 1.05$^{+ 0.22}_{- 0.22}$& marginally resolved\\
    Centaurus A&2& 1.57$\pm$ 0.03&0.58$\pm 0.02$        &0.91$^{+0.03}_{-0.03}$& 0.56$^{+0.03}_{-0.03}$  &$>$24                     &$<$ 2.5                   &$<$ 0.05& over-resolved and unresolved\\
IRAS 13349+2438&4& 0.50$\pm$ 0.02&1.00                  &\ldots                &\ldots                   &$<$ 6.54                  &$<$ 3.3                   &$<$ 6.2& unresolved\\
      IC 4329 A&1& 1.03$\pm$ 0.02&0.56$\pm 0.05$        &0.58$^{+0.05}_{-0.05}$& 0.55$^{+0.05}_{-0.05}$  &30$^{+17}_{- 8}$          &$<$ 2.5                   &$<$ 0.83& resolved and unresolved\\
       Circinus&1&10.23$\pm$ 0.05&0.06$\pm 0.012$       &0.51$^{+0.12}_{-0.10}$& 9.61$^{+0.10}_{-0.12}$  &80.2$^{+ 9.1}_{- 5.4}$    &38.3$^{+ 4.4}_{- 2.6}$    & 0.78$^{+ 0.09}_{- 0.05}$& resolved and unresolved\\
       NGC 5506&2& 0.97$\pm$ 0.08&0.52$\pm 0.05$        &0.50$^{+0.01}_{-0.01}$& 0.37$^{+0.01}_{-0.01}$  &$>$37                     &$<$ 2.5                   &$<$ 0.35& over-resolved and unresolved\\
       NGC 5995&1& 0.33$\pm$ 0.01&0.57$^{+0.06}_{-0.08}$&0.15$^{+0.02}_{-0.03}$& 0.18$^{+0.03}_{-0.02}$  &22.8$^{+ 4.8}_{- 4.2}$    &$<$ 5.0                   &$<$ 2.48& resolved and unresolved\\
       NGC 7469&1& 0.52$\pm$ 0.02&0.37$\pm 0.05$        &0.23$^{+0.03}_{-0.03}$& 0.39$^{+0.03}_{-0.03}$  &46.7$^{+14.5}_{- 8.3}$    &13.6$^{+ 4.2}_{- 2.4}$    & 4.0$^{+ 1.2}_{- 0.7}$& resolved and unresolved\\
       \hline
       NGC 3227&0&0.32$\pm$0.02&$<$ 0.57&$<$ 0.150&\ldots&\ldots&\ldots&\ldots&limit on $F_p$ from non-detection\\
        IC 3639&0&0.54$\pm$0.03&$<$ 0.28&$<$ 0.150&\ldots&\ldots&\ldots&\ldots&limit on $F_p$ from non-detection\\
\end{tabular}
\end{table}
\end{landscape}

\clearpage
\clearpage
\clearpage
\clearpage
\clearpage
\clearpage
\clearpage
\clearpage
\clearpage
\clearpage
\clearpage
\clearpage
\clearpage
\clearpage
\clearpage
\clearpage
\clearpage
\clearpage
\clearpage
\clearpage
\clearpage
\clearpage
\clearpage

\clearpage

\clearpage

%
%
%
%
%
%
%
%
%
\clearpage
\section{Discussion}
%
%
\subsection{Torus scaling relations}

In the most simple conception of an AGN torus, the angular size it subtends at any wavelength simply scales with its innermost radius, i.e. the radius at which the dust sublimates. This radius was originally given by \citet{barvainis1987} as

\begin{equation}
	r_{\rm in, Barvainis} = 1.3 L_{\rm UV, 46}^{1/2} T_{1500}^{-2.8} {\rm pc}
\end{equation}

and depends on $L_{\rm UV, 46}$, the ultraviolet luminosity\footnote{We assume $L_{\rm bol} \approx 1.5 L_{\rm UV}$, from \citet{elvis1994} and \citet{runnoe2012}, with the assumption of a flat powerlaw spectrum in the optical \citep[e.g.][]{kishimoto2008}.} of the AGN in units of $10^{46}$ ergs/s, and the temperature $T_{1500} \equiv T/1500$~K at which grains are destroyed. It also depends on the dust grain size (the larger the grains the higher the opacity) and on the chemistry of the dust (graphite grains sublimate at higher temperature than silicate grains). 

More recently, \citet{kishimoto2007} studied the innermost regions of eight nearby ($z <$ 0.2) type 1 AGNs using NICMOS data from the HST archive. They found that the actual innermost radius, as determined by near-infrared dust reverberation mapping \citep{suganuma2006}, is about a factor of three {\em smaller} than the theoretically expected sublimation radius:

\begin{equation}
	\label{eq:r_in_makoto}
	r_{\rm in} = 0.47 L_{\rm UV, 46}^{1/2} {\rm pc}.
\end{equation}

On the other hand, in the same objects, the interferometrically determined radius of the hottest dust \citep{kishimoto2011,weigelt2012} is larger and lies somewhere between the reverberation radius and the radius given by Barvainis.\footnote{A likely interpretation for this difference is that interferometric measurements are sensitive to the brightness-weighted effective radius while the reverberation signal reacts to the actual innermost radius \citep{kishimoto2011}.} For the following, we will use the empirically derived innermost radius as defined in Eq.~\ref{eq:r_in_makoto}.

Since the angular size $\rho$ of the torus scales with distance $D$ as $\rho = r/D$ and the flux $F$ scales $\propto 1/D^2$, $\rho$ therefore scales as

\begin{equation}
\label{eq:th_prop_sqrt_f}
	\rho \propto \sqrt{F}
\end{equation}

We will make use of this ``intrinsic'' size scale, to separate resolution effects from intrinsic differences in the appearance of the AGN.

In order to compare the mid-infrared-derived radii with data from other wavelengths, we need to estimate bolometric correction factors for the mid-infrared luminosities. We use the prescription from \citet{gandhi2009} that combines their mid-infrared--X-ray relation and the X-ray bolometric correction factors from \citet{marconi2004} and give the inferred bolometric luminosities for all targets as well as the expected inner radii in Table~\ref{tab:radialfitr12}.

%
\begin{table*}
\caption{\label{tab:radialfitr12}Mid-IR luminosity of the sample targets, inferred bolometric luminosity, inner radii using Eq.~\ref{eq:r_in_makoto} and ``Barvainis radius'' at 300 K as well as observed mid-infrared half-light radius in pc.}
\centering
\begin{tabular}{l | l l | l l l}
\hline
Name               & $\log(L_{\rm MIR}/{\rm erg/s})$ & $\log(L_{\rm bol}/{\rm erg/s})$ & $r_{\rm in}$ & $r_{\rm 300 K}$ & $r_{1/2}$\\
                   &                                 &                                 &     [pc]     &  [pc]           &    [pc]\\
\hline\hline
         I Zw 1& 44.9& 46.2& 0.49& 124&                      $<$ 2.7\\
        NGC 424& 43.6& 44.7& 0.09&  23&      1.3$^{+ 0.3}_{- 0.2}$\\
       NGC 1068& 44.0& 45.2& 0.15&  37&      $1.6 \pm 0.1$\\
       NGC 1365& 42.5& 43.3& 0.02& 4.4&      0.3 $\pm$ 0.1\\
IRAS 05189-2524& 44.6& 45.9& 0.36&  89&                      $<$ 3.5\\
     H 0557-385& 44.4& 45.6& 0.25&  63&                      $<$ 1.6\\
IRAS 09149-6206& 44.9& 46.3& 0.53& 132&                      $<$ 2.7\\
   MCG-05-23-16& 43.5& 44.6& 0.08&  19&      1.9$^{+ 0.3}_{- 0.2}$\\
       Mrk 1239& 44.0& 45.2& 0.15&  38&      0.8$^{+ 0.4}_{- 0.5}$\\
       NGC 3281& 43.4& 44.4& 0.06&  15&                      $<$ 0.6\\
       NGC 3783& 43.7& 44.8& 0.09&  23&      2.0$^{+ 0.5}_{- 0.4}$\\
       NGC 4151& 43.0& 44.0& 0.04& 9.4&                     $>$  1.7\\
         3C 273& 45.7& 47.2& 1.54& 387&      8.1$^{+ 1.6}_{- 1.8}$\\
       NGC 4507& 43.7& 44.8& 0.10&  25&      3.0$^{+ 0.4}_{- 0.3}$\\
       NGC 4593& 43.1& 44.1& 0.04&  11&                       \ldots \\
     ESO 323-77& 43.7& 44.8& 0.10&  24&      1.1 $\pm$ 0.2\\
    Centaurus A& 41.8& 42.6& 0.01& 1.8&                      $<$ 0.05\\
IRAS 13349+2438& 45.5& 47.0& 1.22& 305&                      $<$ 6.2\\
      IC 4329 A& 44.2& 45.4& 0.19&  47&                      $<$ 0.8\\
       Circinus& 42.7& 43.6& 0.03& 6.4&      0.8 $\pm$ 0.1\\
       NGC 5506& 43.4& 44.4& 0.06&  16&                      $<$ 0.4\\
       NGC 5995& 44.1& 45.2& 0.16&  50&                      $<$ 2.5\\
       NGC 7469& 43.9& 45.0& 0.12&  30&      4.0$^{+ 1.2}_{- 0.7}$\\
\end{tabular}
\vspace{0.5cm}
\raggedright\\
\end{table*}

%
%
\subsection{Observability constraints and properties of the sample}
\begin{figure*}
	\sidecaption
	{\includegraphics[trim=2cm 2cm 2cm 2cm, width=0.7\hsize]{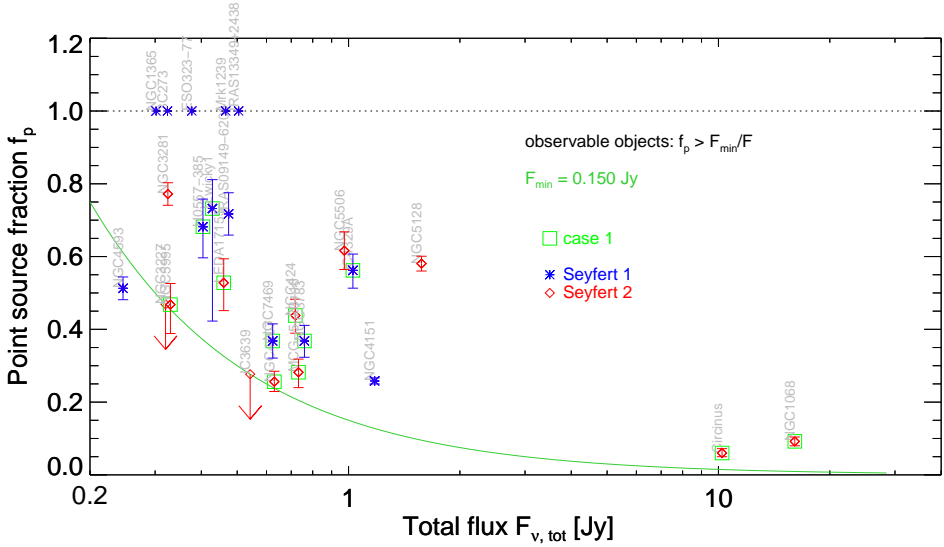}}
	\caption{\label{fig:stat:fp_f}Point source fraction $f_p$ as a function of total flux for all observed objects. Type 1 AGNs are marked with blue stars, type 2 AGNs with red diamonds. The two upper limits are unsuccessful fringe tracks in otherwise stable nights. The green curve denotes the observational limit which is given by the minimum unresolved flux required to track a source.}
\end{figure*}

In order to track fringes on an astronomical object with MIDI, the object's unresolved flux must be greater than some minimum flux $F_{\rm min}$ which we estimate to be $\approx$ 150 mJy at 12 \um, depending on atmospheric conditions. This means that the point source flux $F_p$ has to be at least of this order. In the context of our modeling, we give $F_p$ as a fraction of the total flux, i.e. the point source fraction $f_p$ (see Eq.~\ref{eq:fp} in Section~\ref{sec:radial:model}). This is the model parameter that we can determine best and for all objects with our MIDI observations.

In Fig.~\ref{fig:stat:fp_f} we show $f_p$ as a function of total flux $F_{\rm \nu, tot}$ which demonstrates that the weaker the source is, the higher the point source fraction has to be in order to observe it. This observational limit is drawn as a green line. One object (NGC 4593) has an unresolved flux slightly below 150 mJy. MIDI was able to track fringes on it under favorable conditions. For two more objects (NGC 3227 and IC 3639) we are confident to derive an upper limit to $f_p$ from the fact that the sources could not be tracked (see Section \ref{sec:notrack_limits}).

Our selection criteria only required the flux of a southern AGN in the mid-infrared to be $\gtrsim$ 300 mJy as observed within $\lesssim$ 0.5 arcsec, i.e. within the mid-infrared PSF of a 4-10 m class telescope. While our sample is not complete in any respect, these criteria do not introduce any obvious bias with respect to the flux distribution inside the unresolved source. Especially did we not pick out particularly ``pointy'' objects that would be easier to observe interferometrically since the observability constraint limits our ability to observe weak sources with very shallow intensity profiles, i.e. very low point source fractions. Given these relatively weak selection criteria, it is noteworthy that most sources from our initial sample selection have been observed successfully. We therefore conclude that it lies in the nature of weak AGNs (with fluxes $<$ 1 Jy) that a significant part of the mid-infrared emission on scales~$\lesssim$~0.5~\arcsec originates from scales $\lesssim$~5~mas (0.1 -- 10 pc)!

The median point source fraction in our sample is 70\% for type 1 and 47\% for type 2 objects, respectively. This effect is dominated by the two mid-IR brightest AGNs, that happen to be of type 2, and by the four type 1 objects that are unresolved ($f_p = 1$).

Centaurus~A (NGC~5128) sticks out due its relatively high point source fraction despite its high total flux. In this particular source we have strong evidence that about 50\% of its mid-infrared emission is of non-thermal origin \citep[][and in prep.]{meisenheimer2007,burtscher2010}.

\begin{figure*}
	\sidecaption
	{\includegraphics[trim=2cm 2cm 2cm 2cm, width=0.7\hsize]{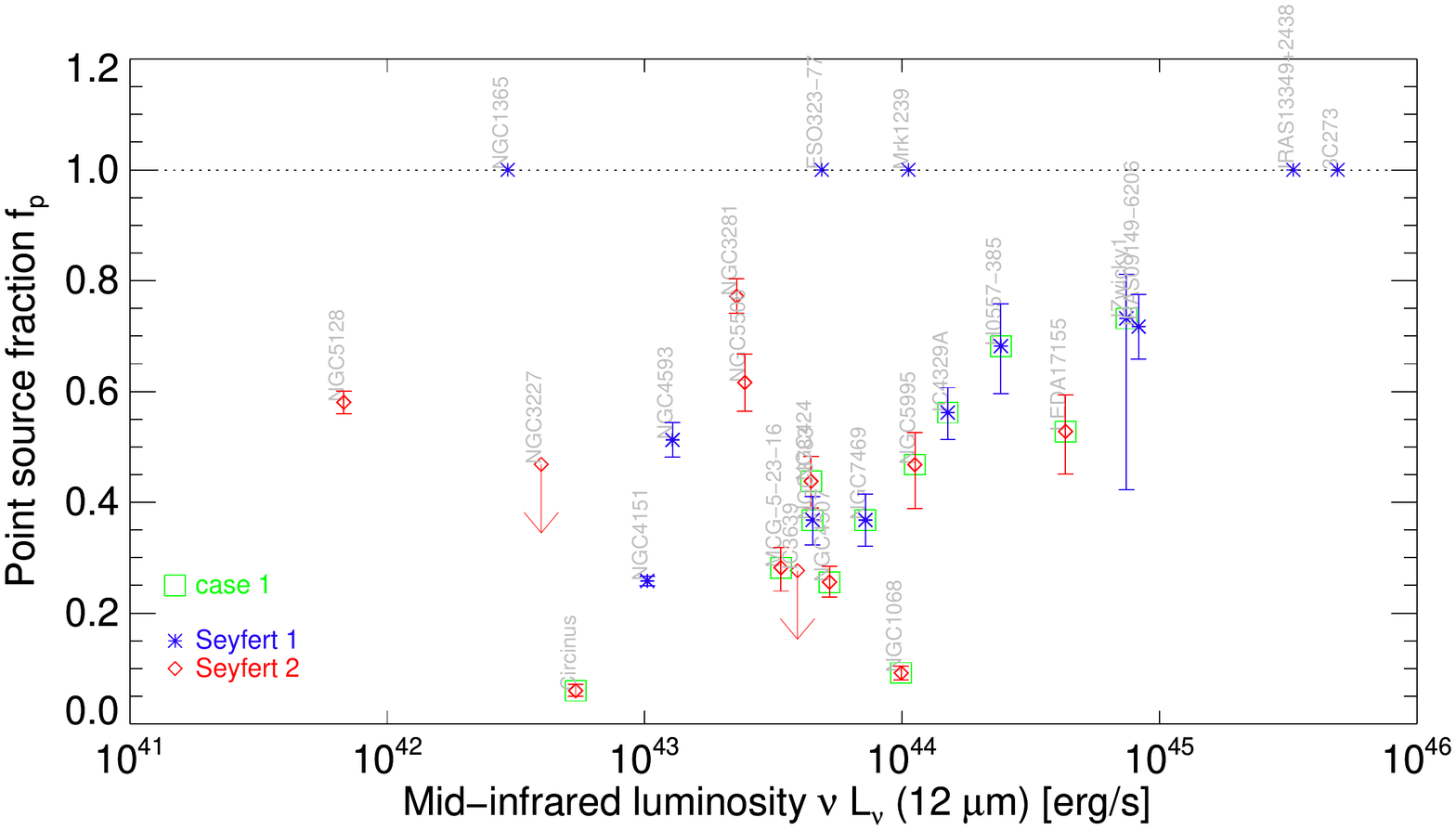}}
	\caption{\label{fig:stat:fp_lum}Point source fraction as a function of 12 \um luminosity. Blue stars are type 1 sources, red diamonds are type 2 sources. A green box denotes objects of model case 1 (resolved Gaussian + point source).}
\end{figure*}

At first sight, the point source fraction seems to increase with the mid-infrared luminosity of the source, Fig.~\ref{fig:stat:fp_lum}. At closer inspection, however, it becomes clear that this effect is simply due to the fact that type 1 sources -- which have higher point source fractions on average in our sample -- are less common in the nearby universe. In fact, in a formal correlation analysis (Spearman's rank correlation) we do not find a significant correlation in either type 1 or type 2 objects separately.

\begin{figure*}
	\sidecaption
	{\includegraphics[trim=2cm 2cm 2cm 2cm, width=0.7\hsize]{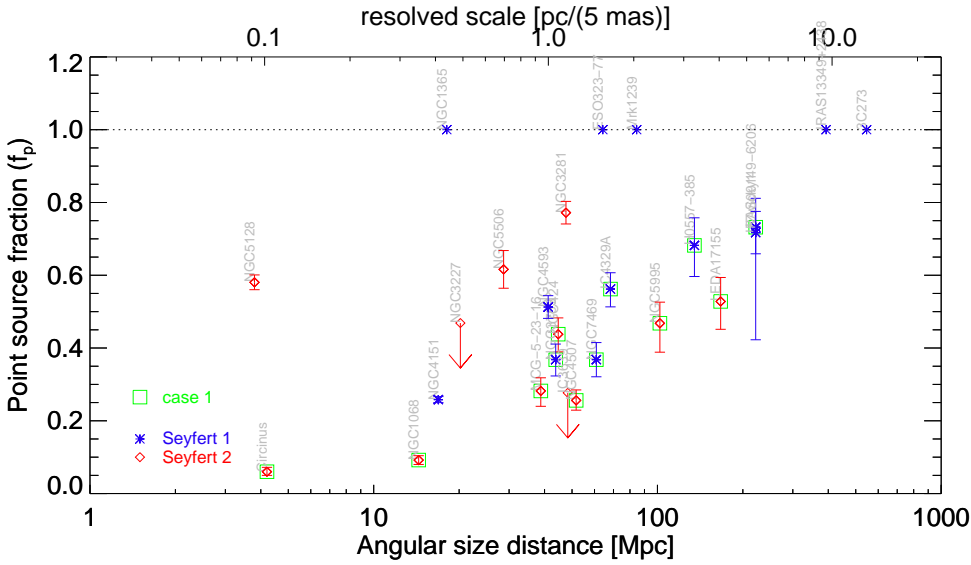}}
	\caption{\label{fig:stat:fp_scale}Point source fraction as a function of distance. Blue stars are type 1 sources, red diamonds are type 2 sources. A green box denotes objects of model case 1 (resolved Gaussian + point source).}
\end{figure*}

Since all but the two brightest targets in the sample are in a very narrow range of flux, luminosity and distance are nearly linearly related, see Fig.~\ref{fig:obs:lum_dist}. For this reason, the plot of point source fraction vs. distance (resolution), Fig.~\ref{fig:stat:fp_scale}, looks similar to the plot of point source fraction vs. luminosity (Fig.~\ref{fig:stat:fp_lum}), except for the two brightest objects. Note that in both plots, the resolved sources (the ones which are marked with green boxes) have the lowest point source fractions for a given distance or luminosity, i.e. these are the maximally extended objects. In both plots, there is only marginal evidence for a correlation. It appears that neither luminosity nor distance (resolution) are a strong predictor for the morphology of AGN tori.

\begin{figure*}
	{\includegraphics[trim=4cm 4cm 4cm 4cm, width=\hsize]{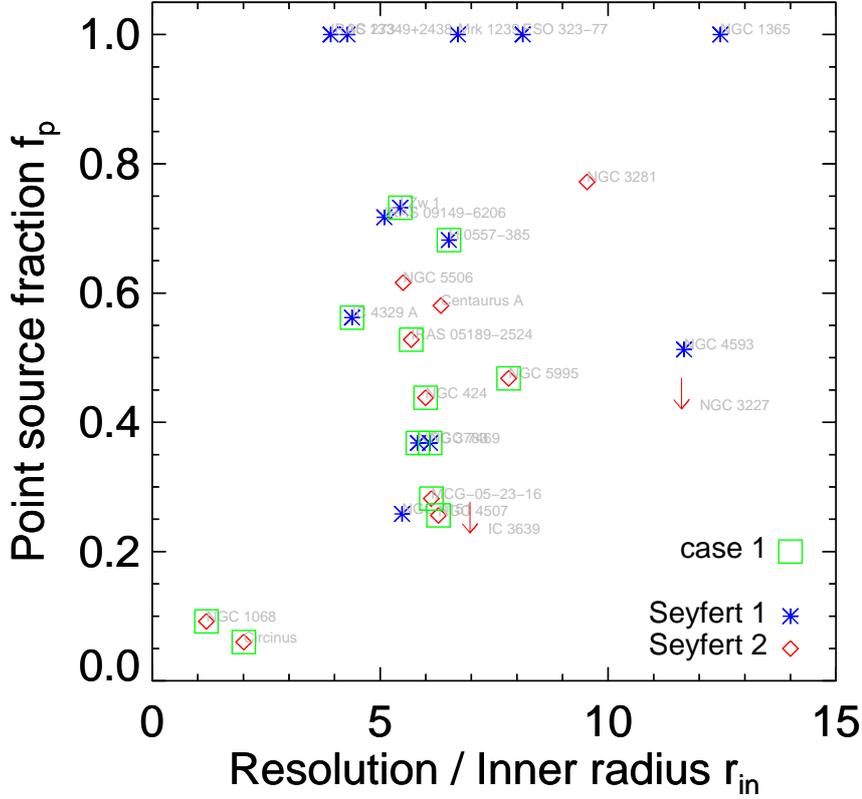}}
	\caption{\label{fig:stat:fp_rin}Point source fraction as a function of intrinsic resolution (it decreases to the right). The intrinsic resolution is the resolution of the interferometer divided by the inner radius of the source, i.e. 5 mas/(2 $\cdot r_{\rm in})$. In the two bright sources, MIDI's resolution almost reaches $r_{\rm in}$, while in the weak sources our resolution corresponds to radii 4 -- 12 times as large as $r_{\rm in}$. Type 1 sources are denoted as blue stars, type 2 sources as red diamonds, the source names are given next to the symbols. As before, objects which we model with case 1 are marked with a green square.}
\end{figure*}

This becomes even clearer when looking at a plot of point source fraction $f_p$ versus ``intrinsic resolution'', i.e. the resolution (in mas) divided by the inner-most radius of the target, Eq.~\ref{eq:r_in_makoto}. From Fig.~\ref{fig:stat:fp_rin} it is first of all evident that we only reach close to the inner-most radius of dust with MIDI in the two brightest objects, NGC~1068 and the Circinus~galaxy. This may well be the reason why NGC~1068 is the only galaxy in which MIDI has detected hot dust \citep{jaffe2004,raban2009}. However, despite a similar intrinsic resolution, no such hot dust component has been found in the Circinus galaxy \citep{tristram2007}.

In the fainter galaxies, the resolution of our observations varies between 4 and 12 $r_{\rm in}$ -- but the point source fraction does not seem to depend on it. In fact, for about half of the objects our resolution is essentially identical at $6 \pm 1 r_{\rm in}$ while $f_p$ varies between 0.25 and 1. There is no clear dependency on optical classification either, except at the extreme ends: The highly resolved, brightest and nearby galaxies are of type 2 and all objects with $f_p = 1$ are type 1 AGNs. In all objects where we interferometrically resolve an emitter (case 1 in the modeling, green boxes in Fig.~\ref{fig:stat:fp_rin}), the resolution is better than 8 $r_{\rm in}$, but there are only few objects with lower resolution. We conclude from this plot that the intrinsic resolution is also {\em not} a strong predictor to explain the differences in $f_p$. This implies that AGN tori do not exhibit a common radial profile, contrary to the suggestion by \citet{kishimoto2009}.

%
%
\subsection{Scaling the bright sources down}
\begin{figure*}
	\centering
	\subfloat{\includegraphics[trim=4cm 1cm 4cm 1cm, width=0.5\hsize]{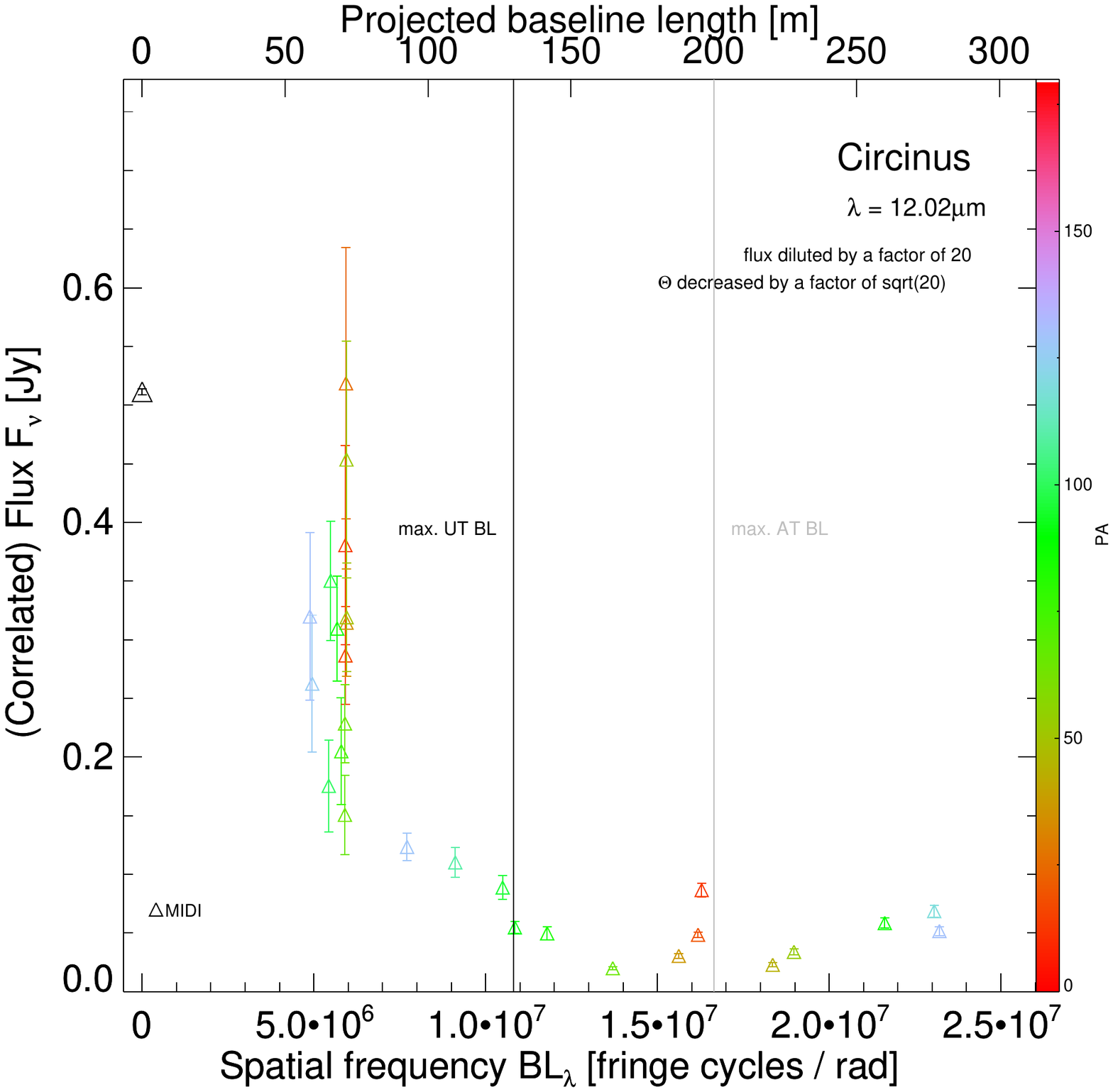}}
	~~~~~~~~~
	\subfloat{\includegraphics[trim=4cm 1cm 4cm 1cm, width=0.5\hsize]{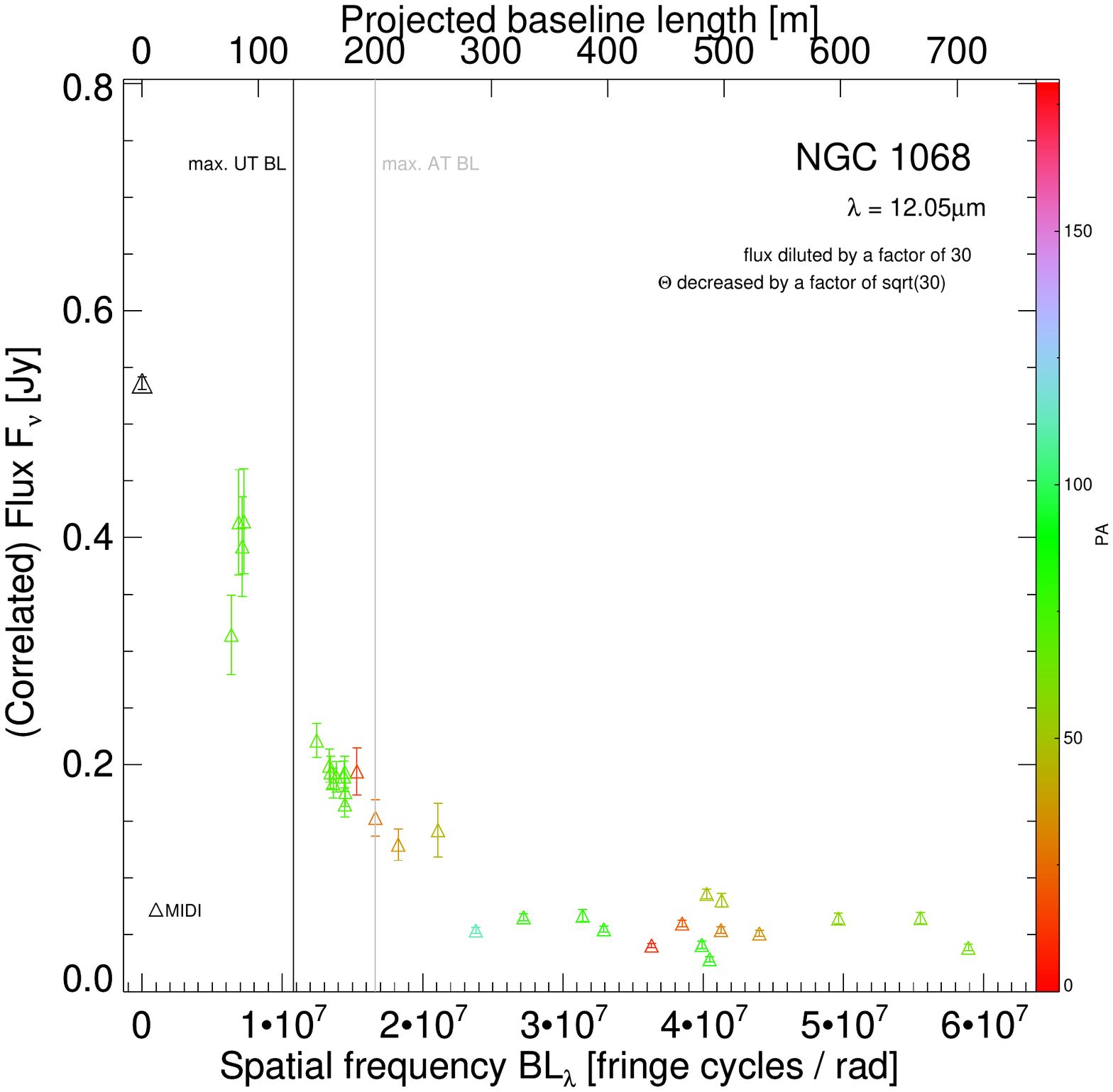}}
	\caption{\label{fig:discuss:LPexpect}Expected radial plot for Large Program sources, if the expectation is based on the two brightest targets, Circinus~galaxy and NGC~1068. For this Figure, these two targets have been ``placed'' at respective distances $\sqrt{20}$ and $\sqrt{30}$ times as far away as they actually are so that their flux matches the median flux of the weak sources. At these distances, baselines much longer than those offered at the VLTI (black / gray: maximum baseline using UTs / ATs) would be needed to resolve the elongated disk component that has been found with MIDI in these two bright and nearby sources.}
\end{figure*}

Can we understand the interferometrically derived properties of this sample of galaxies by scaling down the brightest sources such that their total flux is similar to that of the faint sources? NGC~1068 and the Circinus~galaxy are about a factor of 10 brighter than the next brightest AGN (NGC~4151)\footnote{We exclude the third brightest AGN, Cen A, from this analysis as approximately half of its mid-infrared flux is of non-thermal nature \citep{meisenheimer2007}}. The MIDI data on these two sources were modeled with two Gaussian emitters representing a compact disk (contributing less than 10\% to the total flux) and an extended emission component \citep{tristram2012,raban2009}. We can easily imagine how these sources would appear if they were at distances such that their fluxes matched those of the weak sources (Fig.~\ref{fig:discuss:LPexpect}).

The radial plots (Fig.~\ref{fig:discuss:LPexpect}) would look the same as Figs.~\ref{fig:rad:NGC1068} and \ref{fig:rad:Circinus}, but note how the axis ranges have changed: While the ordinate shows the lower flux level, as intended, the abscissae show that much longer baselines would be needed to collect the same data. When comparing these remote versions of the brightest sources with the weak targets, one should therefore only look at the data to the left of the vertical black lines which denote the maximum baseline length achievable with UTs.

The fluxes at the maximum UT baselines would be 100 mJy for the remote version of the Circinus~galaxy and 200 mJy for the remote NGC~1068, thus comparable to the weak sources. In both sources, however, the radial plots would show strongly falling fluxes with baseline length (as the extended component would be well resolved) and, in both sources there would also be a clear PA-dependence due to the elongation of the source which is not seen such strongly in the weaker targets.\footnote{In NGC~1068, the position-angle dependence of the correlated flux on short baselines is not seen in this plot, because only very limited amount of AT data were available for this plot. New AT observations have been performed in September 2012 that show the large-scale elongation more clearly (Lopez et al. in prep.).} The level of unresolved flux would be relatively low ($\approx$ 20 \%) in Circinus compared to the weak targets while in NGC 1068 it would be $\approx$ 50 \%, close to the median $f_p$ for all the weak type 2 sources (47 \%). Another difference between the scaled-down bright sources and the weak sources are the differential phases: only the Circinus~galaxy and NGC~1068 show a significant, non-zero, signal in the differential phases (see Section \ref{sec:dphases} in the appendix). This is not a signal/noise effect as significant differential phases are observed in NGC~1068 also with the smaller auxiliary telescopes where the signal/noise is comparable to the observations of faint targets with the larger UTs (Lopez et al. in prep.).

Additionally, the unresolved flux of these scaled sources would appear different from the ``point source'' that is seen in many of the weaker targets, both in type~1 sources (NGC~4151, NGC~4593, IC~4329~A, NGC~7469) and in type~2 sources (NGC~4507 and NGC~5506): In the scaled bright sources, a strong decrease of correlated flux with baseline length would be observed and, especially in the case of NGC 1068, there would be no sign for a flat part in the radial plot that would point to a distinct second component.

Lastly, looking at the spectra (Fig.~\ref{fig:spectra} and following) shows that all of the weak sources with $r_{1/2} >$ 5 mas feature either the [S IV] or the [Ne II] line (or both) in their spectra. On the other hand, only 50 \% of the sources with upper limits to their half-light radius show any mid-IR lines and in no correlated flux spectrum are any lines observed. Mid-IR emission lines arise, for example, in star forming regions and in AGN narrow-line regions \citep{sturm2002,groves2008}. However, the two bright sources do not show any lines in their total flux spectra.

We conclude that we cannot fully explain the observed characteristics of the faint sources by simply scaling down the bright sources.

%
%
\subsection{Is there an AGN size-luminosity relation?}

%
%
\begin{figure*}
	\centering
	\subfloat{\includegraphics[trim=7cm 4cm 7cm 4cm, width=0.5\hsize]{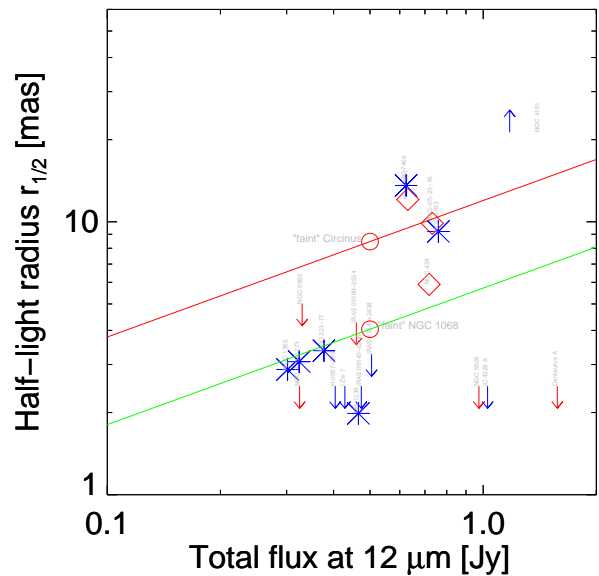}}
	\caption{\label{fig:stat:r12_f}Half-light radius$r_{1/2}$ in observed (angular) units as a function of total flux $F_{\rm \nu, tot}$ at 12 \um. Type 1 sources are shown as blue stars, type 2 sources with red diamonds. The two colored lines denote the expected source sizes using the scaling $r_{1/2} \propto \sqrt{F}$ and normalizations from NGC~1068 (green line) and Circinus (red line) as a reference. The two red circles denote the sizes of the scaled-down versions of the bright sources. They are compatible with the actual sizes of the faint targets, showing that the source sizes indeed scale with $\sqrt{F}$. However, the relation is dominated by large scatter, especially when including the limits. The statistical errors of the MIDI size measurements are smaller than the symbols and therefore not shown.}
\end{figure*}

With different prescriptions used for determining the size and different selection of the sources, the AGN torus size-luminosity relation has been investigated with controversial results: slopes between 0 \citep{kishimoto2011b} and 0.5 \citep{tristram2009} are compatible with the data. Can we solve this controversy using the large amount of data presented here and the significantly increased number of sources?

\begin{figure*}
	\sidecaption
	{\includegraphics[trim=2cm 2cm 2cm 2cm, width=0.7\hsize]{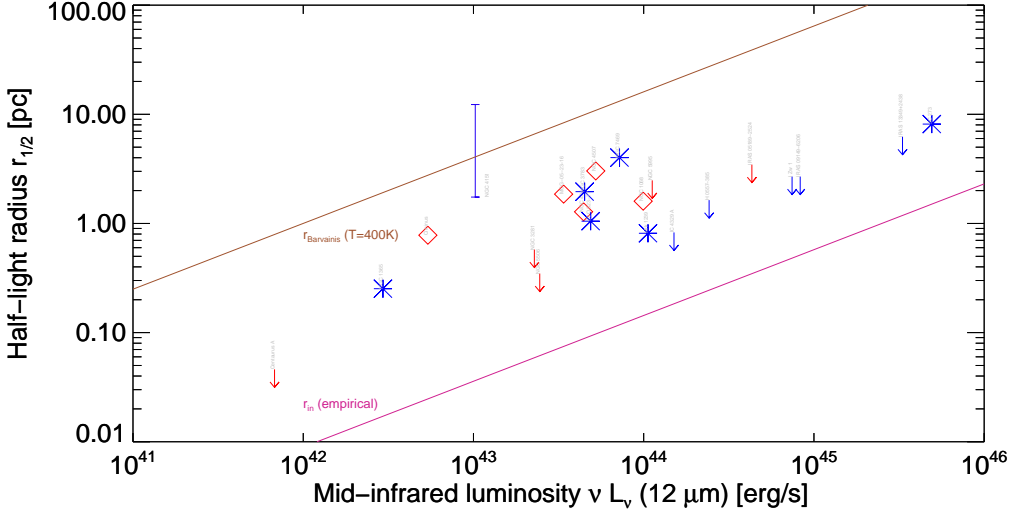}}
	\caption{\label{fig:stat:sl_b}Half-light radius as a function of mid-IR luminosity; blue stars / red diamonds denote type 1/2 AGNs; the error bar for NGC~4151 is not a statistical error but shows the range for $r_{1/2}$ from the upper limit (from single-dish observation) and lower limit from our modeling. The magenta and red lines show the innermost radius as derived empirically (Eq.~\ref{eq:r_in_makoto}) and the outermost radius for 300 K warm dust as given by the Barvainis formula. The half-light radii at 12 \um show a large scatter, but are about 20 times larger than at 2 \um. They are almost always a factor of three smaller than the Barvainis radius. The statistical errors of the MIDI size measurements are smaller than the symbols and therefore not shown. This plot makes use of the bolometric correction factors described in the text.}
\end{figure*}

We would like to caution that, in a flux-limited sample with narrow flux range, a correlation in size vs. luminosity appears even for an uncorrelated set of points simply because more luminous objects are also more distant. This is most clearly seen for the upper limits of Fig.~\ref{fig:stat:r12_f} that turn into a narrow ``correlation'' in Fig.~\ref{fig:stat:sl_b}. We therefore first investigate the size-luminosity plot in observed units (Fig.~\ref{fig:stat:r12_f}). For the resolved sources (denoted with stars and diamonds), we find a correlation between $r_{1/2}$ and flux that is roughly compatible with $\sqrt{F}$. The two colored lines show the expected scaling of source size with flux using the normalizations from the bright sources: Many of the faint sources show sizes that are compatible with the half-light radii of the scaled versions of the Circinus~galaxy and NGC~1068 (albeit all the differences in the detailed characteristics, see above). It is also immediately obvious that the relation is dominated by large scatter, especially if also taking the limits into account.

%
%

\begin{figure*}
	\sidecaption
	{\includegraphics[trim=2cm 2cm 2cm 2cm, width=0.7\hsize]{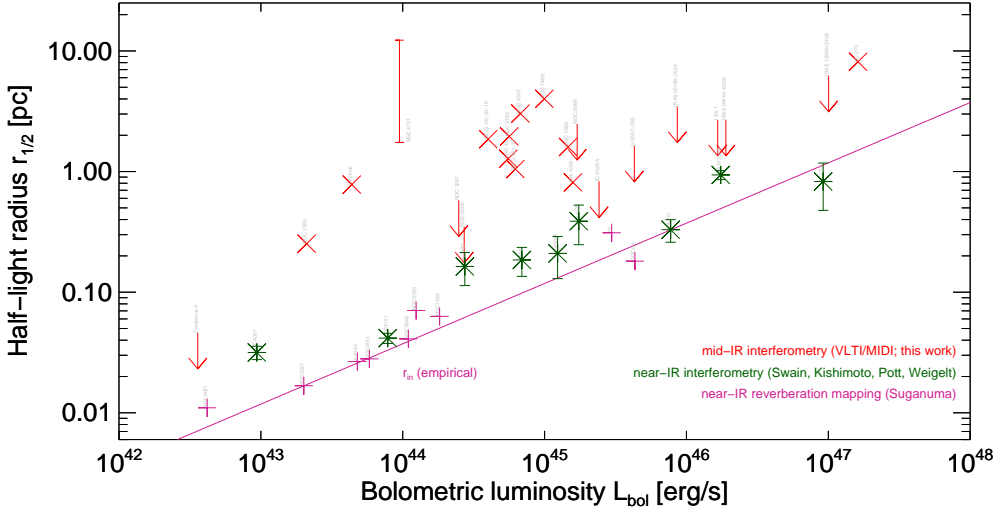}}
	\caption{\label{fig:stat:torus} (Half-light) radius at various pass-bands, probing the structure of the torus, as a function of bolometric luminosity. The red crosses and arrows are our own size estimates from MIDI observations. The green stars are the near-infrared interferometry observations and derived ring-fit radii \citep{kishimoto2011,weigelt2012}. The magenta stars are the UV/optical--near-infrared reverberation mapping radii from \citep{suganuma2006} and references therein; the magenta line a fit to these data. The scatter in the mid-infrared sizes is significantly larger than in the near-infrared. The error bar for NGC~4151 is not a statistical error but denotes the range for $r_{1/2}$ from the upper limit (from single-dish observation) and lower limit from our modeling. The statistical errors of the MIDI size measurements are smaller than the symbols and therefore not shown.}
\end{figure*}

In physical units, we find from the size-luminosity plot (Fig.~\ref{fig:stat:sl_b}) that the half-light radii of the resolved sources are about 20 times larger at 12 \um than at 2 \um, but almost always three times smaller than the radius $r_{\rm Barvainis}$ expected for an optically thin medium (Figs.~\ref{fig:stat:sl_b}, \ref{fig:stat:torus}). Either there is significant absorption, even at mid-IR wavelengths, limiting the extent of the mid-IR emission or there is just no AGN-heated dust at larger scales. The very compact upper limits, on the other hand, show that there is significant amount of mid-IR emitting material on scales as compact as $\approx 4 \times r_{\rm in}$.

It is also evident from this plot that there is essentially no difference in size between type 1 and type 2 sources, contrary to the expectation from clumpy torus models that tori of type 2 sources appear on average about a factor of 2.5 more extended than those of type 1 sources, albeit with large scatter \citep{tristram2011}. This is in agreement with the fact that both type 1 and type 2 tori follow the same mid-infrared -- X-ray relation \citep{horst2008,gandhi2009}.

%
%

In order to explore the scatter in the mid-infrared, we compile all nuclear size measurements of AGN tori from UV/optical -- near-infrared reverberation mapping \citep[][and references therein]{suganuma2006}, near-infrared interferometry \citep{kishimoto2011,weigelt2012} and our own mid-infrared size estimations using MIDI data (Fig.~\ref{fig:stat:torus}). For the mid-infrared we use the bolometric corrections described at the beginning of this Section, for the near-infrared data we use the conversion from \citet{kishimoto2007}, that is based on \citet{elvis1994}. In this plot, color now denotes observing wavelength and technique; the MIDI points are displayed as red crosses, near-IR interferometry measurements are shown in green and the reverberation data in magenta. The correlation is significantly better for the near-infrared data with a highly significant $r$ correlation value of 0.97, while in the mid-infrared, we find $r$ values of 0.77 or 0.56 depending on whether or not the limits are included in the analysis (see Table~\ref{tab:discuss:torus}).\footnote{Some of the scatter in the near-IR data is due to variability which does not play a role in the mid-IR. For NGC~4151, for example, three measurements are available with about 0.5 dex spread both in luminosity and in radius. While it is unclear how exactly the inner radius of dust (as probed by reverberation mapping) changes with luminosity variations \citep{koshida2009,hoenig2011b}, the radius as measured with near-IR interferometry does not seem to undergo any fast changes even in near-IR variable objects \citep{pott2010,kishimoto2011}. In the mid-infrared, so far no changes of the torus structure have been reported either.}

Since the mid-IR sizes are dominated by large scatter, we do not fit a linear slope to this relation as it would highly depend on the selection of sources (only resolved / marginally resolved sources or all sources?). While there is a general trend of increasing source size with luminosity, we conclude that there may not be a meaningful common size--luminosity relation for all AGN tori as their structure appears to be very different from source to source. A probable reason for the large scatter is that the nuclear mid-IR emission of AGNs may consist of at least three distinct components: large-scale emission from AGN-heated dust in the narrow line region or heated by star-formation, compact AGN-heated dust and non-thermal emission. The relative strengths of these components determine the observed nuclear mid-infrared size of the AGN.

\begin{table*}
\caption{\label{tab:discuss:torus}$r$ correlation values for the near-infrared reverberation mapping and for the near- and mid-infrared interferometry observations of AGN tori.}
\centering
\begin{tabular}{l l l l}
Observational technique             & number of sources & $r$ correlation value & $p$ value\\
\hline
optical--near-IR reverberation mapping          & 9     & 0.97                  & $2.2 \cdot 10^{-5}$ \\
near-IR interferometry                         & 9     & 0.97                  & $2.2 \cdot 10^{-5}$ \\
mid-IR interferometry (measurements + limits)  & 21    & 0.77                  & $3.9 \cdot 10^{-5}$\\
mid-IR interferometry (measurements only)       & 11    & 0.56                  & 0.071 \\
\end{tabular}
\vspace{0.5cm}
\raggedright\\
\end{table*}

%
%
%
%
%
%
%
%
%

\section{Summary, Conclusions and Outlook}
We present here the largest sample of AGNs studied with infrared interferometry. In total we present interferometric data for \nsources AGNs and derive limits for another two targets, using new observations from the MIDI AGN Large Program, as well as data from the archive. Since most of the sources in this sample are close to the instrumental sensitivity limit, we developed both an improved observing strategy and modified the data reduction and calibration procedures to be able to reliably analyze a large number of weak targets.

We summarize our results as follows:

\begin{enumerate}

\item {\bf Many of the faint sources show unexpectedly high levels of unresolved flux.} \quad From the empirical scaling of the innermost radius $r_{\rm in}$ with luminosity as $r_{\rm in} \propto \sqrt{L}$ and the structure of the two mid-IR bright sources, we would have expected the faint sources to be highly resolved as well. However, among type 1 sources, the median unresolved flux is 70 \% of the total flux, in type 2 sources it is still 47 \%. This shows that a significant part of the mid-infrared flux of these AGNs originates from scales $\lesssim$ 5 milli-arcseconds (mas; 0.1 -- 10 pc) in diameter. On the one hand, this is the main reason for being able to observe so many AGNs with MIDI at the VLTI. On the other hand, {\em the high level of unresolved flux means that even higher resolution, i.e. longer baselines, are required to resolve the detailed shape of the dust emission.}

\item {\bf The derived half-light radii of the sources are between a few mas and a few tens of mas and correspond to physical radii of tenths of parsecs to almost 10 pc.} There is a large variety of half-light radii, even at similar luminosities (e.g. between $3 \cdot 10^{43}$ and $10^{44}$ erg/s). This is another manifestation of the variety of unresolved flux (point source fraction $f_p$). For most of the sources we are able to derive a half-light radius, based on our models, albeit half of them only as (very compact) limits.

\item {\bf The large variety of nuclear mid-infrared morphologies is not due to resolution.} \quad There is essentially no correlation between point source fraction $f_p$ and luminosity or distance (resolution). Even the correlation with intrinsic resolution, i.e. resolution in units of the innermost radius of dust, is not very strong. While the detailed shape of the nuclear mid-IR emission can only be resolved in objects where the resolution is less than two times the innermost radius of dust, for the weak sources no clear dependence is seen. On the contrary: {\em For the same intrinsic resolution, very different levels of unresolved flux are observed in different sources.}

\item {\bf In many sources, the radial profiles indicate that the unresolved flux originates from a distinct ``point source'' within this size limit.} The rest of the flux is either (12 sources) in a component that we resolve with the interferometer or stems from over-resolved emission that is not observed on any baseline with the interferometer (6 sources). In the other five sources we find only a single component of emission that is marginally resolved in four and unresolved in one.

\item {\bf Simple, axisymmetric radial models provide good fits to most sources, but there is evidence for elongations in nine sources.} The sources with detected elongations are mostly highly resolved ones ($f_p \lesssim 50\%$). In the sources with the best $(u,v)$ coverages, NGC~424, the Circinus~galaxy and NGC~3783, these elongations have been studied in detail and are compatible with emission in the polar direction, possibly in the narrow line region / outflow cone of the AGN \citep{hoenig2012,tristram2012,hoenig2013}. But if the (over-)resolved emission is dust in the polar region, it cannot account for the toroidal obscuration. This leaves the unresolved ``point source flux'' as the only contribution from the torus. However, in the brightest sources, this ``point source'' is resolved into a very thin disk that does not seem to provide the opening angle required for unification. We would like to invite modelers to investigate this {\em apparent contradiction} to the unified model of AGNs.

\item {\bf We see no difference between type 1 and type 2 sources in terms of sizes, radial profiles or elongations.} The only distinction we find between type 1 and type 2 sources is the higher point source fraction of the former. This is in contrast to expectations from clumpy torus models that see type 2 sources more extended \citep{tristram2011} and indicates that {\em torus orientation is not the only factor in creating the different point source fraction of type 1 and type 2 sources}.

\item {\bf While the nuclear mid-infrared size does scale with luminosity in a similar way as the near-infrared size, the mid-infrared sizes show much larger scatter.} We investigate both the angular-size--flux relation and the physical-size--luminosity relation and find that brighter / more luminous AGNs indeed have larger mid-IR sizes, as expected from the established scaling relations in the near-infrared. However, the mid-infrared relation exhibits significantly more scatter than in the near-infrared where the emission is dominated by the hot dust from the innermost radius: {\em The mid-infrared half-light radius is between less than 4 and 20 times larger than the innermost torus radius $r_{\rm in}$.} A plausible reason for the larger scatter in the mid-IR is that at 12 $\um$, the emission is composed of multiple components, possibly of different physical nature. Apart from dust emission from a nuclear disk (as resolved in the two brightest sources), the compact emission can also arise due to non-thermal processes (e.g. the unresolved flux in Centaurus~A). The interferometrically (over-)resolved emission, on the other hand, is likely to arise in the polar region of the AGN or in the circum-nuclear star-forming region.

\end{enumerate}

\subsection{Outlook: Prospects for AGN studies with mid-IR interferometry}

MIDI is now close to decommissioning due to the arrival of the second generation instruments at the VLTI and it seems that most AGN observations possible with MIDI have been performed. 

In the near future, the second generation mid-infrared interferometer MATISSE will observe in the $L$ ($\approx 3.8 \um$), $M$ ($\approx 4.5 \um$) and $N$ (8--13 \um) bands and provide up to six baselines at a time as well as closure phases allowing to reconstruct images. While the $L$ and $M$ bands will provide higher spatial resolution than the $N$ band, AGNs will likely still be best resolved at 12 \um, at least if the existing near- and mid-IR observations are any guidance.

The images that MATISSE will provide will be crucial to unveil the structure of the resolved emission and possibly connect this emission to observations on larger scales. For such observations, a high sensitivity is required as should be provided with an external fringe tracker\footnote{PRIMA's FSU has been demonstrated to work as an external fringe tracker for MIDI \citep{mueller2010,pott2012} and increases the sensitivity for sources with blue infrared colors. Unfortunately AGNs typically have red mid-IR colors and could therefore not yet profit from this external fringe tracker.}. A next-generation fringe tracker for the VLTI, currently under investigation at ESO, will likely boost the sensitivity for fringe tracking and will unleash the full potential of MATISSE for AGN torus studies.

A much increased sensitivity would allow us to observe faint AGNs on the longest baselines of the VLTI platform (200 m), only accessible with the smaller Auxiliary Telescopes. In order to resolve the ``point sources'' in faint AGNs into a possible disk component, however, much longer baselines of the order of 300 -- 700 m are required that will be possible with a new generation of optical--infrared interferometers.

%
%
%
%
%
%
%
%
%
%

\begin{acknowledgements}
The authors thank the anonymous referee whose thorough reading and constructive criticism helped to improve the readability of the paper.
The authors would also like to thank the VLT(I) crew for their support during many years of observations. L.B. acknowledges helpful discussions with Hagai Netzer, Gilles Orban de Xivry, David Rosario, Daniel Asmus, Dieter Lutz and Daniela Klotz. S.F.H. acknowledges support by Deutsche Forschungsgemeinschaft (DFG) in the framework of a research fellowship (Auslandsstipendium).
\end{acknowledgements}

\bibliographystyle{apj}
\bibliography{../../../apj-jour,../../../papers}

\Online
\section{Data reduction and calibration}

\subsection{Correlation losses}

\begin{figure*}
	{\includegraphics[trim=3cm 3cm 3cm 3cm, width=0.5\hsize]{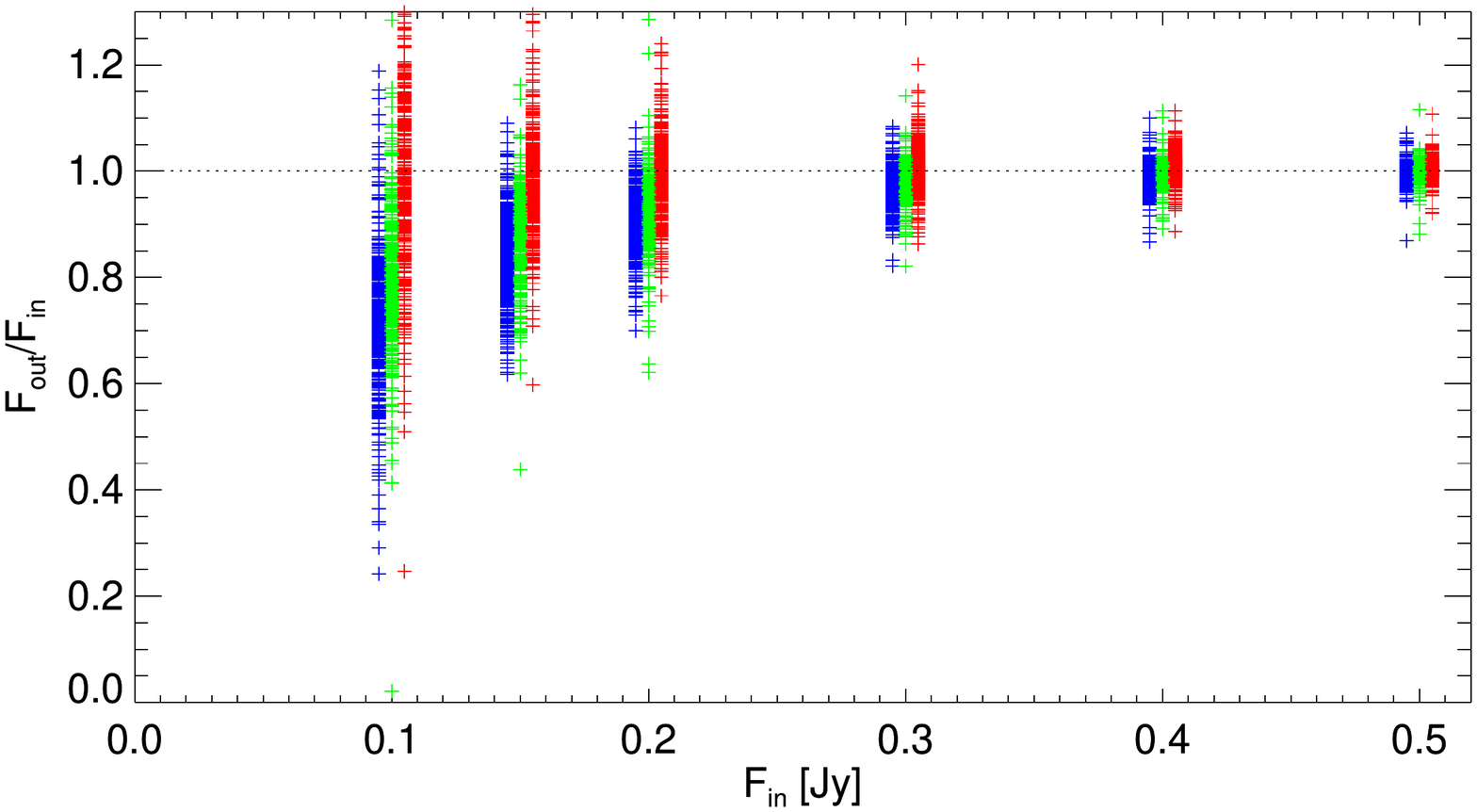}}
	\caption{\label{fig:obs:decorr_fluxes}Correlation losses with EWS 2.0. The blue, green and red crosses denote the recovered flux as a fraction of simulated input flux for 8.5, 10.5 and 12.5 \um. One can see (numerically also from Tab.~\ref{tab:obs:decorr}) that correlation losses do not play a role at 12.5 \um, but have to be corrected at shorter wavelengths for fluxes lower than $\approx$ 200 mJy.}
\end{figure*}

\begin{table*}
\caption{\label{tab:obs:decorr}Correlation losses}
\centering
\begin{tabular}{l l l l}
\hline
&\multicolumn{3}{c}{$F_{\rm out}/F_{\rm in}$ (faint data + faint GD)}\\
Flux [Jy] & 8.5 \um & 10.5 \um & 12.5 \um\\
\hline\hline
0.500&0.995$\pm$0.001&0.998$\pm$0.001&1.006$\pm$0.001\\
0.400&0.988$\pm$0.001&0.994$\pm$0.001&1.007$\pm$0.001\\
0.300&0.969$\pm$0.002&0.979$\pm$0.002&1.004$\pm$0.002\\
0.200&0.909$\pm$0.003&0.929$\pm$0.003&0.990$\pm$0.003\\
0.150&0.839$\pm$0.004&0.876$\pm$0.004&0.978$\pm$0.005\\
0.100&0.719$\pm$0.007&0.804$\pm$0.007&1.009$\pm$0.011\\
\end{tabular}
\vspace{0.2cm}
\end{table*}

Since the accuracy of the group delay estimation depends on the signal to noise of the data and since the group delay estimate is derived from the data itself, a bias is introduced for weak sources that is not calibrated using any standard calibration technique: The jitter of the group delay estimate around its true value is larger for weak sources than for bright sources, causing a reduction in flux for weak sources, the so called ``correlation losses''. To estimate the severity of this unavoidable effect, we simulated observations of weak targets by adding the ``diluted'' signal of a bright calibrator to noise-only datasets and reduced and calibrated these simulated observations as if they were real observations. This revealed significant correlation losses \citep{burtscher2012b}. After tweaking various parameters of the data reduction software, EWS 2.0 is now able to produce estimates of the correlated flux of weak targets that are nearly free of this bias down to a flux level of about 100 mJy at 12 \um. However, at shorter wavelengths, where a given error in group delay estimation translates into a larger phase error, significant losses ($>$ 10\%) still occur for fluxes lower than about 200 mJy (see Fig.~\ref{fig:obs:decorr_fluxes} Tab.~\ref{tab:obs:decorr}).

A different approach to estimate correlation losses for weak sources -- using real observations of weak standard stars -- with similar outcome has been developed by \citet{kishimoto2011b}. See \citet{burtscher2012b} for a comparison of the two methods.

\subsection{Direct flux calibration}

\begin{figure*}
	{\includegraphics[trim=3cm 3cm 3cm 3cm, width=0.5\hsize]{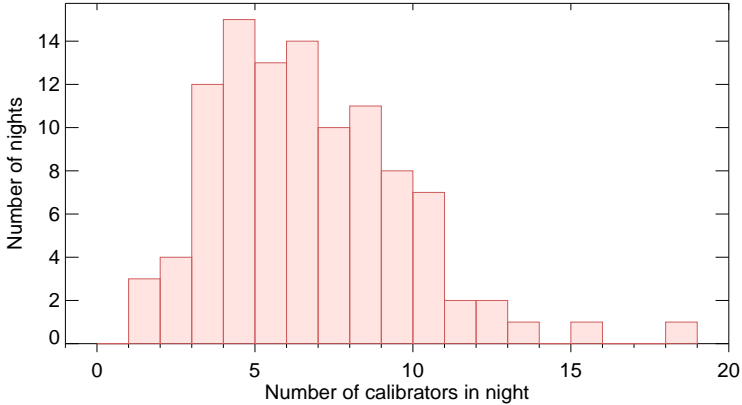}}
	\caption{\label{fig:obs:cal_histogram}Histogram of number of calibrators per night. In most nights enough calibrator measurements were available to reliably determine the variance in the conversion factor.}
\end{figure*}

\begin{figure*}
	{\includegraphics[trim=3cm 3cm 3cm 3cm, width=0.5\hsize]{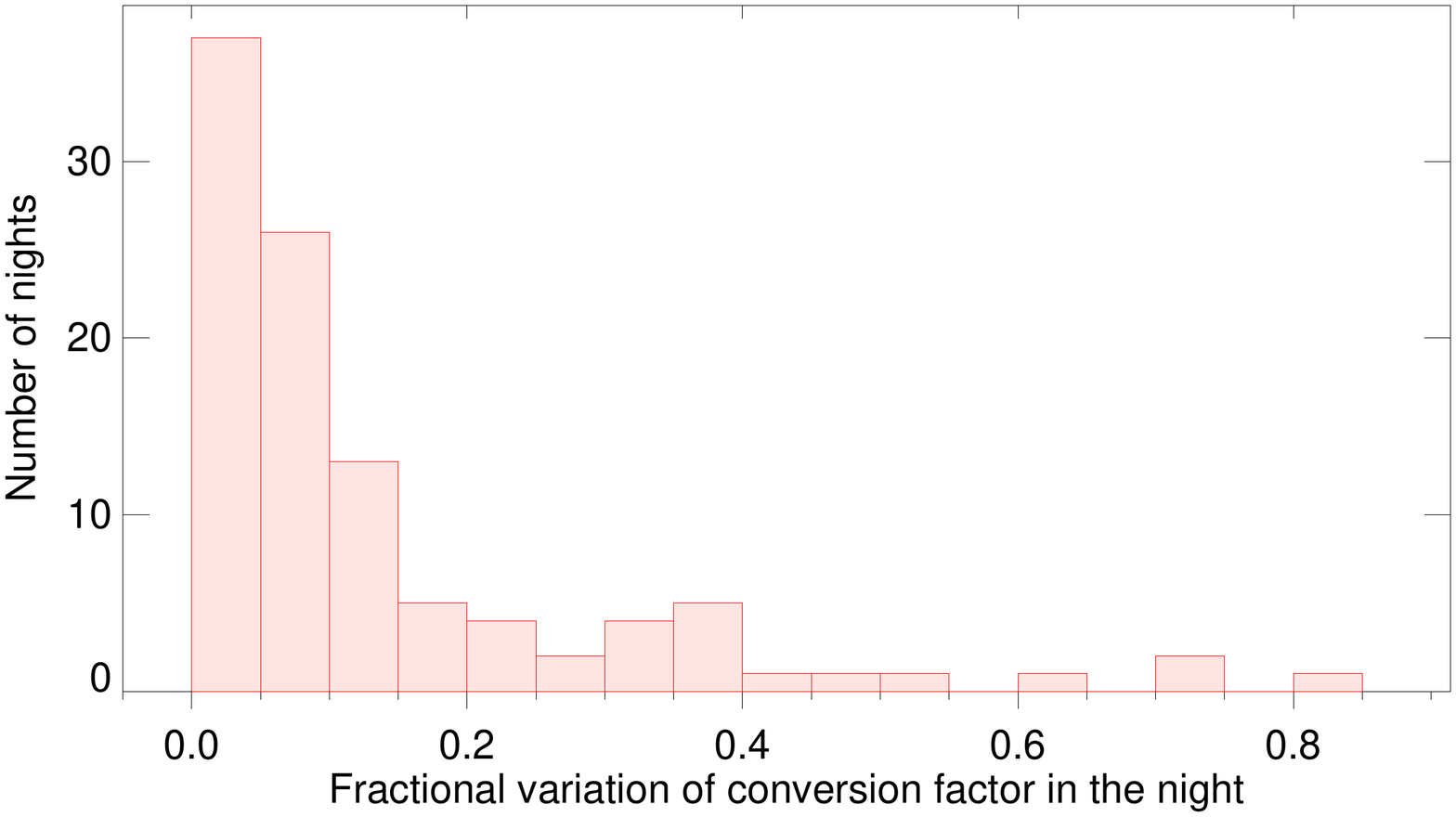}}
	\caption{\label{fig:obs:gainvar_histogram}Histogram of relative Conversion Factor (CF) variations for all nights with five or more calibrator observations. The tail of very large CF variations ($>$ 0.4) originates from early observations (before 2007).}
\end{figure*}

We calibrate the correlated fluxes -- the principal outcome of any MIDI ``fringe track'' observation -- directly with the correlated flux of the calibrator as derived from the flux and diameter of the star taken from the catalog of \citet{vanboekel2004}. The standard calibration is to take also single-dish measurements (in the community referred to as ``photometry'') and compute visibilities, i.e. the ratio correlated flux / ``photometry''. The standard procedure has the advantage that variations in the VLTI's and atmospheric transfer function are negligible since correlated flux and photometry are taken right after each other without a time-consuming acquisition of a new target. However, for the weak targets, the ``photometry'' measurements are of low quality due to an incomplete background subtraction (see Section~\ref{sec:obs:phot}).

When directly calibrating correlated fluxes, variations in the conversion factor (CF)\footnote{The CF is defined as the ratio between the measured count rate on the detector and the known flux of the object, as a function of wavelength.} have to be taken into account. These variations often constitute the largest uncertainty in any MIDI direct flux observation. We investigated a number of possible dependencies, such as variations with airmass, seeing and coherence time $\tau_0$ -- but found no clear correlations \citep{burtscher2012b}. We therefore treat these variations as noise and add the relative variation of the CF, i.e. the standard deviation of the CF as a fraction of the average CF of the night, quadratically to the photon noise of each measurement. We set the minimum error to 5\%, since we estimate remaining systematic errors to be of that order.

In most nights, enough calibrator measurements were available to reliably determine the CF variance. The median number of calibrators per night was 6 with a relatively normal distribution (see Fig. \ref{fig:obs:cal_histogram}). The average (median) CF variation within a night was 13.2 \% (7.6 \%). A histogram of the relative CF variations for all nights with more than five calibrator observations is given in Fig.~\ref{fig:obs:gainvar_histogram}. In nights with clear trends of the CF or when more than one baseline was used, we split the nights into bins and determined the CF variance for each bin separately.

Thus, the CF variations do not prohibit the direct flux calibration method. However, when observing in this mode, it is advisable to observe calibrators as often as possible to allow for an accurate estimate of this error source.


\subsection{Stacking}
While estimating the CF variations from many or all calibrators of the night, we calibrated each target observation only with the CF measurement of the nearest calibrator, which we define as the one that minimizes a penalty $p$:

\begin{equation}
\label{eq:obs:closestcal}
p = \frac{\Delta t}{t_0} + \frac{\Delta \phi}{\phi_0}.
\end{equation}

A difference in angular distance $\Delta \phi = \sqrt{(\Delta {\rm RA} \cdot \cos \delta)^2 + (\Delta \delta)^2}$ of 15$^{\circ}$ was considered as ``bad'' as a difference in time $\Delta t$ of 1 hour.

Since the specific value of CF adopted for each calibration is the dominant part of the flux uncertainty and since the correlated fluxes of our sources turned out not to be extremely sensitive to small changes in $(u,v)$ co-ordinates, we stacked together observations that use the same calibrator, i.e. the same estimate for the CF. For all sources, except NGC~1068 and the Circinus~galaxy, fringe tracks were reduced together when they were less than 1 hour apart and were calibrated with the same star; for the two bright sources, the maximum time difference for two fringe tracks to be reduced together was set to 30 min. Within this time, the projected baseline and position angle change maximally by about 20 meters or 15 degrees and in most cases much less.

\subsection{Quality control}
\begin{figure*}
	{\includegraphics[width=0.8\hsize]{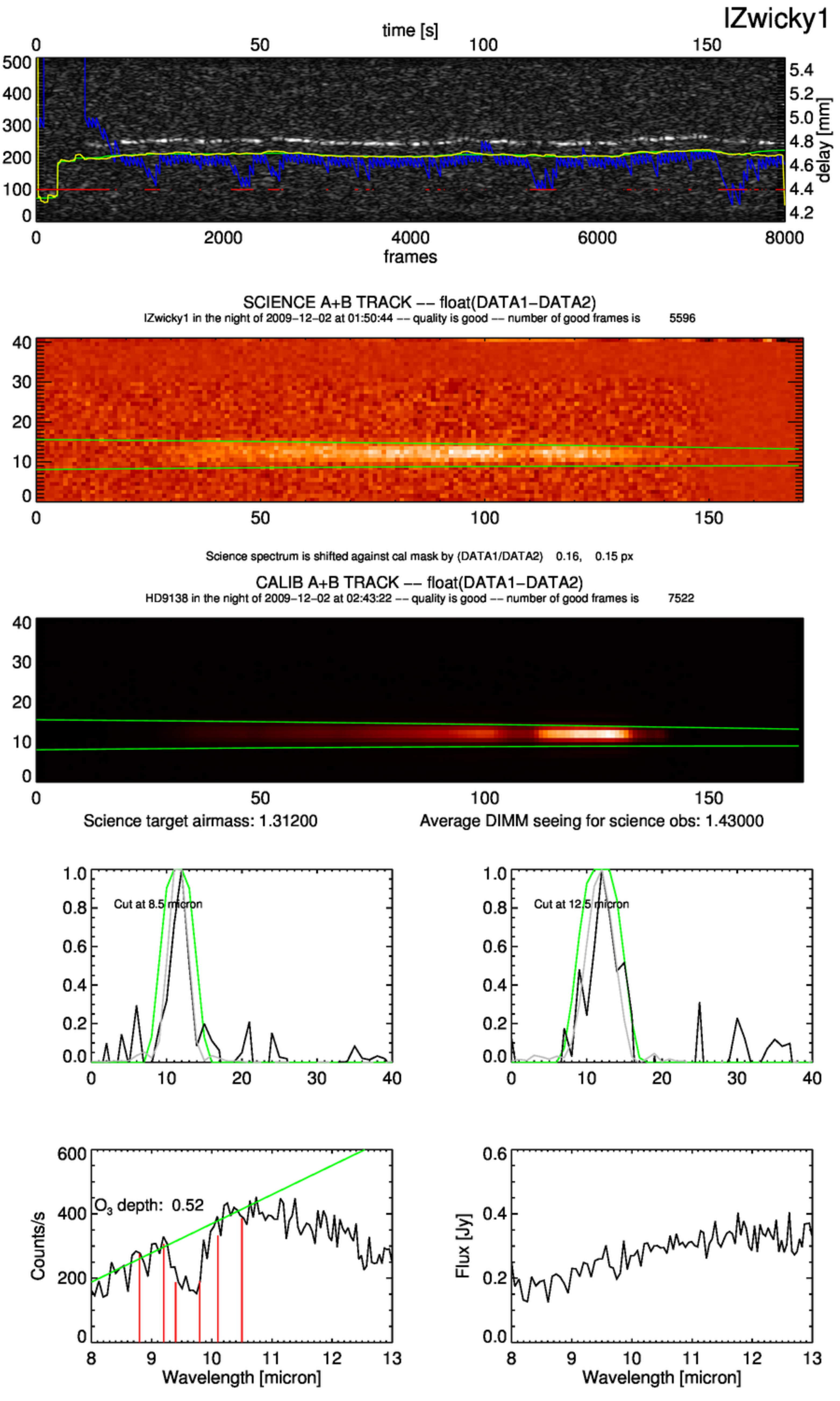}}
	\caption{\label{fig:obs:QC_track}Quality control plot for a fringe track (here observation 2009-12-02/01:50:44). From top to bottom, left to right: Group delay plot, coherent image of the target and the calibrator and (small plots) cuts through the spectrum of the target (black), calibrator (grey) and mask (green) at two different wavelength position. Bottom row: raw spectrum of the target showing the ozone criterion and calibrated spectrum of the target. The MIDI pixel scale in the employed mode is $0\farcs09$/pixel. See text for details.}
\end{figure*}

\begin{figure*}
\caption{\label{fig:spectra}Total and correlated flux spectra for I Zw 1 as well as VISIR photometry and mean photometries: average MIDI total flux spectrum (black, dotted), VISIR spectrum (black solid line), VISIR photometry (black open square), error-weighted means of MIDI total flux and VISIR spectrum (black filled circle). MIDI correlated fluxes are color-coded by baseline with increasing ``blueness'' indicating longer baselines. For typical error bars of the correlated fluxes, see the radial plots, Fig~\ref{fig:rad:IZwicky1} and following. The wavelength shown is restframe wavelength. The positions of bright mid-IR spectral lines are indicated by vertical dotted lines; for a list of detected lines, we refer to Section \ref{sec:spectra}.}
\centering
\includegraphics[trim=4cm 3.5cm 4cm 3.5cm, width=0.75\hsize]{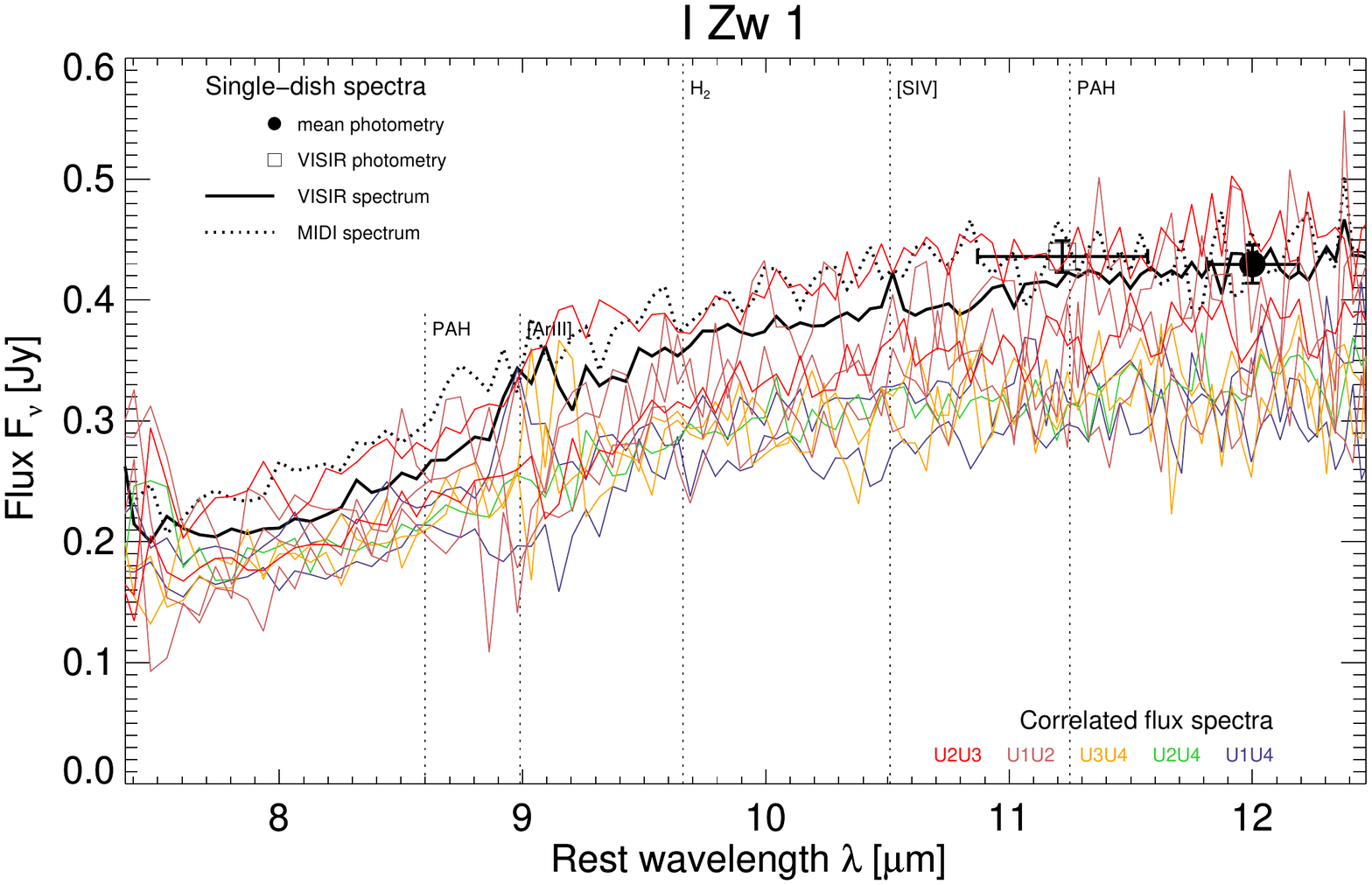}
\end{figure*}

In order to quickly check the quality of each observation, a ``quality control'' (Fig.~\ref{fig:obs:QC_track}) plot has turned out to be useful. 

It shows:

\begin{itemize}
	\item A {\em group delay plot}; each column in it is the Fourier transform of the spectrum extracted from each frame. A fit to the position of maximum intensity returns the position of zero optical path difference (ZOPD). Overlaid are the best-fit position (two iterations in yellow and green, see \citet{burtscher2012b}) as well as the tracking position of the VLTI during the observation (in blue). Red dots or lines denote regions flagged as bad by {\tt EWS}, either because the distance between tracking and true (i.e. found) group delay was too large or because there were jumps in the found group delay position or if the jitter in a region is too large. The number of un-flagged, i.e. ``good'', frames is printed out below the plot. The group delay plot allows a quick assessment of the tracking quality of the observation.
	\item Two {\em coherent images} of the detector. These are generated by omitting the mask extraction step. The extraction mask is generated from a fit to the coherent image of the calibrator. From these images it is easy to estimate the signal-to-noise of the observation at a glance.
	\item Two {\em cuts through the coherent images} showing potential offsets between calibrator and target position on the detector (in that case we shifted the mask) as well as showing the goodness of fit of the mask.
	\item The {\em raw spectrum}, i.e. the result of the coherent integration, of the target. If the flux resulted from outside the earth's atmosphere, it must have the telluric ozone feature at 9.6 \um imprinted on it. We interpolate the flux outside the feature (green line in Fig.~\ref{fig:obs:QC_track}) and compare the measured count rate in the feature with the interpolated value there. If the ratio is larger than 0.76 we reject the observation, because a large part of the flux in this spectrum did not originate outside the atmosphere, i.e. from the target. If the ozone feature test indicates that the observation is good and the number of un-flagged frames is larger than 500 (so that the S/N of this observation is $\gtrsim 1$), the overall judgement of this observation is ``good'', else it is ``bad''.
	\item Finally we plot the {\em calibrated correlated flux}. When flipping through a number of these plots for one source it is easy to see whether there are changes in the correlated flux or not, thereby revealing whether or not there is a resolved component in this object.
\end{itemize}

The fraction of ``good'' observations per source depends mostly on the level of correlated flux and also on the brightness and distance of the AO guide star. It ranges from 61\% in NGC~7469 to 100\% in H 0557-385, IRAS 09149-6206, Mrk 1239 and IRAS 13349+2438 -- all of which are type 1 AGNs. The average (median) fraction of ``good'' observations per source is 82\% (83\%).

From the raw spectrum (in counts/s), the noise can be estimated to be about 50 counts/s (roughly 50 mJy) limiting the ultimate sensitivity of MIDI to about this level. Compared to the sensitivity of the single-dish mid-IR spectrometer and imager VISIR \citep{asmus2011}, the sensitivity of the MIDI correlated fluxes is only about a factor of five lower (1.7 magnitudes higher). The fact that such a high sensitivity can be achieved even after so many reflections in the VLTI beam train is due to the ``lock-in'' amplifier-like observation method in MIDI: The fringe pattern is modulated with a known frequency by instrument-internal piezo-driven mirrors allowing to reject most of the noise which is not modulated with this frequency.

\clearpage
\section{Total and correlated flux spectra}

\begin{figure*}
\caption{Single-dish and correlated flux spectra for NGC 424 (see caption for Fig.~\ref{fig:spectra})}
\centering
\includegraphics[trim=4cm 3.5cm 4cm 3.5cm, width=0.75\hsize]{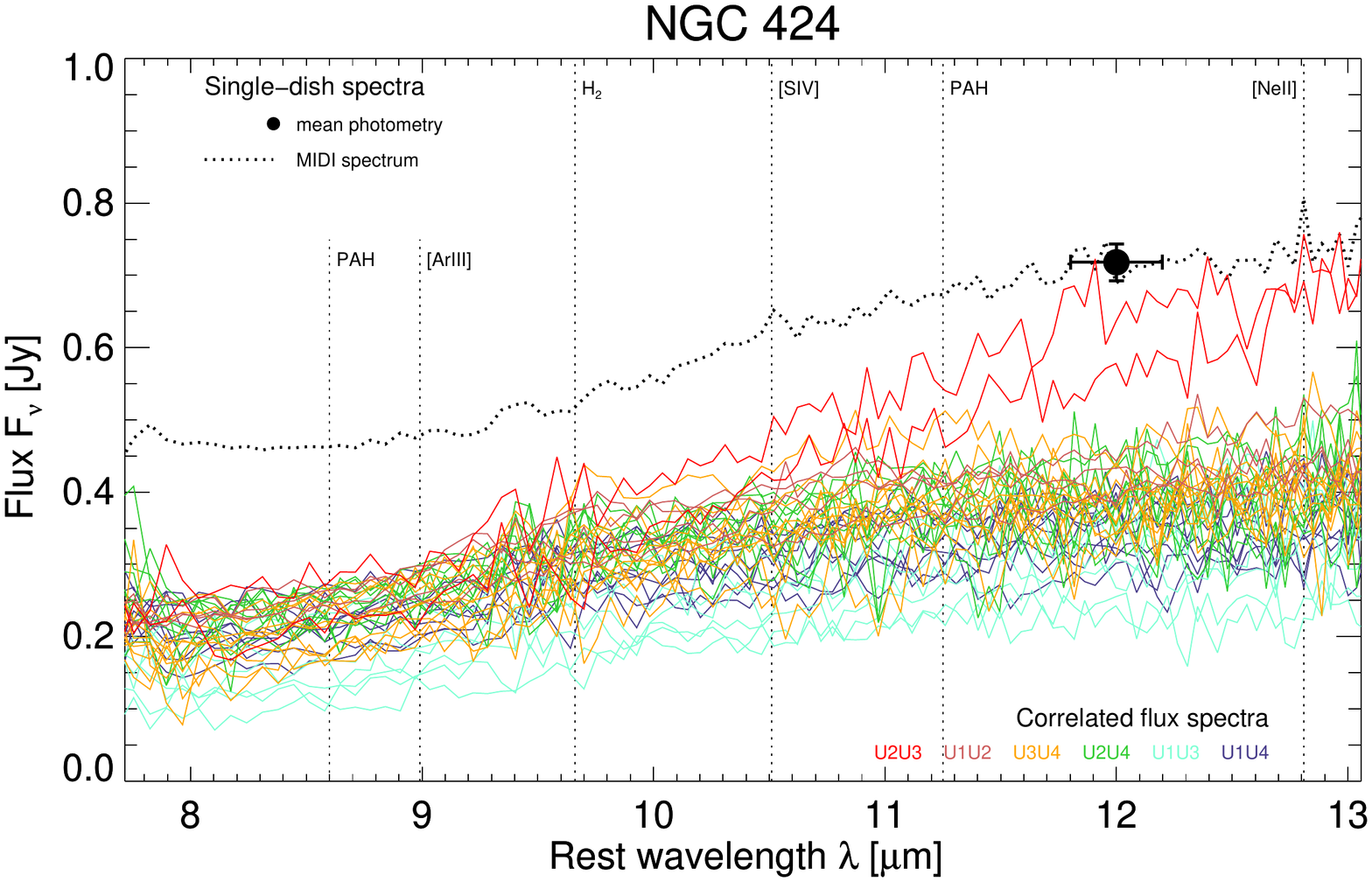}
\end{figure*}

\clearpage

\begin{figure*}
\caption{Single-dish and correlated flux spectra for NGC 1068 (see caption for Fig.~\ref{fig:spectra})}
\centering
\includegraphics[trim=4cm 3.5cm 4cm 3.5cm, width=0.75\hsize]{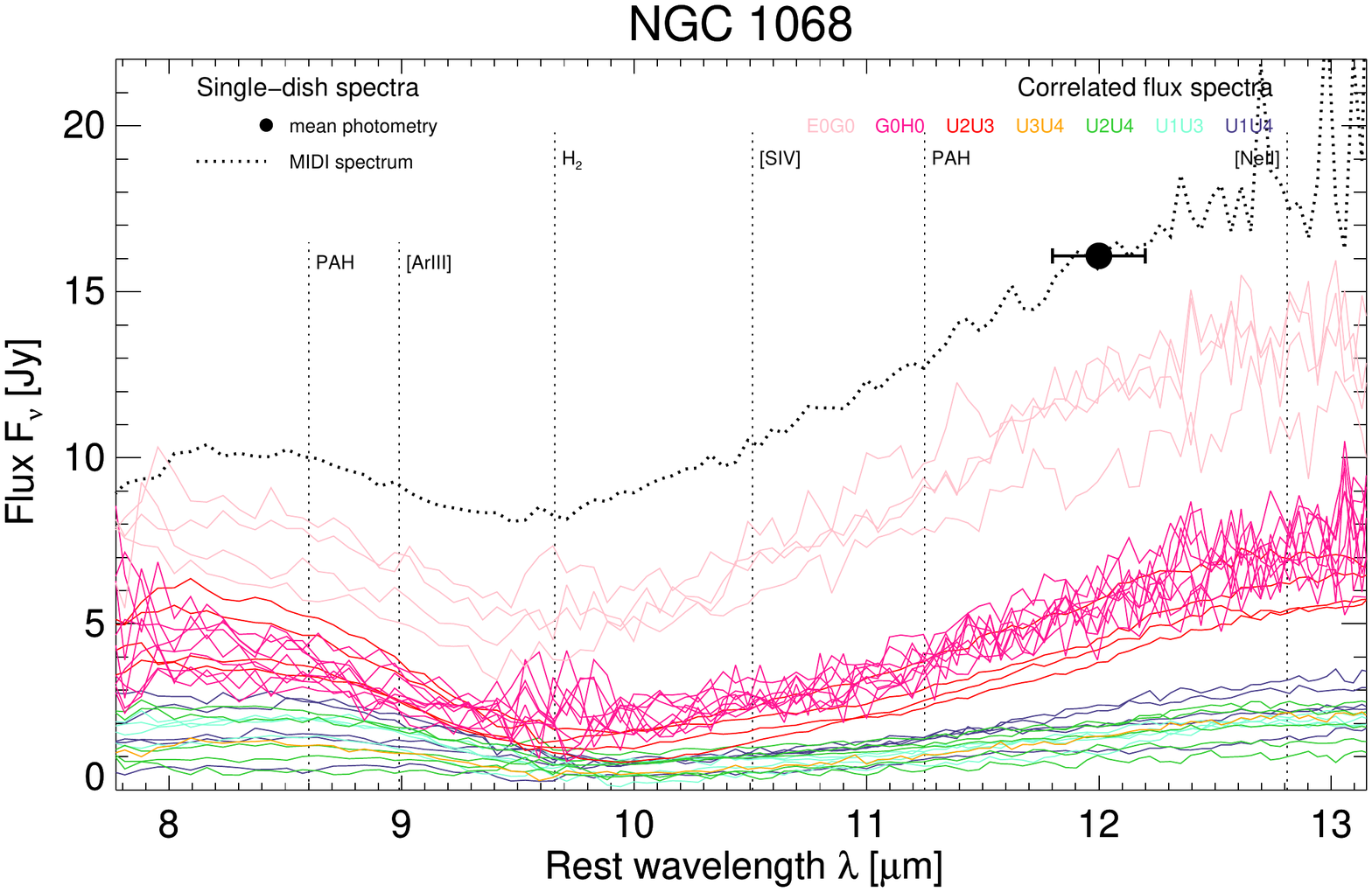}
\end{figure*}

\begin{figure*}
\caption{Single-dish and correlated flux spectra for NGC 1365 (see caption for Fig.~\ref{fig:spectra})}
\centering
\includegraphics[trim=4cm 3.5cm 4cm 3.5cm, width=0.75\hsize]{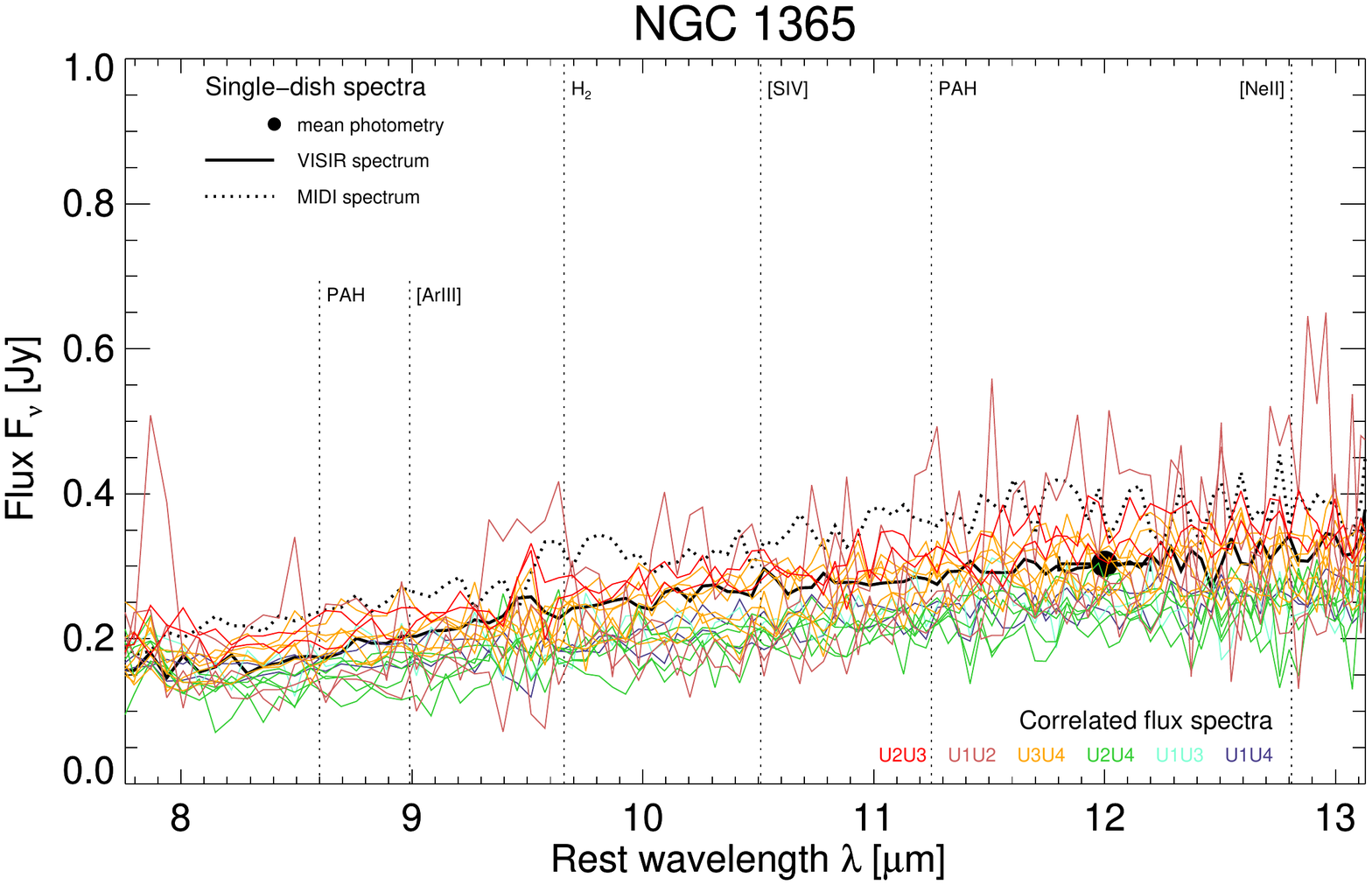}
\end{figure*}

\clearpage

\begin{figure*}
\caption{Single-dish and correlated flux spectra for IRAS 05189-2524 (see caption for Fig.~\ref{fig:spectra})}
\centering
\includegraphics[trim=4cm 3.5cm 4cm 3.5cm, width=0.75\hsize]{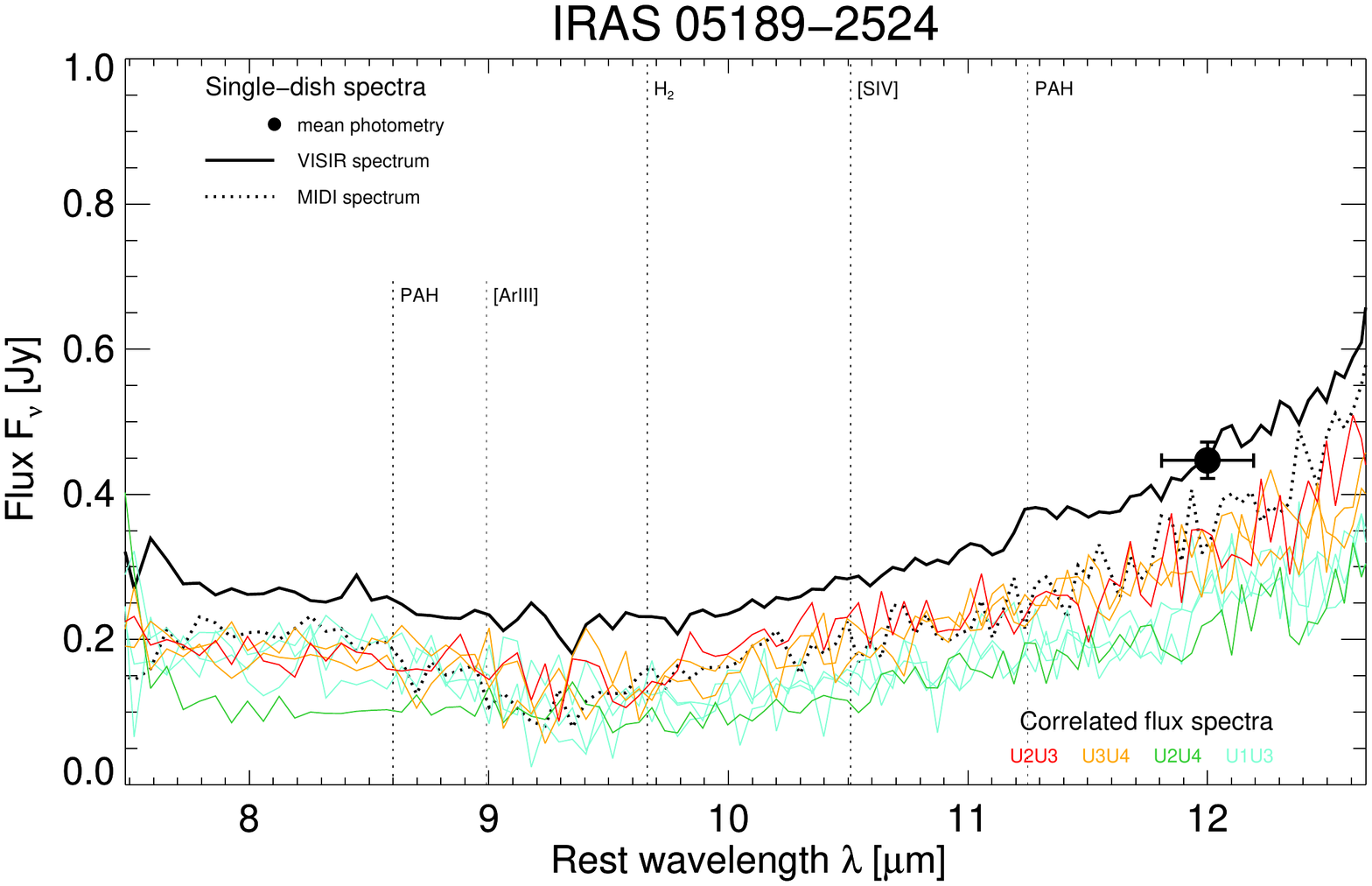}
\end{figure*}

\begin{figure*}
\caption{Single-dish and correlated flux spectra for H 0557-385 (see caption for Fig.~\ref{fig:spectra})}
\centering
\includegraphics[trim=4cm 3.5cm 4cm 3.5cm, width=0.75\hsize]{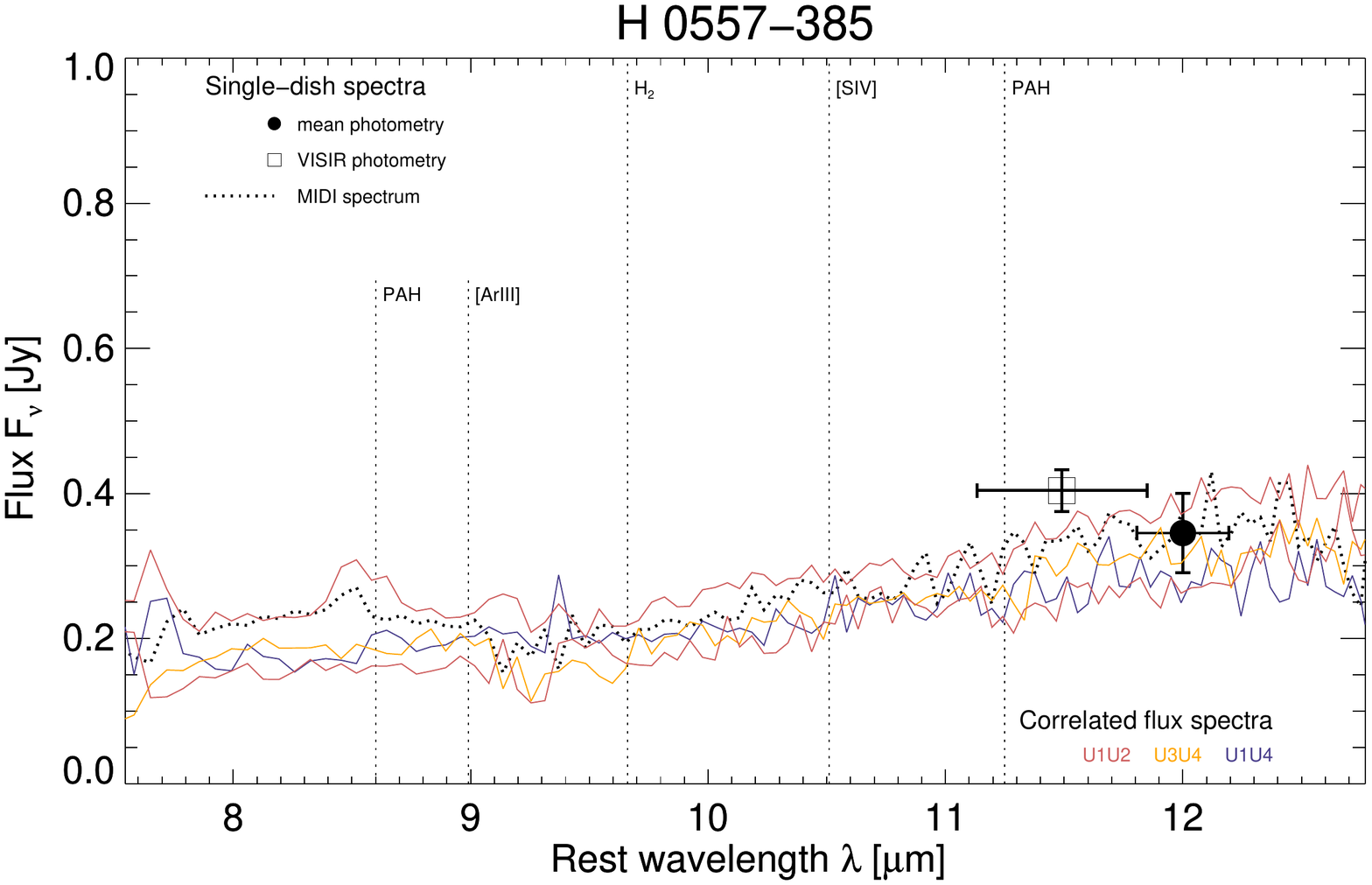}
\end{figure*}

\clearpage

\begin{figure*}
\caption{Single-dish and correlated flux spectra for IRAS 09149-6206 (see caption for Fig.~\ref{fig:spectra})}
\centering
\includegraphics[trim=4cm 3.5cm 4cm 3.5cm, width=0.75\hsize]{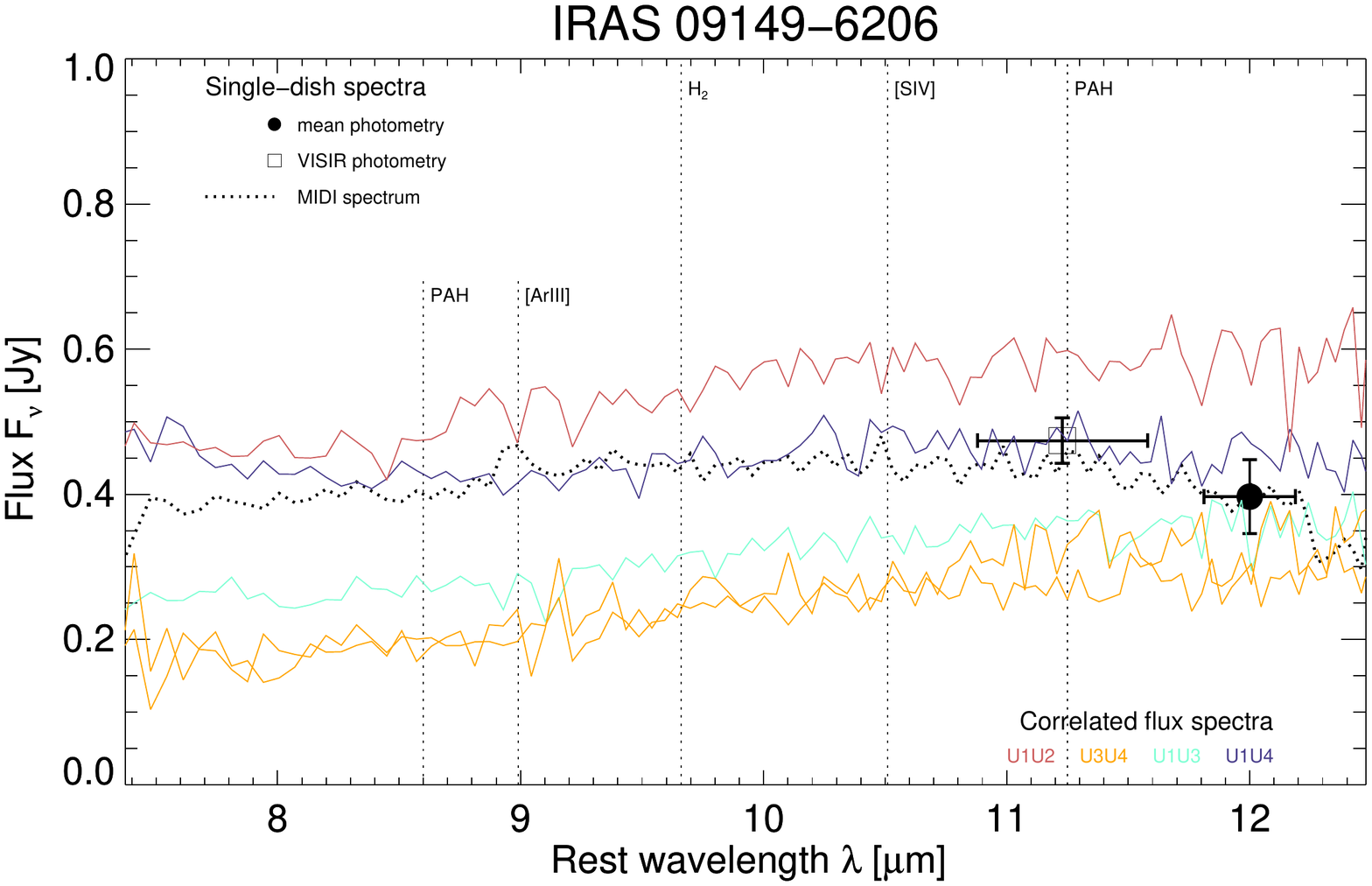}
\end{figure*}

\begin{figure*}
\caption{Single-dish and correlated flux spectra for MCG-05-23-16 (see caption for Fig.~\ref{fig:spectra}).}
\centering
\includegraphics[trim=4cm 3.5cm 4cm 3.5cm, width=0.75\hsize]{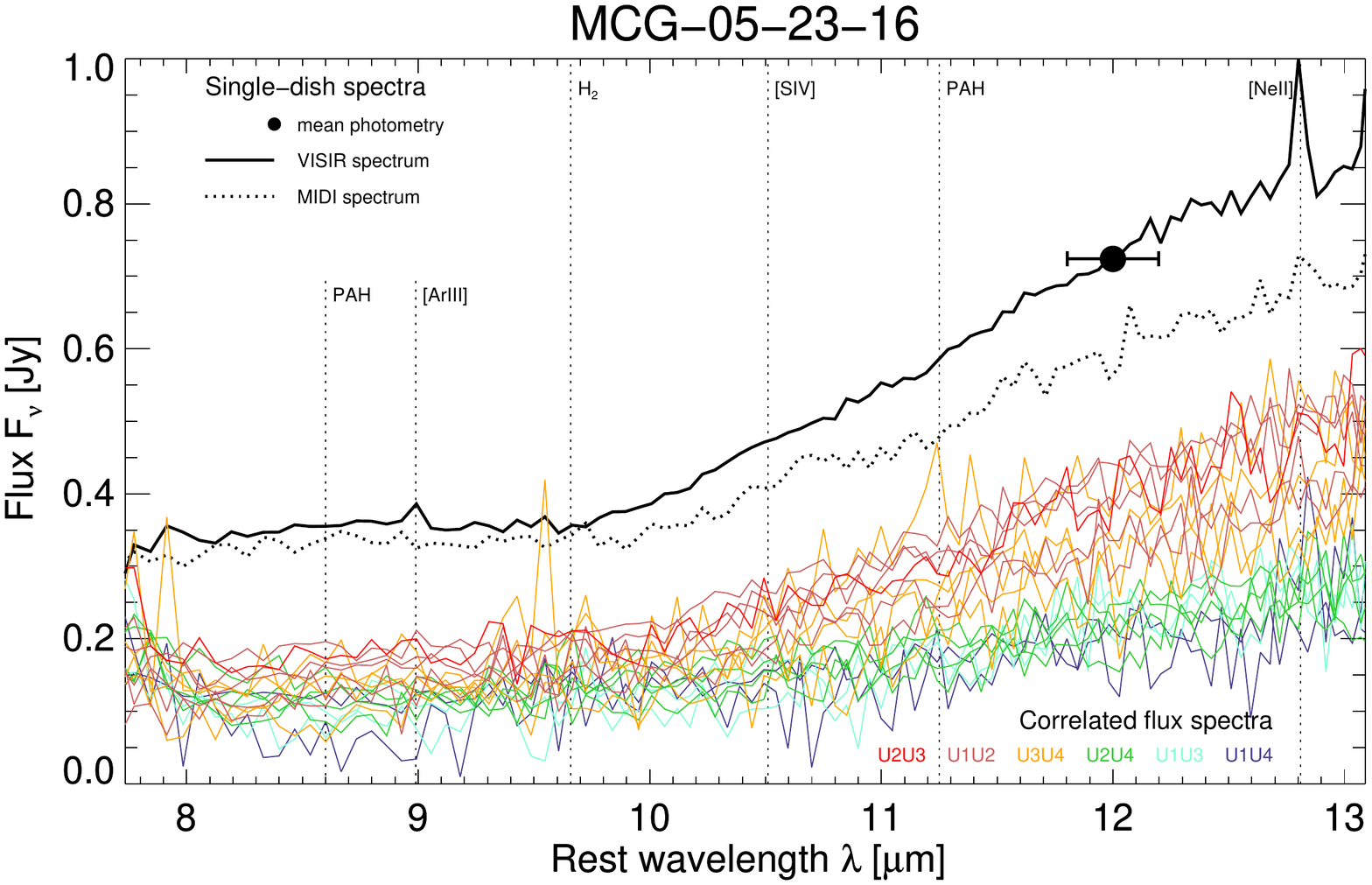}
\end{figure*}

\clearpage

\begin{figure*}
\caption{Single-dish and correlated flux spectra for Mrk 1239 (see caption for Fig.~\ref{fig:spectra}). For unknown reasons, the VISIR spectrum of this source is lower not only than the MIDI total flux but also most of the MIDI correlated flux observations. The only physical explanation for this behavior would be fast variability in the source. However, we have observed the source twice with VISIR (see Tab.~\ref{tab:obs:visir} on page~\pageref{tab:obs:visir}) in between the MIDI observations (see Tab.~\ref{log:midi:mrk1239} on page~\pageref{log:midi:mrk1239}) and found no indications for variability. We show the VISIR spectrum here for completeness, but we only use the MIDI photometry for the fits.}
\centering
\includegraphics[trim=4cm 3.5cm 4cm 3.5cm, width=0.75\hsize]{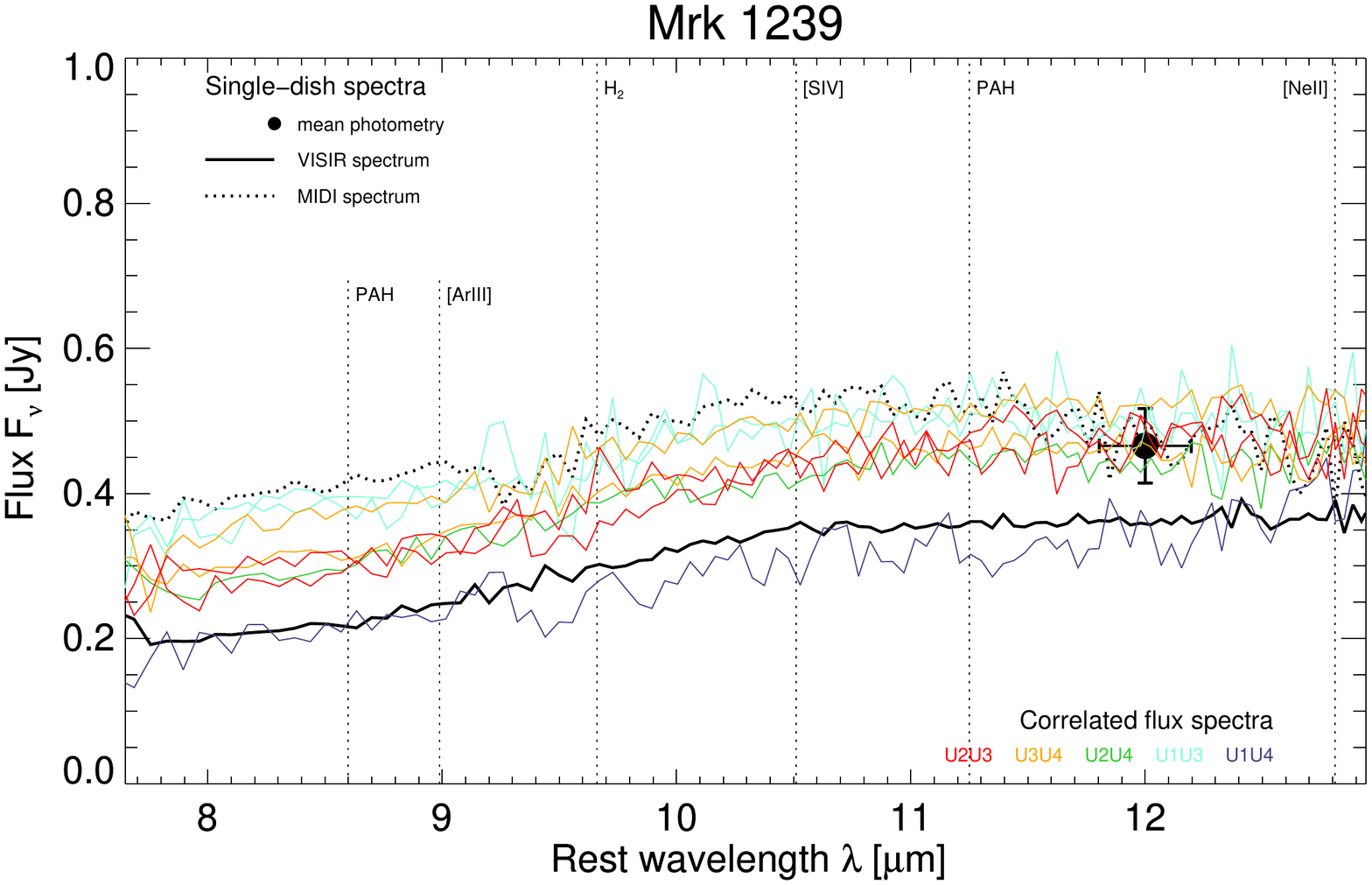}
\end{figure*}

\begin{figure*}
\caption{Single-dish and correlated flux spectra for NGC 3281 (see caption for Fig.~\ref{fig:spectra}). For this source no acceptable MIDI spectrum was observed.}
\centering
\includegraphics[trim=4cm 3.5cm 4cm 3.5cm, width=0.75\hsize]{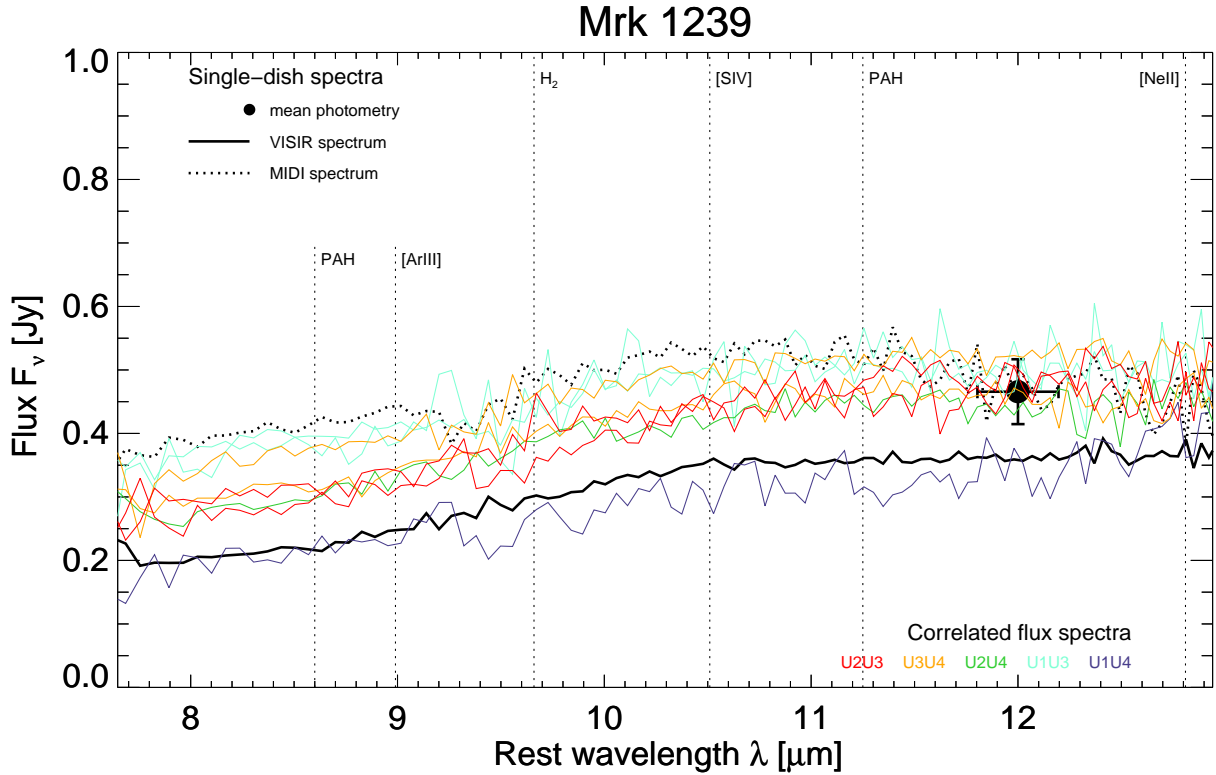}
\end{figure*}

\clearpage

\begin{figure*}
\caption{Single-dish and correlated flux spectra for NGC 3783 (see caption for Fig.~\ref{fig:spectra})}
\centering
\includegraphics[trim=4cm 3.5cm 4cm 3.5cm, width=0.75\hsize]{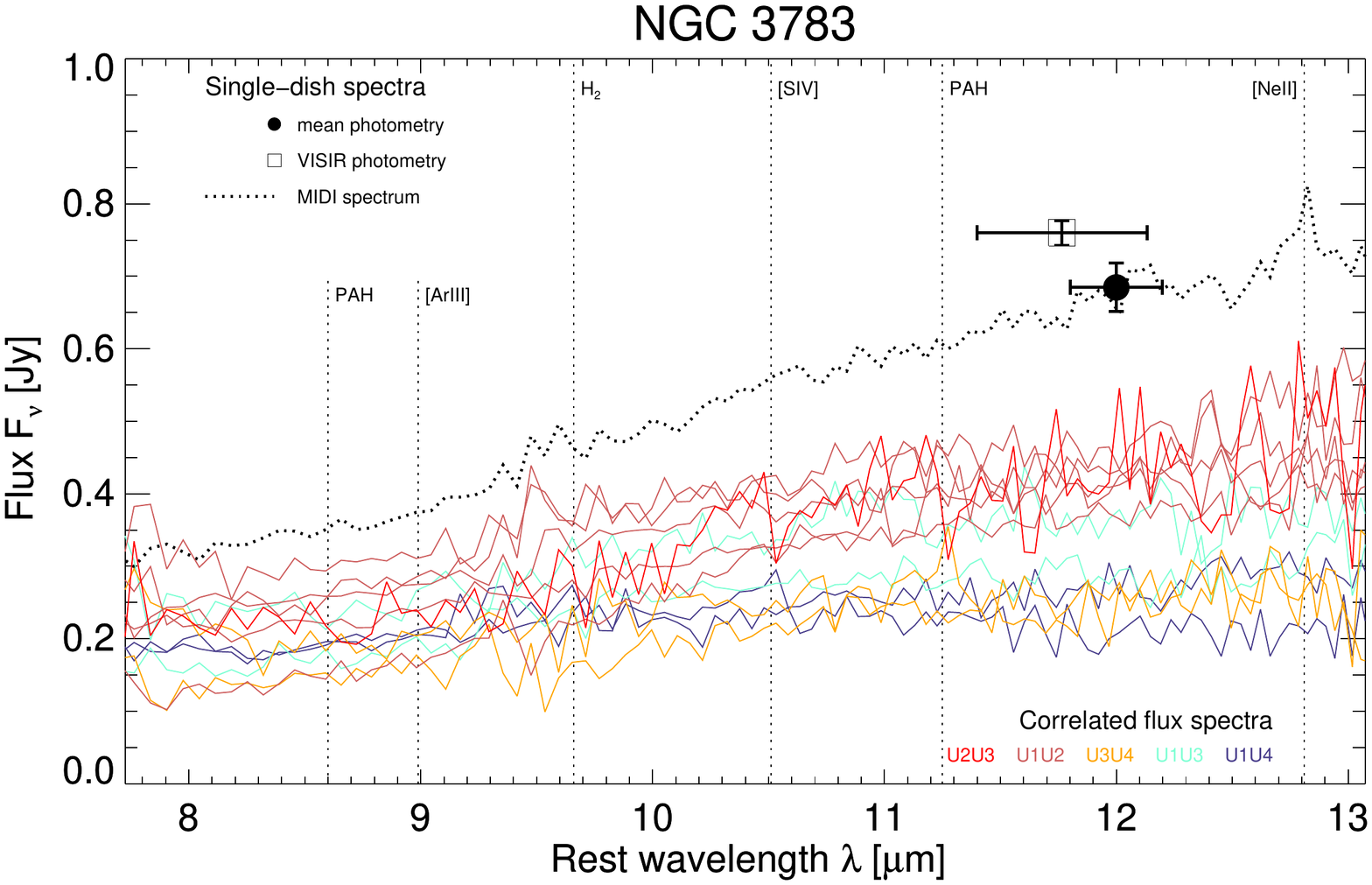}
\end{figure*}

\begin{figure*}
\caption{Single-dish and correlated flux spectra for NGC 4151 (see caption for Fig.~\ref{fig:spectra}). The VISIR photometry has been extracted with a much larger aperture than the MIDI photometry and correlated fluxes. We show it here for completeness, but only use the MIDI photometry for the fits.}
\centering
\includegraphics[trim=4cm 3.5cm 4cm 3.5cm, width=0.75\hsize]{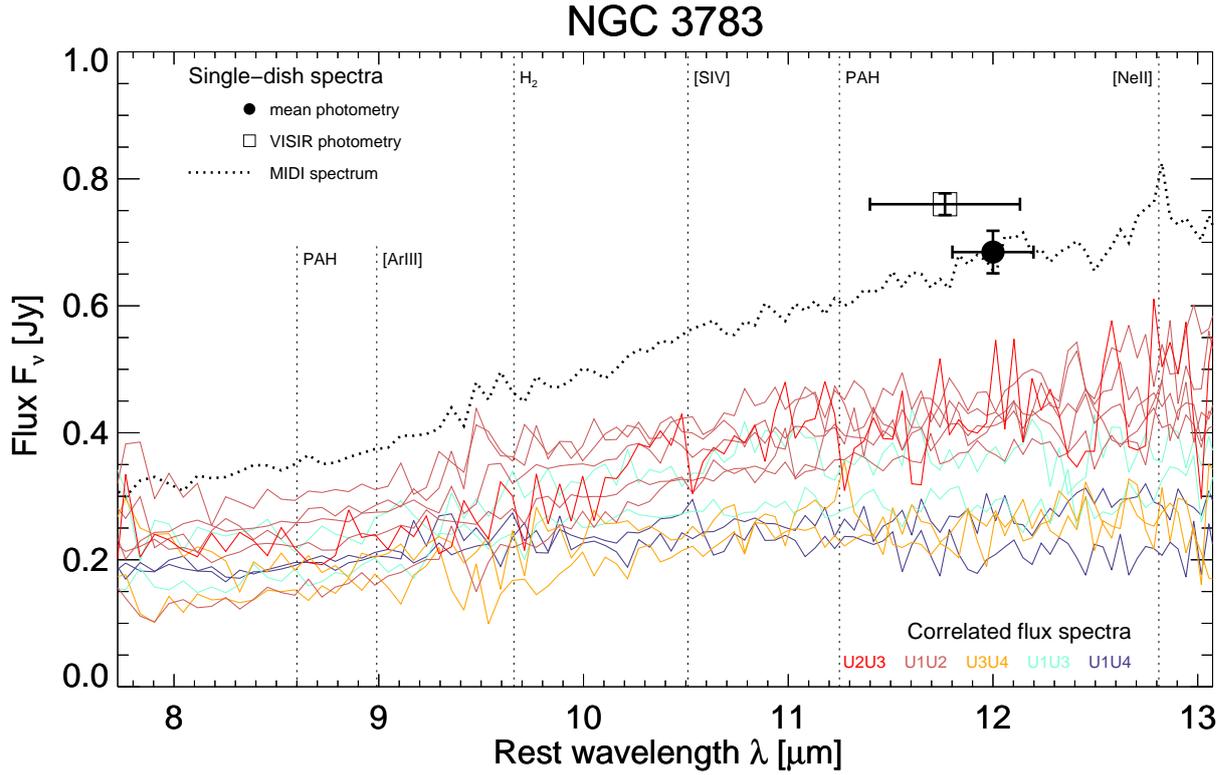}
\end{figure*}

\clearpage

\begin{figure*}
\caption{Single-dish and correlated flux spectra for 3C 273 (see caption for Fig.~\ref{fig:spectra})}
\centering
\includegraphics[trim=4cm 3.5cm 4cm 3.5cm, width=0.75\hsize]{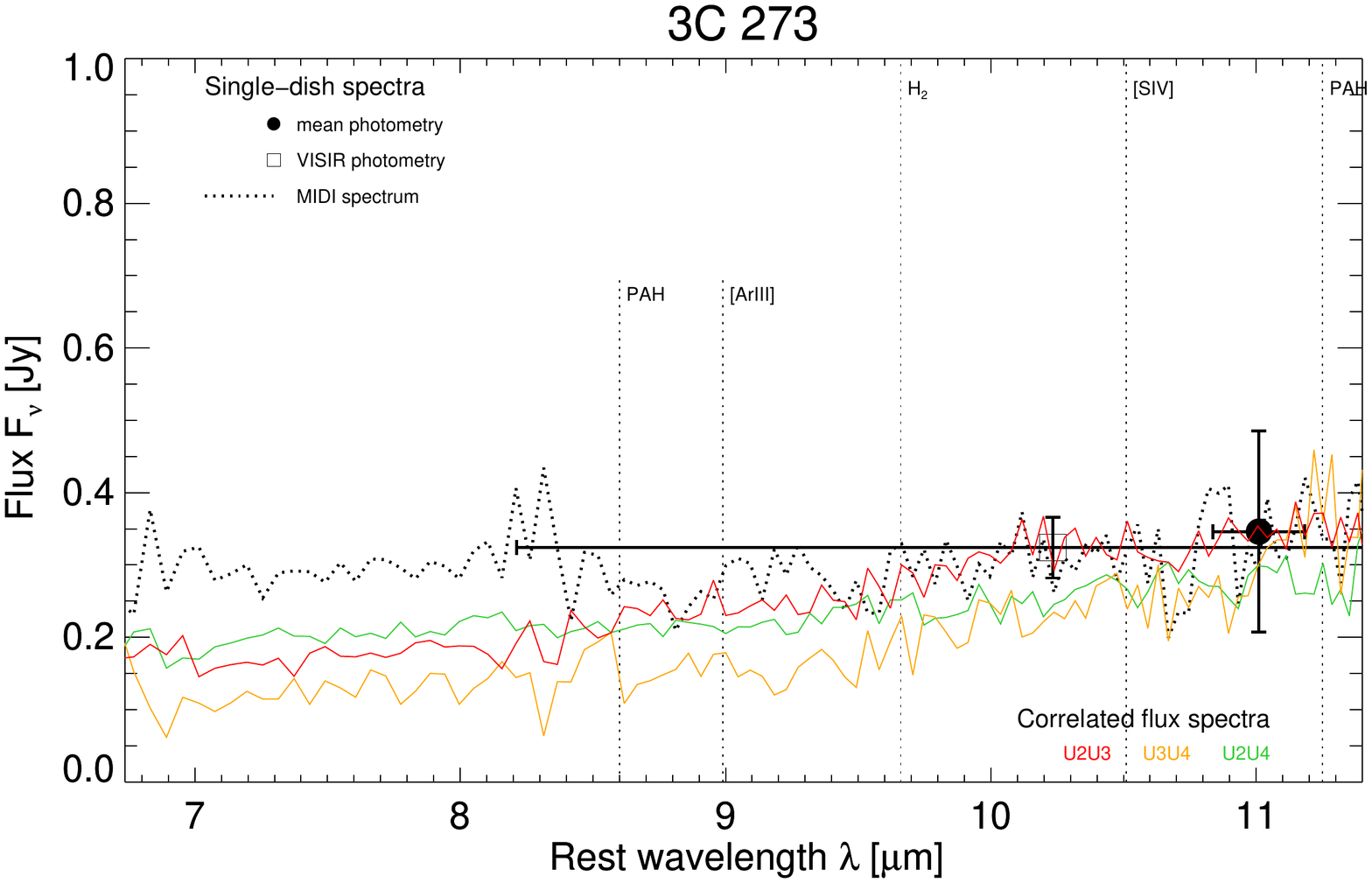}
\end{figure*}

\begin{figure*}
\caption{Single-dish and correlated flux spectra for NGC 4507 (see caption for Fig.~\ref{fig:spectra})}
\centering
\includegraphics[trim=4cm 3.5cm 4cm 3.5cm, width=0.75\hsize]{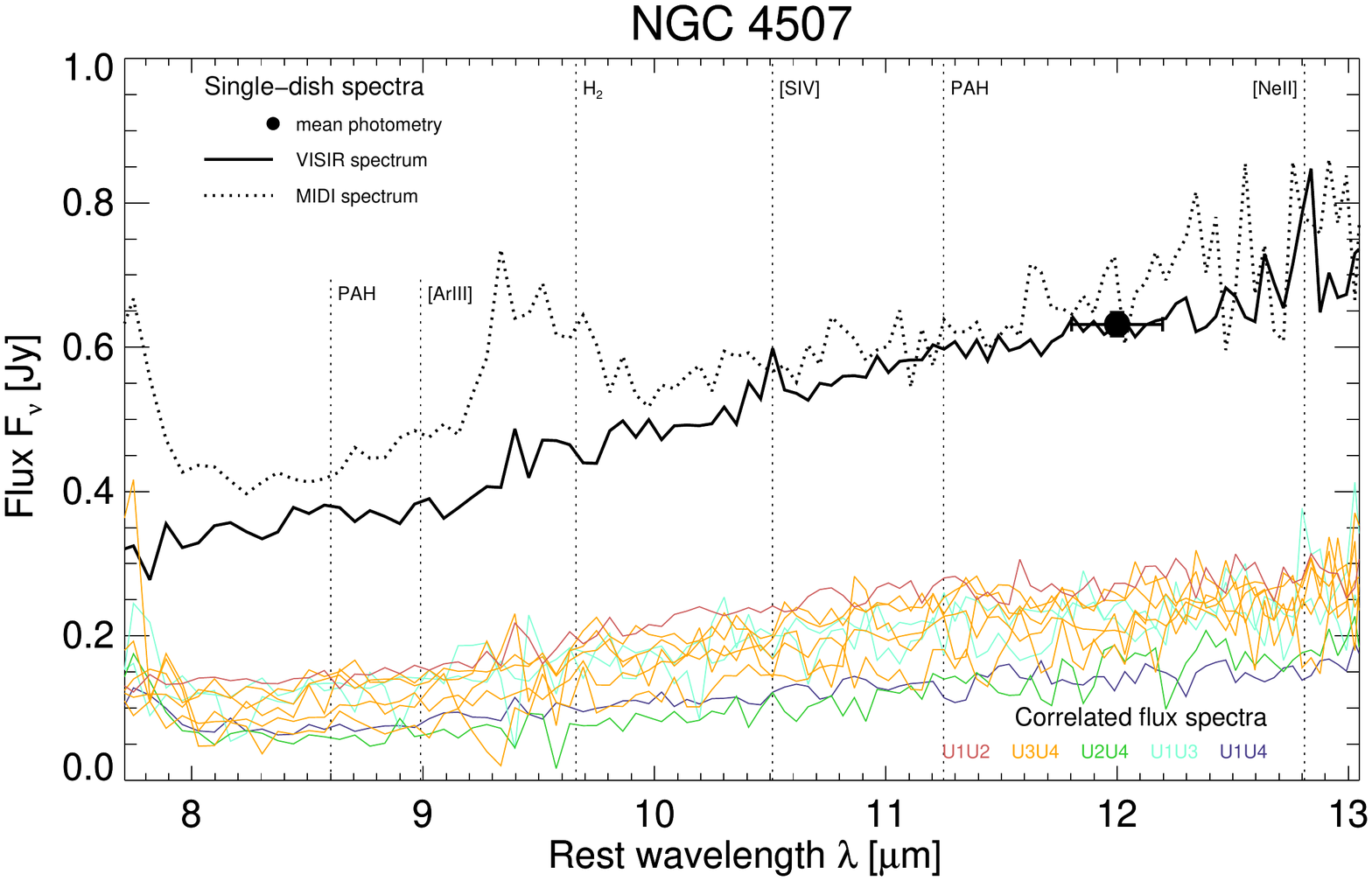}
\end{figure*}

\clearpage

\begin{figure*}
\caption{Single-dish and correlated flux spectra for NGC 4593 (see caption for Fig.~\ref{fig:spectra})}
\centering
\includegraphics[trim=4cm 3.5cm 4cm 3.5cm, width=0.75\hsize]{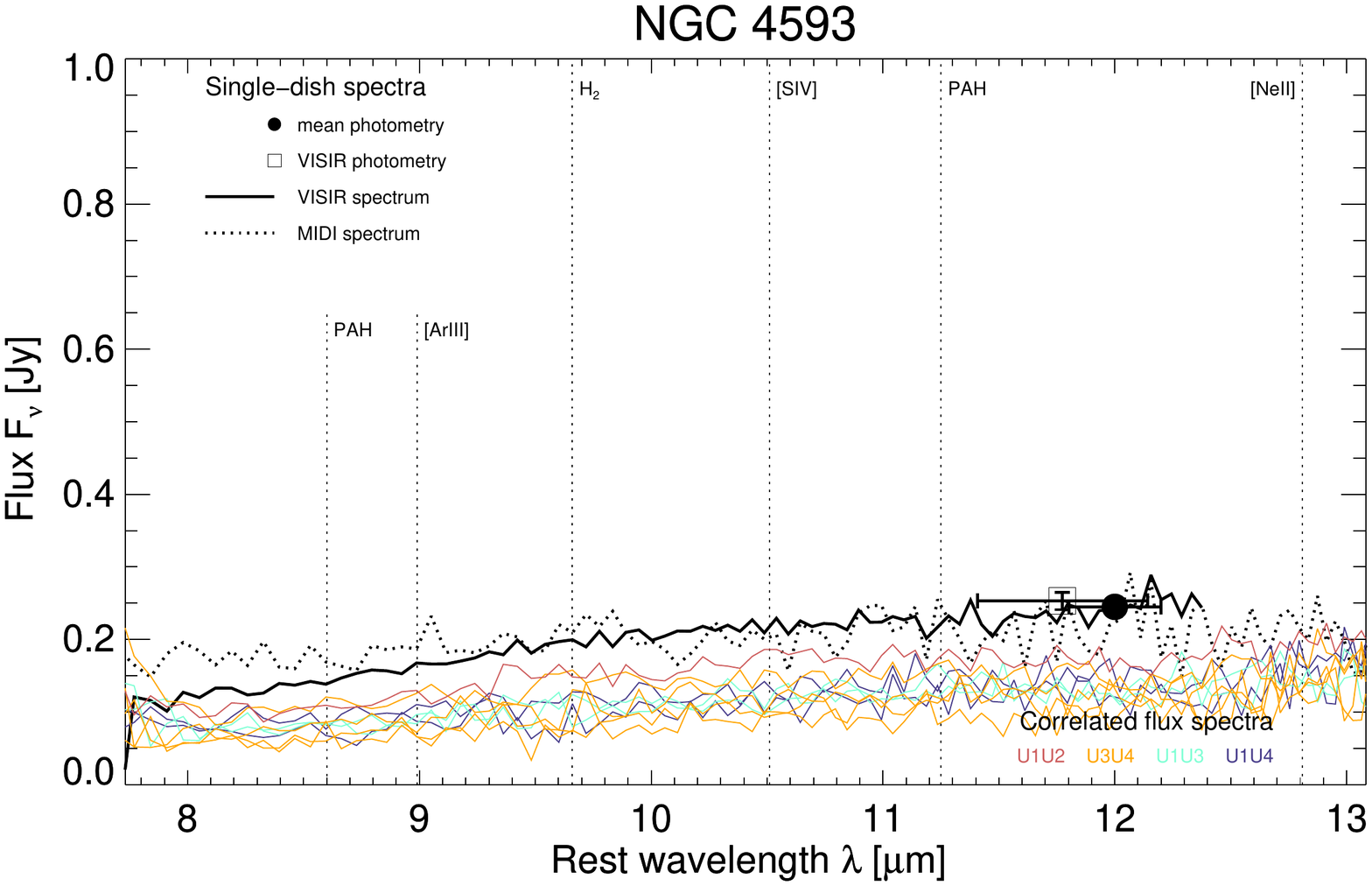}
\end{figure*}

\begin{figure*}
\caption{Single-dish and correlated flux spectra for ESO 323-77 (see caption for Fig.~\ref{fig:spectra})}
\centering
\includegraphics[trim=4cm 3.5cm 4cm 3.5cm, width=0.75\hsize]{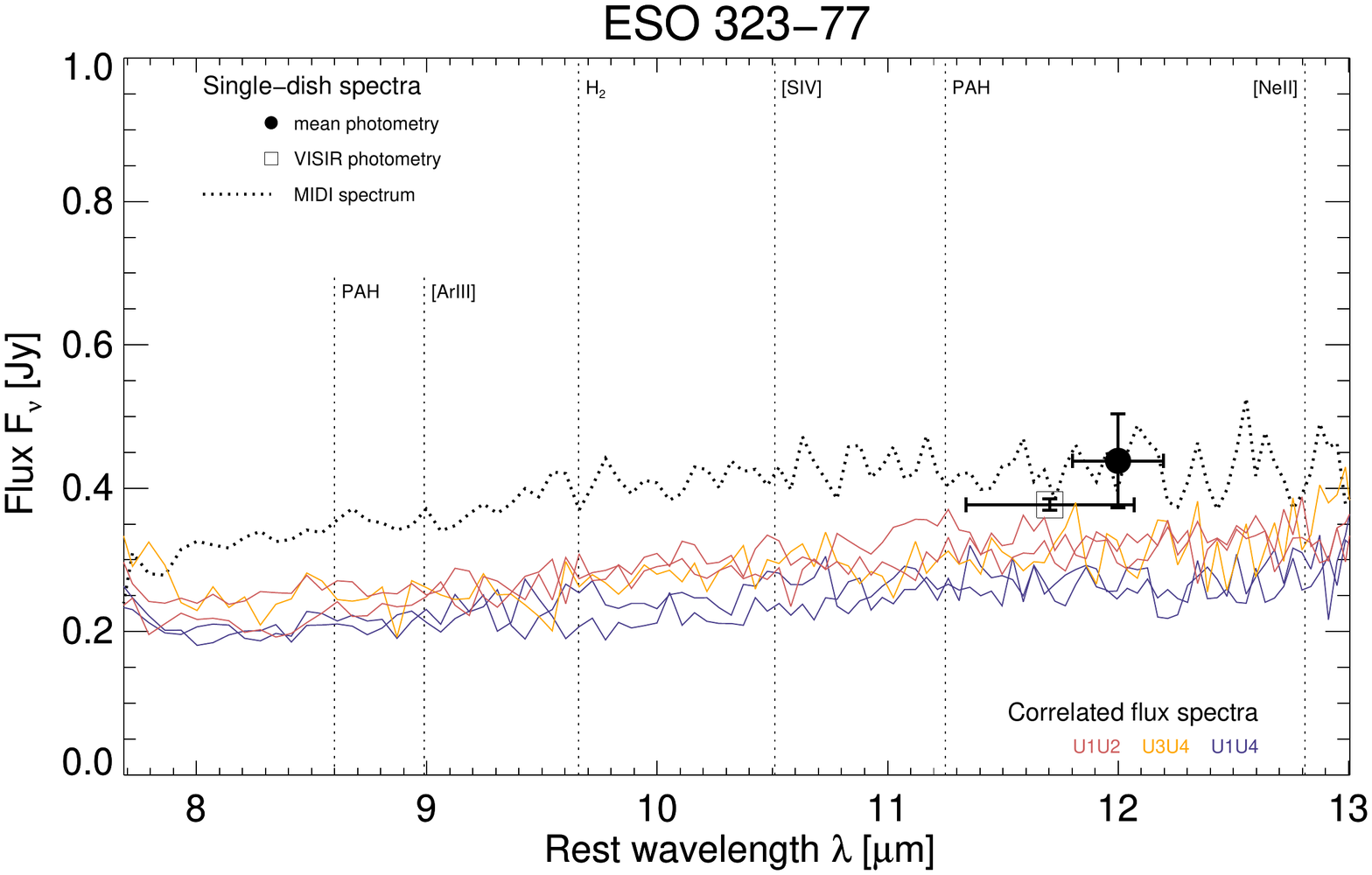}
\end{figure*}

\clearpage

\begin{figure*}
\caption{Single-dish and correlated flux spectra for Centaurus A (see caption for Fig.~\ref{fig:spectra})}
\centering
\includegraphics[trim=4cm 3.5cm 4cm 3.5cm, width=0.75\hsize]{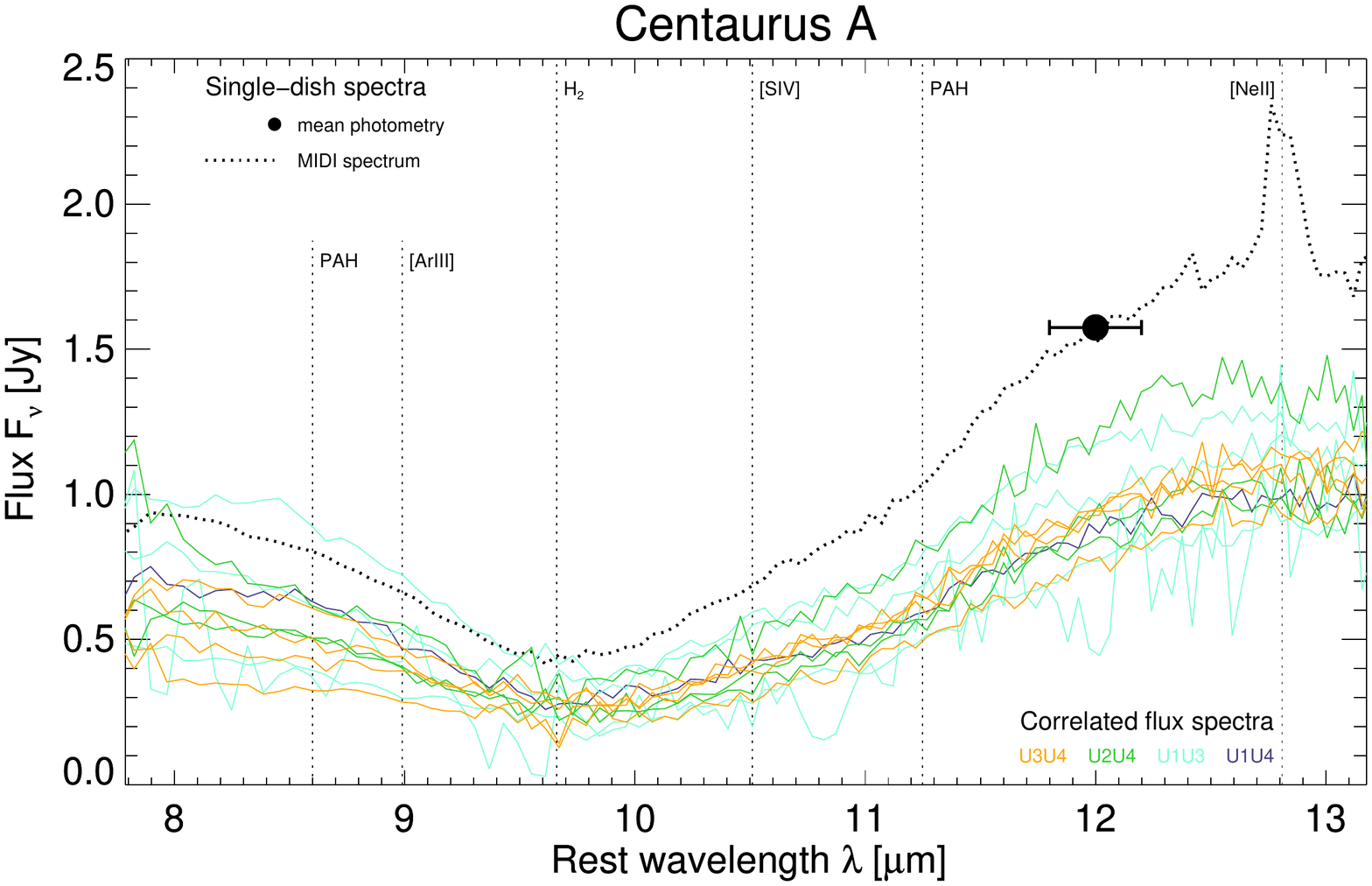}
\end{figure*}

\begin{figure*}
\caption{Single-dish and correlated flux spectra for IRAS 13349+2438 (see caption for Fig.~\ref{fig:spectra})}
\centering
\includegraphics[trim=4cm 3.5cm 4cm 3.5cm, width=0.75\hsize]{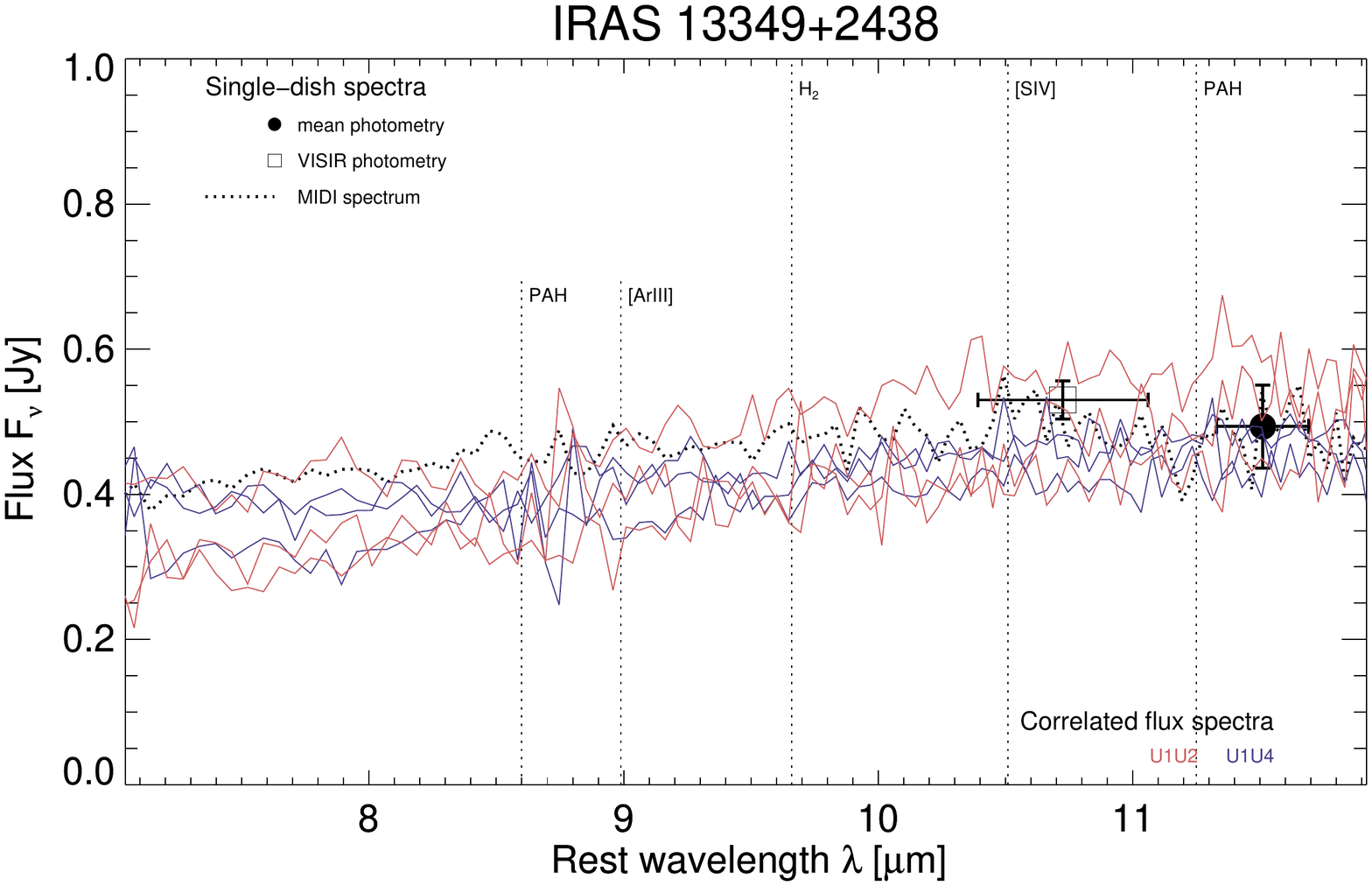}
\end{figure*}

\clearpage

\begin{figure*}
\caption{Single-dish and correlated flux spectra for IC 4329 A (see caption for Fig.~\ref{fig:spectra})}
\centering
\includegraphics[trim=4cm 3.5cm 4cm 3.5cm, width=0.75\hsize]{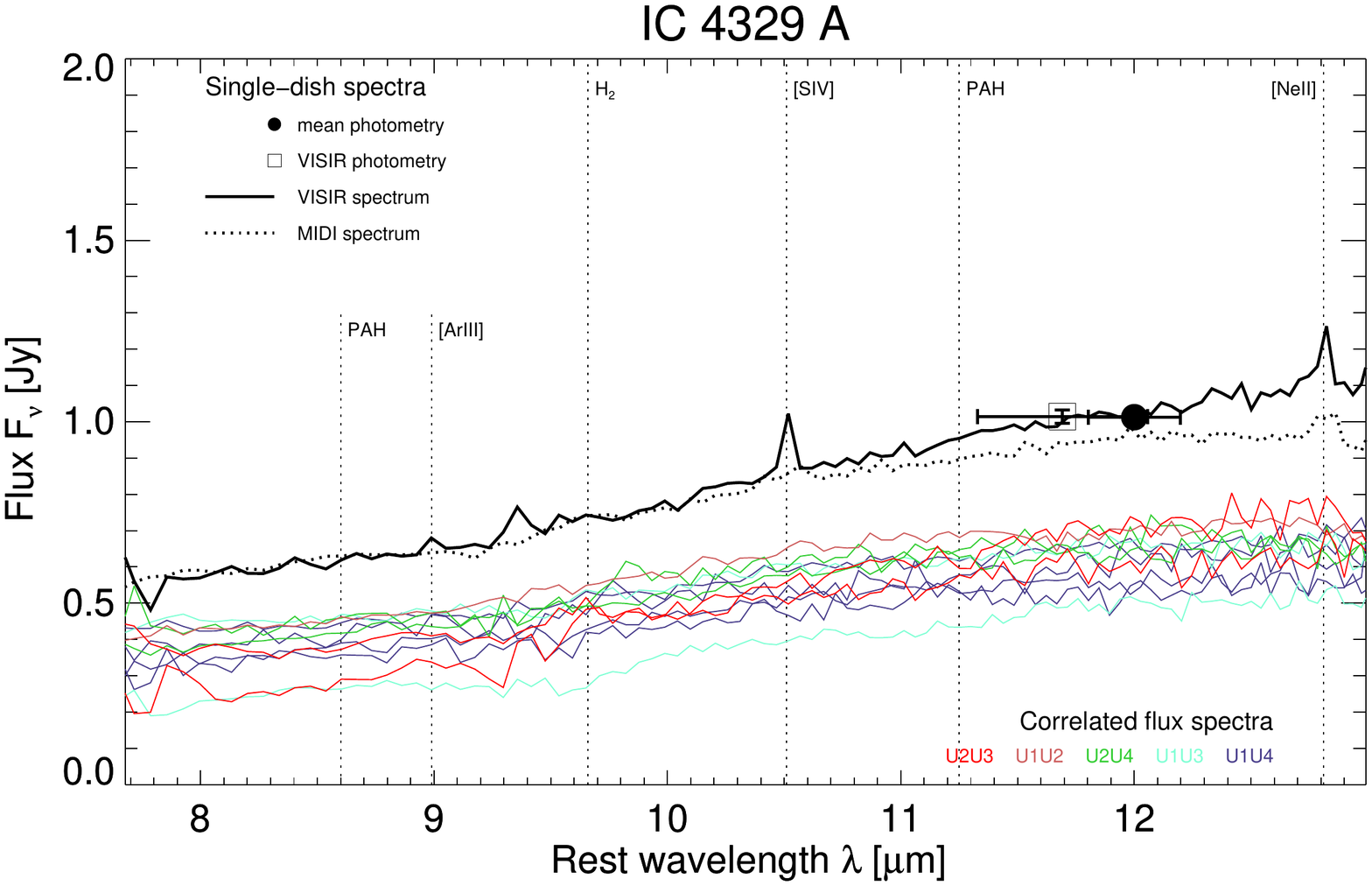}
\end{figure*}

\begin{figure*}
\caption{Single-dish and correlated flux spectra for Circinus (see caption for Fig.~\ref{fig:spectra})}
\centering
\includegraphics[trim=4cm 3.5cm 4cm 3.5cm, width=0.75\hsize]{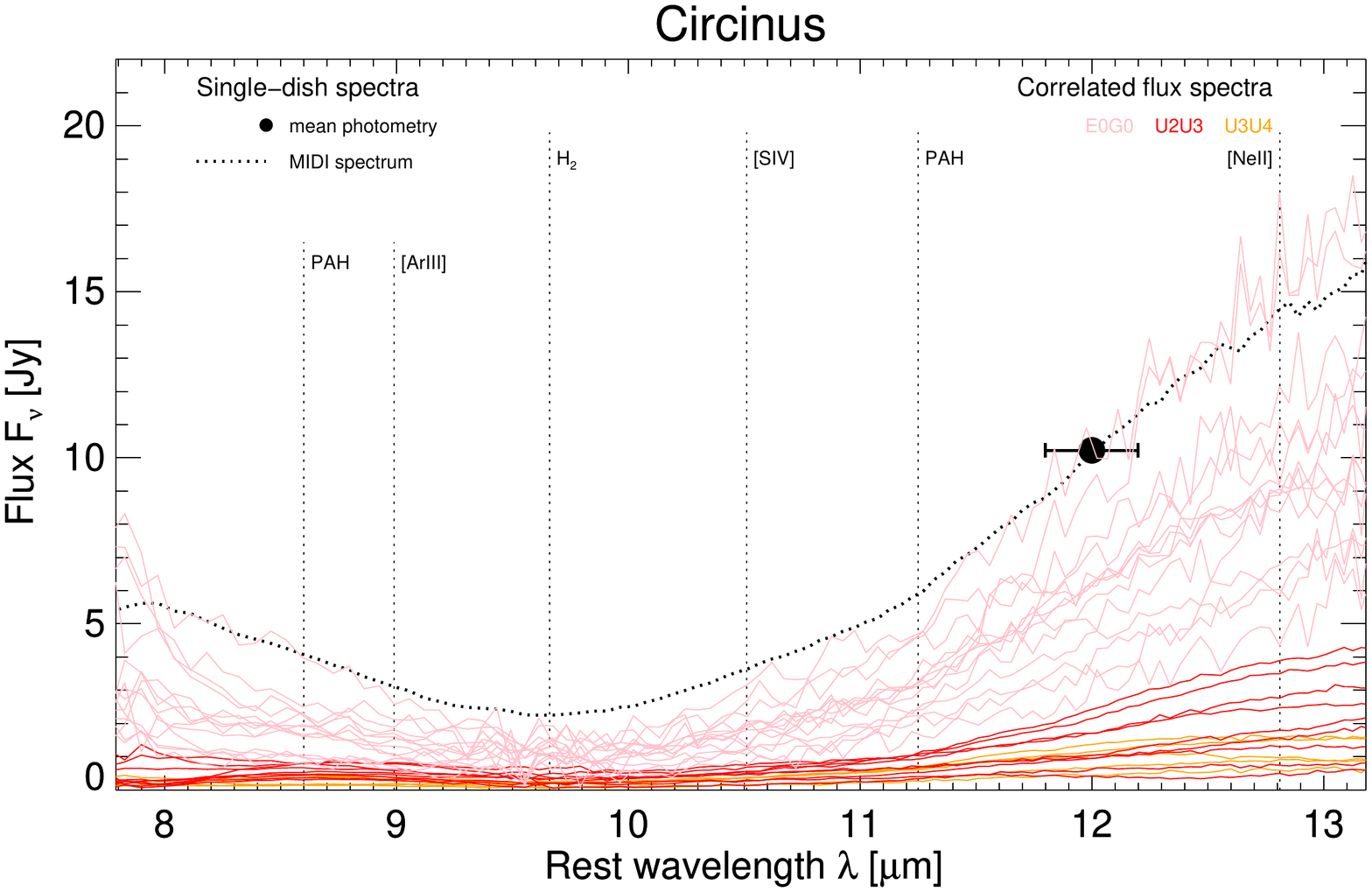}
\end{figure*}

\clearpage

\begin{figure*}
\caption{Single-dish and correlated flux spectra for NGC 5506 (see caption for Fig.~\ref{fig:spectra})}
\centering
\includegraphics[trim=4cm 3.5cm 4cm 3.5cm, width=0.75\hsize]{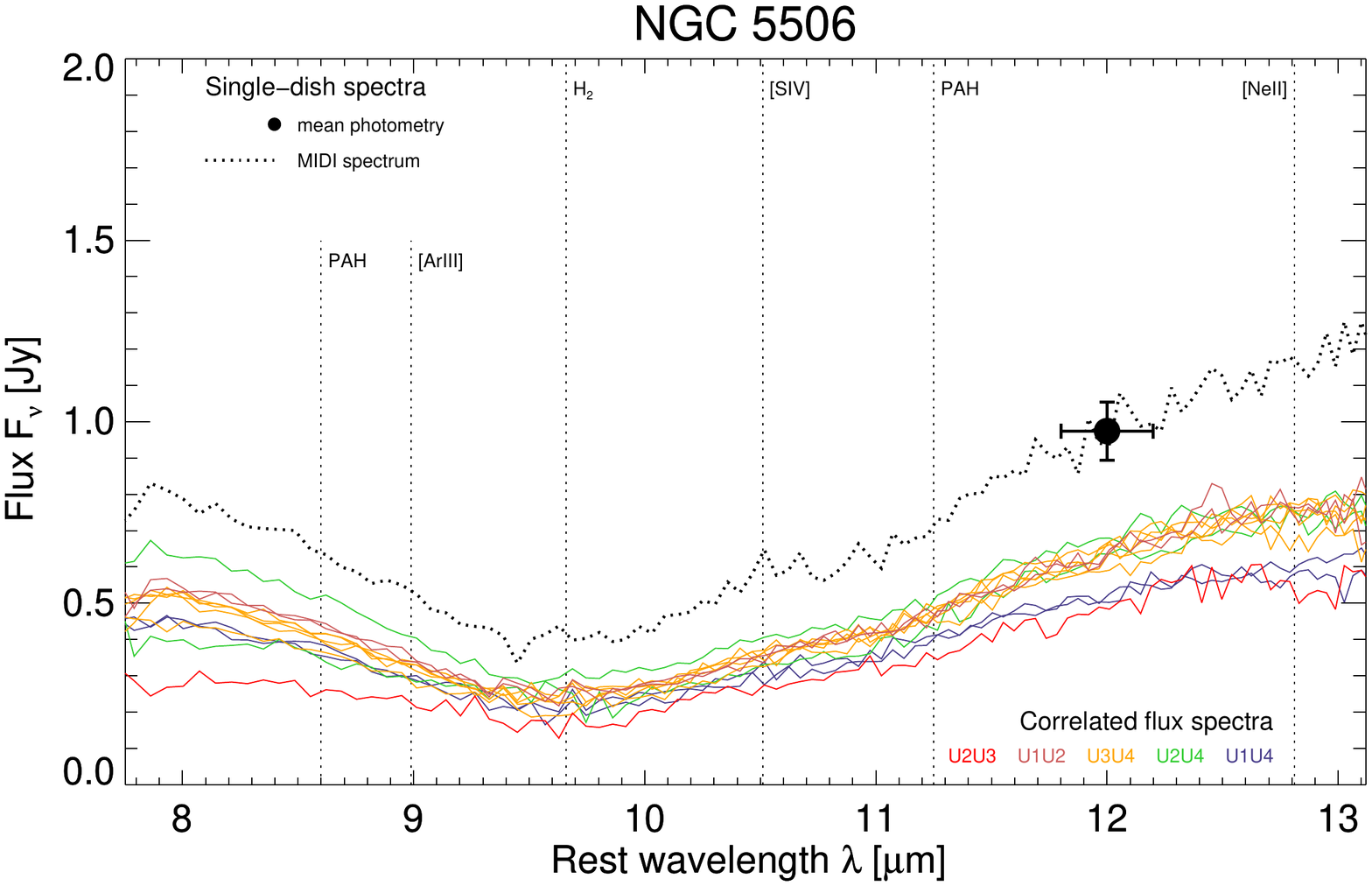}
\end{figure*}

\begin{figure*}
\caption{Single-dish and correlated flux spectra for NGC 5995 (see caption for Fig.~\ref{fig:spectra})}
\centering
\includegraphics[trim=4cm 3.5cm 4cm 3.5cm, width=0.75\hsize]{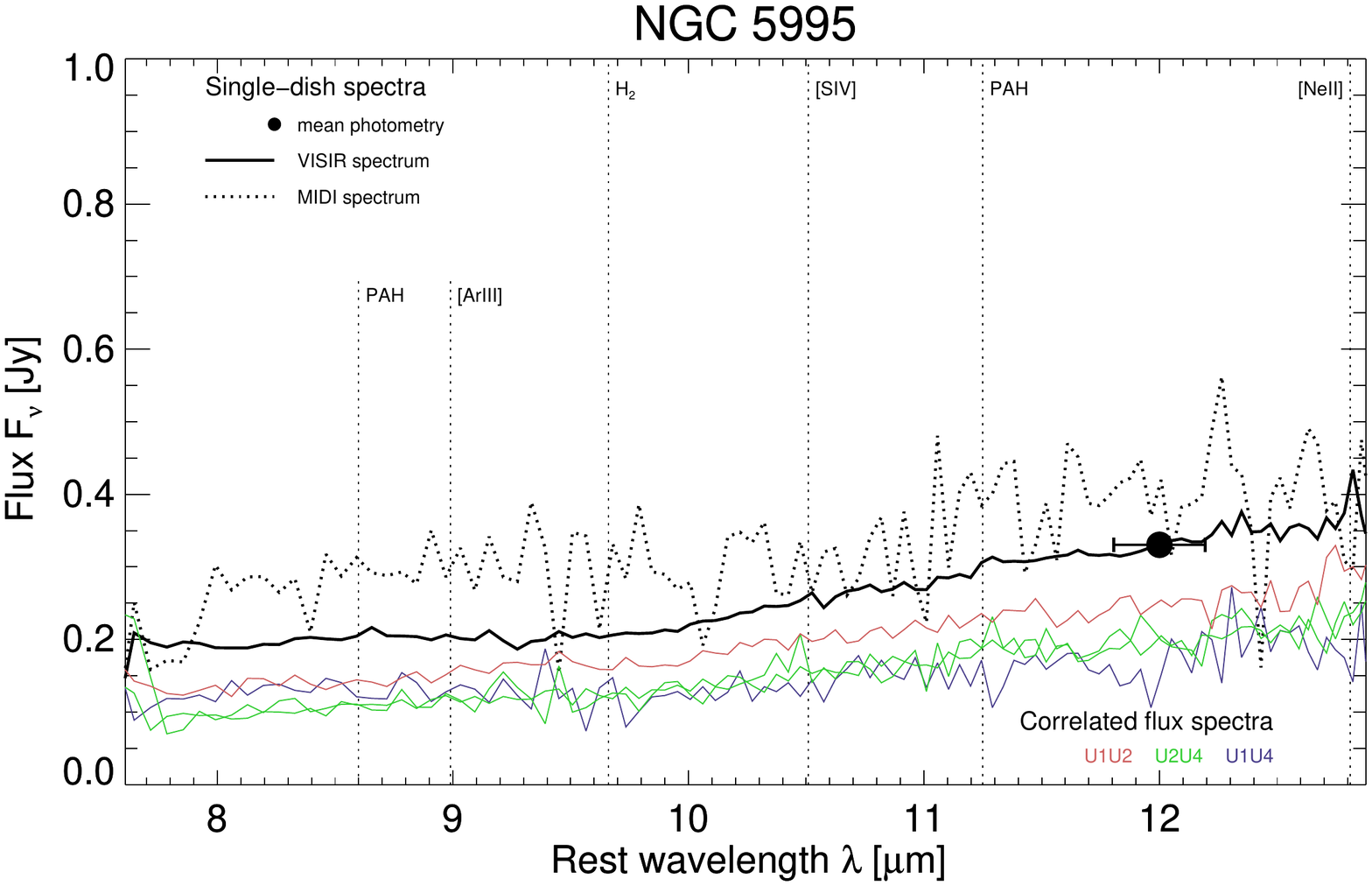}
\end{figure*}

\clearpage

\begin{figure*}
\caption{Single-dish and correlated flux spectra for NGC 7469 (see caption for Fig.~\ref{fig:spectra})}
\centering
\includegraphics[trim=4cm 3.5cm 4cm 3.5cm, width=0.75\hsize]{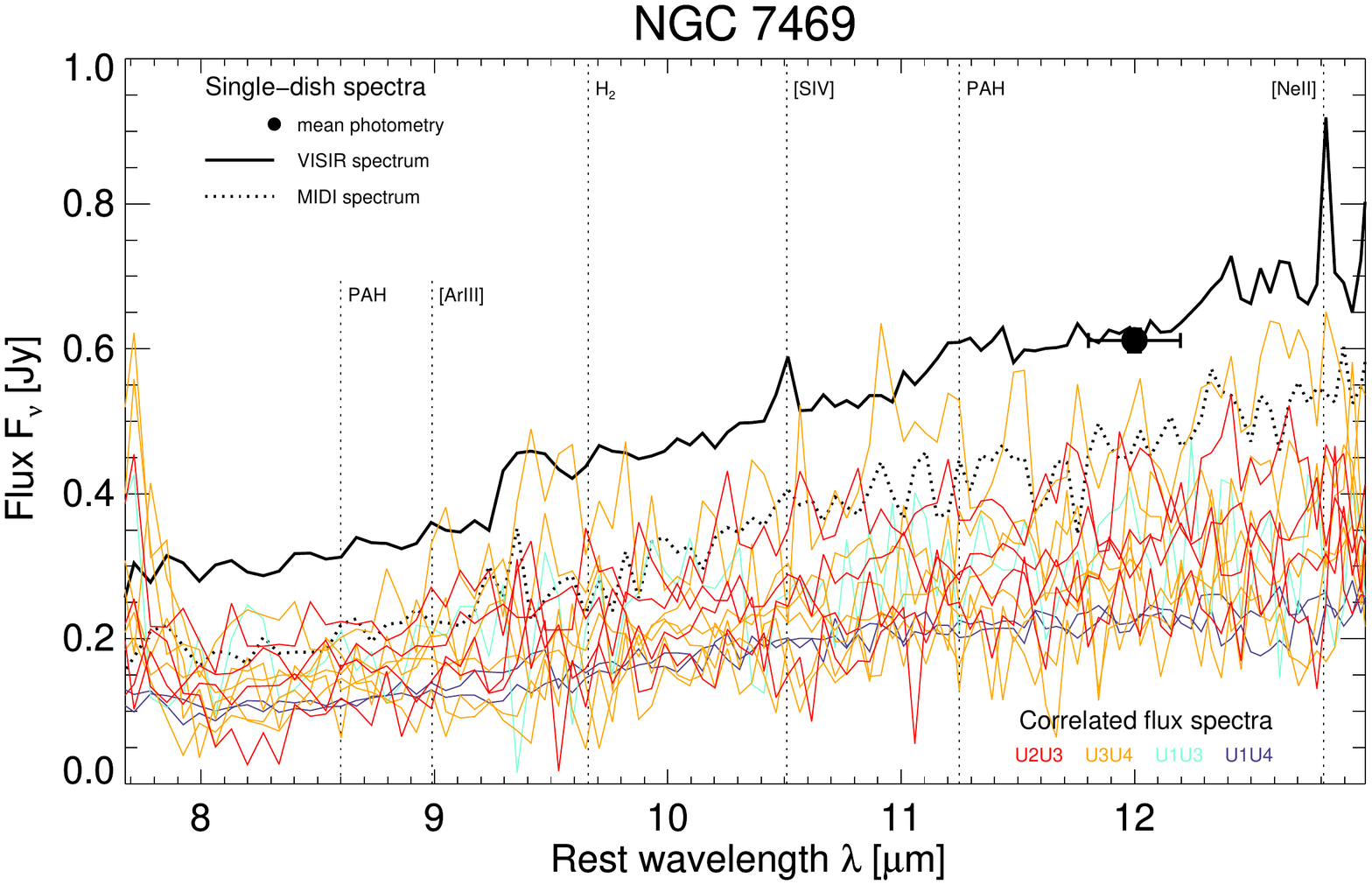}
\end{figure*}

\clearpage
\section{Differential phases}
\label{sec:dphases}

\begin{figure*}
	\centering
	\subfloat{\includegraphics[trim=2cm 4cm 2cm 4cm, width=0.25\hsize]{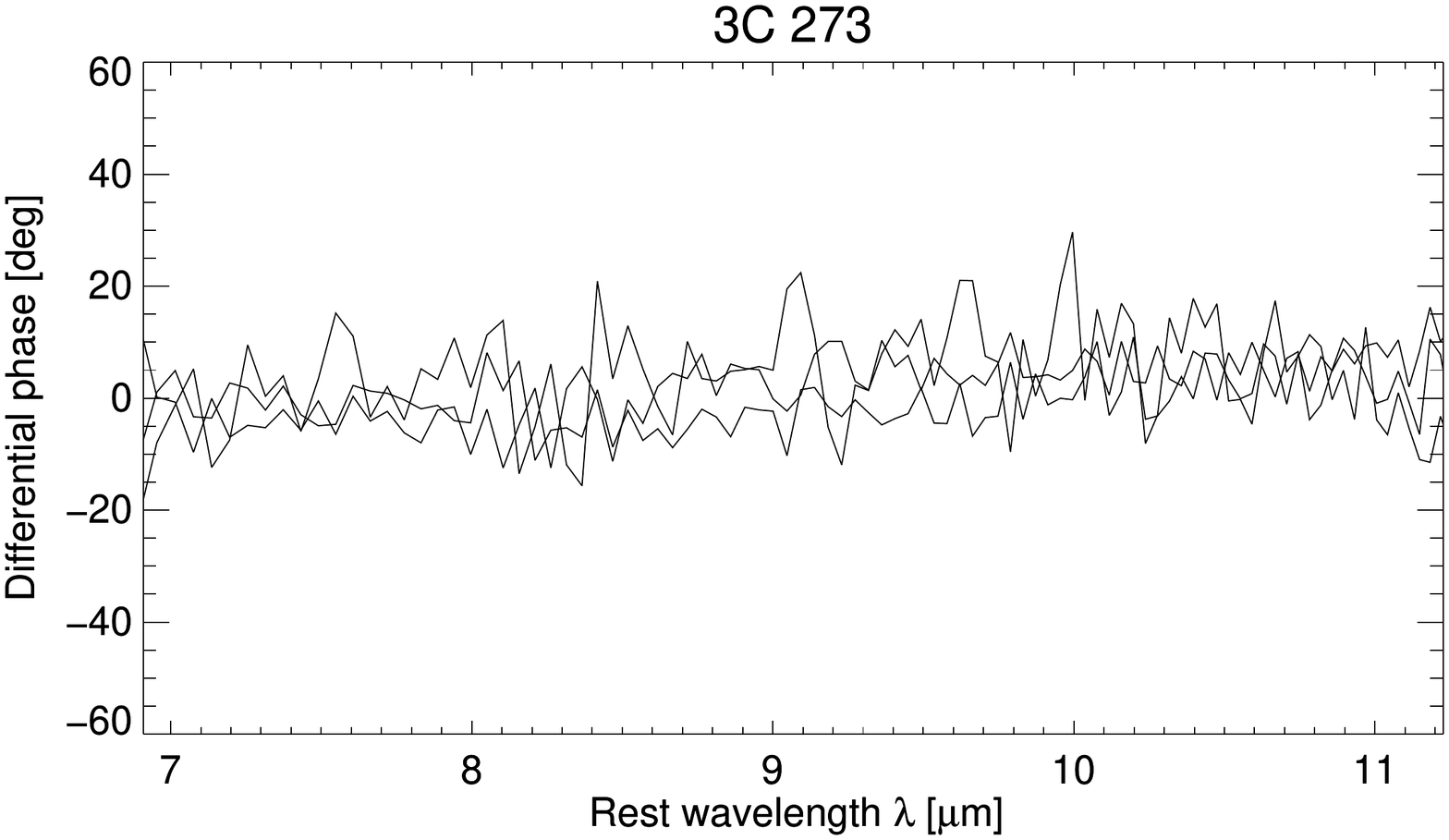}}
	\subfloat{\includegraphics[trim=2cm 4cm 2cm 4cm, width=0.25\hsize]{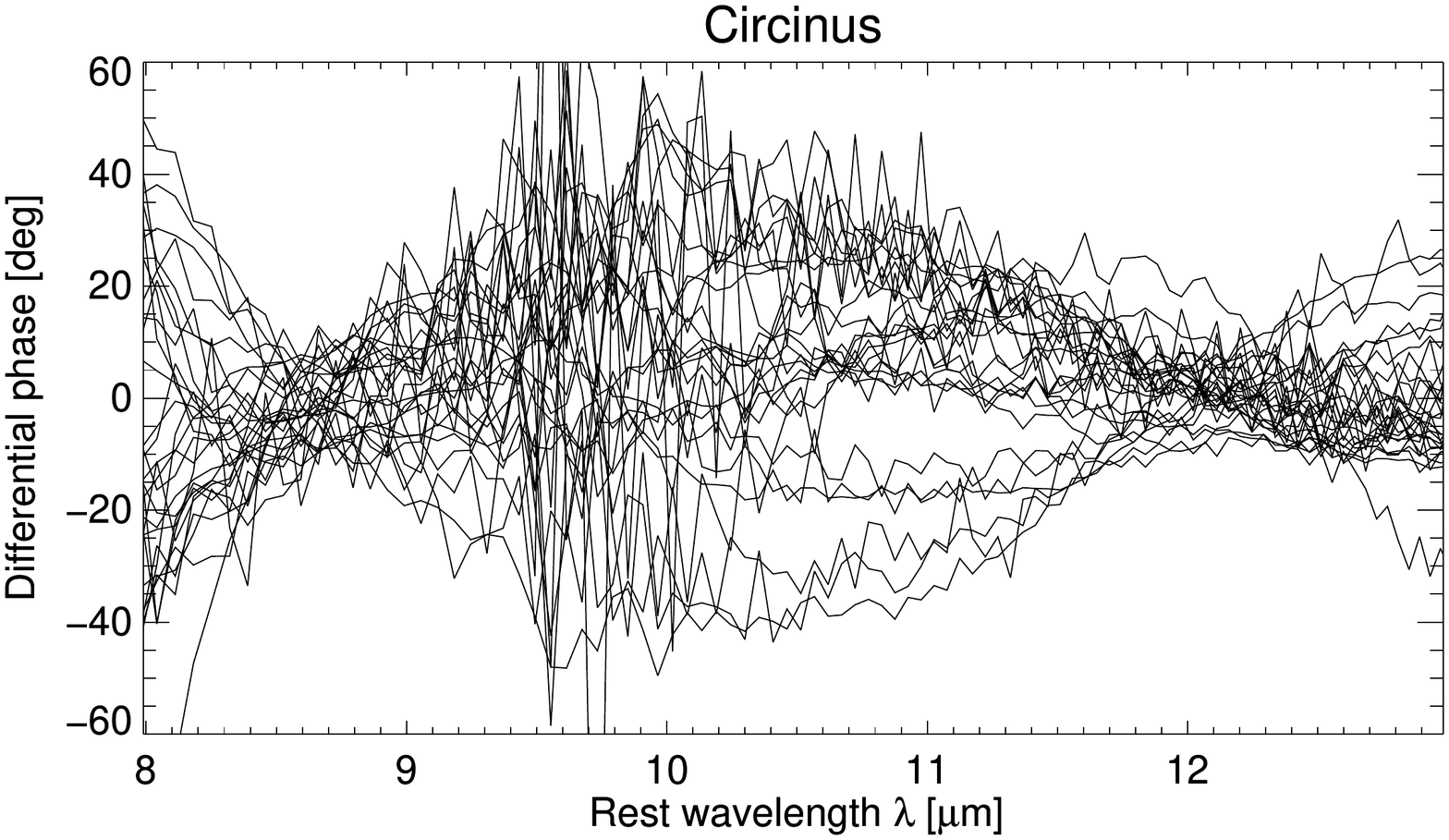}}
	\subfloat{\includegraphics[trim=2cm 4cm 2cm 4cm, width=0.25\hsize]{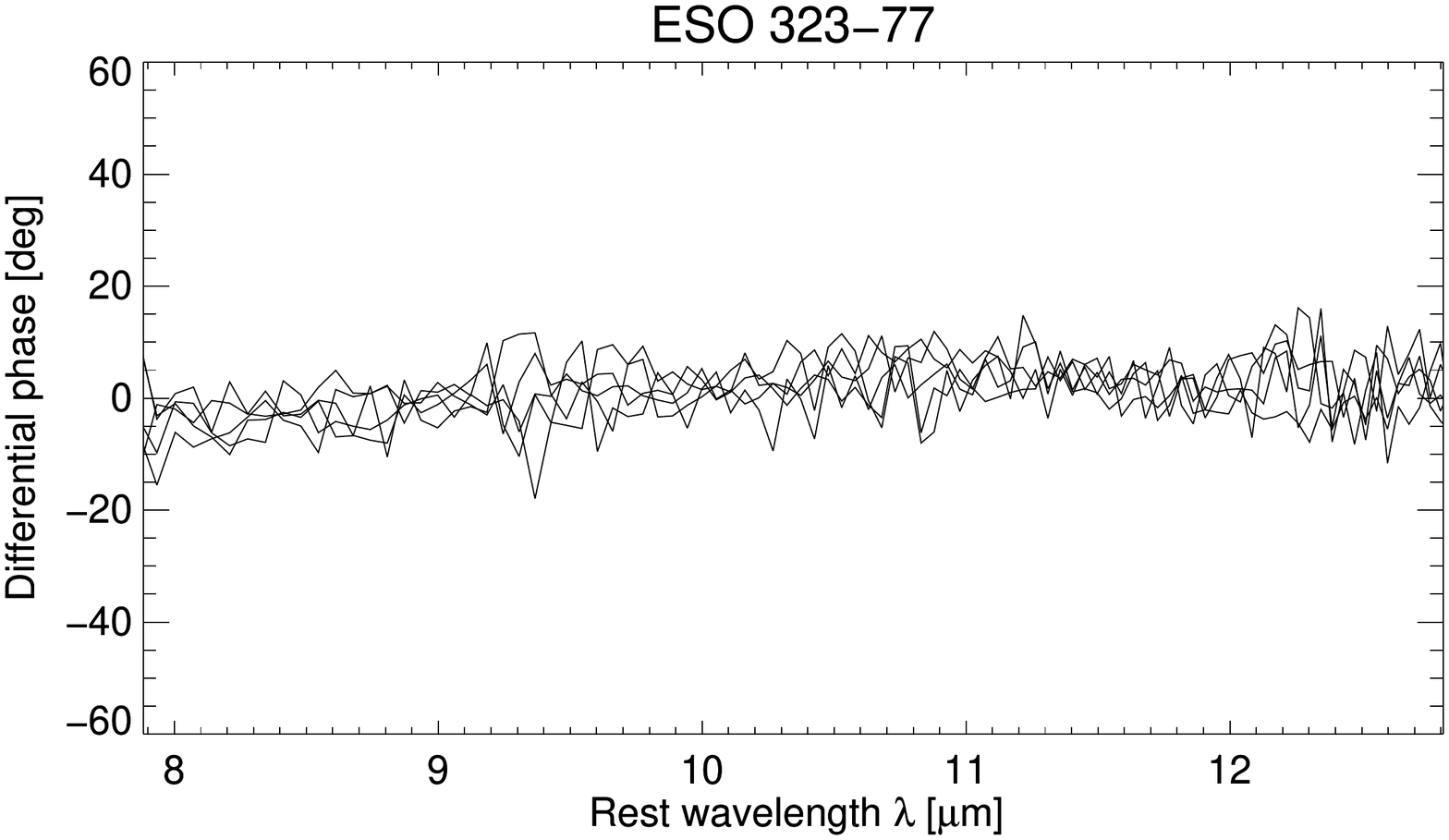}}\\
	\subfloat{\includegraphics[trim=2cm 4cm 2cm 4cm, width=0.25\hsize]{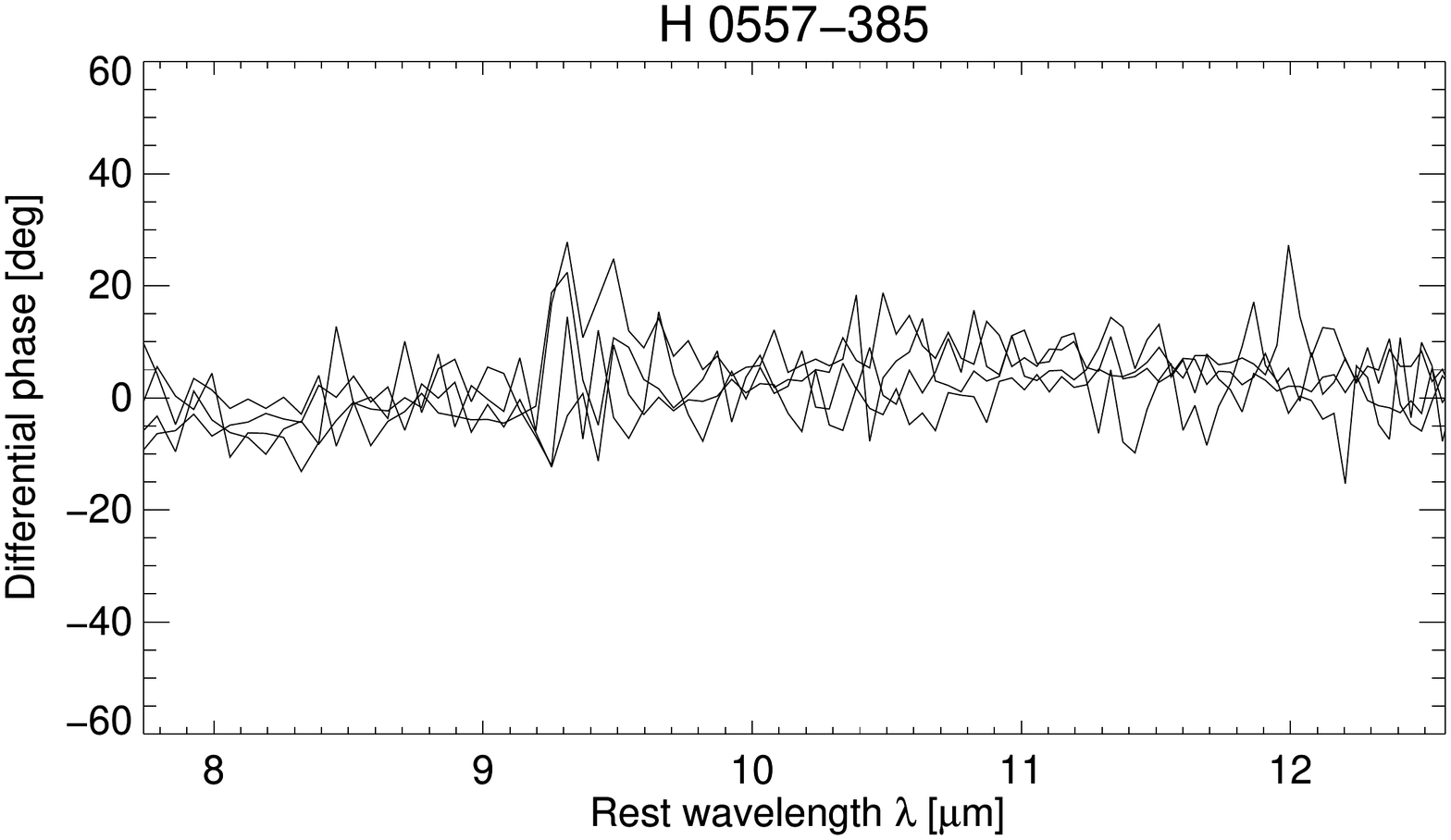}}
	\subfloat{\includegraphics[trim=2cm 4cm 2cm 4cm, width=0.25\hsize]{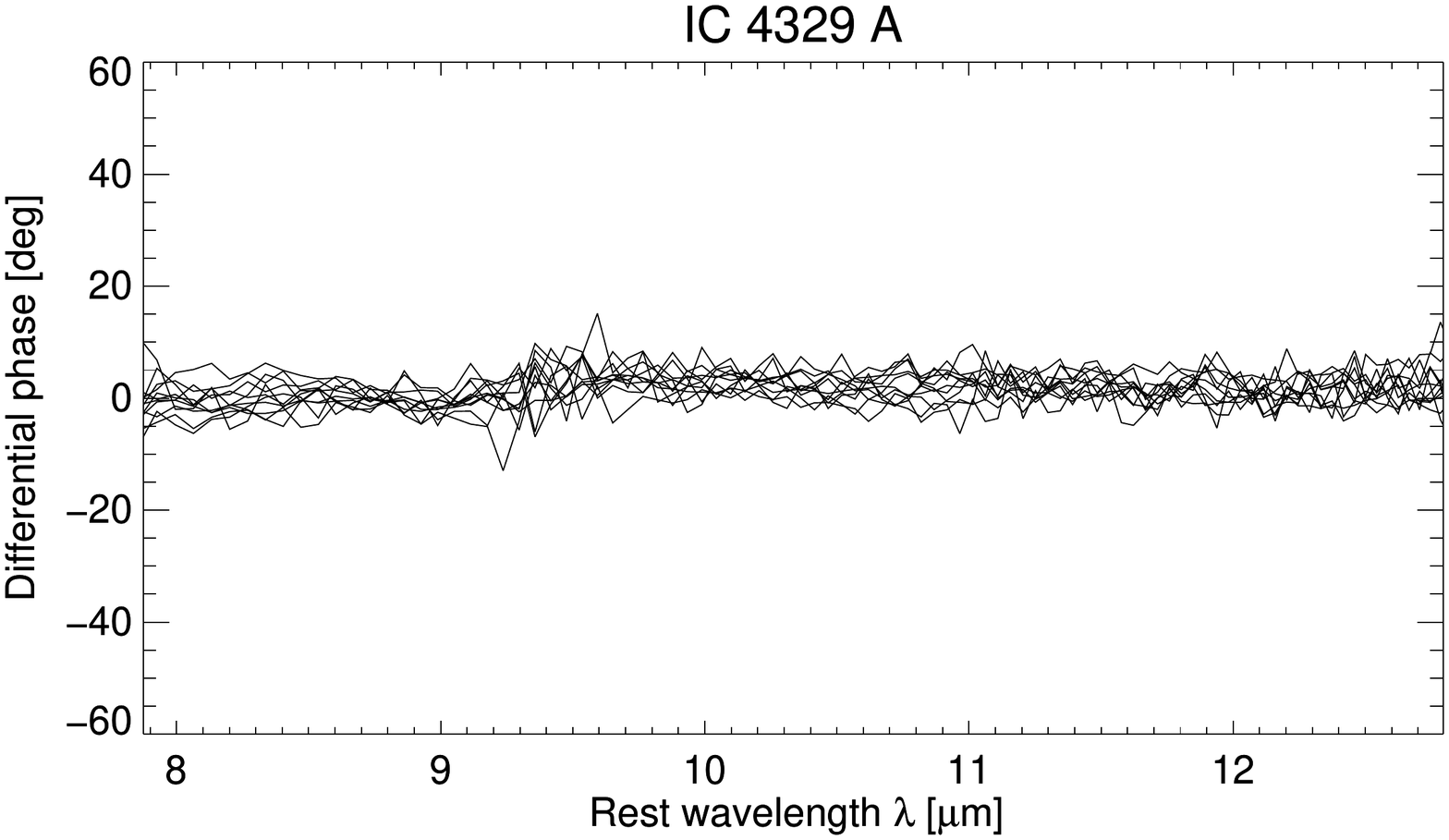}}
	\subfloat{\includegraphics[trim=2cm 4cm 2cm 4cm, width=0.25\hsize]{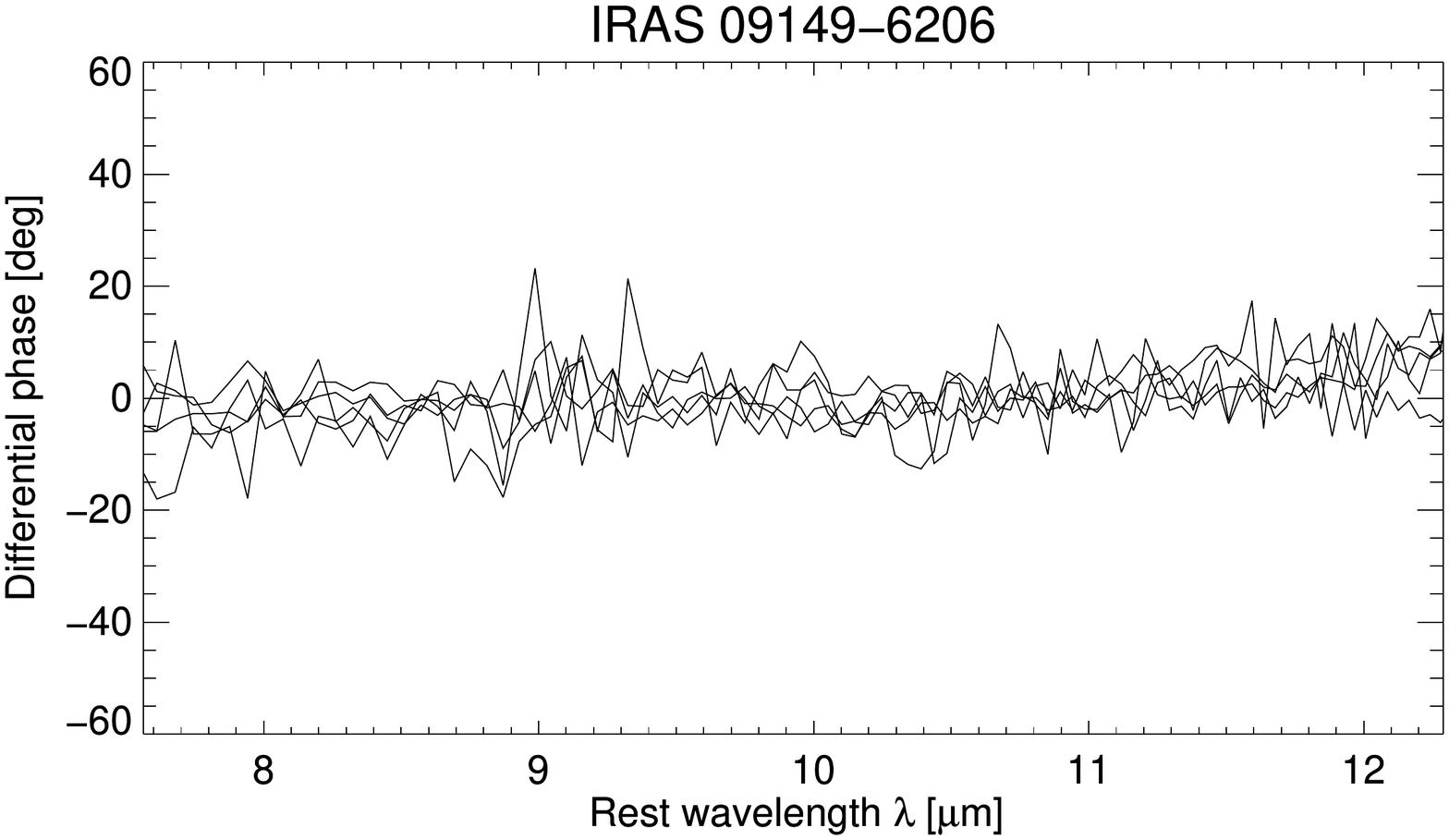}}\\
	\subfloat{\includegraphics[trim=2cm 4cm 2cm 4cm, width=0.25\hsize]{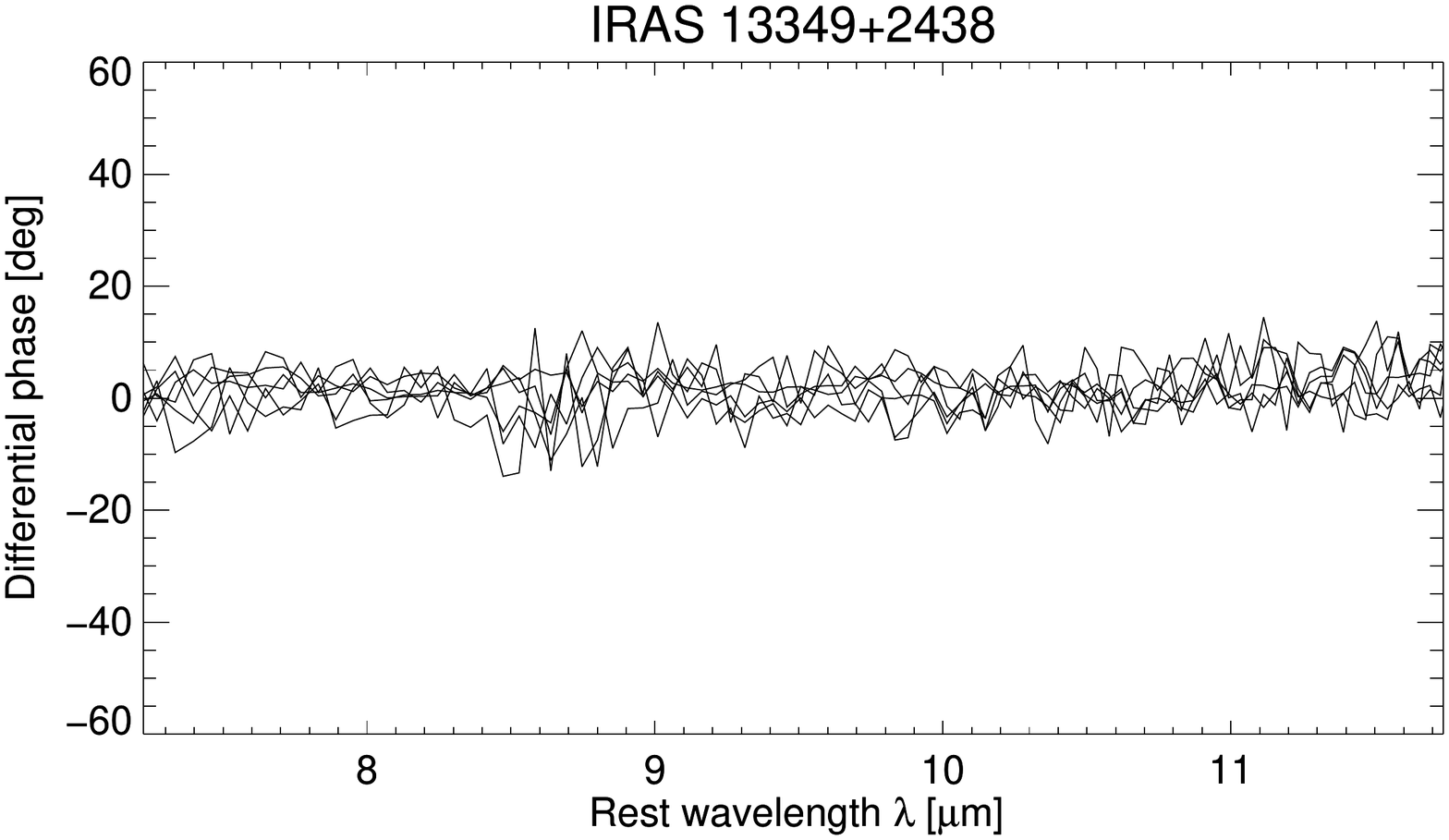}}
	\subfloat{\includegraphics[trim=2cm 4cm 2cm 4cm, width=0.25\hsize]{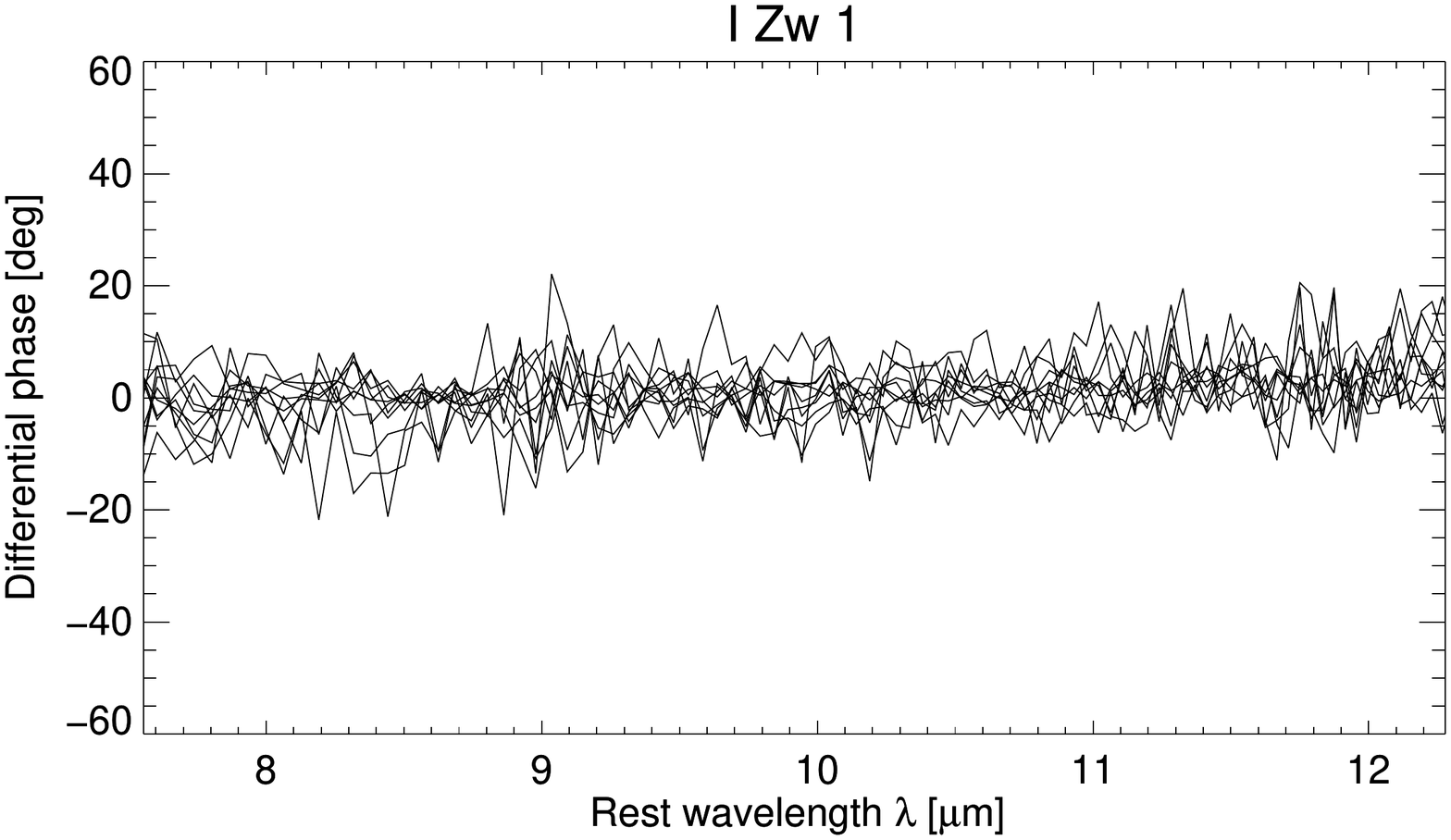}}
	\subfloat{\includegraphics[trim=2cm 4cm 2cm 4cm, width=0.25\hsize]{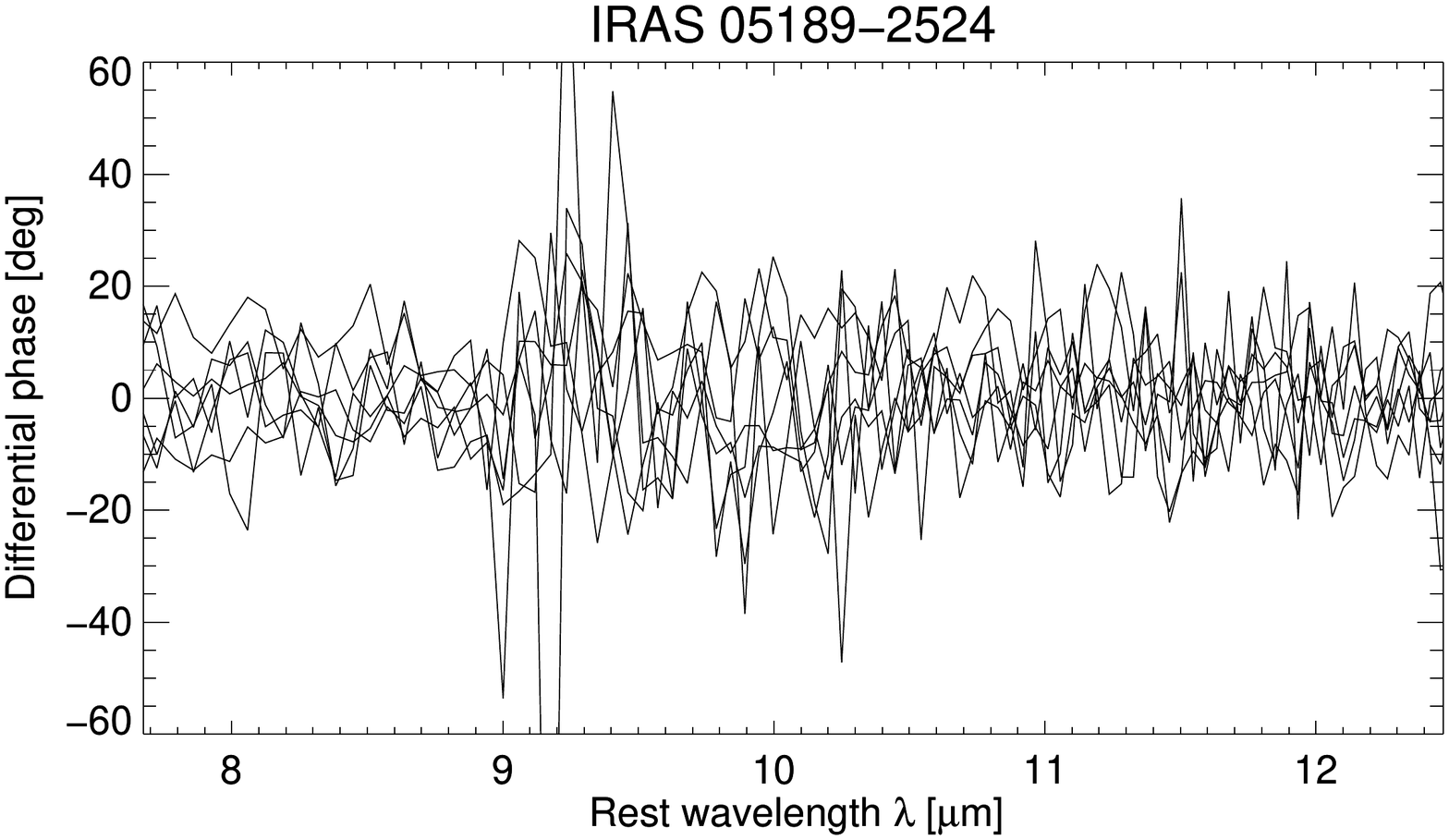}}\\
	\subfloat{\includegraphics[trim=2cm 4cm 2cm 4cm, width=0.25\hsize]{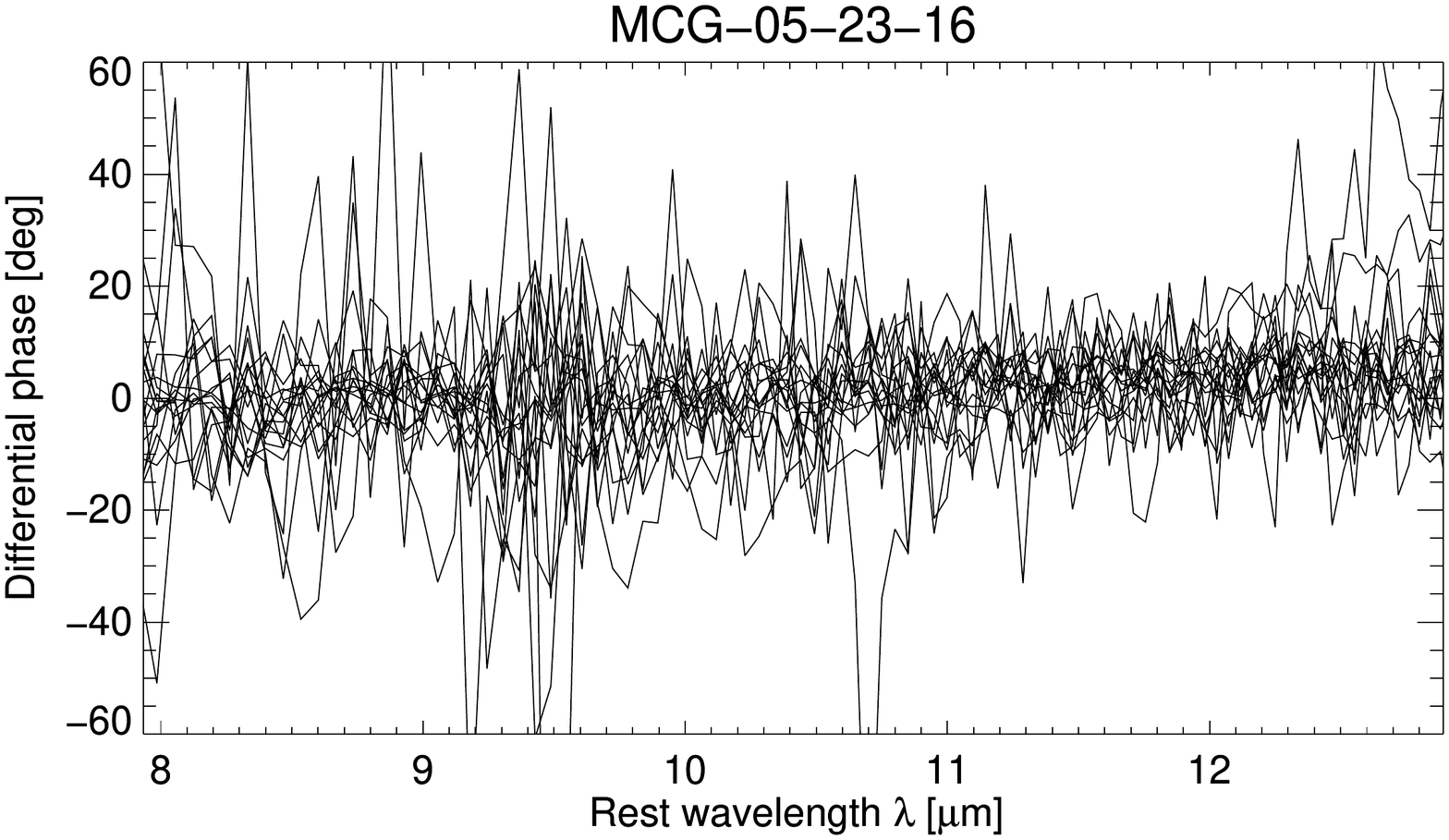}}
	\subfloat{\includegraphics[trim=2cm 4cm 2cm 4cm, width=0.25\hsize]{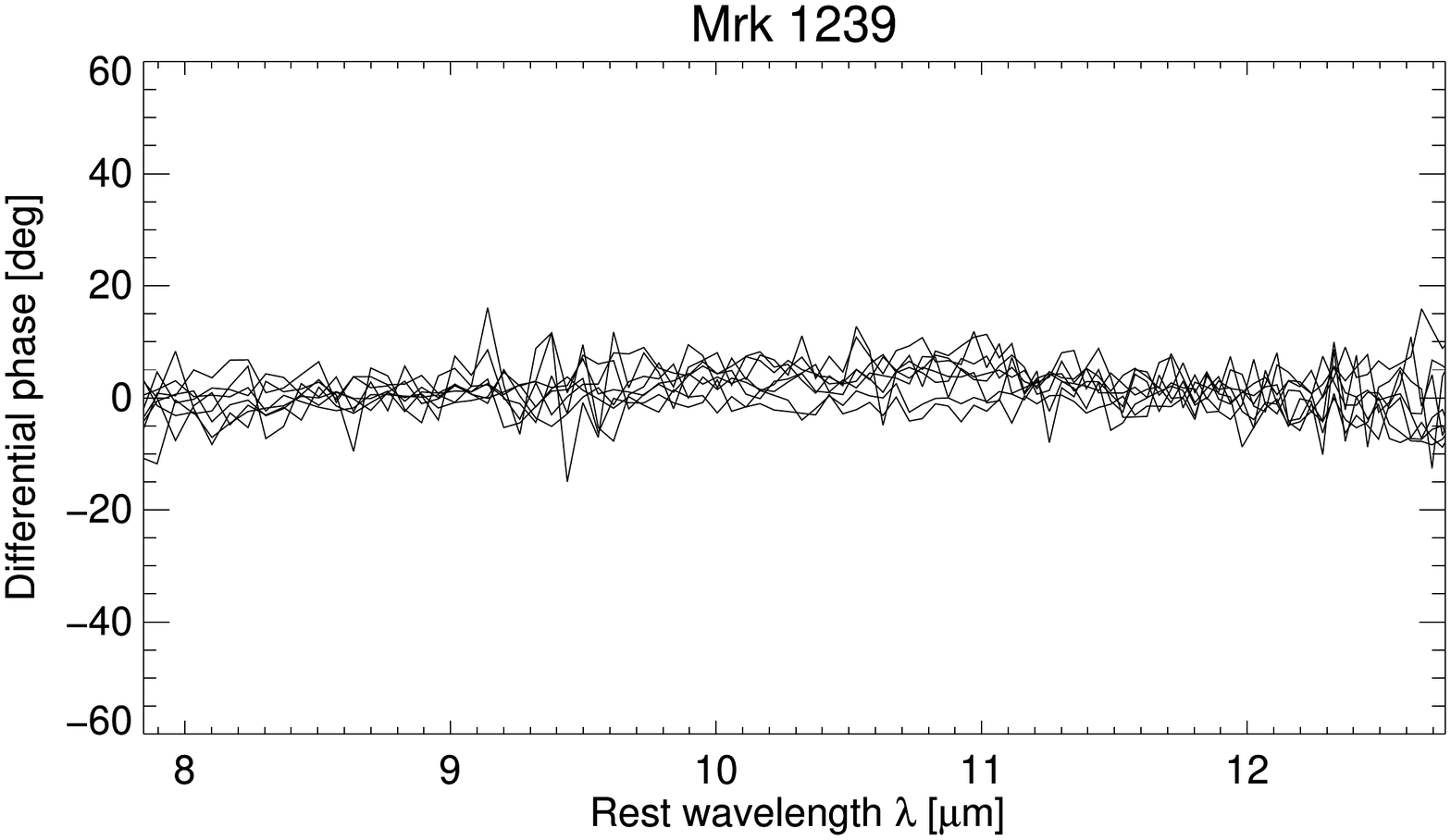}}
	\subfloat{\includegraphics[trim=2cm 4cm 2cm 4cm, width=0.25\hsize]{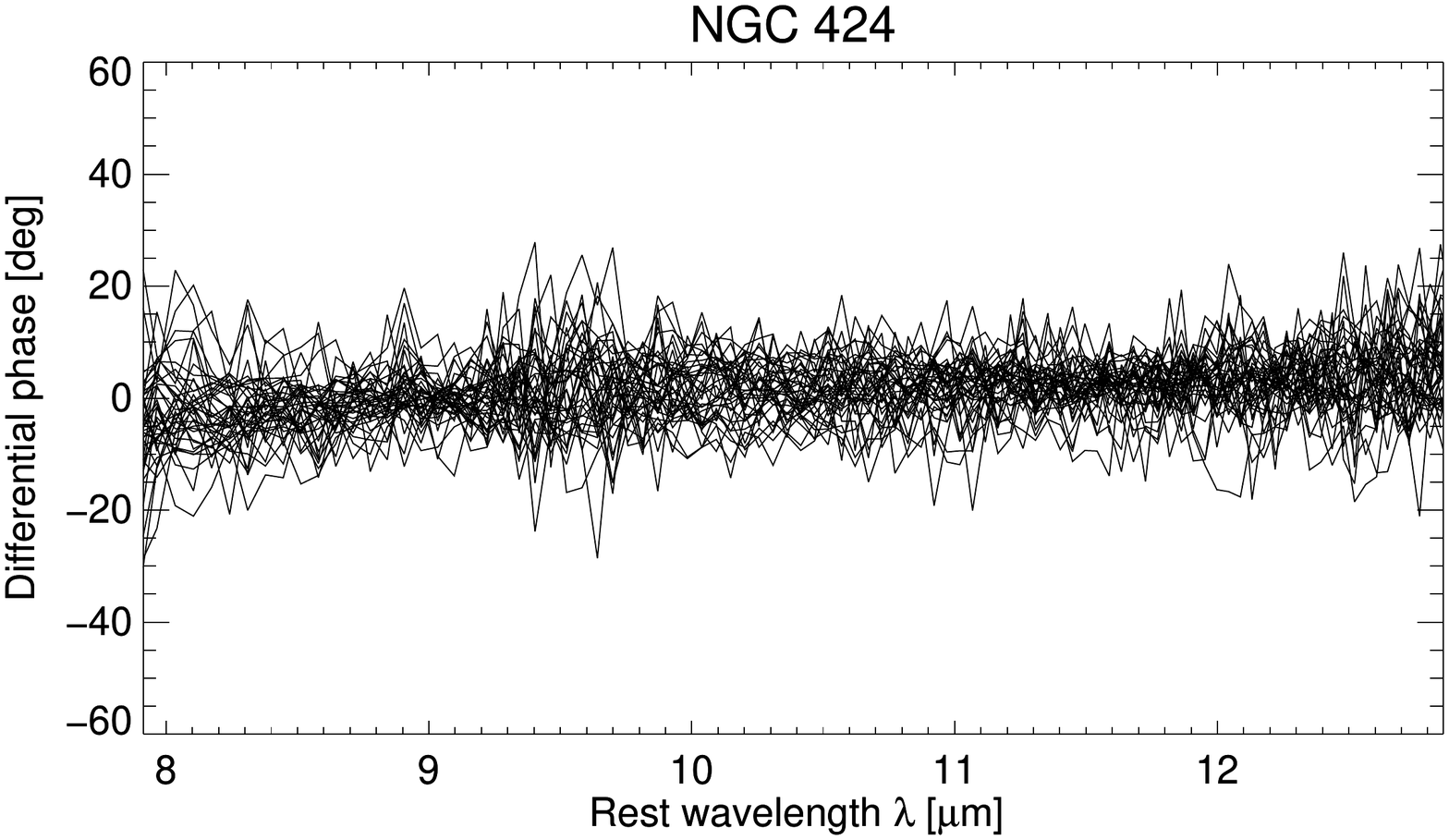}}\\
	\subfloat{\includegraphics[trim=2cm 4cm 2cm 4cm, width=0.25\hsize]{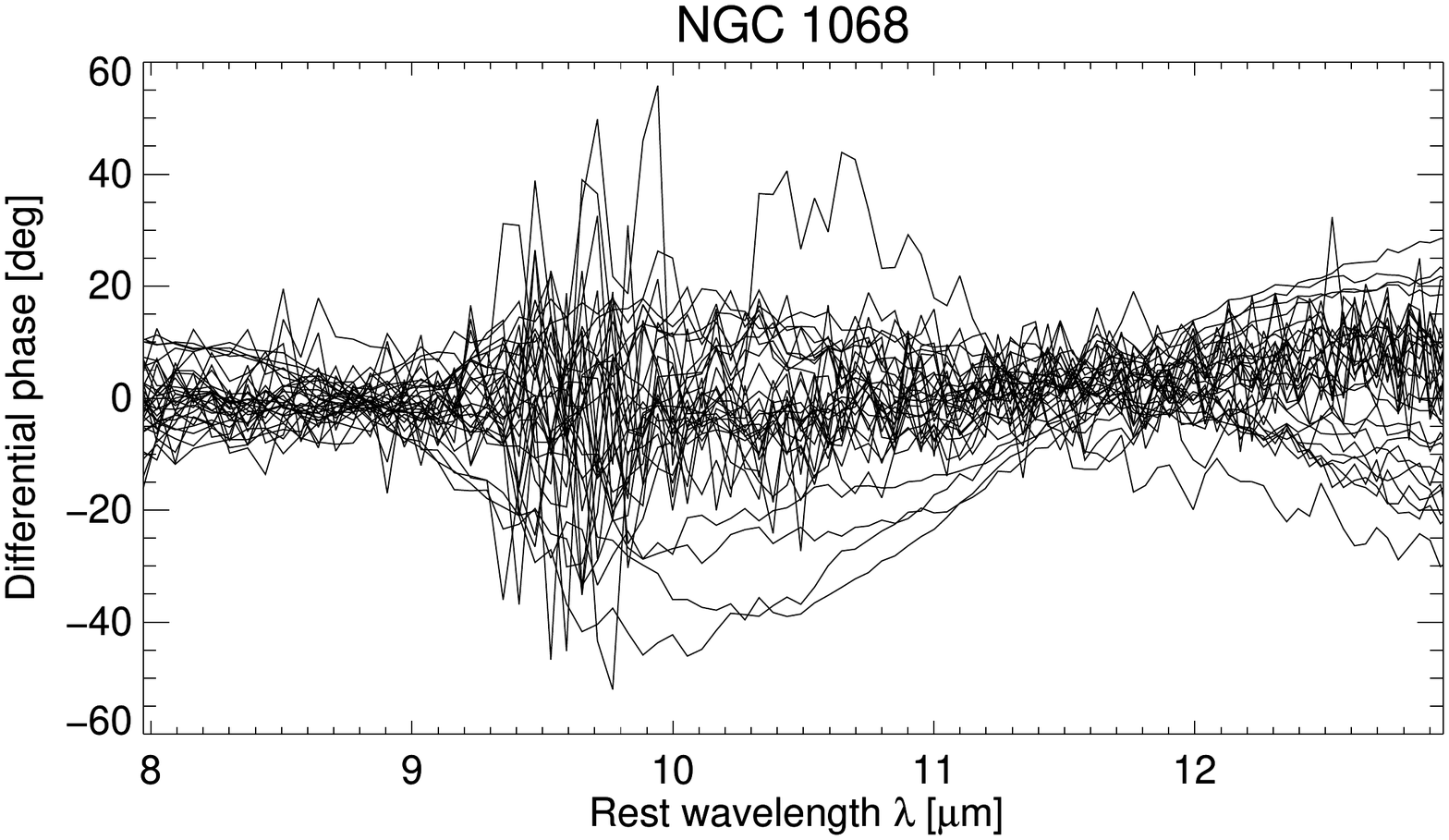}}
	\subfloat{\includegraphics[trim=2cm 4cm 2cm 4cm, width=0.25\hsize]{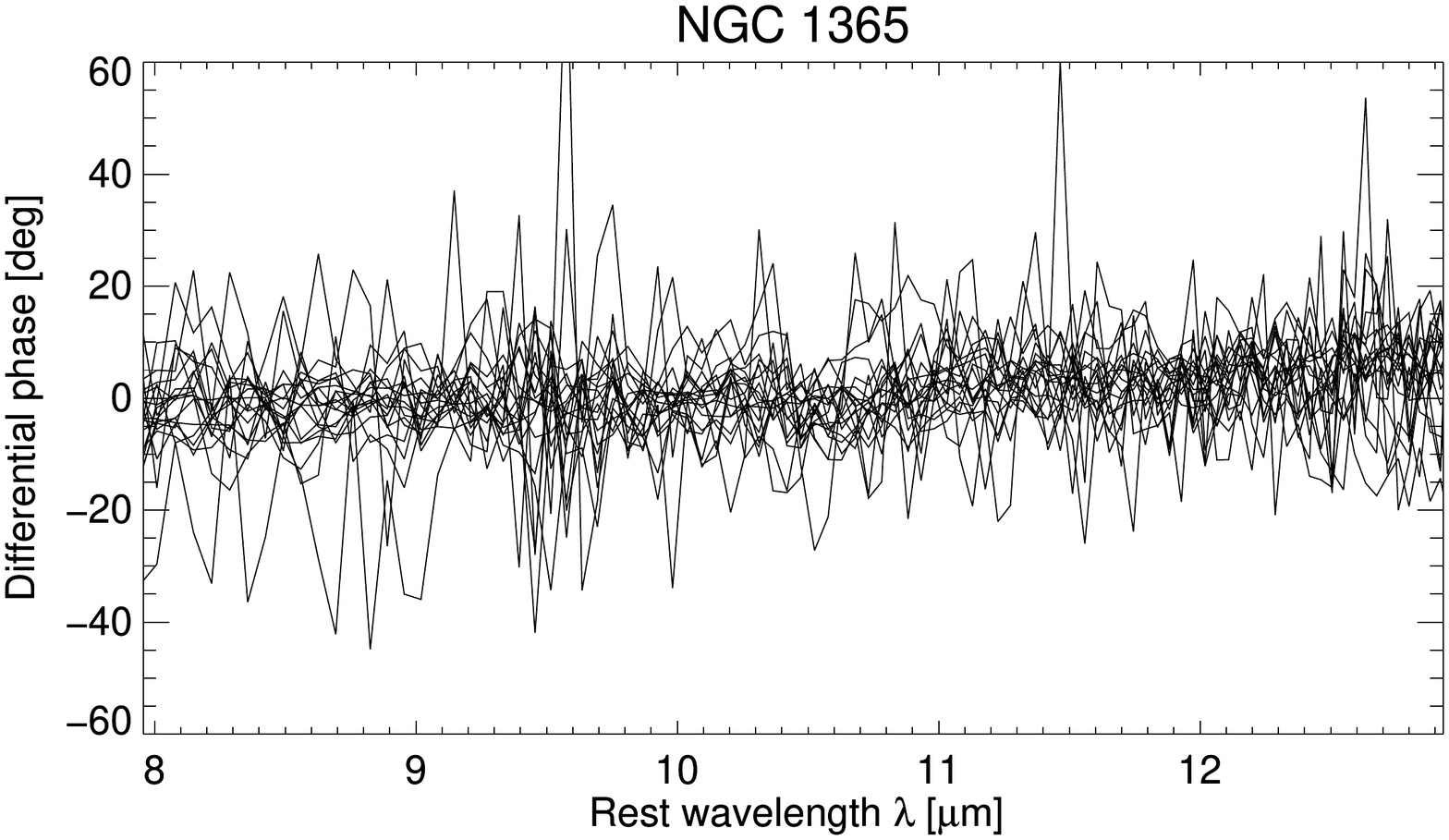}}
	\subfloat{\includegraphics[trim=2cm 4cm 2cm 4cm, width=0.25\hsize]{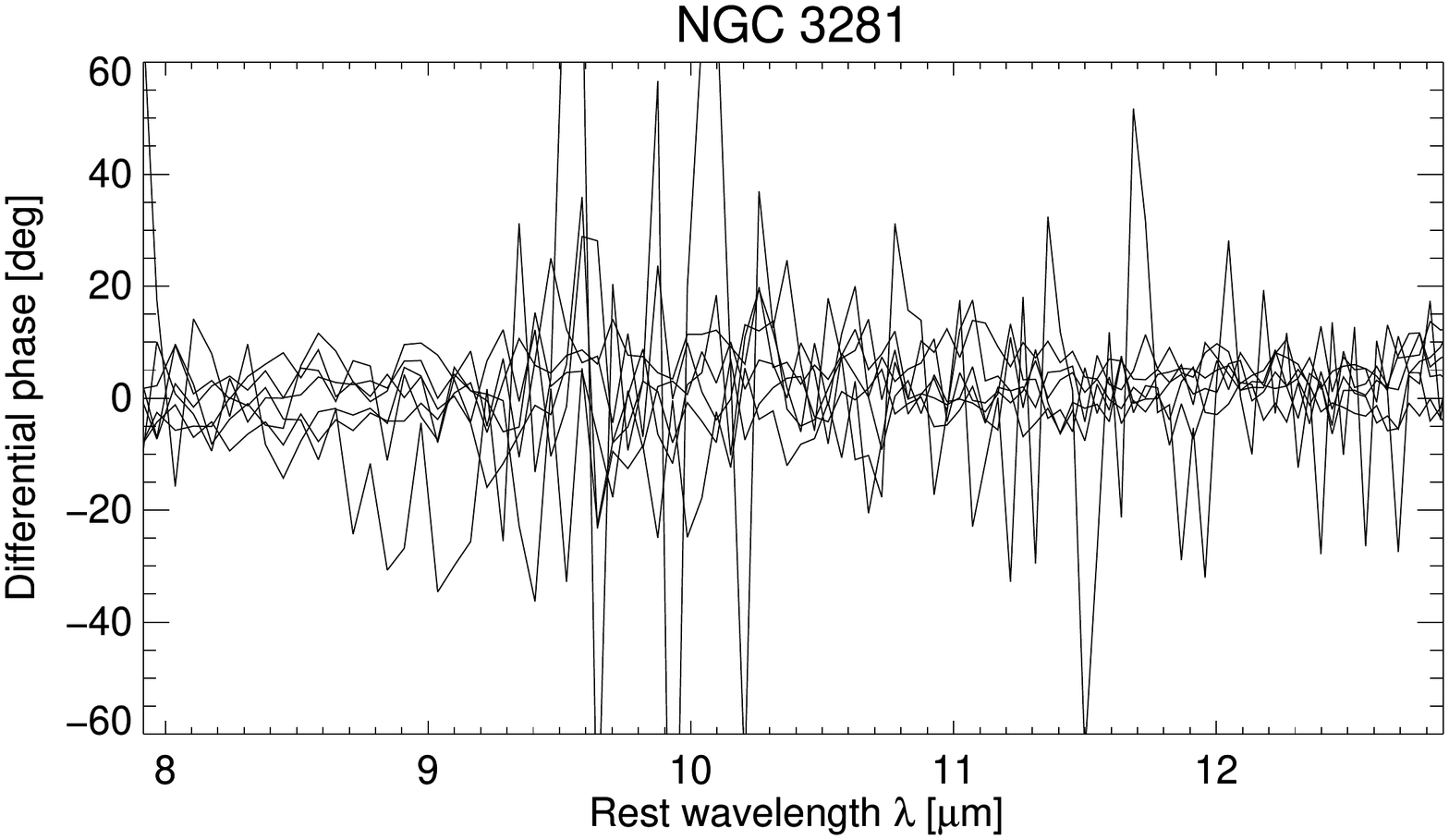}}\\
	\subfloat{\includegraphics[trim=2cm 4cm 2cm 4cm, width=0.25\hsize]{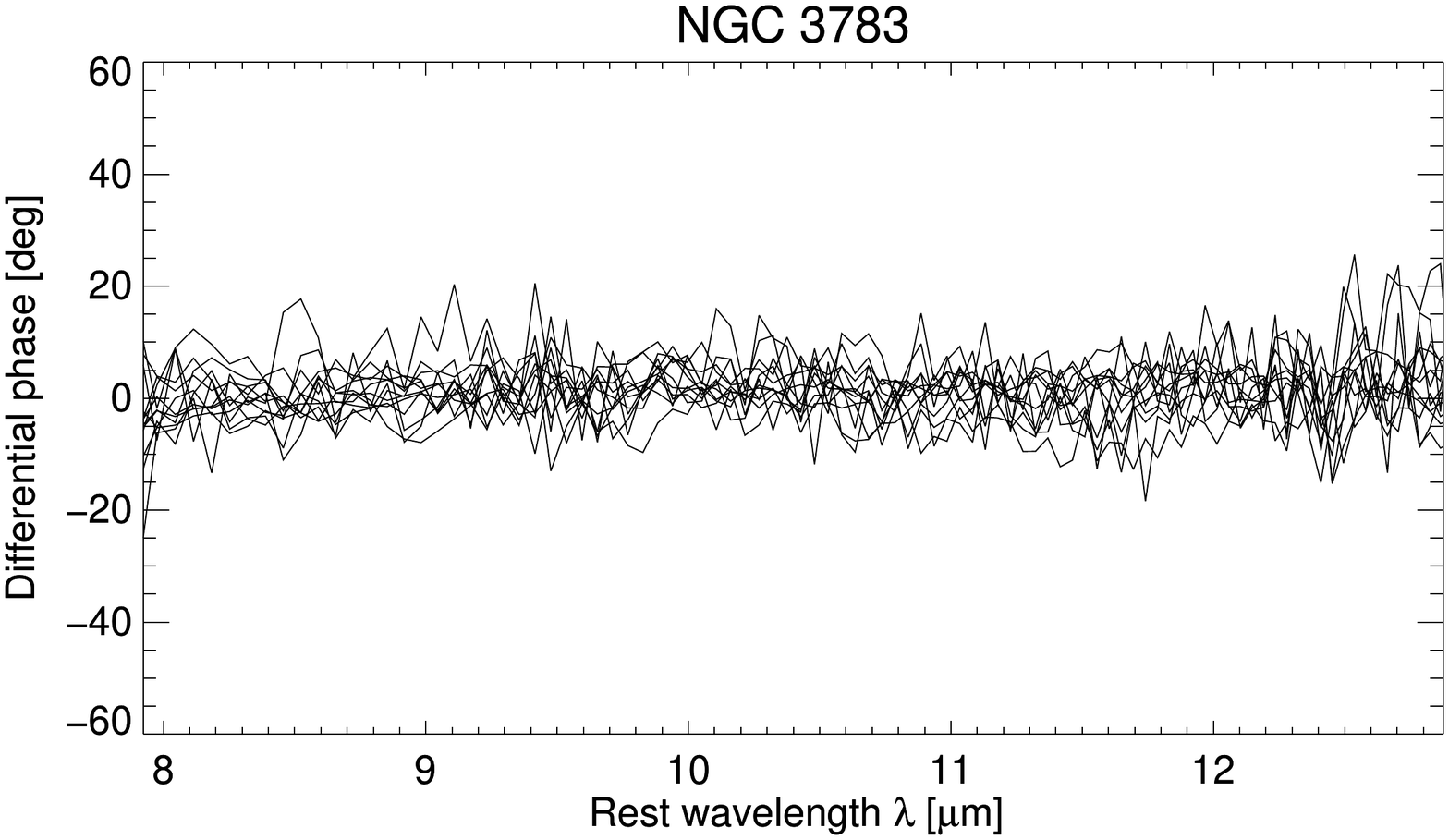}}
	\subfloat{\includegraphics[trim=2cm 4cm 2cm 4cm, width=0.25\hsize]{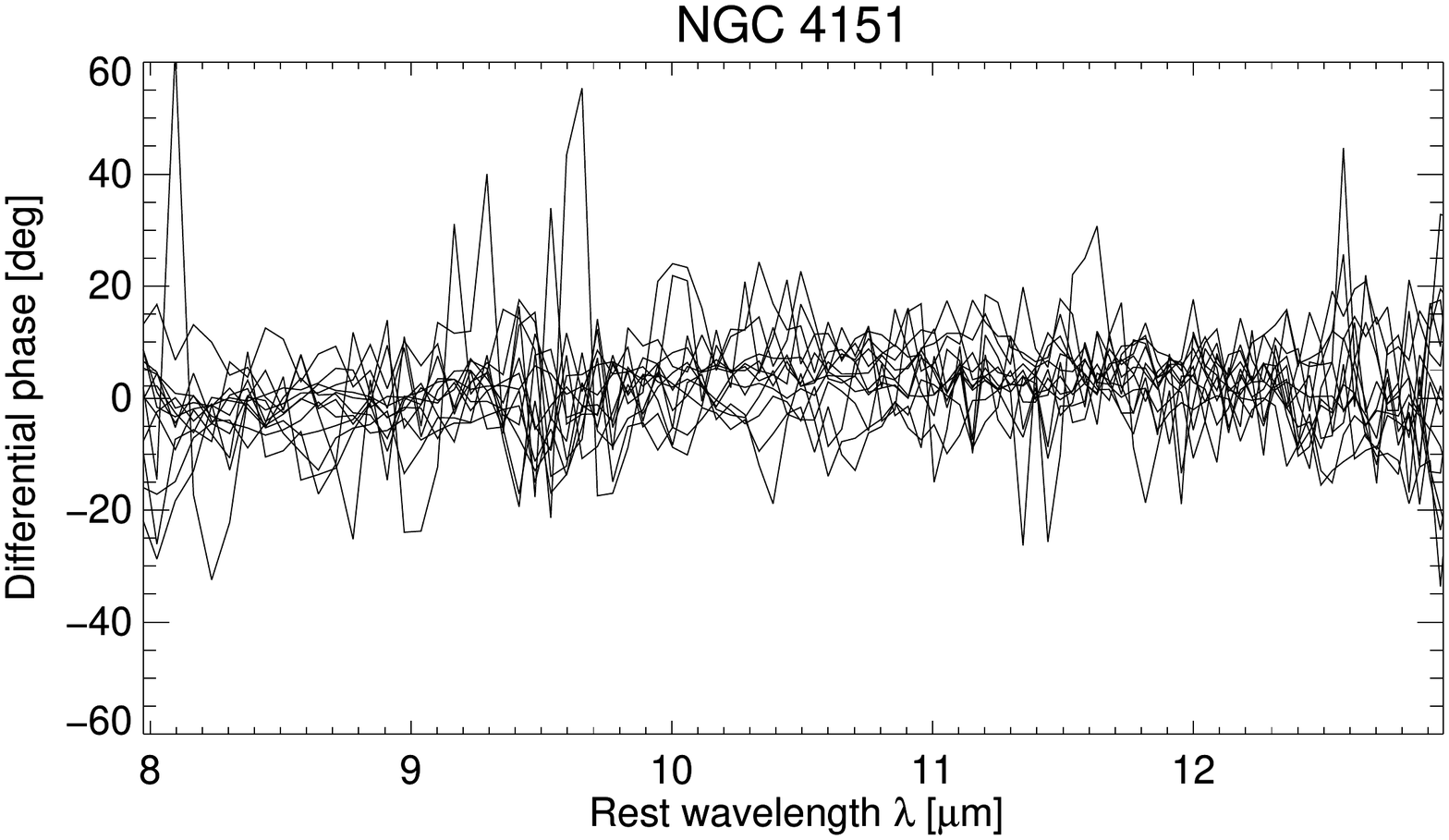}}
	\subfloat{\includegraphics[trim=2cm 4cm 2cm 4cm, width=0.25\hsize]{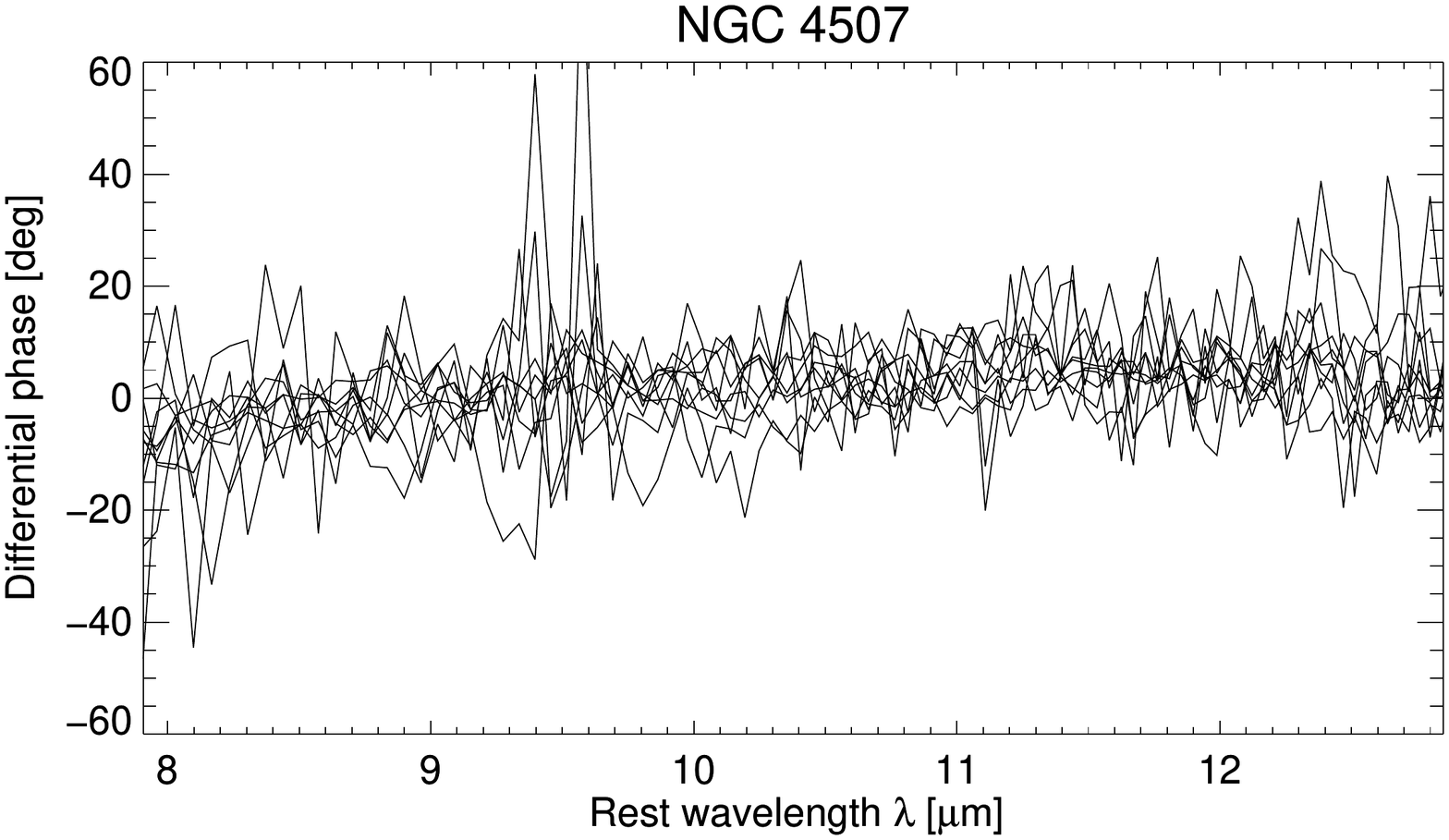}}\\
	\subfloat{\includegraphics[trim=2cm 4cm 2cm 4cm, width=0.25\hsize]{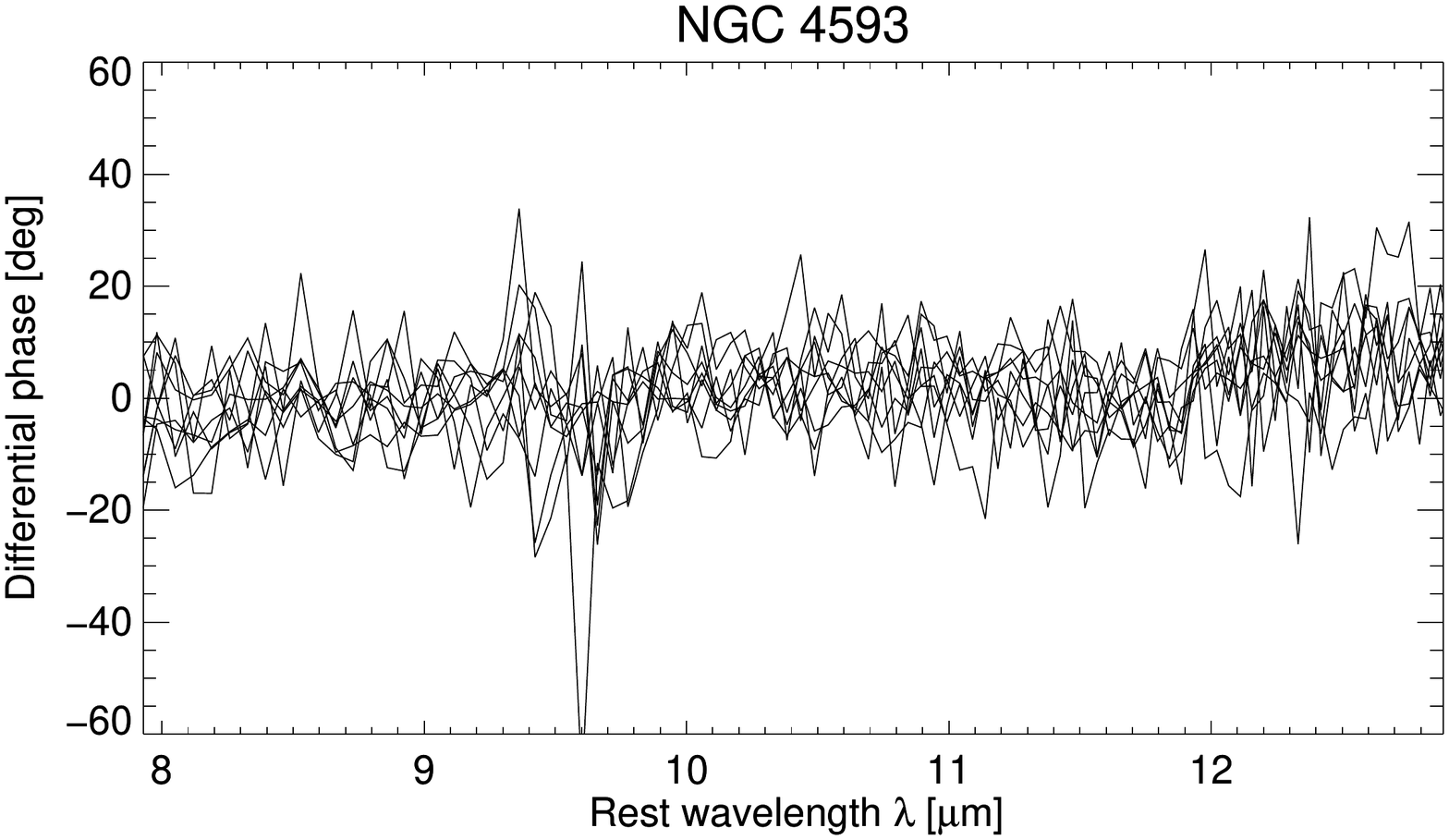}}
	\subfloat{\includegraphics[trim=2cm 4cm 2cm 4cm, width=0.25\hsize]{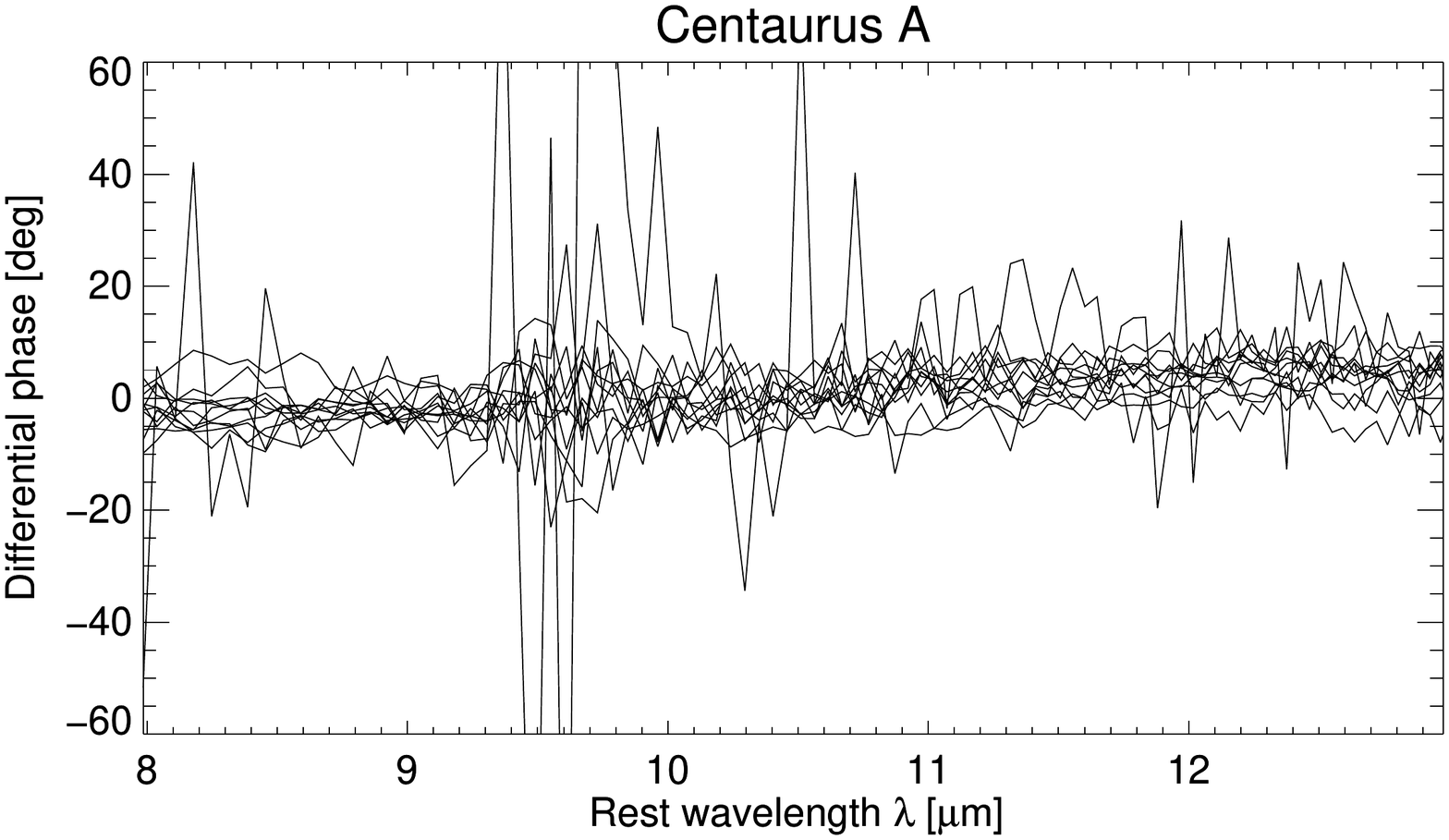}}
	\subfloat{\includegraphics[trim=2cm 4cm 2cm 4cm, width=0.25\hsize]{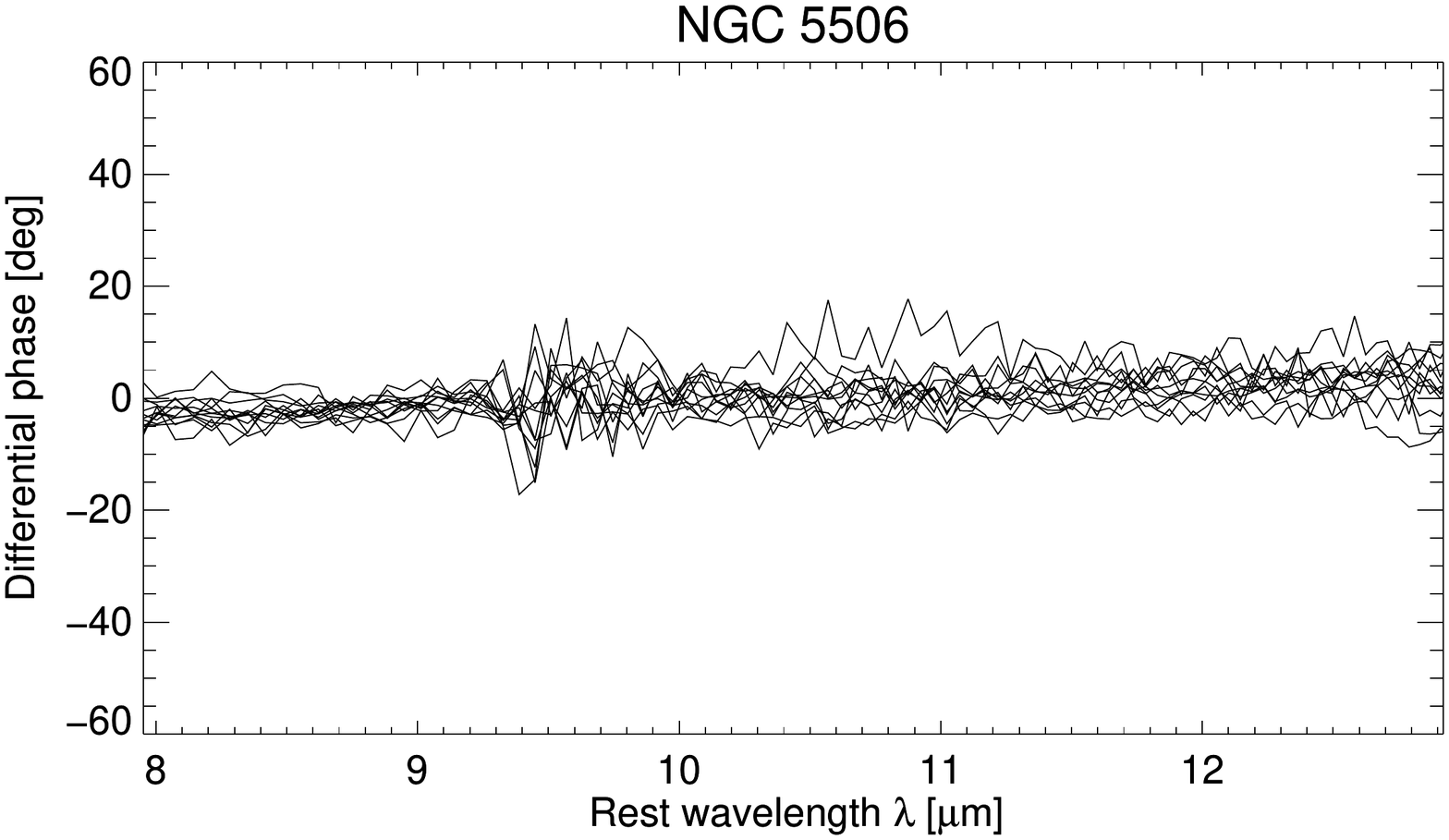}}\\
	\subfloat{\includegraphics[trim=2cm 4cm 2cm 4cm, width=0.25\hsize]{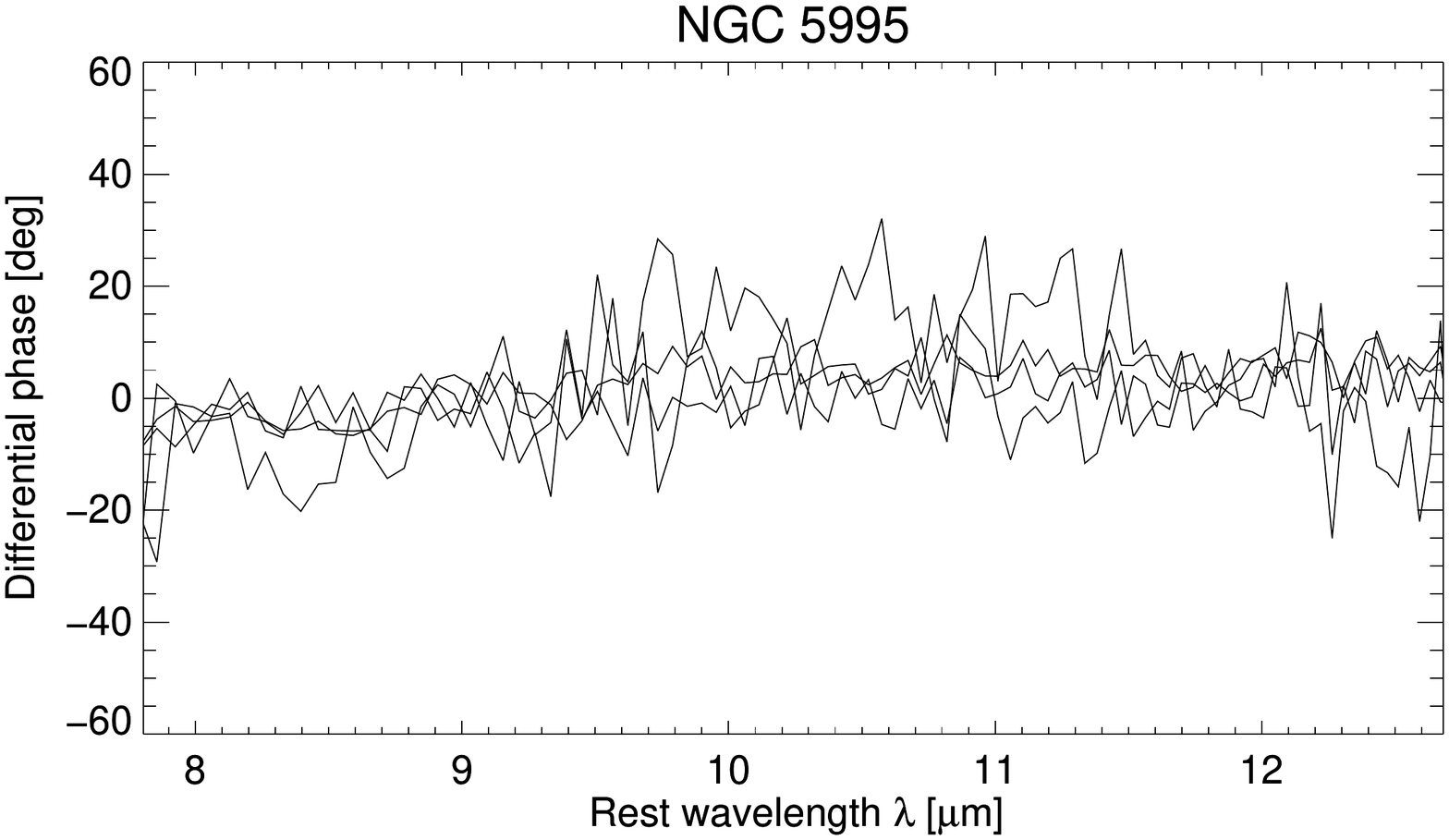}}
	\subfloat{\includegraphics[trim=2cm 4cm 2cm 4cm, width=0.25\hsize]{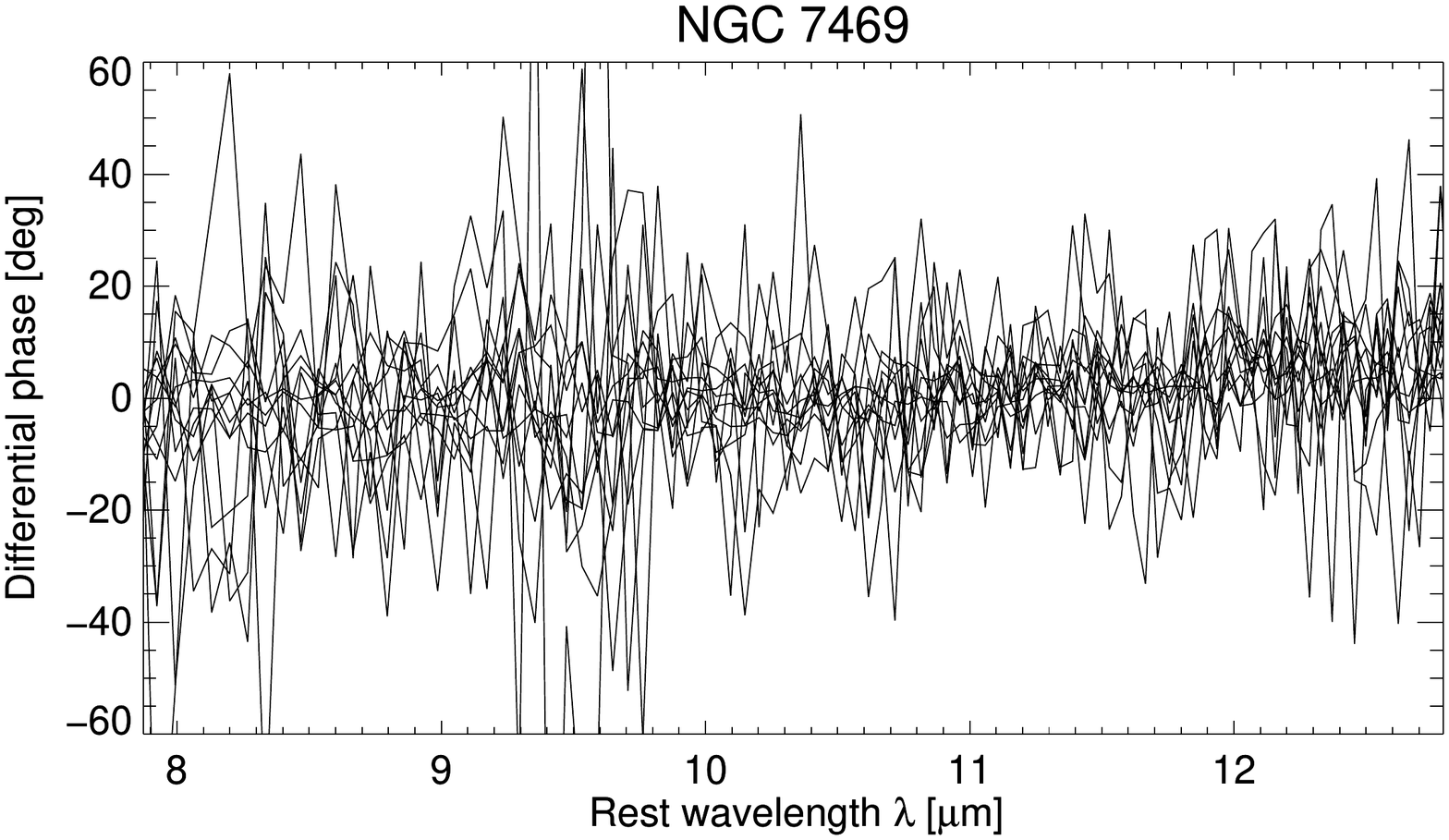}}
	\caption{\label{fig:dphases}Differential phases for all AGNs in this sample}
\end{figure*}

\clearpage
\section{Observing logs}
\subsection{VISIR observations}

\begin{table*}
\caption{\label{tab:obs:visir}VISIR mid-IR spectroscopy of LP sources (program ID 086.B-0919).}
\centering
\begin{tabular}{l c c c c l}
\hline\hline\
Object           &  Setting    & Observing date   & Calibrator &   Program   & Reference \\
                 & ($\um$) &(YYYY-MM-DDTHH:MM)&            &             &           \\ \hline
      I~Zw~1 &    8.5      & 2010-10-20T02:15 & HD211416   & 086.B-0919  & This work \\
                 &    9.8      & 2010-10-20T02:34 & HD211416   &             &           \\
                 &   11.4      & 2010-10-17T02:29 & HD206445   &             &           \\
                 &   12.4      & 2010-10-17T02:47 & HD206445   &             &           \\ \hline
        NGC~1365 &    8.5      & 2010-12-11T05:07 & HD75691    & 086.B-0919  & This work \\
                 &    9.8      & 2010-12-11T05:18 & HD75691    &             &           \\
                 &   11.4      & 2010-12-26T01:19 & HD16815    &             &           \\
                 &   12.4      & 2010-12-26T01:31 & HD16815    &             &           \\ \hline
 IRAS 05189-2524 &    8.5      & 2010-11-22T07:37 & HD29085    & 086.B-0919  & This work \\
                 &    9.8      & 2010-11-22T07:48 & HD29085    &             &           \\
                 &   11.4      & 2010-11-25T07:37 & HD39425    &             &           \\
                 &   12.4      & 2010-11-25T07:48 & HD39425    &             &           \\ \hline 
       Mrk~1239  &    8.5      & 2011-03-13T03:37 & HD81799    & 086.B-0919  & This work \\
                 &    9.8      & 2011-03-13T03:45 & HD81799    &             &           \\
                 &   11.4      & 2011-03-13T03:53 & HD81799    &             &           \\
                 &   12.4      & 2011-03-13T04:01 & HD81799    &             &           \\ 
                 &    8.5      & 2011-02-19T03:48 & HD76110    &             &           \\
                 &    9.8      & 2011-02-19T03:55 & HD76110    &             &           \\
                 &   11.4      & 2011-02-19T04:03 & HD76110    &             &           \\
                 &   12.4      & 2011-02-19T04:11 & HD76110    &             &           \\ \hline
        NGC~3281 &    8.5      & 2011-02-19T05:03 & HD76110    & 086.B-0919  & This work \\
                 &    9.8      & 2011-02-19T05:11 & HD76110    &             &           \\
                 &   11.4      & 2011-02-19T05:19 & HD76110    &             &           \\
                 &   12.4      & 2011-02-19T05:26 & HD76110    &             &           \\ \hline
        NGC~4593 &    8.5      & 2011-02-18T07:18 & HD110014   & 086.B-0919  & This work \\
                 &    9.8      & 2011-02-19T06:06 & HD110014   &             &           \\
                 &   11.4      & 2011-02-19T07:14 & HD110014   &             &           \\ \hline
\end{tabular}
\end{table*}

\subsection{MIDI observations}
\begin{table*}
\caption{\label{tab:obs:log:IZw1} Log of MIDI observations: I Zw 1. The columns are: {\em Time} of fringe track observation. {\em S}tacked with following observation (yes: 1, no: 0), number of frames {\em NDIT}, projected baseline length {\em BL}, projected baseline position angle {\em PA}, {\em g}rism setting ({\em P}RISM or {\em G}RISM), MIDI observing mode, airmass of fringe track, seeing during fringe track observation, goodness of observation ({\em O}: good:1, bad:0, N.A: -1), number of good frames {\em NGOOD}, ozone absorption depth of associated photometry observation {\em o3} (-1: photometry does not exist), number of A/B photometry observations, goodness of photometry observation ({\em O}), time of calibrator observation, difference in airmass between calibrator observation and science observation $\Delta am$. The date is given as date of night begin.}
\centering
\begin{tabular}{c c c c c c c c c c c c c c c c}
\hline\hline
\multicolumn{8}{c}{Fringe track} & \multicolumn{3}{c}{FT QC} & PH & PH QC & \multicolumn{2}{c}{Calibration} \\
time & s & NDIT & BL & PA & g & mode & am & see & O & NGOOD & o3 & A/B & O & caltime & $\Delta$ am \\
\hline
\hline
\multicolumn{15}{l}{2009-12-02: U2U4} \\
01:50:44 &  1 &  8000 &  88 &  79 & P & FT\_OF & 1.3 & 1.4 &  1 &  5596 &  0.52 & 1/1 &  1 & 02:43:22 &      0.1\\
02:02:25 &  1 &  8000 &  87 &  78 & P & FT\_OF & 1.3 & 1.3 &  1 &  6680 &  0.53 & 0/0 & -1 & 02:43:22 &      0.1\\
02:06:02 &  1 &  8000 &  86 &  78 & P & FT\_OF & 1.3 & 1.3 &  1 &  6263 &  0.63 & 0/0 & -1 & 02:43:22 &      0.1\\
02:10:22 &  0 &  8000 &  86 &  78 & P & FT\_OF & 1.4 & 1.2 &  1 &  5849 &  0.59 & 2/2 &  0 & 02:43:22 &      0.1\\
\hline
\multicolumn{15}{l}{2010-01-02: U3U4} \\
\hline
\end{tabular}
\end{table*}

\clearpage
\begin{table}
\caption{Log of MIDI observations: NGC 424}
\centering
\begin{tabular}{c c c c c c c c c c c c c c c c}
\hline\hline
\multicolumn{8}{c}{Fringe track} & \multicolumn{3}{c}{FT QC} & PH & PH QC & \multicolumn{2}{c}{Calibration} \\
time & s & NDIT & BL & PA & g & mode & am & see & O & NGOOD & o3 & A/B & O & caltime & $\Delta$ am \\
\hline
\hline
\multicolumn{15}{l}{2007-10-22: U3U4} \\
04:06:30 &  1 &  8000 &  62 & 113 & P & FT\_OF & 1.0 & 0.8 &  1 &  1762 &  0.59 & 0/0 & -1 & 23:44:01 &     -0.2\\
04:22:17 &  0 &  8000 &  62 & 115 & P & FT\_OF & 1.0 & 0.8 &  1 &  1287 &  0.41 & 0/0 & -1 & 23:44:01 &     -0.2\\
\hline
\multicolumn{15}{l}{2007-10-23: U3U4} \\
01:53:39 &  0 &  8000 &  56 &  94 & P & FT\_OF & 1.1 & 1.2 &  1 &  6154 &  0.49 & 0/0 & -1 & 02:43:50 &     -0.0\\
03:41:12 &  1 &  8000 &  62 & 109 & P & FT\_OF & 1.0 & 0.9 &  1 &  6850 &  0.57 & 0/0 & -1 & 02:43:50 &     -0.1\\
\hline
\end{tabular}
\end{table}

\clearpage

\begin{table}
\caption{Log of observations: NGC 424 (continued)}
\centering
\begin{tabular}{c c c c c c c c c c c c c c c c}
\hline\hline
\multicolumn{8}{c}{Fringe track} & \multicolumn{3}{c}{FT QC} & PH & PH QC & \multicolumn{2}{c}{Calibration} \\
time & s & NDIT & BL & PA & g & mode & am & see & Good? & NGOOD & o3 & A/B & O & caltime & $\Delta$ am \\
\hline
\hline
\multicolumn{15}{l}{2009-08-04: U1U4} \\
05:33:06 &  1 &  8000 & 121 &  24 & P & FT\_OF & 1.4 & 0.5 &  1 &  6855 &  0.51 & 1/1 &  0 & 05:11:14 &     -0.2\\
05:44:45 &  0 &  8000 & 122 &  26 & P & FT\_OF & 1.4 & 0.6 &  1 &  6712 &  0.54 & 1/1 &  0 & 05:11:14 &     -0.3\\
06:25:05 &  1 &  8000 & 125 &  35 & P & FT\_OF & 1.2 & 0.6 &  1 &  7204 &  0.52 & 1/1 &  1 & 06:57:51 &      0.1\\
06:36:31 &  1 &  8000 & 126 &  37 & P & FT\_OF & 1.2 & 0.7 &  1 &  7007 &  0.50 & 1/1 &  1 & 06:57:51 &      0.0\\
07:17:02 &  0 &  8000 & 128 &  45 & P & FT\_OF & 1.1 & 0.7 &  1 &  7500 &  0.56 & 1/1 &  1 & 06:57:51 &     -0.0\\
\hline
\end{tabular}
\end{table}

\clearpage
\begin{table}
\caption{Log of MIDI observations: NGC 1068}
\centering
\begin{tabular}{c c c c c c c c c c c c c c c c}
\hline\hline
\multicolumn{8}{c}{Fringe track} & \multicolumn{3}{c}{FT QC} & PH & PH QC & \multicolumn{2}{c}{Calibration} \\
time & s & NDIT & BL & PA & g & mode & am & see & O & NGOOD & o3 & A/B & O & caltime & $\Delta$ am \\
\hline
\hline
\multicolumn{15}{l}{2003-06-14: U1U3} \\
10:04:07 & -1 & 12000 &  78 & 179 & P & OBS\_F & 0.0 & 0.5 & -1 &    -1 & -1.00 & 1/1 &  0 & 09:37:19 &      0.0\\
10:22:54 & -1 &  9000 &  78 &   3 & P & OBS\_F & 0.0 & 0.6 & -1 &    -1 & -1.00 & 1/1 &  0 & 09:37:19 &      0.0\\
\hline
\multicolumn{15}{l}{2003-06-15: U1U3} \\
09:56:54 & -1 & 14000 &  78 & 179 & P & OBS\_F & 0.0 & 0.5 & -1 &    -1 & -1.00 & 1/1 &  0 & 09:34:22 &      0.0\\
10:10:12 & -1 & 15000 &  78 &   1 & P & OBS\_F & 0.0 & 0.5 & -1 &    -1 & -1.00 & 1/1 &  0 & 09:34:22 &      0.0\\
\hline
\end{tabular}
\end{table}

\clearpage

\begin{table}
\caption{Log of observations: NGC 1068 (continued)}
\centering
\begin{tabular}{c c c c c c c c c c c c c c c c}
\hline\hline
\multicolumn{8}{c}{Fringe track} & \multicolumn{3}{c}{FT QC} & PH & PH QC & \multicolumn{2}{c}{Calibration} \\
time & s & NDIT & BL & PA & g & mode & am & see & Good? & NGOOD & o3 & A/B & O & caltime & $\Delta$ am \\
\hline
\multicolumn{15}{l}{2007-10-07: E0G0} \\
06:04:14 &  1 &  8000 &  15 &  72 & P & FT\_OF & 1.1 & 2.0 &  1 &  6466 &  0.45 & 0/0 & -1 & 06:37:22 &     -0.1\\
06:08:04 &  0 &  8000 &  15 &  72 & P & FT\_OF & 1.1 & 2.0 &  1 &  6593 &  0.42 & 1/1 &  0 & 06:37:22 &     -0.1\\
07:06:35 &  1 &  8000 &  15 &  72 & P & FT\_OF & 1.1 & 1.7 &  1 &  6916 &  0.41 & 0/0 & -1 & 06:37:22 &     -0.1\\
07:10:28 &  0 &  8000 &  15 &  72 & P & FT\_OF & 1.1 & 1.9 &  1 &  6494 &  0.39 & 1/1 &  0 & 06:37:22 &     -0.1\\
08:09:51 &  1 &  8000 &  15 &  71 & P & FT\_OF & 1.2 & 1.9 &  1 &  7070 &  0.43 & 0/0 & -1 & 07:45:02 &     -0.2\\
\hline
\end{tabular}
\end{table}

\clearpage
\begin{table}
\caption{Log of MIDI observations: NGC 1365}
\centering
\begin{tabular}{c c c c c c c c c c c c c c c c}
\hline\hline
\multicolumn{8}{c}{Fringe track} & \multicolumn{3}{c}{FT QC} & PH & PH QC & \multicolumn{2}{c}{Calibration} \\
time & s & NDIT & BL & PA & g & mode & am & see & O & NGOOD & o3 & A/B & O & caltime & $\Delta$ am \\
\hline
\hline
\multicolumn{15}{l}{2006-09-10: U2U3} \\
07:42:55 & -1 &  8000 &  46 &  30 & P & FT\_OF & 1.1 & 1.3 &  0 &    -1 & -1.00 & 1/1 &  0 & 08:28:27 &     -0.0\\
09:36:20 & -1 &  8000 &  45 &  45 & P & FT\_OF & 1.0 & 1.3 &  0 &   103 &  1.24 & 0/0 & -1 & 08:28:27 &     -0.0\\
09:41:27 & -1 &  8000 &   0 & *** & P & FT\_OF & 1.0 & 1.4 &  0 &   563 &  1.17 & 0/0 & -1 & 08:28:27 &     -0.0\\
\hline
\multicolumn{15}{l}{2007-11-23: U3U4} \\
01:53:47 &  0 &  8000 &  54 &  92 & P & FT\_OF & 1.2 & 1.3 &  1 &  5607 &  0.56 & 1/1 &  0 & 02:38:26 &      0.1\\
\hline
\end{tabular}
\end{table}

\begin{table}
\caption{Log of MIDI observations: NGC 1365 (continued)}
\centering
\begin{tabular}{c c c c c c c c c c c c c c c c}
\hline\hline
\multicolumn{8}{c}{Fringe track} & \multicolumn{3}{c}{FT QC} & PH & PH QC & \multicolumn{2}{c}{Calibration} \\
time & s & NDIT & BL & PA & g & mode & am & see & O & NGOOD & o3 & A/B & O & caltime & $\Delta$ am \\
\hline
\hline
\multicolumn{15}{l}{2010-08-28: U1U4} \\
07:28:41 &  1 &  4000 & 125 &  38 & P & FT\_OF & 1.2 & 0.6 &  1 &  3161 &  0.40 & 0/0 & -1 & 07:05:16 &     -0.1\\
07:31:13 &  1 &  8000 & 125 &  39 & P & FT\_OF & 1.2 & 0.6 &  1 &  6312 &  0.51 & 1/1 &  0 & 07:05:16 &     -0.1\\
07:44:43 &  0 & 12000 & 126 &  41 & P & FT\_OF & 1.1 & 0.6 &  1 &  9137 &  0.44 & 1/1 &  1 & 07:05:16 &     -0.1\\
10:08:36 &  1 &  8000 & 128 &  64 & P & FT\_OF & 1.0 & 1.4 &  1 &  6472 &  0.47 & 0/0 & -1 & 10:24:25 &     -0.0\\
10:12:22 &  0 & 12000 & 128 &  64 & P & FT\_OF & 1.0 & 1.3 &  1 & 10401 &  0.51 & 0/0 & -1 & 10:24:25 &     -0.0\\
\hline
\multicolumn{15}{l}{2010-08-29: U2U3} \\
09:50:32 &  0 & 12000 &  45 &  41 & P & FT\_OF & 1.0 & 0.9 &  1 & 10788 &  0.54 & 1/1 &  0 & 09:34:04 &     -0.0\\
10:04:34 &  0 & 12000 &  45 &  43 & P & FT\_OF & 1.0 & 0.7 &  1 & 10640 &  0.54 & 1/1 &  1 & 10:21:56 &     -0.0\\
\hline
\end{tabular}
\end{table}

\clearpage
\begin{table}
\caption{Log of MIDI observations: IRAS 05189-2524}
\centering
\begin{tabular}{c c c c c c c c c c c c c c c c}
\hline\hline
\multicolumn{8}{c}{Fringe track} & \multicolumn{3}{c}{FT QC} & PH & PH QC & \multicolumn{2}{c}{Calibration} \\
time & s & NDIT & BL & PA & g & mode & am & see & O & NGOOD & o3 & A/B & O & caltime & $\Delta$ am \\
\hline
\hline
\multicolumn{15}{l}{2007-11-23: U3U4} \\
03:10:23 & -1 &  8000 &  50 &  95 & P & FT\_OF & 1.2 & 1.2 &  0 &  1250 &  0.77 & 0/0 & -1 & 03:33:48 &      0.2\\
\hline
\multicolumn{15}{l}{2010-01-01: U1U3} \\
02:04:52 &  0 &  8000 & 101 &  23 & P & FT\_OF & 1.0 & 0.7 &  1 &  3738 &  0.54 & 1/1 &  1 & 01:04:54 &     -0.1\\
02:18:30 &  0 &  8000 & 101 &  25 & P & FT\_OF & 1.0 & 0.8 &  1 &  3433 &  0.28 & 0/0 & -1 & 02:41:01 &     -0.0\\
05:21:24 &  1 &  4000 &  97 &  42 & P & FT\_OF & 1.1 & 0.8 &  1 &   753 &  0.60 & 1/1 &  0 & 05:49:16 &     -0.1\\
\hline
\end{tabular}
\end{table}

\clearpage
\begin{table}
\caption{Log of MIDI observations: H 0557-385}
\centering
\begin{tabular}{c c c c c c c c c c c c c c c c}
\hline\hline
\multicolumn{8}{c}{Fringe track} & \multicolumn{3}{c}{FT QC} & PH & PH QC & \multicolumn{2}{c}{Calibration} \\
time & s & NDIT & BL & PA & g & mode & am & see & O & NGOOD & o3 & A/B & O & caltime & $\Delta$ am \\
\hline
\hline
\multicolumn{15}{l}{2009-05-09: U1U2} \\
23:33:33 &  1 &  8000 &  42 &  46 & P & FT\_OF & 1.7 & 0.8 &  1 &  7424 &  0.55 & 1/1 &  1 & 23:07:47 &      0.2\\
23:45:17 &  0 &  8000 &  40 &  46 & P & FT\_OF & 1.8 & 0.7 &  1 &  7234 &  0.54 & 2/1 &  0 & 23:07:47 &      0.3\\
\hline
\multicolumn{15}{l}{2009-08-01: U3U4} \\
09:14:43 &  0 &  8000 &  32 &  60 & P & FT\_OF & 2.0 & 0.7 &  1 &  7379 &  0.39 & 1/1 &  1 & 08:29:02 &      0.9\\
\hline
\multicolumn{15}{l}{2009-08-02: U1U2} \\
09:02:41 &  0 &  8000 &  56 & 167 & P & FT\_OF & 2.1 & 0.7 &  1 &  6995 &  0.39 & 1/1 &  1 & 09:23:42 &      0.3\\
\hline
\multicolumn{15}{l}{2009-08-04: U1U4} \\
09:48:34 &  0 &  8000 & 119 &  17 & P & FT\_OF & 1.6 & 1.0 &  1 &  6862 &  0.46 & 1/1 &  1 & 09:29:46 &     -0.1\\
\hline
\end{tabular}
\end{table}

\clearpage
\begin{table}
\caption{Log of MIDI observations: IRAS 09149-6206}
\centering
\begin{tabular}{c c c c c c c c c c c c c c c c}
\hline\hline
\multicolumn{8}{c}{Fringe track} & \multicolumn{3}{c}{FT QC} & PH & PH QC & \multicolumn{2}{c}{Calibration} \\
time & s & NDIT & BL & PA & g & mode & am & see & O & NGOOD & o3 & A/B & O & caltime & $\Delta$ am \\
\hline
\hline
\multicolumn{15}{l}{2009-03-14: U1U2} \\
00:29:02 &  1 &  8000 &  49 &  12 & P & FT\_OF & 1.4 & 1.3 &  1 &  7009 &  0.53 & 1/1 &  0 & 23:58:48 &     -0.2\\
00:39:49 &  0 &  8000 &  49 &  14 & P & FT\_OF & 1.3 & 1.3 &  1 &  7024 &  0.44 & 1/1 &  1 & 23:58:48 &     -0.2\\
\hline
\multicolumn{15}{l}{2009-05-07: U3U4} \\
00:25:35 &  0 &  8000 &  62 & 126 & P & FT\_OF & 1.3 & 1.2 &  1 &  5261 &  0.54 & 2/1 &  0 & 23:55:01 &      0.2\\
01:45:20 &  1 &  8000 &  62 & 144 & P & FT\_OF & 1.5 & 1.4 &  1 &  6234 &  0.47 & 1/1 &  1 & 23:55:01 &      0.3\\
02:01:43 &  0 &  8000 &  62 & 147 & P & FT\_OF & 1.5 & 1.7 &  1 &  5238 &  0.61 & 2/1 &  1 & 23:55:01 &      0.4\\
\hline
\multicolumn{15}{l}{2009-05-08: U1U3} \\
00:18:43 &  1 &  8000 &  80 &  53 & P & FT\_OF & 1.3 & 0.8 &  1 &  4607 &  0.52 & 1/1 &  0 & 00:04:16 &      0.2\\
00:30:37 &  0 &  8000 &  78 &  55 & P & FT\_OF & 1.3 & 0.7 &  1 &  5297 &  0.49 & 1/1 &  1 & 00:04:16 &      0.2\\
\hline
\multicolumn{15}{l}{2009-05-10: U1U4} \\
23:38:10 &  1 &  8000 & 119 &  77 & P & FT\_OF & 1.3 & 0.9 &  1 &  7455 &  0.51 & 1/1 &  1 & 23:05:50 &      0.1\\
23:49:33 &  0 &  8000 & 117 &  79 & P & FT\_OF & 1.3 & 0.7 &  1 &  7476 &  0.55 & 1/1 &  1 & 23:05:50 &      0.1\\
\hline
\end{tabular}
\end{table}

\clearpage
\begin{table}
\caption{Log of MIDI observations: MCG-05-23-16}
\centering
\begin{tabular}{c c c c c c c c c c c c c c c c}
\hline\hline
\multicolumn{8}{c}{Fringe track} & \multicolumn{3}{c}{FT QC} & PH & PH QC & \multicolumn{2}{c}{Calibration} \\
time & s & NDIT & BL & PA & g & mode & am & see & O & NGOOD & o3 & A/B & O & caltime & $\Delta$ am \\
\hline
\hline
\multicolumn{15}{l}{2005-12-18: U2U3} \\
06:39:51 &  0 & 12000 &  46 &  24 & P & FT\_OF & 1.1 & 0.7 &  1 &  9868 &  0.46 & 2/2 &  0 & 05:07:40 &     -0.6\\
\hline
\multicolumn{15}{l}{2008-04-18: U1U4} \\
00:51:27 &  0 &  8000 & 129 &  62 & P & FT\_OF & 1.0 & 1.5 &  1 &  3235 &  0.68 & 2/2 &  0 & 01:51:16 &     -0.3\\
\hline
\multicolumn{15}{l}{2008-05-17: U1U3} \\
02:33:58 & -1 &  4000 &  80 &  48 & P & FT\_OF & 1.6 & 0.8 &  0 &   202 &  1.20 & 1/1 &  0 & 03:00:50 &      0.2\\
\hline
\end{tabular}
\end{table}

\begin{table}
\caption{Log of MIDI observations: MCG-05-23-16 (continued)}
\centering
\begin{tabular}{c c c c c c c c c c c c c c c c}
\hline\hline
\multicolumn{8}{c}{Fringe track} & \multicolumn{3}{c}{FT QC} & PH & PH QC & \multicolumn{2}{c}{Calibration} \\
time & s & NDIT & BL & PA & g & mode & am & see & O & NGOOD & o3 & A/B & O & caltime & $\Delta$ am \\
\hline
\hline
\multicolumn{15}{l}{2010-05-27: U1U2} \\
00:44:04 &  1 &  8000 &  51 &  39 & P & FT\_OF & 1.2 & 0.8 &  1 &  5709 &  0.57 & 1/1 &  0 & 01:23:55 &     -0.0\\
00:55:21 &  1 &  8000 &  50 &  39 & P & FT\_OF & 1.3 & 0.8 &  1 &  3687 &  0.58 & 0/0 & -1 & 01:23:55 &      0.0\\
01:01:08 &  0 &  8000 &  50 &  40 & P & FT\_OF & 1.3 & 0.8 &  1 &  3578 &  0.61 & 1/1 &  1 & 01:23:55 &      0.1\\
\hline
\multicolumn{15}{l}{2010-05-31: U3U4} \\
23:21:33 &  0 &  8000 &  60 & 123 & P & FT\_OF & 1.1 & 0.9 &  1 &  5245 &  0.74 & 1/1 &  0 & 23:00:38 &     -0.0\\
23:33:05 & -1 &  8000 &  59 & 125 & P & FT\_OF & 1.1 & 1.0 &  0 &   297 &  0.62 & 0/0 & -1 & 23:00:38 &      0.0\\
01:14:34 &  0 &  8000 &  52 & 145 & P & FT\_OF & 1.4 & 0.9 &  1 &  5190 &  0.48 & 1/1 &  0 & 00:42:03 &      0.1\\
01:26:05 & -1 & 16000 &  51 & 148 & P & FT\_OF & 1.5 & 1.0 &  0 &   492 &  0.83 & 0/0 & -1 & 01:49:02 &      0.1\\
01:34:36 &  0 &  8000 &  51 & 150 & P & FT\_OF & 1.5 & 1.0 &  1 &  4187 &  0.69 & 0/0 & -1 & 01:49:02 &      0.1\\
02:15:43 & -1 &  8000 &  49 & 161 & P & FT\_OF & 1.9 & 1.1 &  0 &    -1 & -1.00 & 0/0 & -1 & 01:49:02 &      0.5\\
\hline
\end{tabular}
\end{table}

\clearpage
\begin{table}
\caption{\label{log:midi:mrk1239}Log of MIDI observations: Mrk 1239}
\centering
\begin{tabular}{c c c c c c c c c c c c c c c c}
\hline\hline
\multicolumn{8}{c}{Fringe track} & \multicolumn{3}{c}{FT QC} & PH & PH QC & \multicolumn{2}{c}{Calibration} \\
time & s & NDIT & BL & PA & g & mode & am & see & O & NGOOD & o3 & A/B & O & caltime & $\Delta$ am \\
\hline
\hline
\multicolumn{15}{l}{2005-12-18: U2U3} \\
07:31:59 &  0 & 12000 &  40 &  36 & P & FT\_OF & 1.1 & 0.8 &  1 & 11233 &  0.54 & 2/2 &  1 & 08:15:17 &      0.0\\
08:47:23 &  0 & 12000 &  44 &  42 & P & FT\_OF & 1.1 & 0.9 &  1 & 11196 &  0.52 & 2/2 &  0 & 08:15:17 &     -0.0\\
\hline
\multicolumn{15}{l}{2008-04-19: U1U4} \\
00:35:32 &  0 & 12000 & 127 &  62 & P & FT\_OF & 1.1 & 1.6 &  1 & 10419 &  0.55 & 1/1 &  1 & 01:17:48 &      0.1\\
\hline
\multicolumn{15}{l}{2008-04-20: U1U3} \\
\hline
\end{tabular}
\end{table}

\clearpage
\begin{table}
\caption{Log of MIDI observations: NGC 3281}
\centering
\begin{tabular}{c c c c c c c c c c c c c c c c}
\hline\hline
\multicolumn{8}{c}{Fringe track} & \multicolumn{3}{c}{FT QC} & PH & PH QC & \multicolumn{2}{c}{Calibration} \\
time & s & NDIT & BL & PA & g & mode & am & see & O & NGOOD & o3 & A/B & O & caltime & $\Delta$ am \\
\hline
\hline
\multicolumn{15}{l}{2010-01-01: U1U3} \\
07:25:42 &  0 &  4000 & 102 &  25 & P & FT\_OF & 1.0 & 0.6 &  1 &   955 &  0.65 & 0/0 & -1 & 08:17:07 &     -0.2\\
\hline
\multicolumn{15}{l}{2010-01-02: U3U4} \\
07:50:16 &  1 &  4000 &  61 & 106 & P & FT\_OF & 1.0 & 0.8 &  1 &  2834 &  0.40 & 1/1 &  0 & 07:21:17 &     -0.0\\
08:01:22 &  0 &  4000 &  61 & 107 & P & FT\_OF & 1.0 & 0.7 &  1 &  3234 &  0.47 & 0/0 & -1 & 07:21:17 &     -0.0\\
\hline
\multicolumn{15}{l}{2010-01-30: U2U4} \\
\hline
\end{tabular}
\end{table}

\clearpage
\begin{table}
\caption{Log of MIDI observations: NGC 3783}
\centering
\begin{tabular}{c c c c c c c c c c c c c c c c}
\hline\hline
\multicolumn{8}{c}{Fringe track} & \multicolumn{3}{c}{FT QC} & PH & PH QC & \multicolumn{2}{c}{Calibration} \\
time & s & NDIT & BL & PA & g & mode & am & see & O & NGOOD & o3 & A/B & O & caltime & $\Delta$ am \\
\hline
\hline
\multicolumn{15}{l}{2005-05-27: U1U2} \\
03:39:57 &  0 &  8000 &  44 &  45 & P & OBS\_F & 1.5 & 0.6 &  1 &  6953 &  0.53 & 1/1 &  1 & 02:53:01 &      0.2\\
04:00:21 &  0 &  8000 &  42 &  45 & P & OBS\_F & 1.6 & 0.6 &  1 &  7292 &  0.48 & 1/1 &  1 & 04:28:57 &     -0.4\\
\hline
\multicolumn{15}{l}{2005-05-30: U2U4} \\
00:22:53 & -1 &  8000 &  88 &  87 & P & OBS\_F & 1.0 & 0.8 &  0 &  1134 &  1.20 & 1/1 &  1 & 23:02:55 &      0.0\\
03:01:16 & -1 & 12000 &  68 & 114 & P & OBS\_F & 1.4 & 0.7 &  0 &    -1 & -1.00 & 1/1 &  1 & 02:19:20 &      0.1\\
\hline
\end{tabular}
\end{table}

\clearpage
\begin{table}
\caption{Log of MIDI observations: NGC 4151}
\centering
\begin{tabular}{c c c c c c c c c c c c c c c c}
\hline\hline
\multicolumn{8}{c}{Fringe track} & \multicolumn{3}{c}{FT QC} & PH & PH QC & \multicolumn{2}{c}{Calibration} \\
time & s & NDIT & BL & PA & g & mode & am & see & O & NGOOD & o3 & A/B & O & caltime & $\Delta$ am \\
\hline
\hline
\multicolumn{15}{l}{2007-02-06: U2U3} \\
07:49:59 & -1 &  8000 &  34 &  62 & P & FT\_OF & 2.3 & 0.7 &  0 &   113 &  0.96 & 0/0 & -1 & 06:31:50 &      1.3\\
08:03:42 &  0 &  8000 &  35 &  60 & P & FT\_OF & 2.3 & 0.5 &  1 &  1657 &  0.70 & 0/0 & -1 & 06:31:50 &      1.3\\
\hline
\multicolumn{15}{l}{2008-04-21: U3U4} \\
02:15:00 &  1 &  8000 &  61 & 104 & P & FT\_OF & 2.3 & 0.7 &  1 &  5755 &  0.42 & 1/1 &  1 & 03:02:16 &     -0.0\\
02:29:51 &  0 & 12000 &  60 & 102 & P & FT\_OF & 2.3 & 0.8 &  1 &  8920 &  0.51 & 1/1 &  1 & 03:02:16 &     -0.0\\
\hline
\end{tabular}
\end{table}

\clearpage
\begin{table}
\caption{Log of MIDI observations: 3C 273}
\centering
\begin{tabular}{c c c c c c c c c c c c c c c c}
\hline\hline
\multicolumn{8}{c}{Fringe track} & \multicolumn{3}{c}{FT QC} & PH & PH QC & \multicolumn{2}{c}{Calibration} \\
time & s & NDIT & BL & PA & g & mode & am & see & O & NGOOD & o3 & A/B & O & caltime & $\Delta$ am \\
\hline
\hline
\multicolumn{15}{l}{2007-02-06: U2U3} \\
05:41:06 & -1 &  8000 &  35 &  26 & P & FT\_OF & 1.4 & 0.7 &  0 &   315 &  1.08 & 0/0 & -1 & 06:31:50 &      0.4\\
05:50:44 & -1 &  8000 &  35 &  28 & P & FT\_OF & 1.3 & 0.8 &  0 &    -1 & -1.00 & 0/0 & -1 & 06:31:50 &      0.3\\
06:04:10 &  1 &  8000 &  36 &  30 & P & FT\_OF & 1.3 & 0.7 &  1 &  6769 &  0.50 & 0/0 & -1 & 06:31:50 &      0.3\\
06:08:32 &  0 &  8000 &  36 &  31 & P & FT\_OF & 1.3 & 0.7 &  1 &  6855 &  0.57 & 0/0 & -1 & 06:31:50 &      0.3\\
\hline
\multicolumn{15}{l}{2008-04-21: U3U4} \\
06:52:20 &  1 &  8000 &  30 & 126 & P & FT\_OF & 2.0 & 0.7 &  1 &  5546 &  0.47 & 1/1 &  0 & 06:10:08 &      0.3\\
07:03:47 &  0 &  8000 &  28 & 129 & P & FT\_OF & 2.2 & 0.6 &  1 &  4966 &  0.54 & 1/1 &  0 & 06:10:08 &      0.5\\
\hline
\multicolumn{15}{l}{2009-05-04: U3U4} \\
05:50:15 & -1 &  8000 &  33 & 123 & P & FT\_OF & 1.9 & 0.6 &  0 &    -1 & -1.00 & 1/1 &  1 & 06:23:15 &     -0.7\\
\hline
\multicolumn{15}{l}{2010-01-30: U2U4} \\
09:16:32 &  1 &  8000 &  88 &  81 & P & FT\_OF & 1.1 & 0.9 &  1 &  7040 &  0.59 & 1/1 &  0 & 08:56:39 &      0.1\\
09:28:12 &  0 &  8000 &  87 &  81 & P & FT\_OF & 1.2 & 0.9 &  1 &  7136 &  0.60 & 1/1 &  0 & 08:56:39 &      0.1\\
\hline
\end{tabular}
\end{table}

\clearpage
\begin{table}
\caption{Log of MIDI observations: NGC 4507}
\centering
\begin{tabular}{c c c c c c c c c c c c c c c c}
\hline\hline
\multicolumn{8}{c}{Fringe track} & \multicolumn{3}{c}{FT QC} & PH & PH QC & \multicolumn{2}{c}{Calibration} \\
time & s & NDIT & BL & PA & g & mode & am & see & O & NGOOD & o3 & A/B & O & caltime & $\Delta$ am \\
\hline
\hline
\multicolumn{15}{l}{2010-01-01: U1U3} \\
07:59:57 & -1 &  4000 & 102 &  12 & P & FT\_OF & 1.2 & 0.6 &  0 &    35 &  0.68 & 0/0 & -1 & 08:17:07 &     -0.0\\
08:31:34 &  1 &  4000 & 102 &  17 & P & FT\_OF & 1.1 & 0.6 &  1 &  2145 &  0.56 & 0/0 & -1 & 08:17:07 &     -0.1\\
08:34:09 &  0 &  4000 & 102 &  17 & P & FT\_OF & 1.1 & 0.6 &  1 &  1582 &  0.50 & 0/0 & -1 & 08:17:07 &     -0.1\\
\hline
\multicolumn{15}{l}{2010-01-02: U3U4} \\
08:17:24 &  1 &  4000 &  54 &  90 & P & FT\_OF & 1.2 & 0.7 &  1 &  1227 &  0.50 & 1/1 &  0 & 08:41:24 &     -0.0\\
\hline
\end{tabular}
\end{table}

\clearpage
\begin{table}
\caption{Log of MIDI observations: NGC 4593}
\centering
\begin{tabular}{c c c c c c c c c c c c c c c c}
\hline\hline
\multicolumn{8}{c}{Fringe track} & \multicolumn{3}{c}{FT QC} & PH & PH QC & \multicolumn{2}{c}{Calibration} \\
time & s & NDIT & BL & PA & g & mode & am & see & O & NGOOD & o3 & A/B & O & caltime & $\Delta$ am \\
\hline
\hline
\multicolumn{15}{l}{2010-01-31: U3U4} \\
07:01:27 &  1 &  8000 &  60 & 108 & P & FT\_OF & 1.2 & 0.8 &  1 &  5935 &  0.53 & 2/2 &  0 & 08:12:57 &      0.1\\
07:13:22 & -1 &  8000 &  60 & 108 & P & FT\_OF & 1.1 & 0.8 &  0 &    22 & -1.00 & 0/0 & -1 & 08:12:57 &      0.1\\
07:35:57 &  1 &  8000 &  61 & 108 & P & FT\_OF & 1.1 & 0.9 &  1 &  6766 &  0.69 & 1/1 &  0 & 08:12:57 &      0.0\\
07:47:54 &  0 &  8000 &  62 & 108 & P & FT\_OF & 1.1 & 0.8 &  1 &  6826 &  0.64 & 1/1 &  0 & 08:12:57 &      0.0\\
\hline
\multicolumn{15}{l}{2010-03-26: U1U2} \\
\hline
\end{tabular}
\end{table}

\clearpage
\begin{table}
\caption{Log of MIDI observations: ESO 323-77}
\centering
\begin{tabular}{c c c c c c c c c c c c c c c c}
\hline\hline
\multicolumn{8}{c}{Fringe track} & \multicolumn{3}{c}{FT QC} & PH & PH QC & \multicolumn{2}{c}{Calibration} \\
time & s & NDIT & BL & PA & g & mode & am & see & O & NGOOD & o3 & A/B & O & caltime & $\Delta$ am \\
\hline
\hline
\multicolumn{15}{l}{2009-05-07: U3U4} \\
01:01:10 &  1 &  8000 &  56 &  94 & P & FT\_OF & 1.1 & 1.1 &  1 &  4255 &  0.58 & 0/0 & -1 & 01:27:53 &      0.0\\
01:06:55 &  0 &  8000 &  57 &  95 & P & FT\_OF & 1.1 & 1.0 &  1 &  5408 &  0.51 & 1/1 &  0 & 01:27:53 &      0.0\\
\hline
\multicolumn{15}{l}{2009-05-08: U1U3} \\
04:35:22 & -1 &  8000 &  90 &  46 & P & FT\_OF & 1.1 & 0.5 &  0 &   126 &  1.09 & 0/0 & -1 & 04:15:32 &      0.0\\
\hline
\multicolumn{15}{l}{2009-05-09: U1U2} \\
03:06:43 &  1 &  8000 &  54 &  30 & P & FT\_OF & 1.0 & 1.5 &  1 &  7037 &  0.62 & 1/1 &  0 & 02:49:26 &     -0.0\\
03:18:35 &  0 &  8000 &  53 &  31 & P & FT\_OF & 1.1 & 1.4 &  1 &  6922 &  0.57 & 1/1 &  0 & 02:49:26 &     -0.0\\
04:09:58 &  1 &  8000 &  51 &  37 & P & FT\_OF & 1.1 & 1.1 &  1 &  6947 &  0.56 & 2/1 &  1 & 03:55:51 &     -0.0\\
04:21:28 &  0 &  8000 &  51 &  38 & P & FT\_OF & 1.1 & 1.1 &  1 &  7365 &  0.58 & 0/0 & -1 & 03:55:51 &      0.0\\
\hline
\multicolumn{15}{l}{2009-05-10: U1U4} \\
04:07:36 &  0 &  8000 & 118 &  75 & P & FT\_OF & 1.1 & 1.1 &  1 &  7449 &  0.56 & 1/1 &  1 & 03:52:22 &     -0.0\\
\hline
\multicolumn{15}{l}{2009-05-11: U1U4} \\
00:05:13 &  1 &  8000 & 126 &  36 & P & FT\_OF & 1.2 & 1.4 &  1 &  6139 &  0.54 & 1/1 &  1 & 23:49:25 &     -0.0\\
00:16:41 &  0 &  8000 & 127 &  39 & P & FT\_OF & 1.2 & 1.3 &  1 &  6645 &  0.54 & 1/1 &  0 & 23:49:25 &     -0.1\\
\hline
\end{tabular}
\end{table}

\clearpage
\begin{table}
\caption{Log of MIDI observations: Centaurus A}
\centering
\begin{tabular}{c c c c c c c c c c c c c c c c}
\hline\hline
\multicolumn{8}{c}{Fringe track} & \multicolumn{3}{c}{FT QC} & PH & PH QC & \multicolumn{2}{c}{Calibration} \\
time & s & NDIT & BL & PA & g & mode & am & see & O & NGOOD & o3 & A/B & O & caltime & $\Delta$ am \\
\hline
\hline
\multicolumn{15}{l}{2008-04-16: U1U3} \\
04:05:48 &  0 & 12000 &  99 &  31 & P & FT\_OF & 1.1 & 2.7 &  1 &  2209 &  0.29 & 2/2 &  0 & 03:39:41 &      0.0\\
\hline
\multicolumn{15}{l}{2008-04-18: U1U3} \\
02:54:31 &  0 &  8000 & 101 &  22 & P & FT\_OF & 1.1 & 1.0 &  1 &  7374 &  0.41 & 1/1 &  0 & 03:14:48 &      0.0\\
04:51:38 &  1 &  8000 &  96 &  38 & P & FT\_OF & 1.1 & 1.1 &  1 &  7447 &  0.39 & 1/1 &  1 & 04:25:30 &     -0.0\\
05:03:06 &  0 &  8000 &  95 &  39 & P & FT\_OF & 1.1 & 1.0 &  1 &  7420 &  0.41 & 1/1 &  1 & 04:25:30 &      0.0\\
\hline
\end{tabular}
\end{table}

\clearpage
\begin{table}
\caption{Log of MIDI observations: IRAS 13349+2438}
\centering
\begin{tabular}{c c c c c c c c c c c c c c c c}
\hline\hline
\multicolumn{8}{c}{Fringe track} & \multicolumn{3}{c}{FT QC} & PH & PH QC & \multicolumn{2}{c}{Calibration} \\
time & s & NDIT & BL & PA & g & mode & am & see & O & NGOOD & o3 & A/B & O & caltime & $\Delta$ am \\
\hline
\hline
\multicolumn{15}{l}{2009-05-09: U1U2} \\
01:51:33 &  0 &  8000 &  34 &  28 & P & FT\_OF & 1.6 & 1.4 &  1 &  5820 &  0.53 & 0/0 & -1 & 01:31:59 &     -0.1\\
02:30:36 &  0 &  8000 &  38 &  33 & P & FT\_OF & 1.5 & 1.4 &  1 &  5392 &  0.45 & 3/1 &  0 & 02:12:09 &     -0.1\\
04:40:22 &  0 &  8000 &  49 &  39 & P & FT\_OF & 1.7 & 1.1 &  1 &  6986 &  0.47 & 1/1 &  1 & 02:12:09 &      0.1\\
\hline
\multicolumn{15}{l}{2009-05-10: U1U4} \\
02:40:05 &  1 &  8000 & 116 &  71 & P & FT\_OF & 1.5 & 0.8 &  1 &  7464 &  0.48 & 1/1 &  1 & 02:24:31 &     -0.1\\
02:51:43 &  0 &  8000 & 118 &  70 & P & FT\_OF & 1.5 & 0.8 &  1 &  7515 &  0.49 & 1/1 &  1 & 02:24:31 &     -0.1\\
\hline
\multicolumn{15}{l}{2009-05-11: U1U4} \\
01:06:50 &  1 &  8000 &  89 &  76 & P & FT\_OF & 1.8 & 1.3 &  1 &  6746 &  0.46 & 1/1 &  0 & 00:50:15 &     -0.1\\
01:18:14 &  0 &  8000 &  93 &  75 & P & FT\_OF & 1.7 & 1.2 &  1 &  6857 &  0.47 & 1/1 &  1 & 00:50:15 &     -0.1\\
04:11:10 &  0 &  8000 & 129 &  63 & P & FT\_OF & 1.6 & 1.1 &  1 &  7226 &  0.44 & 1/1 &  1 & 04:28:23 &      0.0\\
\hline
\end{tabular}
\end{table}

\clearpage
\begin{table}
\caption{Log of MIDI observations: IC 4329 A}
\centering
\begin{tabular}{c c c c c c c c c c c c c c c c}
\hline\hline
\multicolumn{8}{c}{Fringe track} & \multicolumn{3}{c}{FT QC} & PH & PH QC & \multicolumn{2}{c}{Calibration} \\
time & s & NDIT & BL & PA & g & mode & am & see & O & NGOOD & o3 & A/B & O & caltime & $\Delta$ am \\
\hline
\hline
\multicolumn{15}{l}{2007-02-06: U2U3} \\
08:43:55 &  1 &  8000 &  46 &  35 & P & FT\_OF & 1.0 & 0.7 &  1 &  7362 &  0.61 & 0/0 & -1 & 09:08:18 &      0.0\\
08:49:02 &  0 &  8000 &  46 &  36 & P & FT\_OF & 1.0 & 0.7 &  1 &  7403 &  0.57 & 1/0 &  0 & 09:08:18 &      0.0\\
\hline
\multicolumn{15}{l}{2008-04-19: U1U4} \\
02:47:35 &  1 &  8000 & 125 &  44 & P & FT\_OF & 1.1 & 0.9 &  1 &  7032 &  0.55 & 1/1 &  1 & 03:25:05 &      0.0\\
02:58:55 &  0 &  8000 & 126 &  46 & P & FT\_OF & 1.1 & 1.1 &  1 &  6724 &  0.59 & 1/1 &  1 & 03:25:05 &      0.0\\
\hline
\end{tabular}
\end{table}

\clearpage
\begin{table}
\caption{Log of MIDI observations: Circinus}
\centering
\begin{tabular}{c c c c c c c c c c c c c c c c}
\hline\hline
\multicolumn{8}{c}{Fringe track} & \multicolumn{3}{c}{FT QC} & PH & PH QC & \multicolumn{2}{c}{Calibration} \\
time & s & NDIT & BL & PA & g & mode & am & see & O & NGOOD & o3 & A/B & O & caltime & $\Delta$ am \\
\hline
\hline
\multicolumn{15}{l}{2004-02-11: U2U3} \\
06:55:11 &  1 & 12000 &  43 &  19 & P & OBS\_F & 1.5 & 1.1 &  1 &  9054 &  0.39 & 1/1 &  0 & 07:36:34 &      0.0\\
07:06:46 &  0 & 12000 &  43 &  21 & P & OBS\_F & 1.5 & 0.9 &  1 & 10847 &  0.34 & 1/1 &  1 & 07:36:34 &     -0.0\\
\hline
\multicolumn{15}{l}{2004-06-02: U2U3} \\
05:50:45 &  0 &  8000 &  29 &  92 & P & OBS\_F & 1.7 & 0.7 &  1 &  7345 &  0.38 & 1/1 &  1 & 06:42:24 &     -0.4\\
07:58:04 &  0 &  8000 &  20 & 129 & P & OBS\_F & 2.5 & 0.7 &  1 &  7319 &  0.26 & 1/1 &  0 & 06:42:24 &      0.4\\
\hline
\end{tabular}
\end{table}

\clearpage

\begin{table}
\caption{Log of observations: Circinus (continued)}
\centering
\begin{tabular}{c c c c c c c c c c c c c c c c}
\hline\hline
\multicolumn{8}{c}{Fringe track} & \multicolumn{3}{c}{FT QC} & PH & PH QC & \multicolumn{2}{c}{Calibration} \\
time & s & NDIT & BL & PA & g & mode & am & see & Good? & NGOOD & o3 & A/B & O & caltime & $\Delta$ am \\
\hline
\hline
\multicolumn{15}{l}{2008-04-25: E0G0} \\
00:11:26 &  1 &  8000 &  15 &  14 & P & FT\_OF & 1.9 & 0.8 &  1 &  5628 &  0.37 & 1/1 &  0 & 23:30:59 &     -0.4\\
00:23:42 &  1 &  8000 &  15 &  17 & P & FT\_OF & 1.8 & 0.8 &  1 &  6170 &  0.36 & 1/1 &  0 & 23:30:59 &     -0.4\\
00:35:58 &  0 & 12000 &  15 &  20 & P & FT\_OF & 1.8 & 0.8 &  1 &  9192 &  0.35 & 1/1 &  1 & 23:30:59 &     -0.5\\
00:55:33 &  1 & 12000 &  15 &  25 & P & FT\_OF & 1.7 & 0.8 &  1 &  6832 &  0.42 & 1/1 &  1 & 02:15:06 &      0.3\\
01:09:14 &  1 & 12000 &  15 &  28 & P & FT\_OF & 1.6 & 0.7 &  1 & 10599 &  0.28 & 1/1 &  0 & 02:15:06 &      0.3\\
\hline
\hline
\end{tabular}
\end{table}

\clearpage

\begin{table}
\caption{Log of observations: Circinus (continued)}
\centering
\begin{tabular}{c c c c c c c c c c c c c c c c}
\hline\hline
\multicolumn{8}{c}{Fringe track} & \multicolumn{3}{c}{FT QC} & PH & PH QC & \multicolumn{2}{c}{Calibration} \\
time & s & NDIT & BL & PA & g & mode & am & see & Good? & NGOOD & o3 & A/B & O & caltime & $\Delta$ am \\
\hline
\hline
\multicolumn{15}{l}{2009-04-26: E0G0} \\
00:15:00 &  1 & 12000 &  15 &  16 & P & FT\_OF & 1.9 & 1.3 &  1 & 10030 &  0.36 & 1/1 &  0 & 01:03:58 &      0.3\\
00:28:53 &  1 & 12000 &  15 &  19 & P & FT\_OF & 1.8 & 1.2 &  1 & 10279 &  0.42 & 1/1 &  1 & 01:03:58 &      0.2\\
00:42:38 &  0 & 12000 &  15 &  22 & P & FT\_OF & 1.7 & 0.8 &  1 & 11145 &  0.41 & 1/1 &  1 & 01:03:58 &      0.2\\
01:35:15 &  1 & 12000 &  15 &  35 & P & FT\_OF & 1.5 & 1.2 &  1 &  9884 &  0.39 & 1/1 &  1 & 01:03:58 &     -0.0\\
01:48:49 &  1 & 12000 &  15 &  38 & P & FT\_OF & 1.5 & 1.0 &  1 & 11173 &  0.42 & 1/1 &  1 & 01:03:58 &     -0.1\\
\hline
\end{tabular}
\end{table}

\clearpage
\begin{table}
\caption{Log of MIDI observations: NGC 5506}
\centering
\begin{tabular}{c c c c c c c c c c c c c c c c}
\hline\hline
\multicolumn{8}{c}{Fringe track} & \multicolumn{3}{c}{FT QC} & PH & PH QC & \multicolumn{2}{c}{Calibration} \\
time & s & NDIT & BL & PA & g & mode & am & see & O & NGOOD & o3 & A/B & O & caltime & $\Delta$ am \\
\hline
\hline
\multicolumn{15}{l}{2010-03-26: U1U2} \\
09:20:14 &  0 &  8000 &  56 &  35 & P & FT\_OF & 1.4 & 0.8 &  1 &  7473 &  0.43 & 1/1 &  1 & 08:57:57 &      0.2\\
09:31:35 &  0 &  8000 &  56 &  35 & P & FT\_OF & 1.5 & 0.9 &  1 &  7306 &  0.41 & 1/1 &  1 & 09:55:03 &      0.0\\
\hline
\multicolumn{15}{l}{2010-03-27: U1U4} \\
07:36:50 &  1 &  8000 & 129 &  63 & P & FT\_OF & 1.1 & 0.8 &  1 &  7470 &  0.40 & 1/1 &  0 & 08:10:22 &     -0.0\\
07:48:08 &  0 &  8000 & 129 &  63 & P & FT\_OF & 1.1 & 0.6 &  1 &  7488 &  0.44 & 1/1 &  0 & 08:10:22 &      0.0\\
\hline
\end{tabular}
\end{table}

\clearpage
\begin{table}
\caption{Log of MIDI observations: NGC 5995}
\centering
\begin{tabular}{c c c c c c c c c c c c c c c c}
\hline\hline
\multicolumn{8}{c}{Fringe track} & \multicolumn{3}{c}{FT QC} & PH & PH QC & \multicolumn{2}{c}{Calibration} \\
time & s & NDIT & BL & PA & g & mode & am & see & O & NGOOD & o3 & A/B & O & caltime & $\Delta$ am \\
\hline
\hline
\multicolumn{15}{l}{2010-03-26: U1U2} \\
08:20:08 &  1 &  8000 &  55 &  27 & P & FT\_OF & 1.0 & 0.6 &  1 &  7438 &  0.65 & 1/1 &  0 & 08:57:57 &     -0.2\\
08:31:29 &  0 &  8000 &  56 &  28 & P & FT\_OF & 1.0 & 0.8 &  1 &  7483 &  0.53 & 1/1 &  0 & 08:57:57 &     -0.2\\
\hline
\multicolumn{15}{l}{2010-03-27: U1U4} \\
05:00:11 & -1 &  8000 &  99 &  35 & P & FT\_OF & 1.4 & 0.7 &  0 &  6219 &  0.79 & 1/1 &  1 & 05:35:11 &      0.4\\
05:12:07 &  0 &  8000 & 101 &  38 & P & FT\_OF & 1.4 & 0.7 &  1 &  6053 &  0.57 & 1/1 &  0 & 05:35:11 &      0.3\\
\hline
\multicolumn{15}{l}{2011-04-17: U2U4} \\
09:29:36 &  1 &  8000 &  70 &  90 & P & FT\_OF & 1.3 & 1.0 &  1 &  5702 &  0.52 & 0/0 & -1 & 09:49:07 &     -0.0\\
09:33:32 &  1 &  8000 &  69 &  90 & P & FT\_OF & 1.3 & 1.0 &  1 &  6212 &  0.47 & 0/0 & -1 & 09:49:07 &     -0.0\\
09:37:03 &  1 &  8000 &  68 &  91 & P & FT\_OF & 1.3 & 0.9 &  1 &  6142 &  0.58 & 0/0 & -1 & 09:49:07 &      0.0\\
09:40:37 &  1 &  8000 &  67 &  91 & P & FT\_OF & 1.4 & 0.9 &  1 &  6463 &  0.55 & 0/0 & -1 & 09:49:07 &      0.0\\
09:57:34 &  1 &  8000 &  63 &  92 & P & FT\_OF & 1.4 & 1.0 &  1 &  5860 &  0.43 & 0/0 & -1 & 09:49:07 &      0.1\\
10:01:07 &  0 &  8000 &  62 &  92 & P & FT\_OF & 1.5 & 1.0 &  1 &  6056 &  0.57 & 0/0 & -1 & 09:49:07 &      0.1\\
10:04:40 &  1 &  8000 &  61 &  93 & P & FT\_OF & 1.5 & 1.3 &  1 &  6425 &  0.49 & 0/0 & -1 & 10:15:59 &      0.0\\
10:08:05 &  0 &  8000 &  60 &  93 & P & FT\_OF & 1.5 & 1.4 &  1 &  3524 &  0.56 & 0/0 & -1 & 10:15:59 &      0.0\\
\hline
\end{tabular}
\end{table}

\clearpage
\begin{table}
\caption{Log of MIDI observations: NGC 7469}
\centering
\begin{tabular}{c c c c c c c c c c c c c c c c}
\hline\hline
\multicolumn{8}{c}{Fringe track} & \multicolumn{3}{c}{FT QC} & PH & PH QC & \multicolumn{2}{c}{Calibration} \\
time & s & NDIT & BL & PA & g & mode & am & see & O & NGOOD & o3 & A/B & O & caltime & $\Delta$ am \\
\hline
\hline
\multicolumn{15}{l}{2006-09-10: U2U3} \\
06:09:20 & -1 &  8000 &  46 &  46 & P & FT\_OF & 1.3 & 1.1 &  0 &    -1 & -1.00 & 1/1 &  0 & 05:00:21 &      0.2\\
\hline
\multicolumn{15}{l}{2007-11-23: U3U4} \\
01:07:41 & -1 &  8000 &  52 & 106 & P & FT\_OF & 1.3 & 1.1 &  0 &   460 &  1.07 & 0/0 & -1 & 00:00:39 &      0.2\\
01:12:56 & -1 &  8000 &  51 & 106 & P & FT\_OF & 1.3 & 1.1 &  0 &  1093 &  0.91 & 0/0 & -1 & 00:00:39 &      0.2\\
\hline
\multicolumn{15}{l}{2010-04-28: U2U4} \\
\hline
\end{tabular}
\end{table}

\clearpage

\begin{table}
\caption{Log of MIDI observations: NGC 7469 (continued)}
\centering
\begin{tabular}{c c c c c c c c c c c c c c c c}
\hline
\multicolumn{8}{c}{Fringe track} & \multicolumn{3}{c}{FT QC} & PH & PH QC & \multicolumn{2}{c}{Calibration} \\
time & s & NDIT & BL & PA & g & mode & am & see & O & NGOOD & o3 & A/B & O & caltime & $\Delta$ am \\
\hline
\multicolumn{15}{l}{2011-08-17: U3U4} \\
06:12:23 &  1 &  8000 &  60 & 107 & P & FT\_OF & 1.2 & 1.0 &  1 &  6874 &  0.57 & 0/0 & -1 & 06:26:11 &      0.0\\
06:16:09 &  1 &  8000 &  60 & 107 & P & FT\_OF & 1.2 & 0.9 &  1 &  7236 &  0.45 & 0/0 & -1 & 06:26:11 &      0.0\\
06:36:38 &  0 &  8000 &  58 & 106 & P & FT\_OF & 1.2 & 1.0 &  1 &  5741 &  0.53 & 0/0 & -1 & 06:26:11 &      0.1\\
06:40:53 &  0 &  8000 &  58 & 106 & P & FT\_OF & 1.2 & 1.3 &  1 &  4471 &  0.43 & 0/0 & -1 & 06:52:40 &      0.1\\
07:29:31 & -1 &  8000 &  52 & 106 & P & FT\_OF & 1.3 & 1.3 &  0 &  1698 &  0.87 & 0/0 & -1 & 07:39:28 &      0.1\\
\hline
\end{tabular}
\end{table}

\end{document}